\documentclass[preprint,amsmath,amssymb,aps,superscriptaddress,prd,
showpacs,floatfix,nofootinbib]{revtex4-1}

\pdfoutput=1

\usepackage{amsmath,amssymb,amsfonts}
\usepackage{graphicx}
\usepackage{dcolumn}% Align table columns on decimal point
\usepackage{bm}
\usepackage{hyperref}
\usepackage{footmisc}
\usepackage{longtable}

\newcommand{\be}{\begin{equation}}
\newcommand{\ee}{\end{equation}}
\newcommand{\ba}{\begin{eqnarray}}
\newcommand{\ea}{\end{eqnarray}}
\newcommand{\SuperField}[1]{\hat{#1}}
\usepackage{url}

\allowdisplaybreaks
\begin{document}

\title {$Z^\prime$ mass limits and the naturalness of supersymmetry}

\author{P. Athron}
 \email{peter.athron@monash.edu}
 \affiliation{
ARC Centre of Excellence for Particle Physics at the Terascale,
School of Physics, Monash University, Melbourne VIC 3800, Australia
}

\author{D. Harries}
 \email{dylan.harries@adelaide.edu.au}
 \affiliation{
ARC Centre of Excellence for Particle Physics at the Terascale,
Department of Physics, The University of Adelaide, Adelaide,
South Australia 5005, Australia}

\author{A. G. Williams}
 \email{anthony.williams@adelaide.edu.au}
 \affiliation{
ARC Centre of Excellence for Particle Physics at the Terascale,
Department of Physics, The University of Adelaide, Adelaide,
South Australia 5005, Australia}

\date{\today}

\begin{abstract}

The discovery of a 125 GeV Higgs boson and rising lower bounds on the
masses of superpartners have lead to concerns that supersymmetric
models are now fine-tuned.  Large stop masses, required for a $125$
GeV Higgs, feed into the electroweak symmetry breaking conditions
through renormalization group equations forcing one to fine-tune these
parameters to obtain the correct electroweak vacuum expectation value.
Nonetheless, this fine-tuning depends crucially on our assumptions about
the supersymmetry breaking scale.  At the same time, $U(1)$ extensions
provide the most compelling solution to the $\mu$ problem, which is also
a naturalness issue, and allow the tree-level Higgs mass to be raised
substantially above $M_Z$.  These very well-motivated supersymmetric models
predict a new $Z^\prime$ boson which could be discovered at the LHC, and the
naturalness of the model requires that the $Z^\prime$ boson mass
should not be too far above the TeV scale.  Moreover, this fine-tuning
appears at the tree level, making it less dependent on assumptions
about the supersymmetry breaking mechanism.  Here we study this fine-tuning
for several $U(1)$ supersymmetric extensions of the Standard Model and
compare it to the situation in the MSSM where the most direct tree-level
fine-tuning can be probed through chargino mass limits.  We show
that future LHC $Z^\prime$ searches are extremely important for
challenging the most natural scenarios in these models.

\end{abstract}

\preprint{ADP-15-4/T906}
\maketitle

\section{\label{sec:intro}Introduction}

The discovery of an approximately 125 GeV Higgs
\cite{Aad:2012tfa, Chatrchyan:2012ufa} at the Large Hadron Collider
(LHC) has interesting implications for physics beyond the Standard
Model (SM) and supersymmetry (SUSY).  On the one hand, it provides a
light Higgs boson, as expected from supersymmetry, and
can be fitted in the minimal supersymmetric standard model (MSSM).  On
the other hand, the Higgs mass is slightly heavier than the constrained
version of the MSSM (cMSSM) can accommodate naturally
\cite{Cassel:2011zd, Ghilencea:2012gz}.

In the MSSM the Higgs mass causes a naturalness problem because at
tree level it has an upper bound of the mass of the $Z$ boson,
$M_Z$.  The dominant higher-order corrections to the Higgs mass come
from stops, and to obtain a $125$ GeV Higgs they need to be rather
heavy.  Heavy stops will provide a large contribution to the low-energy
value of $m_{H_u}^2$, the soft breaking mass for the up-type Higgs
scalar, through the evolution of the renormalization group equations
(RGEs) from the grand unification (GUT) scale to the electroweak (EW)
scale.  This affects the SUSY prediction of the electroweak vacuum expectation
value (VEV), $v$, or $M_Z$.  This naturalness problem motivates both
further examination of nonminimal SUSY models that can raise the Higgs mass
without the need for heavy stops and alternative possibilities for how the
soft breaking parameters get generated, which might set them at lower
energies, reducing the influence the stops have on $m_{H_u}^2$.

In addition to that naturalness issue, often referred to as the little
hierarchy problem, the MSSM also suffers from the $\mu$ problem.  This
is also a naturalness problem since there should be a natural explanation
of how the $\mu$ superpotential parameter can be set to the same scale
as the soft breaking masses.

$U(1)$ extensions of the MSSM provide a very elegant solution to this
$\mu$ problem \cite{Fayet:1977yc, Kim:1983dt, Suematsu:1994qm,
  Cvetic:1995rj, Cvetic:1996mf, Jain:1995cb, Nir:1995bu,
  Cvetic:1997ky} and also raise the Higgs mass with new $F$ and
$D$ terms.  Nonetheless, as was recently demonstrated in the context
of the exceptional supersymmetric standard model (E$_6$SSM)
\cite{King:2005jy, King:2005my, Athron:2010zz}, such models can still
suffer from naturalness problems with the mass of the new $Z^\prime$
associated with the break down of the new $U(1)$ appearing in the
electroweak symmetry breaking (EWSB) conditions at tree level
\cite{Athron:2013ipa}.  Despite this the constrained version of the
E$_6$SSM (cE$_6$SSM) \cite{Athron:2009ue, Athron:2009bs} was still
found to be significantly less tuned than the cMSSM.  Tree-level
fine-tuning from the $Z^\prime$ mass was also considered
previously \cite{Drees:1985js}.

However, this comparison of fine-tuning depends crucially upon the
assumptions of these gravity mediated SUSY breaking motivated
constrained models and, in particular, the universality
constraints being applied at the GUT scale.  As mentioned above, given
the findings at the LHC, it is worth considering other possibilities,
which may allow the soft masses to be set at lower energies.  As the
scale at which the parameters fulfill some breaking inspired
constraints is lowered the stop masses contribute less to the fine-tuning.

At the same time in $U(1)$ extensions, lowering the UV boundary scale
for the RGE evolution also allows even larger $F$-term contributions
to the Higgs mass, so long as one only requires $\lambda$, the
coupling between the Singlet Higgs, $S$ and the up- and down-type
Higgs bosons, $H_u$ and $H_d$, to remain perturbative up to the UV
scale and not all of the way up to the GUT scale.

However, the tuning from the $Z^\prime$ mass limit does not disappear
as the UV boundary condition is lowered.  This tuning appears in the
EWSB conditions at tree level and is quite difficult to avoid without
introducing a pure gauge singlet \cite{Athron:2014pua}.

In this paper we investigate how big this tuning is if we bring this
scale all the way down to 20 TeV, effectively minimizing the contribution
from the stops.  We find that the $Z^\prime$ limit is enough to already
require moderate fine-tuning in the E$_6$SSM.  We also show this is
comparable to the situation in the MSSM defined at the same scale if
charginos could be ruled out below $700$ GeV.  We then show how this
tuning from the $Z^\prime$ mass looks for different $U(1)$ extensions,
finding that the current severity depends upon the charges but that
$Z^\prime$ limits are important in constraining the most natural
scenarios of these models.  Therefore, the $Z^\prime$ constraint
is amongst the most important in terms of tuning and attacking natural
supersymmetry experimentally and the next run of the LHC will be
crucial in this respect.

Finally, we make a case study, for a few benchmarks, of the impact of
raising the high-scale boundary condition, $M_X$, at which the SUSY
breaking parameters must be fixed by some SUSY breaking mechanism.  We
show that which model has less fine-tuning depends on $M_X$.  We also
see rather complicated behavior in the tuning for the E$_6$SSM points
due to the combination of different sources of tuning.

As mentioned earlier, the-fine tuning of the cE$_6$SSM was recently studied
\cite{Athron:2013ipa} and there it was revealed that the associated
$Z^\prime$ boson leads to a new source of fine-tuning since its mass
appears in the EWSB conditions.

However in this study we will examine this source of fine-tuning in more
detail by considering low-energy constructions where the usual fine-tuning
problem from the Higgs is minimized.  We will also consider
alternate charges for the extra $U(1)$ symmetry to relax the focus on
the E$_6$SSM and demonstrate that this is quite a generic result.

To quantify the fine-tuning we will employ the traditional
Barbieri-Giudice measure \cite{Ellis:1986yg, Barbieri:1987fn}.  This
has been used extensively within the literature
e.g. Refs.~\cite{deCarlos:1993yy, deCarlos:1993ca, Chankowski:1997zh,
  Agashe:1997kn, Wright:1998mk, Kane:1998im, BasteroGil:1999gu,
  Feng:1999zg, Allanach:2000ii, Dermisek:2005ar, Barbieri:2005kf,
  Allanach:2006jc, Gripaios:2006nn, Dermisek:2006py, Barbieri:2006dq,
  Kobayashi:2006fh, Perelstein:2012qg, Antusch:2012gv,Cheng:2012pe,
  CahillRowley:2012rv, Ross:2012nr, Basak:2012bd, Kang:2012sy,
  Athron:2013ipa, Boehm:2013qva, Miller:2013jra, Binjonaid:2014oga,
  Miller:2014jza}.

A number of alternative measures have also been applied in the
literature \cite{Anderson:1994dz, Anderson:1994tr, Anderson:1995cp,
Anderson:1996ew, Ciafaloni:1996zh, Chan:1997bi, Barbieri:1998uv,
Giusti:1998gz, Casas:2003jx, Casas:2004uu, Casas:2004gh,
Casas:2006bd, Kitano:2005wc, Athron:2007ry, Athron:2007qr,
Baer:2012up} with varying motivations.  A very different approach is
to work within a Bayesian analysis.  There the concept of naturalness
is automatically incorporated since in models where one must fine-tune
parameters to fit measured values of the observables, the region with
high likelihood will occupy a tiny prior volume \cite{Allanach:2007qk,
Cabrera:2008tj, Ghilencea:2012gz, Ghilencea:2012qk, Fichet:2012sn,
Kim:2013uxa} suppressing the posterior.  Indeed in the MSSM and
the next-to-MSSM (NMSSM) if one transforms GUT scale parameters to the
VEVs, the inverse of the Jacobian for this transformation looks quite like
the derivatives that appear in the traditional fine-tuning measure
\cite{Allanach:2007qk, Cabrera:2008tj, Kim:2013uxa}.  If one thinks
more generally, then a model without fine-tuning is one where the
parameterization is such that all the parameters are observables
\cite{Fichet:2012sn, Kim:2013uxa}.  This provides a quite general
definition of fine-tuning as $1/|J|$ where $|J|$ is the determinant of
the Jacobian for the coordinate transformation between the parameters
and the observables.  Interestingly this means the tuning is the ratio
of the infinitesimal observable space volume element to the
infinitesimal parameter space element and coincides with the measure
proposed in Ref.~\cite{Athron:2007ry} when the interval of variation is
taken to zero.

While this approach has many merits here we will employ the traditional
measure of fine-tuning because it is both simple to apply and easy to
compare with previous results due to it's widespread use.  Fortunately
the derivatives which appear in these tunings are also similar to the
Bayesian motivated measure so there should not be too large a
discrepancy between the two approaches.

The structure of this paper is as follows.  In Sec.~\ref{sec:model}
we review the models we consider.  In Sec.~\ref{sec:ewsb} we
specify the EWSB conditions of the models, with particular focus on
how the $Z^\prime$ mass influences the prediction of $M_Z$.  Then in
Sec.~\ref{sec:tuningmeasure} we introduce our fine-tuning measure
and our approach to evaluating it to obtain the individual
sensitivities.  The results are then given in Sec.~\ref{sec:results}.

\section{\label{sec:model}$U(1)$ extensions and the
E$_6$SSM}

In this paper we consider $U(1)$ extensions of the MSSM where the gauge
group at low energies is \be SU(3)_C\times SU(2)_W\times U(1)_Y\times
U(1)^\prime.  \ee $U(1)^\prime$ is the new gauge group beyond
that of the SM and MSSM.  The minimal superfield content of $U(1)$
extensions which solve the $\mu$ problem should be ordinary left-handed
quark $\hat{Q}_i$ and lepton $\hat{L}_i$
($i=1,2,3$) superfields along with right-handed superfields
$\hat{u}^c_i$, $\hat{d}^c_i$, $\hat{e}^c_i$ ($i=1,2,3$) for the
up-type (s)quarks, down-type (s)quarks and charged (s)leptons
respectively and three Higgs superfields, up-type $\hat{H}_u$,
down-type $\hat{H}_d$ and a singlet under the SM gauge group
$\hat{S}$.

Here we will refer to $U(1)$ extensions of the MSSM, which solve the
$\mu$ problem, as the USSM \cite{Cvetic:1995rj, Jain:1995cb,
  Nir:1995bu, Cvetic:1996mf, Cvetic:1997ky}.  The couplings for the
$U(1)^\prime$ gauge group should allow the following renormalizable
superpotential terms required in the USSM,
\begin{equation}
 W_{USSM} = y^U_{ij} \hat{u}^c_i \hat{H}_u \cdot \hat{Q}_j + y^D_{ij}
\hat{d}^c_i \hat{Q}_j \cdot \hat{H}_d + y^E_{ij} \hat{e}^c_i \hat{L}_j
\cdot \hat{H}_d + \lambda \hat{S} \hat{H}_d \cdot \hat{H}_u ,
\end{equation}
with $i,j \in \{1,2,3\}$.  For the $SU(2)$ dot product we follow the
convention $\hat{A} \cdot \hat{B} \equiv \epsilon_{\alpha \beta}
\hat{A}^\alpha \hat{B}^\beta = \hat{A}^2 \hat{B}^1 - \hat{A}^1 \hat{B}^2$.

The $U(1)^\prime$ charges should allow for cancellations of gauge
anomalies.  The most elegant way to do this is to use an extra $U(1)$
gauge symmetry that can be obtained from the break down of the $E_6$
gauge symmetry which is anomaly free and have all matter fields that fill the
three generations of $27$-plet representations of $E_6$ survive down to
low energies.  Such models are often referred to in the literature as
$E_6$ inspired, and we will adopt this here.

The breaking of $E_6$ into $SO(10)$ gives rise to $E_6 \to
SO(10) \times U(1)_{\psi}$, and the subsequent breaking of $SO(10)$
into $SU(5)$ gives $SO(10) \to SU(5) \times U(1)_{\chi}$ (this is
reviewed in e.g. Ref.~\cite{Langacker:2008yv}).  The extra $U(1)$ gauge
symmetry at low energies should then be a linear combination of these
in the $E_6$ inspired case, \be U(1)^\prime = U(1)_{\chi} \cos\theta
+ U(1)_{\psi} \sin\theta .  \ee In Table \ref{tab:E6charges} the
charges for several popular $E_6$ inspired $U(1)$ extensions are
shown.

\begin{table}[h]
\centering
\begin{ruledtabular}
\begin{tabular}{cccccccccccccc}
 & $\hat{Q}$ & $\hat{u}^c$ & $\hat{d}^c$ & $\hat{L}$ & $\hat{e}^c$ &
$\hat{N}^c$ & $\hat{S}$ & $\hat{H}_2$ & $\hat{H}_1$ & $\hat{D}$ &
$\hat{\overline{D}}$ & $\hat{H}'$ & $\hat{\overline{H'}}$ \\[1mm]
\hline
$\sqrt{\frac{5}{3}}Q_i^Y$ & $\frac{1}{6}$ & $-\frac{2}{3}$ & $\frac{1}{3}$ &
$-\frac{1}{2}$ & $1$ & $0$ & $0$ & $\frac{1}{2}$ & $-\frac{1}{2}$ &
$-\frac{1}{3}$ & $\frac{1}{3}$ & $-\frac{1}{2}$ & $\frac{1}{2}$ \\[1mm]
$2\sqrt{6}Q_i^\psi$ & $1$ & $1$ & $1$ & $1$ & $1$ & $1$ & $4$ & $-2$ & $-2$ &
$-2$ & $-2$ & $1$ & $-1$\\[1mm]
$2\sqrt{10}Q_i^\chi$ & $-1$ & $-1$ & $3$ & $3$ & $-1$ & $-5$ & $0$ & $2$ &
$-2$ & $2$ & $-2$ & $3$ & $-3$\\[1mm]
$\sqrt{40}Q_i^N$ & $1$ & $1$ & $2$ & $2$ & $1$ & $0$ & $5$ & $-2$ & $-3$ &
$-2$ & $-3$ & $2$ & $-2$ \\[1mm]
\end{tabular}
\end{ruledtabular}
\caption{The $U(1)_Y$, $U(1)_\psi$, $U(1)_\chi$ and $U(1)_N$ charges of the
chiral superfields in the $E_6$ model.  The specific case of $U(1)_N$,
corresponding to the E$_6$SSM, is obtained for $\theta = \arctan \sqrt{15}$.}
\label{tab:E6charges}
\end{table}

$U(1)$ and $E_6$ inspired extensions of the MSSM have been studied
very widely in the literature \cite{Gunion:1989we, Gunion:1992hs,
Binetruy:1985xm, Ellis:1986yg, Ibanez:1986si, Gunion:1986ky,
Haber:1986gz, Baer:1987eb, Gunion:1987jd, Grifols:1986vr,
Ellis:1986ip, Morris:1987fm, Drees:1987tp, Ma:1995xk,
Suematsu:1997tv, Suematsu:1997qt, Suematsu:1997au, Keith:1996fv,
Keith:1997zb, Gherghetta:1996yr, Demir:1998dk, Langacker:1998tc,
Hambye:2000bn, Ma:2000jf} (or, for reviews, see Refs.~\cite{Hewett:1988xc,
Langacker:2008yv}).  There has also been a lot of work recently including
investigations of the neutralino sector \cite{Hesselbach:2001ri,
Barger:2005hb, Choi:2006fz, Barger:2007nv}; the relic density of dark matter
\cite{Kalinowski:2008iq}; GUT scale family symmetries which can explain the
hierarchy of masses in the fermion sector and their associated mixings
\cite{Stech:2008wd}; neutrino physics \cite{Kang:2004ix}; explanations of
the matter-antimatter asymmetry of the Universe though EW baryogenesis or
leptogenesis \cite{Hambye:2000bn, Ma:2000jf, Kang:2004pp}; decays of the
$Z^\prime$ boson \cite{Kang:2004bz, Baumgart:2006pa, Athron:2011wu,
Chang:2011be}; dipole moments \cite{GutierrezRodriguez:2006hb}; anomaly
mediated SUSY breaking with $D$-term contributions \cite{Asano:2008ju} and
the (extended) Higgs sectors \cite{Daikoku:2000ep, Ham:2008xf}.

Here we will focus most on the special case where the gauge symmetry
is $U(1)_N$, under which the right-handed neutrino $\hat{N}^c$ does not
participate in gauge interactions.  This is the case in the E$_6$SSM
\cite{King:2005jy, King:2005my, Athron:2010zz}, and closely related variants
\cite{Howl:2007zi, Braam:2009fi, Braam:2010sy, Hall:2011zq, Nevzorov:2012hs,
Athron:2014pua}.  Since the right-handed neutrino has no gauge symmetry
protecting it's mass from becoming extremely heavy such models may explain
the tiny observed masses of neutrinos via the see-saw mechanism and the
baryon asymmetry in the Universe via leptogenesis \cite{Hambye:2000bn,
King:2008qb, King:2008gw}.  Recently it has also been studied in the context
of electroweak baryogenesis \cite{Chao:2014hya}.

The gauge coupling running in the E$_6$SSM at the two-loop level leads
to unification more precisely than in the MSSM \cite{King:2007uj} or,
in slightly modified scenarios, two-step unification can take place
\cite{Howl:2007hq, Howl:2007zi}.  If the exotic particles are light in
these models this can open up nonstandard decays of the SM-like
Higgs boson \cite{Hall:2010ix, Nevzorov:2013tta, Athron:2014pua}.

The correct relic density could be obtained entirely through an almost
decoupled ``inert'' neutralino sector \cite{Hall:2009aj}.  However,
this is no longer phenomenologically viable due to limits from direct
detection of dark matter \cite{2011PhRvL.107m1302A, 2012PhRvL.109r1301A,
Akerib:2013tjd} and due to a significant suppression of the decay of the
lightest Higgs boson into SM states, due to a new channel into inert
singlinos opening up.

There are still several remaining options.  One may specialize to
scenarios known as the EZSSM \cite{Hall:2011zq} where the inert
singlinos that cause these problems are entirely decoupled and the
relic abundance is fitted with a binolike candidate with a novel
mechanism involving back-scattering into a heavier inert
Higgsino.  Another well motivated scenario admits two possible dark
matter candidates \cite{Nevzorov:2012hs}, where one will be an inert
singlino and the other will have a similar composition to MSSM
neutralinos.  The simplest phenomenologically viable solution in that
case is to make the singlinos extremely light hot dark matter
candidates, in which case the lightest ordinary neutralino accounts
for almost all of the observed relic abundance.

The impact of gauge kinetic mixing in the case where both of the extra
$U(1)$ symmetries appearing from the breakdown of $E_6$ are present at
low energy was studied in Ref.~\cite{Rizzo:2012rf}.  The E$_6$SSM was also
included in studies looking at how first- or second-generation sfermion
masses can be used to constrain the GUT scale parameters \cite{Miller:2012vn}
and the renormalization of VEVs \cite{Sperling:2013eva, Sperling:2013xqa}.
The particle spectrum and collider signatures of the cE$_6$SSM have been
studied in a series of papers, \cite{Athron:2009ue, Athron:2009bs,
Athron:2011wu,Athron:2012sq}.  The threshold corrections to the
$\overline{DR}$ gauge and Yukawa couplings in the E$_6$SSM have also been
calculated and their numerical impact in the constrained version examined
\cite{Athron:2012pw}.

With three generations of matter $27$-plet representations of $E_6$
surviving to low energies, the low-energy matter content in each generation,
after integrating out the heavy right-handed neutrinos, includes, \be
(\hat{Q}_i, \, \hat{u}^c_i, \, \hat{d}^c_i, \, \hat{L}_i, \, \hat{e}^c_i)
+ (\hat{D}_i, \, \hat{\overline{D}}_i) + (\hat{S}_{i}) + (\hat{H}^u_i)
+ (\hat{H}^d_i) ,\ee
where the $\hat{S}_{i}$, $\hat{H}^u_i$ and $\hat{H}^d_i$ have the quantum
numbers of a SM singlet, and up-, down-type Higgs fields, respectively, and
the $\hat{D}_i$ and $\hat{\overline{D}}_i$ are $SU(3)_C$ triplets that reside
in the same $SU(5)$ multiplets as these Higgs-like states.

If one wishes to maintain gauge coupling unification this set of states should
be augmented by two extra $SU(2)$ doublet states $H^\prime$ and
$\overline{H}^\prime$ belonging to other $27^\prime$ and $\overline{27}^\prime$
multiplets that must be incomplete at low energies.

The full superpotential for $E_6$ inspired models coming from
$27 \otimes 27 \otimes 27$ decomposition of the fundamental $E_6$
representation will then be
\be
W_{E_6} = W_0 + W_1 + W_2 , \label{Eq:e6-superpot}
\ee
where
\begin{eqnarray}
W_0 &=& \lambda_{ijk} \hat{S}_i \hat{H}^d_{j} \cdot\hat{H}^u_{k} +
\kappa_{ijk} \hat{S}_i \hat{D}_j \hat{\overline{D}}_k + h^N_{ijk} \hat{N}^c_i
\hat{H}^u_{j} \cdot \hat{L}_k \nonumber \\
& & + y^U_{ijk} \hat{u}^c_i \hat{H}^u_{j} \cdot \hat{Q}_k + y^D_{ijk}
\hat{d}^c_i \hat{Q}_k \cdot \hat{H}^d_{j} + y^E_{ijk} \hat{e}^c_i \hat{L}_k
\cdot\hat{H}^d_{j} , \\
W_1 &=& g^Q_{ijk} \hat{D}_i \hat{Q}_j \cdot \hat{Q}_k + g^q_{ijk}
\hat{\bar{D}}_i \hat{d}^c_j \hat{u}^c_k , \\
W_2 &=& g^N_{ijk} \hat{N}^c_i \hat{D}_j \hat{d}^c_k + g^E_{ijk} \hat{e}^c_i
\hat{D}_j \hat{u}^c_k + g^D_{ijk} \hat{Q}_i \cdot \hat{L}_j \hat{\bar{D}}_k .
\label{Eq:e6-superpot-parts}
\end{eqnarray}
Nonetheless, while this model is very elegant so far, the
superpotential of Eq.~(\ref{Eq:e6-superpot}) contains dangerous terms which
can induce proton decay and lead to large flavor changing neutral
currents (FCNCs).  There are a number of approaches to suppress these
terms, involving the use of different discrete symmetries.  Here for
the purposes of renormalization group running we will simply include
the following unsuppressed superpotential terms, which follows the
approach taken in work on the cE$_6$SSM \cite{Athron:2009ue, Athron:2009bs},
%
%% \begin{align}
%%   \begin{split}
    \ba
    W &\approx& y_{\tau} \SuperField{L}_3 \cdot \SuperField{H}_d
      \SuperField{e}^c_3 + y_b \SuperField{Q}_3\cdot\SuperField{H}_d
      \SuperField{d}_3^c + y_t \SuperField{H}_u\cdot\SuperField{Q}_3
      \SuperField{u}_3^c\nonumber\\
    &&
    + \lambda_i \SuperField{S} \SuperField{H}_i^d
    \cdot \SuperField{H}_i^u  + \kappa_i \SuperField{S} \SuperField{D}_i
    \SuperField{\overline{D}}_i + \mu' \SuperField{H}'\cdot
    \SuperField{\overline{H'}},
    \ea
  %% \end{split}
  \label{SuPot_RGE}
%%\end{align}
%
where we denote by $\SuperField{H}_d^3 \equiv \SuperField{H}_d$,
$\SuperField{H}_u^3 \equiv \SuperField{H}_u$ and $\SuperField{S}_3
\equiv \SuperField{S}$ the third-generation Higgs and SM singlet
fields that are assumed to acquire nonzero VEVs.  In addition to the
terms coming from the $27\otimes 27 \otimes 27$ interactions given in
Eq.~(\ref{Eq:e6-superpot}), this superpotential also contains a
bilinear term $\mu' \SuperField{H}'\cdot \SuperField{\overline{H'}}$,
arising from $27^\prime \otimes \overline{27}^\prime$, which is
invariant with respect to the low-energy SM gauge group and the
additional $U(1)'$ symmetry and also anomaly free.  This term is
responsible for setting the masses of the components of the
superfields $\SuperField{H}'$, $\SuperField{\overline{H'}}$, included
to ensure gauge coupling unification, but it is not involved in the
process of EWSB.  Consequently, the impact on the fine-tuning of the
value of $\mu'$ is much smaller than that coming from other sectors,
and so can be safely neglected in our study. In all of the scans we
present below the value of $\mu'$ is fixed to $\mu' = 5$ TeV.

\section{\label{sec:ewsb}Electroweak Symmetry
Breaking}

The Higgs scalar potential for the $E_6$ models considered can be written as
\cite{King:2005jy}
\begin{equation} \label{eq:E6VeffOneLoop}
V = V_F + V_D + V_{\textrm{soft}} + \Delta V \, ,
\end{equation}
where
\begin{align}
V_F &= \lambda^2 |S|^2 (|H_d|^2 + |H_u|^2) + \lambda^2 |H_d \cdot H_u|^2 ,
\label{eq:E6VFterms} \\
V_D &= \frac{\bar{g}^2}{8} \left (|H_u|^2 - |H_d|^2 \right )^2 +
\frac{g_2^2}{2} |H_d^\dagger H_u|^2 + \frac{g_1'^2}{2} (Q_1 |H_d|^2 +
Q_2 |H_u|^2 + Q_S |S|^2)^2 , \label{eq:E6VDterms} \\
V_{\textrm{soft}} &= m_S^2 |S|^2 + m_{H_d}^2 |H_d|^2 + m_{H_u}^2 |H_u|^2 +
\Big [\lambda A_\lambda S H_d \cdot H_u + \textrm{H.c.} \Big ] .
\label{eq:E6Vsoft}
\end{align}
In these expressions $g_2$, $g' = \sqrt{3/5} g_1$, and $g_1'$ are the
$SU(2)$, (non-GUT normalized) $U(1)_Y$ and $U(1)'$ gauge couplings,
respectively, and $\bar{g}^2 = g_2^2 + g'^2$.  The charges $Q_1$, $Q_2$ and
$Q_S$ are effective $U(1)'$ charges for $H_d$, $H_u$ and $S$, respectively,
and $\lambda \equiv \lambda_3$.  In the case of the $U(1)_\chi$ model, $V_F$
may also contain an elementary $\mu$ term, as occurs in the MSSM.  The term
$\Delta V$ contains the Coleman-Weinberg contributions to the effective
potential.  For the purposes of this study, we include in $\Delta V$ only the
one-loop contributions from the top quark and stop squarks,
\begin{equation}
\Delta V = \frac{3}{32\pi^2} \left [ m_{\tilde{t}_1}^4 \left ( \ln
\frac{m_{\tilde{t}_1}^2}{Q^2} - \frac{3}{2} \right ) + m_{\tilde{t}_2}^4
\left ( \ln \frac{m_{\tilde{t}_2}^2}{Q^2} - \frac{3}{2} \right ) - 2m_t^4
\left ( \ln \frac{m_t^2}{Q^2} - \frac{3}{2} \right ) \right ] .
\label{Eq:DeltaV}
\end{equation}
Explicit expressions for the running $\overline{\textrm{DR}}$ top mass $m_t$
and stop masses $m_{\tilde{t}_{1,2}}$ are given below.

Demanding that the Higgs fields $H_1$, $H_2$ and the singlet $S$ acquire real
VEVs of the form
\begin{equation} \label{eq:E6vevs}
\langle H_d \rangle = \frac{1}{\sqrt{2}} \begin{pmatrix} v_1 \\ 0
\end{pmatrix} , \;
\langle H_u \rangle = \frac{1}{\sqrt{2}} \begin{pmatrix} 0 \\ v_2
\end{pmatrix} , \;
\langle S \rangle =\frac{s}{\sqrt{2}} ,
\end{equation}
at the physical minimum leads to the minimization conditions
\begin{subequations} \label{eq:E6EWSBConditions}
\begin{align}
&f_1 = m_{H_d}^2 v_1 + \frac{\lambda^2}{2} (v_2^2 + s^2) v_1-
\frac{\lambda A_\lambda}{\sqrt{2}} s v_2 - \frac{\bar{g}^2}{8}
(v_2^2 - v_1^2) v_1 + D_{H_d} v_1 + \frac{\partial \Delta V}{\partial v_1} = 0 ,
\label{eq:E6EWSBcondition1} \\
&f_2 = m_{H_u}^2 v_2 + \frac{\lambda^2}{2} (v_1^2 + s^2) v_2 -
\frac{\lambda A_\lambda}{\sqrt{2}} s v_1 + \frac{\bar{g}^2}{8}
(v_2^2 - v_1^2) v_2 + D_{H_u} v_2 + \frac{\partial \Delta V}{\partial v_2} = 0 ,
\label{eq:E6EWSBcondition2} \\
&f_3 = m_S^2 s + \frac{\lambda^2}{2} (v_2^2 + v_1^2) s -
\frac{\lambda A_\lambda}{\sqrt{2}} v_2 v_1 + D_S s +
\frac{\partial \Delta V}{\partial s} = 0 . \label{eq:E6EWSBcondition3}
\end{align}
\end{subequations}
The quantities $D_{H_d}$, $D_{H_u}$ and $D_S$ appearing above are $U(1)'$
$D$-term contributions that are absent in the MSSM and NMSSM and are given by
\begin{equation} \label{eq:E6Dterms}
D_\phi \equiv \frac{g_1'^2}{2} \left (Q_1 v_1^2 + Q_2 v_2^2 + Q_S s^2
\right ) Q_\phi .
\end{equation}
We also include these $U(1)'$ $D$-term contributions in the diagonalized stop
masses,
\begin{align}
m_{\tilde{t}_{1,2}}^2 &= \frac{1}{2} \bigg \{ m_{Q_3}^2 + m_{u_3}^2 +
\frac{\bar{g}^2}{8} \left ( v_1^2 - v_2^2 \right ) + D_Q + D_u + 2 m_t^2
\nonumber \\
& \quad {} \mp \sqrt{\left [ m_{Q_3}^2 - m_{u_3}^2 + \frac{1}{8} \left (g_2^2
- g_1^2 \right ) \left (v_1^2 - v_2^2 \right ) + D_Q - D_u \right ]^2 +
4 m_t^2 X_t^2} \bigg \} ,
\end{align}
where $m_t^2 = y_t^2 v_2^2 / 2$, $X_t = A_t -
\frac{\lambda s v_1}{\sqrt{2} v_2}$, $m_{Q_3}^2$, $m_{u_3}^2$ are soft
breaking scalar masses and $A_t$ is a soft trilinear coupling.  By definition
we take $m_{\tilde{t}_1}$ to correspond to the lighter of the two states.

As was noted in Ref.~\cite{Athron:2013ipa}, the first two of the conditions in
Eq.~(\ref{eq:E6EWSBConditions}) may be rewritten in the form
\begin{equation} \label{eq:E6MZequation}
\frac{M_Z^2}{2} = -\frac{\lambda^2 s^2}{2} + \frac{\tilde{m}_{H_d}^2 -
\tilde{m}_{H_u}^2 \tan^2\beta}{\tan^2\beta - 1} +
\frac{D_{H_d} - D_{H_u} \tan^2\beta}{\tan^2\beta - 1} ,
\end{equation}
\begin{equation} \label{eq:E6sin2bequation}
\sin 2\beta = \frac{\sqrt{2} \lambda A_{\lambda} s}
{\tilde{m}_{H_d}^2 + \tilde{m}_{H_u}^2 + \lambda^2 s^2 + D_{H_d} + D_{H_u}} ,
\end{equation}
with $M_Z^2 = \bar{g}^2 v^2 / 4$, $v^2 = v_1^2 + v_2^2$ and $\tan\beta =
v_2 / v_1$ and where we have for convenience absorbed the effects of the
loop corrections into the soft masses,
\begin{align*}
\tilde{m}_{H_d}^2 &= m_{H_d}^2 + \frac{1}{v_1} \frac{\partial \Delta V}
{\partial v_1} , \\
\tilde{m}_{H_u}^2 &= m_{H_u}^2 + \frac{1}{v_2} \frac{\partial \Delta V}
{\partial v_2} .
\end{align*}
Written in the form of Eq.~(\ref{eq:E6MZequation}), we see the
potential new source of fine-tuning alluded to above, in the form of
the third term on the right-hand side.  For large values of the VEV
$s$, the $D$-term contributions can be quite a bit larger than
$M_Z^2$.  In particular, recent experimental limits \cite{Aad:2014cka}
require that the $Z'$ mass be large, with for example bounds of
$M_{Z'} \gtrsim 2.51$ TeV in $U(1)_\psi$ models and $M_{Z'} \gtrsim 2.62$ TeV
in $U(1)_\chi$ models.  To satisfy these limits typically requires large
values of the singlet VEV $s$.  For example, $s \gtrsim 6$ TeV is required in
the E$_6$SSM with $U(1)' = U(1)_N$, so that $|D_{H_1}| , |D_{H_2}| \gg M_Z^2$
for $E_6$ models with $Q_S \neq 0$.  As a result the remaining terms on the
right-hand side of Eq.~(\ref{eq:E6MZequation}) must be tuned to cancel
this very large contribution to $M_Z$.  Moreover, because this is a large
tree-level fine-tuning, it may negate the improvement in naturalness that is
associated with having a reduced need for heavy superpartners.  In $U(1)$
extended models for which $Q_S\neq 0$, the importance of the $Z^\prime$ mass
to the fine-tuning in these models can be made even clearer by writing
Eq.~(\ref{eq:E6MZequation}) in the form given in Ref.~\cite{Athron:2013ipa},
\begin{equation}\label{eq:originalE6MZequation}
c(\theta , \tan\beta) \frac{M_Z^2}{2} = -\frac{\lambda^2 s^2}{2} +
\frac{\tilde{m}_{H_d}^2 - \tilde{m}_{H_u}^2 \tan^2\beta}{\tan^2\beta - 1}
+ d(\theta , \tan\beta) \frac{M_{Z^\prime}^2}{2} ,
\end{equation}
where
\begin{align}
c(\theta , \tan\beta) &= 1 - \frac{4}{\left (\tan^2\beta - 1 \right )}
\frac{g_1'^2}{\bar{g}^2} \left (Q_1 - Q_2 \tan^2\beta \right )
\left (Q_1 \cos^2\beta + Q_2 \sin^2\beta \right ) , \label{eq:cdefn} \\
d(\theta , \tan\beta) &= \frac{Q_1 - Q_2 \tan^2\beta}
{Q_S \left (\tan^2\beta - 1\right )} . \label{eq:ddefn}
\end{align}
Written like so, it is evident that the fine-tuning contribution coming
from the new $D$ terms depends both on the $U(1)'$ charges and the
$Z^\prime$ mass, and that the tuning can be expected to increase with
$M_{Z^\prime}$.  As shall be shown below, the exact size of this tuning
then depends strongly on the choice of $U(1)'$ charges, via the
coefficient $d$.

The extra $U(1)^\prime$ gauge symmetry may mix with the $U(1)_Y$ gauge
symmetry associated with hypercharge through gauge kinetic mixing,
\be
\mathcal{L}_{mix}^{kin}=-\frac{\sin\chi}{2}F^{Y}_{\mu\nu}F^{N}_{\mu\nu}\, ,
\label{Eq:GKM}
\ee where $F_{\mu\nu}^Y$ and $F_{\mu\nu}^{N}$ are field strengths
associated with the $U(1)_Y$ and $U(1)_{N}$, respectively.  The gauge
kinetic mixing can have a significant impact on the phenomenology
\cite{Rizzo:1998ut,Salvioni:2009mt,Krauss:2012ku} and may reduce the
$Z^\prime$ mass limit.

However, if the extra $U(1)$ gauge symmetry appears from the
breakdown of $E_6$, then $\sin\chi$ should be zero at the GUT scale.
Nonetheless, even if this term is zero at the outset, it will still be
radiatively generated if the trace of the $U(1)$ charges, $\sum_i Q_i^Y
Q^\prime_i$, is nonzero.  In the cases studied here, the trace of the
charges over states in the complete $27$-plets vanishes, but to be
consistent with single-step gauge coupling unification, we also
included $\hat{H}^\prime$ and $\hat{\overline{H^\prime}}$ which lead
to a nonzero value for $\sum_i Q_i^Y Q^\prime_i $.  The value induced
by this at the EW scale though is rather small as can be
seen\footnote{The specific incomplete multiplets we consider here
  correspond to the third of the four possible embeddings referred to in
  Ref.~\cite{Rizzo:1998ut}.} in Fig.~3 of Ref.~\cite{Rizzo:1998ut}, and this
was also checked
with two-loop RGEs in the E$_6$SSM \cite{King:2005jy,Athron:2009bs}.
For this reason and due to the huge expansion in the number of terms
in the two-loop RGEs when one allows for gauge kinetic mixing, we will
neglect this in our analysis here and throughout this paper.

In general, though, it is possible for gauge kinetic mixing to be much larger,
which can be the case if one considers an additional $5 + \overline{5}$ pair of
$SU(5)$ multiplets \cite{Rizzo:1998ut} or which has been looked at in the
$U(1)_{B-L}$ \cite{Salvioni:2009mt,Krauss:2012ku}.  In such a case, this will
impact the results in two ways, firstly by altering the $Z^\prime$ limit from
experiment and secondly by altering the charges which appear in the EWSB
condition, which can be seen from examining
Eqs.~(\ref{eq:originalE6MZequation})--(\ref{eq:ddefn}).

\section{\label{sec:tuningmeasure}The Fine-Tuning
Measure}

As stated above, to quantify the resulting fine-tuning we apply the
traditional Barbieri-Giudice measure \cite{Ellis:1986yg, Barbieri:1987fn}.
A specific model is characterized by a set of $n$ model parameters $\{p_i\}$
and is defined at some input scale $M_X$. For a given parameter $p$ in this
set, one computes an associated sensitivity,
\begin{equation} \label{eq:bgmeasure}
\Delta_p = \left | \frac{\partial \ln M_Z^2}{\partial \ln p} \right
| = \left | \frac{p}{M_Z^2} \frac{\partial M_Z^2}{\partial p} \right | .
\end{equation}
The coefficient $\Delta_p$ measures the fractional variation in $M_Z^2$
resulting from a given variation in the parameter $p$.  The overall fine-tuning
is then taken to be
$\Delta = \underset{i}{\max} \, \{ \Delta_{p_i} \}$.

The sensitivity $\Delta_p$ may be calculated directly from the expression
for $M_Z^2$ in terms of the $p_i$ for a particular model, which leads to a
so-called master formula for calculating the fine-tuning.  A master formula
for the E$_6$SSM, obtained from the tree-level scalar potential, was
presented in Ref.~\cite{Athron:2013ipa}.  In order to derive the expression
presented there, the fact that $s \gg v$ was made use of to neglect certain
$O(v^2)$ terms in the EWSB conditions, greatly simplifying the final result.
For the purposes of exploring a wider class of $E_6$ inspired models, we have
derived the master formula without neglecting any $O(v^2)$ terms.  The more
complete tree-level master formula is somewhat complicated.  This is
because, unlike in the MSSM, even at tree level it is not possible to
solve explicitly for the VEVs $v_1$, $v_2$ in terms of the Lagrangian
parameters.  It may be written in the form \be \Delta_p = |C|^{-1} \times
\frac{|p|}{M_Z} \bigg | \sum_{q} \tilde{\Delta}_q \frac{\partial q}
{\partial p} \bigg | , \label{eq:E6masterformula}
\ee where the sum is over all low-energy running parameters appearing
in the tree-level EWSB conditions, i.e., $q \in \{\lambda , A_\lambda ,
m_{H_d}^2 , m_{H_u}^2 , m_S^2 , g_1 , g_2 , g_1'\}$.  Expressions for the
quantity $C$ and the $\tilde{\Delta}_q$ appearing above are given in
Appendix \ref{app:masterformula}.  It should be noted that the effects
of $U(1)$ mixing are neglected in deriving Eq.~(\ref{eq:E6masterformula}).

However, it is well known in the MSSM that radiative corrections can
significantly change (indeed, reduce) the fine-tuning \cite{Cassel:2010px}.
It is, therefore, important when studying the fine-tuning to include loop
corrections to the effective potential in the fine-tuning measure.  To do so
it is most convenient to work in terms of the EWSB conditions
Eq.~(\ref{eq:E6EWSBConditions}), rather than Eq.~(\ref{eq:E6MZequation}).
The general procedure that we use is as follows (this method has also
previously been applied in the NMSSM; see, for example,
Ref.~\cite{Ellwanger:2011mu}).  For a model in which $m$ fields develop real
VEVs (e.g. $m=2$ in the MSSM, $m=3$ in the NMSSM and in the $E_6$
models considered), we require that the $m$ minimization conditions,
\begin{equation} \label{eq:EWSBconditions}
f_1 = f_2 = \dotsb = f_m = 0 ,
\end{equation}
continue to hold under an arbitrary variation in a model parameter
$p \rightarrow p + \delta p$, so that the variations $\delta f_i$ satisfy
\begin{equation} \label{eq:EWSBvariations}
\delta f_1 = \delta f_2 = \dotsb = \delta f_m = 0 .
\end{equation}
Each $f_i$ is a function of the VEVs $v_j$ and $l$ running parameters
$q_k$ evaluated at the scale of EWSB, $f_i = f_i(v_j , q_k)$.  Thus for
each $f_i$ we find that
\begin{equation} \label{eq:EWSBchainrule}
\sum_{j=1}^m \frac{\partial f_i}{\partial v_j}
\frac{\partial v_j}{\partial p} + \sum_{k=1}^l \frac{\partial f_i}
{\partial q_k} \frac{\partial q_k}{\partial p} = 0 .
\end{equation}
The quantities $\frac{\partial f_i}{\partial v_j}$ are the elements of
the CP-even Higgs squared mass matrix $M_h^2$ of the model before
rotating into the mass eigenstate basis.  When evaluated for all $n$
model parameters, the above system of equations can be concisely
expressed as
\begin{equation} \label{eq:tuningsystem}
M_h^2 \begin{pmatrix}
\frac{\partial v_1}{\partial p_1} & \cdots &
\frac{\partial v_1}{\partial p_n} \\
\vdots & \ddots & \vdots \\
\frac{\partial v_m}{\partial p_1} & \cdots & \frac{\partial v_m}{\partial p_n}
\end{pmatrix} = -\begin{pmatrix}
\frac{\partial f_1}{\partial q_1} & \cdots & \frac{\partial f_1}
{\partial q_l} \\
\vdots & \ddots & \vdots \\
\frac{\partial f_m}{\partial q_1} & \cdots & \frac{\partial f_m}{\partial q_l}
\end{pmatrix}
\begin{pmatrix}
\frac{\partial q_1}{\partial p_1} & \cdots & \frac{\partial q_1}
{\partial p_n} \\
\vdots & \ddots & \vdots \\
\frac{\partial q_l}{\partial p_1} & \cdots & \frac{\partial q_l}{\partial p_n}
\end{pmatrix} .
\end{equation}
The quantities forming the first matrix on the right-hand side, along
with $M_h^2$, are easily calculated by differentiating the conditions
in Eq.~(\ref{eq:E6EWSBConditions}) with respect to the VEVs and the
running parameters.  The remaining derivatives $\partial q_k /\partial
p$ must be determined using the RGEs.  Once these have been obtained,
it is straightforward to solve for the $\partial v_i / \partial p$.
The sensitivities $\Delta_p$ are then simply linear combinations of
the $\partial v_i / \partial p$ and $\partial q_k / \partial p$.  The
effects of radiative corrections may be easily included by including
the Coleman-Weinberg potential contributions $\Delta V$ in the EWSB
conditions.  Here we use the one-loop corrections given in
Eq.~(\ref{Eq:DeltaV}).

Evaluating the derivatives $\partial q_k / \partial p$ must, in general, be
done by numerically integrating the two-loop RGEs.  This is time consuming
and presents an obstacle to doing large scans of the parameter space.  For
studying models defined at low energies, as we do here, we can take advantage
of the fact that the running is over much smaller scales than when
evolving up to the GUT scale.  This makes it possible to use approximate
analytic solutions to the RGEs that exhibit good accuracy over the range of
scales considered.  Given the two-loop RG equation for a parameter $q$,
\begin{equation} \label{eq:rge}
\frac{\textrm{d}q}{\mathrm{d}t} \equiv \beta_q = \frac{1}{16\pi^2}
\beta_q^{(1)} + \frac{1}{(16\pi^2)^2} \beta_q^{(2)} , \qquad t \equiv
\ln \frac{Q}{M_X} ,
\end{equation}
a Taylor series expansion of the solution may be used to obtain the
parameter at the scale $Q$,
\begin{align} \label{eq:rgeapproxsoln}
q(Q) &= q(M_X) + \int_0^t \beta_q(t') \; \textrm{d}t' \approx q(M_X) +
\frac{t}{16\pi^2} \left ( \beta_q^{(1)} + \frac{\beta_q^{(2)}}{16\pi^2}
\right ) + \frac{t^2}{32\pi^2} \frac{\textrm{d}\beta_q^{(1)}}{\textrm{d}t}
+ O(t^2) .
\end{align}
Expanded to this order, we obtain the leading log (LL) and next-to-LL
(NLL) contributions at two-loop order.  The $O(t^2)$ terms
not displayed above are formally of three-loop order and are
neglected. The derivative of the one-loop $\beta$ function is given by
\begin{equation} \label{eq:1loopbetaderivative}
\frac{\textrm{d}\beta_q^{(1)}}{\textrm{d}t} = \frac{1}{16\pi^2} \sum_{q_k}
\beta_{q_k}^{(1)} \frac{\partial \beta_q^{(1)}}{\partial q_k} ,
\end{equation}
where the sum is over all running parameters appearing in
$\beta_q^{(1)}$.  The $\beta$ functions appearing on the right-hand side of
Eqs.~(\ref{eq:rgeapproxsoln}) and (\ref{eq:1loopbetaderivative}) are
evaluated at the scale $M_X$, giving a simple analytic expression for the
parameters at the scale of EWSB in terms of the model parameters at $M_X$.
Explicit results for the relevant series expansions in the MSSM and $E_6$
models are presented in Appendix \ref{app:rges}.

\section{\label{sec:results}Results}

Using the approach outlined above, we are able to scan the low-energy
parameter space of the MSSM and E$_6$SSM and calculate the fine-tuning
in each.  To do so, we implemented the above expressions for computing
the fine-tuning in a modified version of the E$_6$SSM spectrum
generator that was used in Ref.~\cite{Athron:2013ipa}.  This code
implemented two-loop RGEs for all parameters except the soft scalar
masses.  In order to properly include the fine-tuning impact of the
$SU(3)$ gaugino soft mass $M_3$, we have extended the original code to
make use of the two-loop RGEs generated by SARAH \cite{Staub:2009bi,
  Staub:2010jh, Staub:2012pb, Staub:2013tta} and FlexibleSUSY
\cite{Athron:2014yba}, which also makes use of SOFTSUSY
\cite{Allanach:2001kg, Allanach:2013kza}.  The CP-even Higgs masses
are calculated including the leading one-loop effective potential
contributions given in Ref.~\cite{Athron:2009bs} and for the light Higgs we
use the leading two-loop\footnote{Two-loop corrections calculated for
  a nonminimal SUSY model may now also be obtained from SARAH
  \cite{Goodsell:2014pla,Goodsell:2015ira}.  However, this was not
  available when the numerical work for this paper was carried out, and
  such corrections go beyond the required precision for studying fine-tuning
  here.} contributions from Ref.~\cite{King:2005jy} which are a
generalization of the corrections in the MSSM and NMSSM calculated
using effective field theory techniques \cite{Carena:1995wu,
  Ellwanger:1999ji}.  To scan over the MSSM parameter space, the
equivalent MSSM fine-tuning expressions were implemented into a
modified version of SOFTSUSY 3.3.10 \cite{Allanach:2001kg}.  For
consistency with the results produced in the $E_6$ models, and for
computational speed, for our main scans only the dominant one- and
two-loop corrections to the CP-even Higgs masses were included.
Finally, in all of the results below the fine-tuning was evaluated at
the scale $Q = M_{\textrm{SUSY}} =
\sqrt{m_{\tilde{t}_1}m_{\tilde{t}_2}}$, where $m_{\tilde{t}_{1,2}}$
are the running $\overline{\textrm{DR}}$ stop masses evaluated at $Q =
M_{\textrm{SUSY}}$.

As discussed in the Introduction, many recent papers interested in
natural supersymmetry have focused on light stops, with much
theoretical effort to find models where it is easier to get a $125$
GeV Higgs boson and light stops simultaneously and much experimental
effort to search for light stops.  This is entirely appropriate since
there are many good reasons to expect the soft masses to be set at
high energies.  However, that is not the only possibility and the fine-tuning
problem depends strongly on the RG evolution from the GUT
scale, as the soft Higgs masses that appear in the EWSB conditions
pick up contributions from the soft squark masses.

\begin{figure}[h]
\begin{center}
\resizebox{!}{6.1cm}{%
\includegraphics{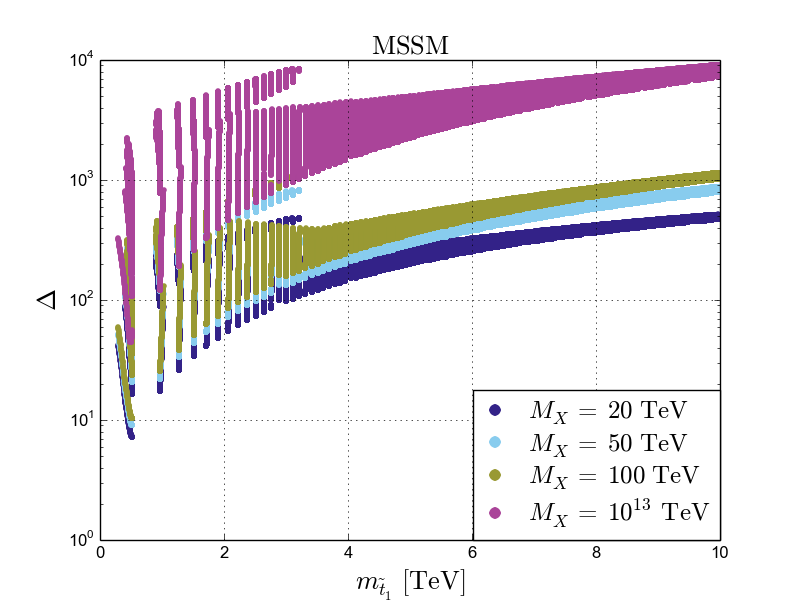}}
\resizebox{!}{6.1cm}{%
\includegraphics{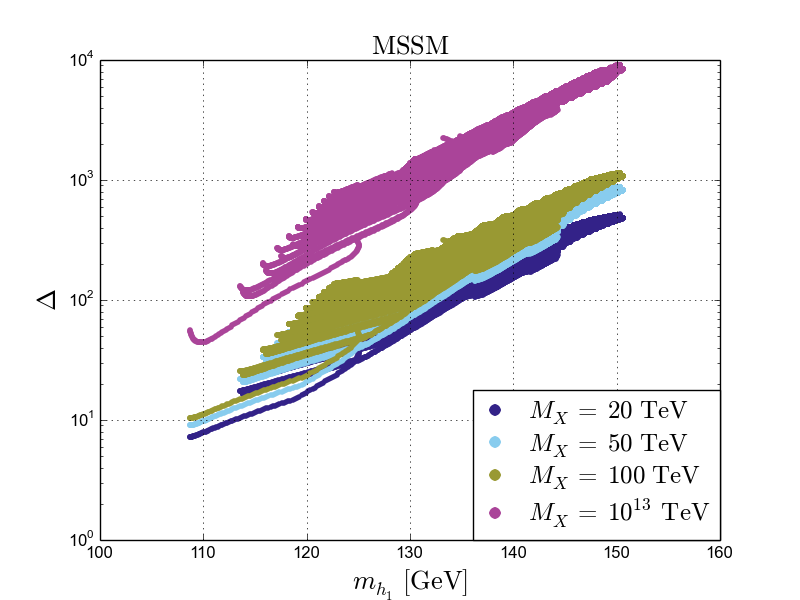}}
\caption{Left panel: Scatter plot of fine-tuning in the MSSM as a
function of the lightest stop mass, $m_{\tilde{t}_1}$, for the cutoff scales
(from bottom to top) $M_X=20$ TeV, $M_X=50$ TeV, $M_X=100$ TeV and
$M_X=10^{16}$ GeV.  Right panel: Scatter plot of fine-tuning in the
MSSM as a function of the lightest Higgs mass, $m_{h_1}$, for the cutoff
scales (from bottom to top) $M_X=20$ TeV, $M_X=50$ TeV, $M_X=100$ TeV and
$M_X=10^{16}$ GeV.}
\label{Fig:mssmstoptuning}
\end{center}
\end{figure}

To illustrate this, in the left panel of Fig.~\ref{Fig:mssmstoptuning}
we show the variation in fine-tuning for $M_X = 20$ TeV, $50$ TeV,
$100$ TeV and $10^{16}$ GeV when we scan over the stop masses and
mixing, with $500 \textrm{ GeV} \leq m_{Q_3} , m_{u_3} \leq 10
\textrm{ TeV}$ and $-3810\textrm{ GeV} \leq A_t \leq -20 \textrm{
  GeV}$.  The remaining parameters we fix such that at $M_{\rm SUSY}$
they have the values $\mu = -97.5$, $B = -84.8$, $M_1 = 92.1$,
$M_2 = 95.9$, $M_3 = 352$, $A_b = -117.9$, $A_\tau =
-7.8$, $m_{L_i} = 400$, $m_{e_i} = 204$, $m_{Q_{1,2}} =
438$, $m_{u_{1,2}} = 436$ and $m_{d_i} = 438$ GeV ($i = 1 , 2
, 3$).  Here we denote by $M_1$, $M_2$ and $M_3$ the soft gaugino
masses for $U(1)$, $SU(2)$ and $SU(3)$, while $A_b$ and $A_\tau$ are
soft trilinear couplings and the $m_\phi$ are soft scalar masses for
the indicated fields.  The soft bilinear coupling $B$ is defined such
that at tree level the mass of the CP-odd MSSM Higgs boson reads
$m_A^2 = 2 B \mu / \sin 2\beta$.  All off-diagonal couplings and
scalar masses are set to zero, as are the first- and second-generation
Yukawa couplings and soft trilinears.  Although we should stress that
making this choice will lead to a spectrum which is in conflict with
the LHC limits, doing so ensures that fine-tuning due to the other
parameters is small, so that we avoid washing out the fine-tuning
impact of the stops when the tuning is small\footnote{For models in
  which the spectrum is heavier, when the stop masses are small the
  fine tuning reaches a lower bound imposed by other heavier
  parameters.} as can be the case when the stop masses are less than 1
TeV.  Note that the Higgs mass is also allowed to vary in this scan,
as shown in the right panel of Fig.~\ref{Fig:mssmstoptuning}.  This
illustrates the tuning problem which people have been worrying about
since the discovery of the 125 GeV Higgs boson as we see that raising
the stop masses is also pushing up the Higgs mass, meaning that
heavier Higgs masses require more fine-tuning.  However, for a low
value of the UV scale this tuning is not so severe unless the stops
are very heavy, and a 125 GeV Higgs can be obtained without much
tuning in this unrealistic case where we have minimized other sources
of tuning.  On the other hand, the tuning becomes more severe as we
increase the cutoff such that for $M_X = 10^{16}$ GeV a lightest stop
mass of 1--3 TeV can result in a fine-tuning of $\approx$ 100 --
1000 and the minimum tuning we find\footnote{Note that in the
  calculation of the Higgs mass there is a significant theoretical
  error, even with leading two-loop corrections, which should be
  considered when thinking about what the results imply for the
  minimum fine tuning in the model consistent with the recent
  discovery of a 125 GeV Higgs.} for a $125$ GeV Higgs is $\approx
200$, as shown in Fig.~\ref{Fig:mssmstoptuning}.

Since the stop mass does not have such a large impact on the fine-tuning
when the cutoff scale is very low we can use this to see more
clearly the impact of the $Z^\prime$ mass on fine-tuning.  To do so we
select a fixed low cutoff of $M_X=20$ TeV and compare the fine tuning
between the MSSM and E$_6$SSM for two different values of the
$Z^\prime$ mass.  We choose to look at $M_{Z^\prime} = 2.5$ TeV, which is
just above the current limits, and $M_{Z^\prime} = 4.5$ TeV, which
should be in reach in run II at the LHC \cite{Godfrey:2013eta} and
then compare the fine tuning calculated in each case to the tuning in
the MSSM.  For this, we have performed a six-dimensional parameter
space scan in both the MSSM and E$_6$SSM, varying those parameters
most relevant for the fine tuning and the Higgs mass.  Therefore, the
set of parameters which we vary includes $\mu$, $B$ and $\tan\beta$
for the MSSM, and $\lambda$, $A_\lambda$ and $\tan\beta$, for the
E$_6$SSM, which appear at tree level in the EWSB conditions of the
models.  While the RGE contribution from large stop masses to the fine
tuning is small for such a low cutoff scale, the stop contributions to
the effective potential can play a significant role in reducing the
fine tuning.  For this reason it is still important to properly treat
the tuning associated with stop contributions to the one-loop
effective potential, and so we also scan over the soft masses
$m_{Q_3}^2$, $m_{u_3}^2$ and the stop mixing $A_t$.  The relevant
parameters and ranges that were scanned over are summarized in Table
\ref{tab:scanranges}.  In addition to this we also repeat each scan
for three different values of $M_2$ to allow more variation in the
chargino masses.

In this case, we now consider realistic scenarios, where the
parameters that are not scanned over are set to values which keep the
associated states comfortably above their experimental limits.  So in
both the MSSM and E$_6$SSM, all other soft scalar masses are set
to $5$ TeV.  We require a valid spectrum with no
tachyonic states to exclude points which would have an unrealistic minimum,
for example due to the appearance of charge or color breaking (CCB) minima.
We work in the third family approximation, taking the first- and
second-generation Yukawa couplings to be zero, and we also assume that their
associated soft trilinears vanish.  Similarly, we take $A_b = A_\tau = 0$ GeV.
The $U(1)$ gaugino soft mass $M_1$ was fixed to $M_1 = 300$ GeV, and we fix
$M_3 = 2000$ GeV.  Additionally, in the E$_6$SSM the $U(1)_N$ gaugino soft mass
$M_1'$ is held fixed with $M_1' = M_1 = 300$ GeV, and $\mu^\prime = 5$ TeV.

\begin{table}[h]
\centering
\begin{ruledtabular}
\begin{tabular}{cc}
MSSM & E$_6$SSM \\
\hline
$2 \leq \tan\beta \leq 50$ & $2 \leq \tan\beta \leq 50$ \\
$-1 \textrm{ TeV } \leq \mu \leq 1 \textrm{ TeV}$ & $-3 \leq \lambda \leq 3$\\
$-1 \textrm{ TeV } \leq B \leq 1 \textrm{ TeV}$ & $-10 \textrm{ TeV } \leq
A_\lambda \leq 10 \textrm{ TeV}$ \\
$ 200 \textrm{ GeV } \leq m_{Q_3} \leq 2000 \textrm { GeV}$ & $200
\textrm{ GeV } \leq m_{Q_3} \leq 2000 \textrm { GeV}$ \\
$200 \textrm{ GeV } \leq m_{u_3} \leq 2000 \textrm { GeV}$ & $200
\textrm{ GeV } \leq m_{u_3} \leq 2000 \textrm { GeV}$ \\
$-10 \textrm{ TeV } \leq A_t \leq 10 \textrm { TeV}$ & $-10
\textrm{ TeV } \leq A_t \leq 10 \textrm { TeV}$ \\
$M_2 = 100 \textrm{, } 1050 \textrm{, } 2000 \textrm{ GeV}$ &
$M_2 = 100 \textrm{, } 1050 \textrm{, } 2000 \textrm{ GeV}$ \\
\end{tabular}
\end{ruledtabular}
\caption{The parameters scanned over and the ranges of values used in
the MSSM and the E$_6$SSM models.}
\label{tab:scanranges}
\end{table}

\begin{figure}[h]
\begin{center}
\resizebox{!}{10cm}{%
\includegraphics{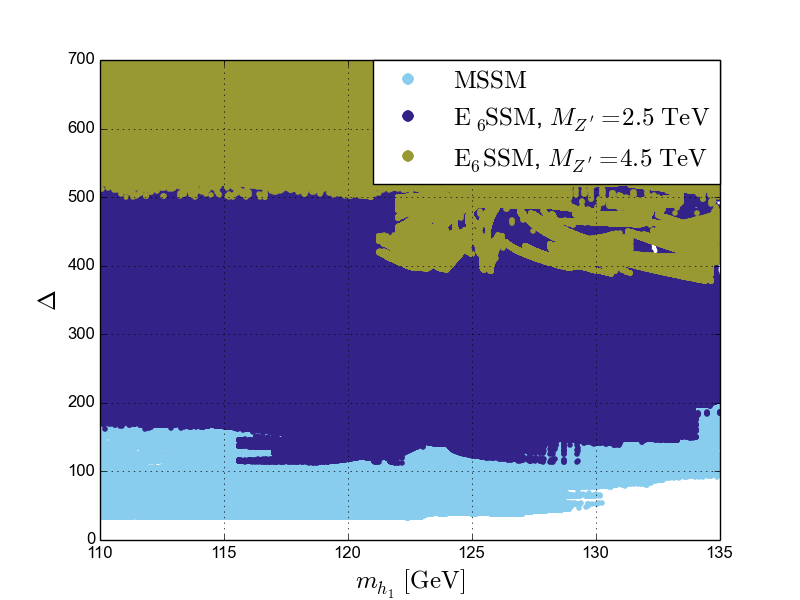}}
\caption{Scatter plot of fine-tuning vs lightest Higgs mass for the
MSSM (light blue, bottom band), E$_6$SSM with $M_{Z^\prime} = 2.5$ TeV
(dark blue, middle band) and E$_6$SSM with $M_{Z^\prime} = 4.5$ TeV
(dark yellow, top band).  Note that there are points for which
the fine-tuning in the MSSM and E$_6$SSM with $M_{Z'} = 2.5$ TeV is larger
than is visible on this plot and those below; however, these points are
obscured by the overlaid data for the E$_6$SSM with $M_{Z'} = 4.5$ TeV, and
it is the lower bound on the achievable tuning that is of interest here.}
\label{Fig:e6ssmvsmssm}
\end{center}
\end{figure}

In Fig.~\ref{Fig:e6ssmvsmssm}, results from the scan are plotted showing
the tuning for each case against the lightest Higgs mass.  As expected,
the dependence on the Higgs mass is now quite weak, while the minimum
tuning in the model for the E$_6$SSM is increased by the mass of the
$Z^\prime$ boson.  So in the case of a very low cutoff the tuning
required to get a 125 GeV Higgs is not so large.  However, the tuning
from the $Z^\prime$ mass appears already at tree level and is,
therefore, not suppressed when the cutoff scale is low.  In our scan we
find that, for the points satisfying the current limit on the mass of
the $Z^\prime$ boson and having an approximately 125 GeV Higgs, the
minimum fine-tuning that can be achieved is
$\Delta_{\textrm{min}} \approx 121$.  If run II of the LHC further
pushes up the limit on the $Z^\prime$ mass to be above $4.5$ TeV then
the fine-tuning in the model will be greater than at least
$\Delta_{\textrm{min}} \approx 394$ for a Higgs mass between $124.5$
and $125.5$ GeV.

This demonstrates two important points about these $U(1)$
extensions--first, that limits on the $Z^\prime$ mass play an
incredibly important role in constraining natural scenarios in such
models and, second, that the tuning from the $Z^\prime$ limits in
these models depends less on assumptions about SUSY breaking than the
tuning required by the $125$ GeV Higgs measurement which concerns
people in the MSSM.

There is another limit which plays a similar role.  Chargino limits
directly constrain the $\mu$ parameter (or effective $\mu$ parameter
in these $U(1)$ extensions).  The LEP bound \cite{Kraan:2005vy} on
chargino masses, excluding $m_{\tilde{\chi}^\pm_1} \lesssim 104$ GeV, implies
that $|\mu|$ should only be greater than $\sim 100$ GeV, which is not
substantially larger than $M_Z$.  Consequently the bound from LEP is not
high enough to have an impact on the fine-tuning obtained in the models and
parameter space regions that we have studied, as we have checked explicitly.
Significantly larger lower bounds on the $\mu$ parameter, and therefore on the
fine-tuning, may arise from chargino limits coming from LHC searches.  However,
the chargino limits from the LHC depend on whether there are light sleptons or
sneutrinos and the mass difference between the lightest chargino and lightest
neutralino.  Current limits placed by CMS and ATLAS extend up to
$m_{\tilde{\chi}^\pm_1} \approx$ 700--740 GeV if there are light
sleptons \cite{Khachatryan:2014qwa, Aad:2014nua} with much weaker
bounds if there are no light sleptons or sneutrinos.\footnote{Useful
summary plots of these limit may be found on the public pages of
ATLAS, \url{https://atlas.web.cern.ch/Atlas/GROUPS/PHYSICS/CombinedSummaryPlots/SUSY/ATLAS_SUSY_EWSummary/ATLAS_SUSY_EWSummary.png}
and CMS \url{http://cms.web.cern.ch/sites/cms.web.cern.ch/files/styles/large/public/field/image/Image_03_exclusion_Combined.png?itok=8FMBpu_1}.}

\begin{figure}[h]
\begin{center}
\resizebox{!}{10cm}{%
\includegraphics{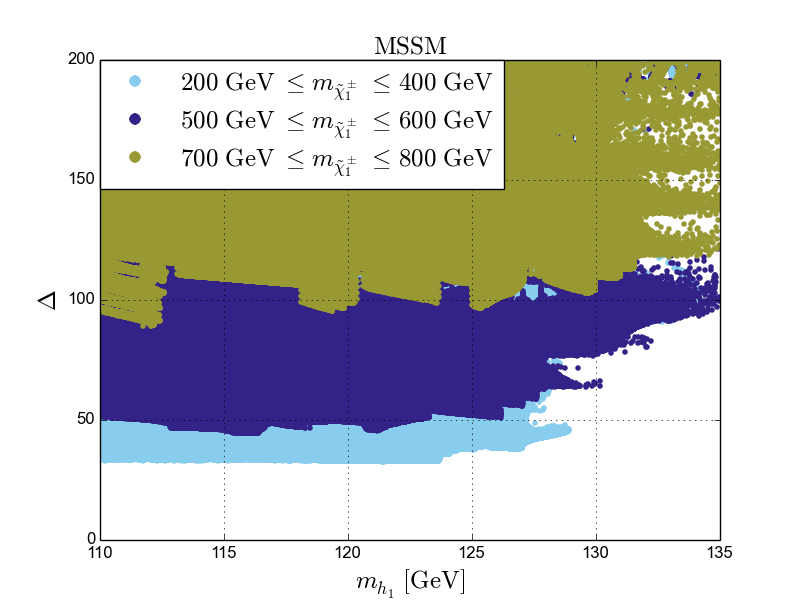}}%%
\caption{Scatter plot of fine-tuning vs lightest Higgs mass in the MSSM with
$200 \textrm{ GeV} \leq m_{\tilde{\chi}^\pm_1} \leq 400$ GeV shown in light
blue (bottom band), $500 \textrm{ GeV} \leq m_{\tilde{\chi}^\pm_1} \leq 600$
GeV in dark blue (middle band), and $700 \textrm { GeV} \leq
m_{\tilde{\chi}^\pm_1} \leq 800$ GeV in dark yellow (top band).}
\label{Fig:chargino-plot}
\end{center}
\end{figure}

Nonetheless, for the MSSM the impact of potential chargino mass limits is
shown in Fig.~\ref{Fig:chargino-plot}.  There we see that if the
full parameter space with $m_{\tilde{\chi}^\pm_1} < 700$ GeV was excluded, the
impact would be to make the tuning in the MSSM with a $20$ TeV cutoff
similar to that of the E$_6$SSM with the same cutoff and a
$Z^\prime$ mass just larger than current limits.  In the E$_6$SSM,
while raising the chargino limit can have the same impact in
principle, due to current limits on the $Z^\prime$ mass already
imposing a significant degree of tuning, chargino masses do not make
much of a noticeable change.

\begin{figure}
\begin{center}
\resizebox{!}{6.1cm}{%
\includegraphics{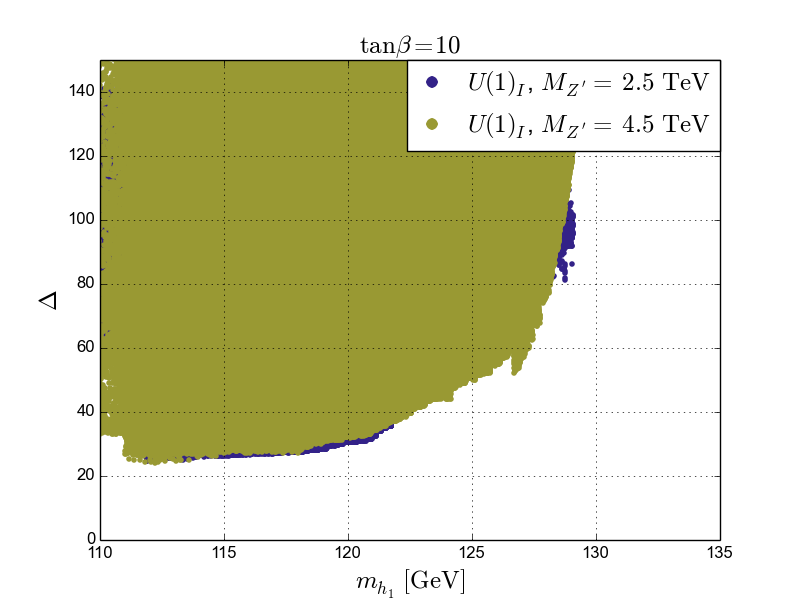}}
\resizebox{!}{6.1cm}{%
\includegraphics{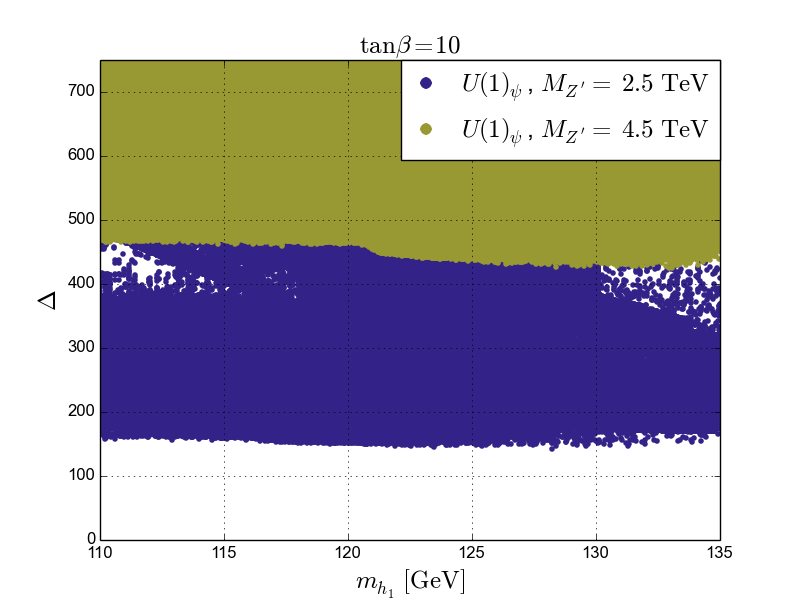}}\\
\resizebox{!}{6.1cm}{%
\includegraphics{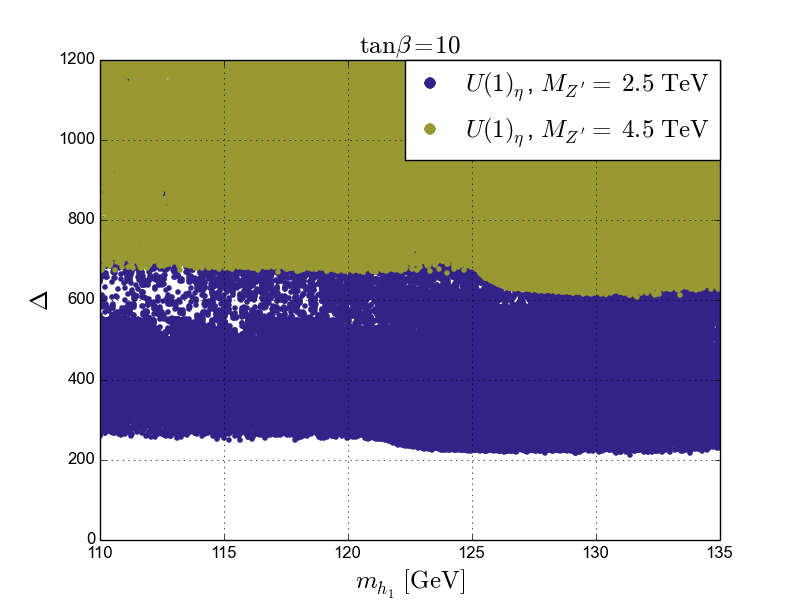}}%%
\caption{Top left panel: Scatter plot of the fine-tuning vs lightest
Higgs mass in the $U(1)_I$ model.  Top right panel: Scatter plot of
the fine-tuning vs lightest Higgs mass in the $U(1)_\psi$ model.
Bottom panel: Scatter plot of the fine-tuning vs lightest Higgs mass
in the $U(1)_\eta$ model.  In each plot points with $M_{Z^\prime} = 2.5$ TeV
are shown in dark blue (bottom band), and points with $M_{Z^\prime} = 4.5$
TeV are shown in dark yellow (top band).}
\label{Fig:othere6modelsvsmssm}
\end{center}
\end{figure}

The exact level of tuning from the $Z^\prime$ depends on the charges
of the extra $U(1)$ gauge symmetry it is associated with.  In
Fig.~\ref{Fig:othere6modelsvsmssm} we look at the fine-tuning for
other $U(1)$ extensions for the same $Z^\prime$ masses as we did for the
E$_6$SSM.  To simplify the analysis we fix $\tan\beta = 10$, but scan
over the remaining parameters as in Table \ref{tab:scanranges} and fix
the rest to the same values we did in the scan carried out for
Fig.~\ref{Fig:e6ssmvsmssm}.  In order to more clearly identify the lower bound
on the obtainable tuning in each model, the parameter values for points in
these main grid scans with a low fine-tuning were then used as the starting
points for smaller scans about those values.  In these smaller scans the
parameters were more finely varied to populate the low fine-tuning regions.

As can be seen in Fig.~\ref{Fig:othere6modelsvsmssm}, the severity of the
tunings varies quite a bit.  This is because the charges appear as
coefficients in front of the $Z^\prime$ mass in the EWSB condition.  These
charges change the value of the coefficient $d$ in
Eq.~(\ref{eq:originalE6MZequation}).  The values of the coefficient $d$ in
each model, for $\tan\beta = 10$, is $\{-0.01 , 0.40 , 0.50 , 0.81 \}$ for
$\{U(1)_I , U(1)_N , U(1)_\psi , U(1)_\eta \}$ and this determines which of
the models are most tuned.

Interestingly, the coefficient $d$ is very small (and negative) in the
case of the $U(1)_I$.  This allows a dramatic reduction in the fine-tuning
from the $U(1)_I$ symmetry.  This is a result of the $H_u$
charge associated with $U(1)_I$ vanishing, which means that the
$D$ terms to the lightest Higgs which is predominantly $H_u$ at large
$\tan\beta$ are suppressed, making it difficult to raise the Higgs
mass in the same way as happens in the other models and explaining why
heavier Higgs values in this model can't be obtained.  Therefore, the
fine-tuning behaviour in this model is closer to that of the MSSM, and
in this case raising the $Z^\prime$ mass limit to $4.5$ TeV will have
little impact on naturalness.  From naively estimating the tuning,
using the $d$ coefficient one can estimate that $Z^\prime$ limits
need to be around $15$ TeV before they will raise the tuning in this
model.

Finally we want to emphasize that while in Fig.~\ref{Fig:e6ssmvsmssm}
the E$_6$SSM looks more fine-tuned than the MSSM this depends on the
high scale boundary, $M_X$, where the parameters are assumed to be set
by some SUSY breaking mechanism.  Indeed in Ref.~\cite{Athron:2013ipa} a
constrained version of the E$_6$SSM, with the high scale boundary at
the GUT scale, is considered and there the cE$_6$SSM was found to be
less tuned than the cMSSM.  Since a 125 GeV Higgs can be achieved in
the E$_6$SSM with lighter stops, then if the cutoff is large, the
larger stop masses of the MSSM can make that model more fine-tuned due
to large RGE effects.

\begin{figure}
\begin{center}
\resizebox{!}{5.5cm}{%
\includegraphics{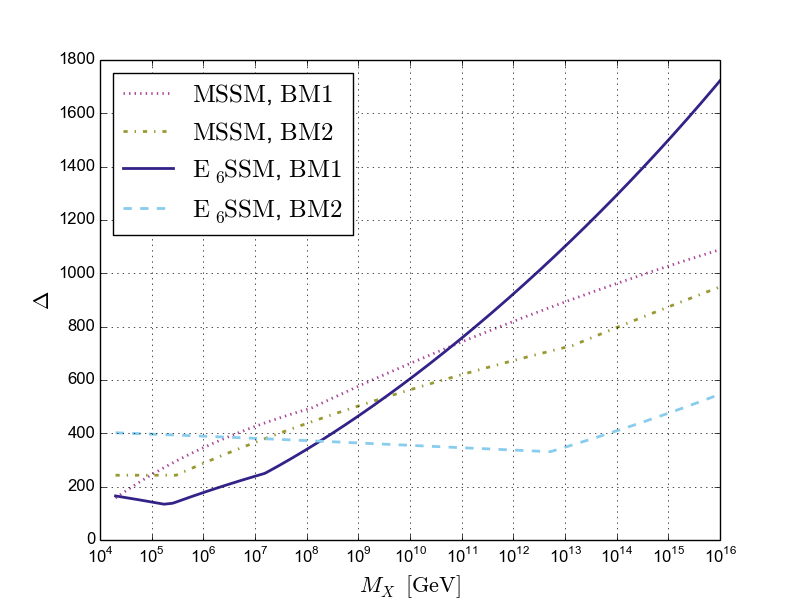} } \\
\resizebox{!}{5.5cm}{%
\includegraphics{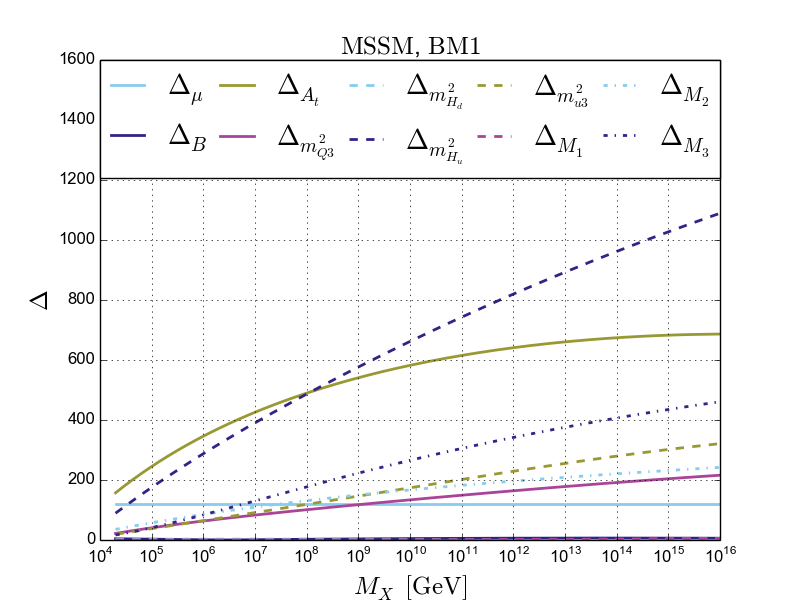}}
\resizebox{!}{5.5cm}{%
\includegraphics{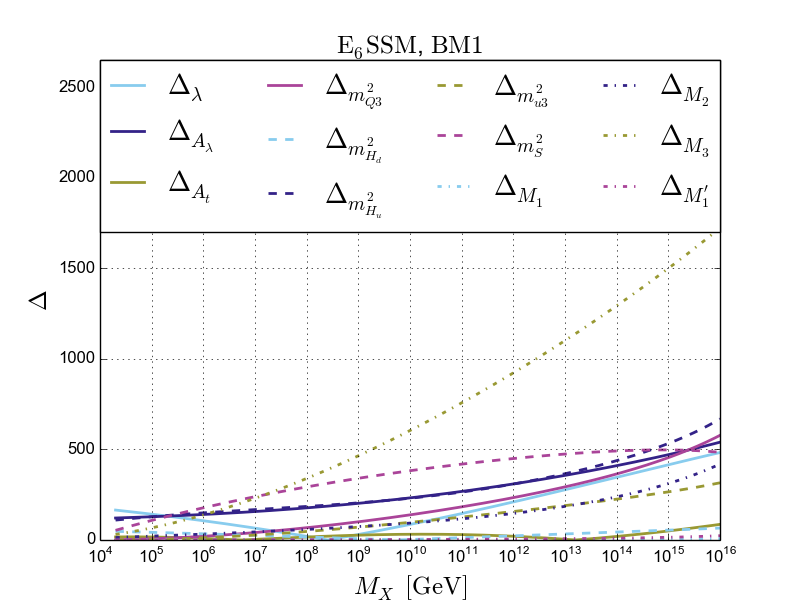}} \\
\resizebox{!}{5.5cm}{%
\includegraphics{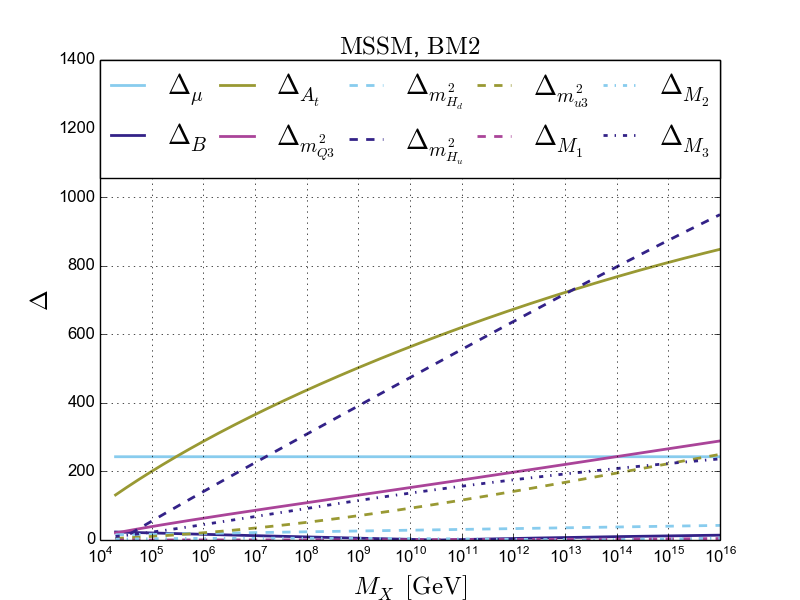}}
\resizebox{!}{5.5cm}{%
\includegraphics{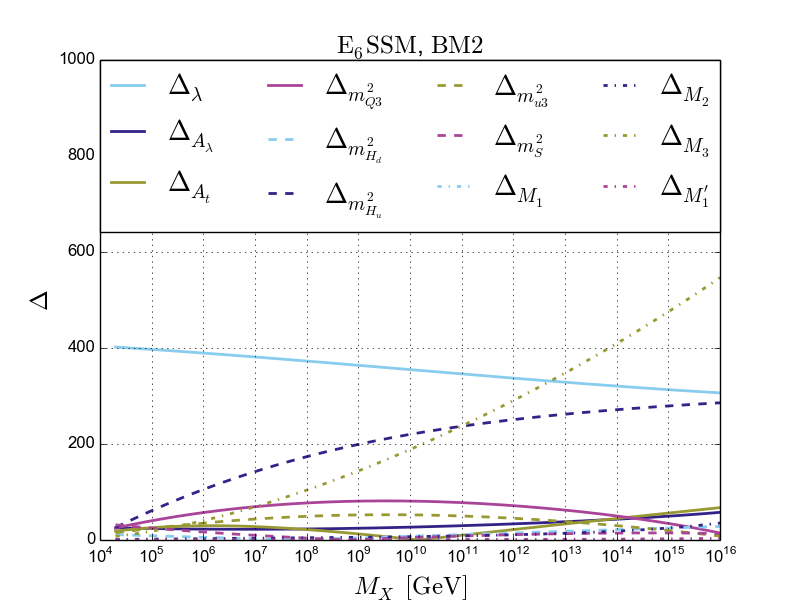}}%%
\caption{Top panel: Scatter plot of the fine-tuning as a function
of the cutoff scale $M_X$ for the four benchmark points given in
Table \ref{tab:benchmarks}.  Middle left panel:  Individual
sensitivities for MSSM BM1 plotted against the high scale $M_X$ which give
the overall tuning shown by the dotted line in the top panel.
Middle right panel: Individual sensitivities for E$_6$SSM BM1 plotted
against the high scale $M_X$ which give the overall tuning shown by the solid
line in the top panel.  Bottom left panel: Individual sensitivities
for MSSM BM2 plotted against the high scale $M_X$ which give the overall
tuning shown by the dash-dotted line in the top panel.
Bottom right panel: Individual sensitivities for E$_6$SSM BM2 plotted
against the high scale $M_X$ which give the overall tuning shown by the dashed
line in the top panel.}
\label{Fig:BMs-varyMX}
\end{center}
\end{figure}

To further illustrate this point, we looked at how the tuning varies
with $M_X$ for low tuning benchmarks in the MSSM and E$_6$SSM.  These
benchmarks are defined in Table \ref{tab:benchmarks} and the results
are shown in Fig.~\ref{Fig:BMs-varyMX}.  Since the behavior is quite
complicated we now discuss these in detail as it provides some insight
into the many differences in the tuning between the two models.

\begingroup
\squeezetable
\begin{table}[h]
\centering
\begin{ruledtabular}
\begin{tabular}{ccccc}
& MSSM BM1 & MSSM BM2 & E$_6$SSM BM1 & E$_6$SSM BM2 \\
\hline
$\tan\beta(M_Z)$ & $10$ & $10$ & $10$ & $10$ \\
$s(M_{\mathrm{SUSY}})$ [GeV] & $\cdots$ & $\cdots$ & $6700$ & $6700$ \\
$\kappa_{1,2,3}(M_{\mathrm{SUSY}})$ & $\cdots$ & $\cdots$ & $0.6$ & $0.52$ \\
$\lambda_{1,2}(M_{\mathrm{SUSY}})$ & $\cdots$ & $\cdots$ & $0.2$ & $0.13$ \\
$\mu_{\mathrm{eff}}(M_{\mathrm{SUSY}})$ [GeV] & $689.7$ & $1013.5$ &
$1093.3$ & $1313.0$ \\
$B_{\mathrm{eff}}(M_{\mathrm{SUSY}})$ [GeV] & $345.7$ & $1032.5$ &
$3792.7$ & $817.8$ \\
$A_\tau(M_{\mathrm{SUSY}})$ [GeV] & $0$ & $-5057.9$ & $0$ & $-88.5$ \\
$A_b(M_{\mathrm{SUSY}})$ [GeV] & $0$ & $-5707.2$ & $0$ & $-1720.7$\\
$A_t(M_{\mathrm{SUSY}})$ [GeV] & $-3335.7$ & $-2734.8$ & $-1100$ & $-1103.2$ \\
$m_{L_{1,2}}^2(M_{\mathrm{SUSY}})$ [GeV$^2$] & $2.5\times 10^7$ &
$6.35\times 10^6$ & $2.5\times 10^7$ & $4.94\times 10^6$ \\
$m_{L_3}^2(M_{\mathrm{SUSY}})$ [GeV$^2$] & $2.5\times 10^7$ &
$6.22\times 10^6$ & $2.5\times 10^7$ & $4.90\times 10^6$ \\
$m_{e_{1,2}}^2(M_{\mathrm{SUSY}})$ [GeV$^2$] & $2.5\times 10^7$ &
$6.27\times 10^6$ & $2.5\times 10^7$ & $5.21\times 10^6$ \\
$m_{e_3}^2(M_{\mathrm{SUSY}})$ [GeV$^2$] & $2.5\times 10^7$ &
$6.03\times 10^6$ & $2.5\times 10^7$ & $5.11\times 10^6$ \\
$m_{Q_{1,2}}^2(M_{\mathrm{SUSY}})$ [GeV$^2$] & $2.5\times 10^7$ &
$7.37\times 10^6$ & $2.5\times 10^7$ & $5.76\times 10^6$ \\
$m_{Q_3}^2(M_{\mathrm{SUSY}})$ [GeV$^2$] & $4.45\times 10^6$ &
$3.97\times 10^6$ & $4.50\times 10^5$ & $3.61\times 10^6$ \\
$m_{u_{1,2}}^2(M_{\mathrm{SUSY}})$ [GeV$^2$] & $2.5\times 10^7$ &
$7.30\times 10^6$ & $2.5\times 10^7$ & $5.54\times 10^6$ \\
$m_{u_3}^2(M_{\mathrm{SUSY}})$ [GeV$^2$] & $4.0 \times 10^6$ &
$6.60\times10^5$& $5.86\times 10^5$ & $2.04\times 10^6$ \\
$m_{d_{1,2}}^2(M_{\mathrm{SUSY}})$ [GeV$^2$] & $2.5\times 10^7$ &
$7.30\times 10^6$ & $2.5\times 10^7$ & $5.88\times 10^6$ \\
$m_{d_3}^2(M_{\mathrm{SUSY}})$ [GeV$^2$] & $2.5\times 10^7$ &
$7.03\times 10^6$ & $2.5\times 10^7$ & $5.78\times 10^6$ \\
$m_{H_d}^2(M_{\mathrm{SUSY}})$ [GeV$^2$] & $1.82\times 10^6$ &
$8.96\times 10^6$ & $4.06\times 10^7$ & $1.04\times 10^7$ \\
$m_{H_u}^2(M_{\mathrm{SUSY}})$ [GeV$^2$] & $-3.60\times 10^5$ &
$-9.35\times 10^5$ & $5.0\times 10^5$ & $-2.66\times 10^5$ \\
$m_S^2(M_{\mathrm{SUSY}})$ [GeV$^2$] & $\cdots$ & $\cdots$ &
$-3.10\times 10^6$ & $-3.17 \times 10^6$ \\
$M_1(M_{\mathrm{SUSY}})$ [GeV] & $300$ & $260.8$ & $300$ & $173.4$ \\
$M_2(M_{\mathrm{SUSY}})$ [GeV] & $2000$ & $479.2$ & $1050$ & $281.4$ \\
$M_3(M_{\mathrm{SUSY}})$ [GeV] & $2000$ & $1312.3$ & $2000$ & $1200$ \\
$M_1^\prime(M_{\mathrm{SUSY}})$ [GeV] & $\cdots$ & $\cdots$ & $300$ & $175.2$ \\
\hline
$M_{Z^\prime}$ [GeV] & $\cdots$ & $\cdots$ & $2473.2$ & $2512.7$ \\
$m_{h_1}$ [GeV] & $124.3$ & $124.4$ & $125.0$ & $126.2$ \\
$m_{\tilde{t}_1}$ [GeV] & $1942.1$ & $861.6$ & $993.8$ & $1665.0$ \\
$m_{\tilde{t}_2}$ [GeV] & $2220.1$ & $2023.9$ & $1174.8$ & $2094.4$ \\
$m_{\tilde{g}}$ [GeV] & $2259.8$ & $1472.9$ & $2290.0$ & $1407.4$ \\
\hline
$\Delta(M_X=20\textrm{ TeV})$ & $157.3$ & $242.8$ & $165.3$ & $402.1$ \\
$\Delta(M_X=10^{16}\textrm{ GeV})$ & $1089.0$ & $949.0$ & $1722.3$ & $546.7$
\end{tabular}
\end{ruledtabular}
\caption{Parameters for the MSSM and E$_6$SSM benchmark points.  In the
E$_6$SSM, we define $\mu_{\mathrm{eff}} \equiv \lambda s/\sqrt{2}$ and
$B_{\mathrm{eff}} = A_\lambda$.  The soft masses $m_{H_d}^2$, $m_{H_u}^2$
and $m_S^2$ are those that satisfy the EWSB conditions including one-loop
Coleman-Weinberg corrections involving the top and stops.  For E$_6$SSM BM1
(BM2) we also set $\mu^\prime = 5000.0$ $(897.9)$ GeV, $B\mu^\prime = 5000.0$
$(-4.21\times 10^5)$ GeV$^2$, $A_{\kappa_{1,2,3}} = 0$ $(-1389.2)$ GeV,
$A_{\lambda_{1,2}} = 0$ $(-52.9)$ GeV, $m_{D_{1,2,3}}^2 = 2.5\times 10^7$
$(4.81\times 10^6)$ GeV$^2$, $m_{\overline{D}_{1,2,3}}^2 = 2.5\times 10^7$
$(4.90\times 10^6)$ GeV$^2$, $m_{H^d_{1,2}}^2 = 2.5\times 10^7$
$(4.46\times 10^6)$ GeV$^2$, $m_{H^u_{1,2}}^2 = 2.5\times 10^7$
$(4.81\times 10^6)$ GeV$^2$, $m_{S_{1,2}}^2 = 2.5\times 10^7$
$(5.28\times 10^6)$ GeV$^2$, $m_{H'}^2 = 2.5\times 10^7$ $(4.94\times 10^6)$
GeV$^2$ and $m_{\overline{H'}}^2 = 2.5\times 10^7$ $(4.87\times 10^6)$ GeV$^2$.}
\label{tab:benchmarks}
\end{table}
\endgroup

In the top panel one can see that the MSSM BM1 tuning (dotted curve)
steadily climbs as the cutoff scale is increased, as one would expect
when the tuning originates from large soft masses entering from the
RGEs.  The panel on the middle left confirms this, showing that the
largest tuning contributions come from $\Delta_{A_t}$ and
$\Delta_{m_{H_u}^2}$ with the former being the larger sensitivity
until $M_X\approx 10^8$ GeV at which point $\Delta_{m_{H_u}^2}$ takes
over, leading to the small kink in overall tuning which can be seen in
the dotted curve in the top panel.  In this case we have chosen a
point with large mixing, which is known to reduce the MSSM tuning.  We
found this does not eliminate the tuning as there is still a strong
sensitivity to $A_t$, but we did find that large mixing lead to less
fine-tuning overall for the points we examined.

Comparing the MSSM tunings to the E$_6$SSM tunings one can see that
which point is more fine-tuned depends on the scale at which the
parameters are defined.  This illustrates that any statement about
which model is more tuned depends on the high scale boundary, $M_X$.

For E$_6$SSM BM1 the fine-tuning is shown by the solid curve in the top
panel of Fig.~\ref{Fig:BMs-varyMX} and the individual sensitivities
are given in the middle right panel.  The tuning actually reduces
initially as the cutoff is is increased from $20$ TeV.  This occurs
because the largest sensitivity is initially $\Delta_\lambda$ (shown
in solid light blue in the middle right panel).  This contains some terms
proportional to $M_Z^{\prime \, 2}$, which provide the dominant
contribution to this sensitivity at very low $M_X$.  However, as $M_X$
is increased contributions from the soft masses become more important
and these actually start to cancel the large contribution to
$\Delta_\lambda$ coming from $M_{Z^\prime}$ until $\Delta_\lambda$
passes through zero.  At the same time though these large soft masses
also cause other sensitivities to grow, in particular $\Delta_{M_3}$.
The fine-tuning rises with $M_X$ once $M_X \gtrsim 10^5 - 10^6$ GeV, but
remains lower than that of the other points, until $M_X \approx 10^8$
GeV.  Eventually the $\Delta_{M_3}$ sensitivity leads to this point
being the most fine-tuned of the four shown in
Fig.~\ref{Fig:BMs-varyMX}.

Although the gluino mass and $M_3(M_{SUSY})$ have similar values to
those in the MSSM BM1 point, in the E$_6$SSM $M_3(M_X)$ is larger due
to the altered RGE running from exotic matter\footnote{This altered
RGE running is a result of the exotic matter introduced to keep the
extra $U(1)$ anomaly free.}.  This is why this E$_6$SSM BM1 has a
larger tuning at larger values of $M_X$, coming from
$\Delta_{M_3}$.

Interestingly other sensitivities are suppressed by this effect since
at the same time larger $M_3$ at higher scales reduces the soft squark
masses at $M_X$.  Therefore, the stop mass contributions are
ameliorated, compared to the MSSM, both by allowing lighter stops at
$M_{SUSY}$ and by the modified RGE running.  Nonetheless the stops
still do lead to $\Delta_{m_{H_u}^2}$ increasing with the cutoff
through the usual mechanism\footnote{Wherein $m_{H_u}^2(M_{SUSY})$
receives a positive contribution to it's mass from $m_{H_u}^2(M_X)$
and a negative contribution from $m_{Q_3}^2(M_X)$ and
$m_{u_3}^2(M_X)$, allowing heavy stop masses to cause fine-tuning.  In this
case $m_{H_u}^2(M_{SUSY})$ is held fixed so as the
cutoff increases the values of these soft masses at $M_X$ will be
larger and there will be a bigger cancellation between them,
increasing the sensitivity of $M_Z$ to both $m_{H_u}^2$ and the soft
scalar masses for the stops.}.

By contrast the tuning for E$_6$SSM BM2 is very different, as is shown
by the dashed line in the top panel, with the individual sensitivities given
in the bottom right panel.  This point was chosen as it had a much lighter
gluino mass that is just above the experimental limit of $1.4$ TeV
\cite{Aad:2014lra}.  At $20$ TeV this benchmark is not amongst the
lowest tuned points, since at that scale the tree-level tuning from
$M_{Z^\prime}$ dominates.  However, the reduction in $M_3$ means that
$\Delta_{M_3}$ is substantially lower and only becomes the dominant
tuning at a much larger scale of $M_X \gtrsim 10^{12} - 10^{13}$ GeV,
giving a tuning at $10^{16}$ GeV of $\approx 546$, which is far below
that of the other three benchmark points.

In addition to this, the soft parameters in E$_6$SSM BM2 follow a pattern
similar to that found in the constrained model.  With the exception of the
parameters $m_{Q_3}^2$, $m_{u_3}^2$, $m_{H_d}^2$, $m_{H_u}^2$, and $M_3$,
the values of which are given in Table \ref{tab:benchmarks}, the soft masses
at the SUSY scale correspond to the values that result in the cE$_6$SSM with
$m_0 = 2.2$ TeV, $M_{1/2} = 1003$ GeV, $A_0 = 500$ GeV,
$\kappa_{1,2,3}(M_X) = 0.1923$, $\lambda(M_X) = 0.2646$ and
$\lambda_{1,2}(M_X) = 0.1$.  This leads to a significant reduction in the
contributions to the RG running of $m_{H_u}^2$ and $m_{Q_3}^2$ coming from
terms of the form $g_1^2\Sigma_1$ and, to a lesser extent, $g_1'^2\Sigma_1'$.
Here we define for the E$_6$SSM (see also
Eqs.~(\ref{eq:USSMGaugeSigmaTerm1}-\ref{eq:USSMGaugeSigmaTerm1Pr}) for
general $U(1)$ inspired models)
\begin{align*}
\Sigma_1 &= \sum_{i=1}^3 \left ( m_{Q_i}^2 - 2 m_{u_i}^2 + m_{d_i}^2 +
m_{e_i}^2 - m_{L_i}^2 + m_{H_i^u}^2 - m_{H_i^d}^2 + m_{\overline{D}_i}^2
- m_{D_i}^2 \right ) -m_{H'}^2 + m_{\overline{H'}}^2 , \\
\Sigma_1' &= \sum_{i=1}^3 \left ( 6 m_{Q_i}^2 + 3 m_{u_i}^2 + 6 m_{d_i}^2
+ m_{e_i}^2 + 4 m_{L_i}^2 - 4 m_{H_i^u}^2 - 6 m_{H_i^d}^2 + 5 m_{S_i}^2
- 9 m_{\overline{D}_i}^2 - 6 m_{D_i}^2 \right ) \\
& \quad {} + 4 m_{H'}^2 - 4 m_{\overline{H'}}^2 .
\end{align*}
In the unconstrained case, this contribution acts to drive up the values of
$m_{Q_3}^2$ and $m_{H_u}^2$, and thus the associated tuning sensitivities,
at the cutoff scale $M_X$.  In the case of E$_6$SSM BM2, on the other hand,
the reduced splitting between the soft masses leads to a much smaller
contribution from these terms.  Together with the reduction in $M_3$ described
above, this allows to maintain the observed low fine-tuning at very large
values of $M_X$.

MSSM benchmark BM2 (dash-dotted in top panel, individual sensitivities in
bottom left panel) is designed to be similar to E$_6$SSM BM2, for a reasonable
comparison.  However, from the individual sensitivities one can see that the
behavior is quite similar to MSSM BM1, though in this case
$\Delta_{m_{H_u}^2}$ becomes the largest tuning at a higher $M_X$ and does not
reach such large values, since more of the tuning is from the mixing in this
case.

\section{\label{sec:conclusion}Conclusions}

Prior to stringent experimental constraints on the mass of the
lightest Higgs boson and squarks in supersymmetric models, a simple
picture of a natural SUSY model emerged from theoretical reasoning,
with soft masses set to similar values at the GUT scale through local
gravity interactions with the hidden sector.  Through the use of
renormalization group running, one can then see that at the EW scale
the stops enter the EWSB condition for $M_Z$; therefore, it was
expected that these masses should not be much bigger than $100$ GeV.
However, to disturb this elegant picture, first LEP placed constraints
on the Higgs mass, requiring it to be above 114.4 GeV
\cite{Barate:2003sz, Schael:2006cr}, which already introduced
significant tuning for constrained models since heavy stops are
required to raise the lightest Higgs mass above its tree-level
bound. Then, recently, this problem got much worse since the LHC measured the
Higgs mass to be around $125$ GeV.

$U(1)$ extensions motivated by the $\mu$ problem, $E_6$ GUT theories
and the connection to string theory contain both $F$- and $D$- term
contributions to the light Higgs mass which can raise the tree-level
mass, evading the need for large radiative corrections to increase it.
However, such models come with their own fine-tuning problem, where the
$Z^\prime$ mass appears in the EWSB condition for $M_Z$ at tree level.
While in a previous study of the constrained E$_6$SSM it was found
that the tuning is less severe than the MSSM, it was still significant.

In light of such difficulties it is worth considering whether the
simple picture which emerged is wrong in some way and if there are
other possibilities that allow naturalness.  Or to phrase this in a
more challenging manner are there ways to constrain the naturalness of
these models that do not rely upon assumptions about how SUSY is
broken?

We have investigated this question here in the context of the MSSM and
$U(1)$ extensions.  Since the RG evolution links the soft masses
together and causes these problems from stop and gluino masses the
most conservative approach to placing naturalness limits is to choose
a low cutoff.  We find that in the MSSM the most direct way to
constrain naturalness in the model without making assumptions about
the SUSY breaking scale is through limits on the chargino
masses.  Current LHC limits on charginos are not model independent and
thereby leave many gaps where one can have light charginos.

In contrast we find that in $U(1)$ extensions of the MSSM there is an
additional way to constrain the naturalness of the models, which is
through the $Z^\prime$ mass limit.  We find when we impose a low
cutoff of $20$ TeV for setting the soft masses, the lowest tuning in
the E$_6$SSM compatible with a $Z^\prime$ mass of $2.5$ TeV was
$\Delta \approx 121$, while if the LHC run II can place a limit of
$4.5$ TeV on $M_Z^\prime$ then the tuning would be approximately $394$.
By comparison the current situation in the MSSM only requires a tuning
of around $38$.  This should be interpreted as saying that in
the most conservative limits one can place on naturalness in these
models, the tuning in the E$_6$SSM is worse.  However, if there are no
charginos below $700$ GeV then the situation in the two models would
be the same.

This should also be contrasted with what happens as we raise the high
scale boundary, $M_X$.  We showed that for our benchmark points, which
one is more tuned depends very strongly on $M_X$.  The E$_6$SSM tuning
is sufficiently complicated by the interplay of these different
sources of tension in the EWSB conditions that a small reduction in
fine-tuning can even occur for a moderate increase in $M_X$.  However,
as $M_X$ increases towards the scale where the gauge couplings unify,
the familiar tunings do dominate, though with tunings from the gluino
mass appearing to be more significant relative to those from soft
scalar masses.

We also looked at the tuning in different $U(1)$ extensions for fixed
$\tan\beta = 10$.  We found that in every case except for the
$U(1)_I$ the fine-tuning was much worse for the larger $Z^\prime$
mass, further emphasizing the importance of this in $U(1)$ extensions.
The $U(1)_I$ model showed the least tuning due to
the vanishing charge of the $H_u$ state.  This model is quite
interesting in the sense that it provides a solution to the
$\mu$ problem while avoiding the large tuning (with current limits)
from the $Z^\prime$ mass.  However, one should remember we are looking
at conservative limits on naturalness here and there is no solution to
the usual tuning coming from the large stops needed to get a $125$ GeV
Higgs in this model, which will be a problem as the UV scale is raised.

\begin{acknowledgments}
We would like to thank R.~Nevzorov for many useful
discussions and comments and F.~Staub for very helpful responses
regarding the gauge kinetic mixing contributions in SARAH.  DH would
also like to thank A.~Ismail and T.~Rizzo for their helpful comments
regarding fine-tuning in the MSSM.  PA would like to thank J.H.~Park,
A.~Voigt and D.~St\"{o}ckinger for useful discussion regarding the use of
FlexibleSUSY in this project.  This work was supported by the University
of Adelaide and by the ARC Centre of Excellence for Particle Physics at
the Terascale.
\end{acknowledgments}

\appendix

\section{\label{app:masterformula}FINE-TUNING MASTER
FORMULA}

To write down the tree-level master formula, it is convenient to define the
quantities
\begin{equation}
z_i =\epsilon_{ijk} \frac{\partial f_j}{\partial s}
\frac{\partial f_k}{\partial \tan\beta}
\label{eq:masterformulacoefficients}
\end{equation}
with $f_1 , f_2 , f_3$ as given in Eq.~(\ref{eq:E6EWSBConditions}). The
relevant partial derivatives are
\ba
\frac{\partial f_1}{\partial \tan\beta} &=& -\frac{2 M_Z}{\bar{g}}
\, \cos^2\beta \left \{ \frac{\lambda A_\lambda s}{\sqrt{2}} \, \cos\beta
+ \sin\beta \left [ m_{H_d}^2 + \frac{s^2}{2} \left ( \lambda^2 +
g_1'^2 Q_1 Q_S \right ) \right . \right . \nonumber \\
& & \left . \left . + M_Z^2 \left ( \frac{5}{2}-\frac{4 \lambda^2}{\bar{g}^2}
- \frac{4 g_1'^2}{\bar{g}^2} Q_1 Q_2 + \frac{6 g_1'^2}{\bar{g}^2} Q_1^2
\right ) \right ] + 3 M_Z^2 \, \sin^3\beta \left [
\frac{2 \lambda^2}{\bar{g}^2} - 1 \right . \right . \nonumber \\
& & \left . \left . + \frac{2 g_1'^2}{\bar{g}^2} \left ( Q_1 Q_2
- Q_1^2 \right ) \right ] \right \} , \nonumber \\
\frac{\partial f_1}{\partial s} &=& \frac{2 M_Z}{\bar{g}} \left [ s \left (
\lambda^2 + g_1'^2 Q_1 Q_S \right ) \, \cos\beta -
\frac{\lambda A_\lambda}{\sqrt{2}} \, \sin\beta \right ] , \nonumber \\
\frac{\partial f_2}{\partial \tan\beta} &=& \frac{2 M_Z}{\bar{g}} \,
\cos^2\beta \left \{ \frac{\lambda A_\lambda s}{\sqrt{2}} \, \sin\beta
+ \cos\beta \left [ m_{H_u}^2 + \frac{s^2}{2} \left ( \lambda^2 +
g_1'^2 Q_2 Q_S \right ) \right . \right . \nonumber \\
& & \left . + M_Z^2 \left ( \frac{5}{2}-\frac{4 \lambda^2}{\bar{g}^2}
- \frac{4 g_1'^2}{\bar{g}^2} Q_1 Q_2 + \frac{6 g_1'^2}{\bar{g}^2} Q_2^2
\right ) \right ] + 3 M_Z^2 \, \cos^3\beta \left [
\frac{2 \lambda^2}{\bar{g}^2} - 1 \right . \nonumber \\
& & \left . \left . + \frac{2 g_1'^2}{\bar{g}^2} \left ( Q_1 Q_2 - Q_2^2
\right ) \right ] \right \} , \nonumber \\
\frac{\partial f_2}{\partial s} &=& \frac{2 M_Z}{\bar{g}} \left [
s \left ( \lambda^2 + g_1'^2 Q_2 Q_S \right ) \, \sin\beta -
\frac{\lambda A_\lambda}{\sqrt{2}} \, \cos\beta \right ] , \nonumber \\
\frac{\partial f_3}{\partial \tan\beta} &=& \frac{2 M_Z^2}{\bar{g}^2} \,
\cos^2\beta \left [ g_1'^2 Q_S s \left ( Q_2 - Q_1 \right ) \, \sin 2\beta
- \sqrt{2} \lambda A_\lambda \, \cos 2\beta \right ] , \nonumber \\
\frac{\partial f_3}{\partial s} &=& m_S^2 +
\frac{2 \lambda^2 M_Z^2}{\bar{g}^2} + \frac{g_1'^2}{2} Q_S \left [
\frac{4 M_Z^2}{\bar{g}^2} \left ( Q_1 \, \cos^2\beta + Q_2 \, \sin^2\beta
\right ) + 3 Q_S s^2 \right ] . \nonumber
\ea
For a running parameter $q$ appearing in the tree-level EWSB conditions, the
corresponding contribution to the individual sensitivity can then be written
\be
\tilde{\Delta}_q = z_1 \frac{\partial f_1}{\partial q} +
z_2 \frac{\partial f_2}{\partial q} + z_3 \frac{\partial f_3}{\partial q} .
\label{eq:tuningContribution}
\ee
It is straightforward to compute the appropriate derivatives directly from the
EWSB conditions, Eq.~(\ref{eq:E6EWSBConditions}).  Similarly, the quantity
$C$ appearing in Eq.~(\ref{eq:E6masterformula}) is given by
\be
C = \frac{1}{2} \left ( z_1 \frac{\partial f_1}{\partial M_Z} +
z_2 \frac{\partial f_2}{\partial M_Z} + z_3 \frac{\partial f_3}{\partial M_Z}
\right ) ,
\label{eq:tuningDenominator}
\ee
with
\ba
\frac{\partial f_1}{\partial M_Z} &=& \frac{2}{\bar{g}} \, \cos\beta
\left ( m_{H_d}^2 + \frac{\lambda^2 s^2}{2} + \frac{g_1'^2}{2} Q_1 Q_S s^2
+ \frac{6 g_1'^2}{\bar{g}^2} Q_1^2 M_Z^2 \right ) - \sqrt{2}
\frac{\lambda A_\lambda s}{\bar{g}} \, \sin\beta \nonumber \\
& & + \frac{3 M_Z^2}{\bar{g}} \, \cos\beta \, \cos 2\beta +
\frac{6}{\bar{g}^3} M_Z^2 \, \sin\beta \, \sin 2\beta \left [ \lambda^2 +
g_1'^2 \left ( Q_1 Q_2 - Q_1^2 \right ) \right ] , \nonumber \\
\frac{\partial f_2}{\partial M_Z} &=& \frac{2}{\bar{g}} \, \sin\beta
\left ( m_{H_u}^2 + \frac{\lambda^2 s^2}{2} + \frac{g_1'^2}{2} Q_2 Q_S s^2
+ \frac{6 g_1'^2}{\bar{g}^2} Q_2^2 M_Z^2 \right ) - \sqrt{2}
\frac{\lambda A_\lambda s}{\bar{g}} \, \cos\beta \nonumber \\
& & - \frac{3 M_Z^2}{\bar{g}} \, \sin\beta \, \cos 2\beta +
\frac{6}{\bar{g}^3} M_Z^2 \, \cos\beta \, \sin 2\beta \left [ \lambda^2 +
g_1'^2 \left ( Q_1 Q_2 - \tilde{Q_2}^2 \right ) \right ] , \nonumber \\
\frac{\partial f_3}{\partial M_Z} &=& \frac{4 M_Z}{\bar{g}^2}\left [
\lambda^2 s - \frac{\lambda A_\lambda}{\sqrt{2}} \, \sin 2\beta + g_1'^2 Q_S s
\left ( Q_1 \, \cos^2\beta + Q_2 \, \sin^2\beta \right ) \right ] . \nonumber
\ea

\section{\label{app:rges}RGE CONTRIBUTIONS}

Provided that one does not run over too large a range of scales, the
solutions to the RG equations for a model can be reasonably well
approximated by a Taylor series, Eq.~(\ref{eq:rgeapproxsoln}).  For a
parameter $p$, this reads
\begin{equation*}
q(Q) = q(M_X) + \frac{t}{16\pi^2} \left ( \beta_q^{(1)} +
\frac{\beta_q^{(2)}}{16\pi^2} \right ) + \frac{t^2}{(16\pi^2)^2}
b_q^{(2)}(M_X) ,
\end{equation*}
where we have for convenience defined
\begin{equation*}
b_q^{(2)}(M_X) = \frac{1}{2!} \left . \sum_{q_k} \beta_{q_k}^{(1)}
\frac{\partial \beta_q}{\partial q_k} \right |_{M_X} .
\end{equation*}
We have constructed the necessary series solutions in both the MSSM
and the $U(1)$ extended models.  Due to the smallness of the first- and
second-generation Yukawa couplings, we neglect them in our
calculations.  The corresponding soft SUSY breaking trilinears are
likewise omitted.  Additionally, all soft mass matrices are assumed
diagonal, and the gaugino masses are taken to be real.

In the MSSM, the relevant parameters for the fine-tuning calculation are
$\mu$, $B$, $m_{H_u}^2$, $m_{H_d}^2$ at tree level.  For the
renormalization group running of the relevant parameters
SOFTSUSY uses the one- and two-loop RGEs from
\cite{Barger:1993gh, Martin:1993zk}. The corresponding $O(t^2)$
contributions are
\begin{subequations} \label{eq:MSSMTreeLevelCoeffs}
\begin{align}
b_\mu^{(2)} &= \frac{\mu}{2} \Bigg [ 45 y_t^4 + 45 y_b^4 + 9 y_\tau^4
+ 30 y_t^2 y_b^2 + 6 y_t^2 y_\tau^2 + 18 y_b^2 y_\tau^2 - 32 g_3^2 ( y_t^2
+ y_b^2 ) \nonumber \\
& \quad {} - 12 g_2^2 ( 3 y_t^2 + 3 y_b^2 + y_\tau^2 ) - \frac{4}{5} g_1^2
( 11 y_t^2 + 8 y_b^2 + 6 y_\tau^2 ) + 3 g_2^4 - \frac{189}{25} g_1^4 +
\frac{18}{5} g_1^2 g_2^2 \Bigg ] , \label{eq:MSSMmub2} \\
b_B^{(2)} &= 72 y_t^4 A_t + 72 y_b^4 A_b + 16 y_\tau^4 A_\tau + 12 y_t^2 y_b^2
( A_t + A_b ) + 12 y_\tau^2 y_b^2 ( A_b + A_\tau ) \nonumber \\
& \quad {} - 32 g_3^2 y_t^2 ( A_t - M_3 ) - 32 g_3^2 y_b^2 ( A_b - M_3 )
\nonumber \\
& \quad {} - 18 g_2^2 y_t^2 ( A_t - M_2 ) - 18 g_2^2 y_b^2 ( A_b - M_2 )
- 6 g_2^2 y_\tau^2 ( A_\tau - M_2 ) \nonumber \\
& \quad {} - \frac{26}{5} g_1^2 y_t^2 ( A_t - M_1 ) - \frac{14}{5} g_1^2 y_b^2
( A_b - M_1 ) - \frac{18}{5} g_1^2 y_\tau^2 ( A_\tau - M_1 ) \nonumber \\
& \quad {} + 12 g_2^4 M_2 + \frac{396}{25} g_1^4 M_1 , \label{eq:MSSMBb2} \\
b_{m_{H_d}^2}^{(2)} &= 72 y_b^4 \left ( m_{H_d}^2 + m_{Q_3}^2 + m_{d_3}^2 +
2 A_b^2 \right ) \nonumber \\
& \quad {} + 6 y_t^2 y_b^2 \left ( m_{H_u}^2 + m_{H_d}^2 + 2 m_{Q_3}^2 +
m_{u_3}^2 + m_{d_3}^2 + ( A_t + A_b )^2 \right ) \nonumber \\
& \quad {} + 12 y_\tau^2 y_b^2 \left ( 2 m_{H_d}^2 + m_{Q_3}^2 + m_{d_3}^2 +
m_{L_3}^2 + m_{e_3}^2 + ( A_\tau + A_b )^2 \right ) \nonumber \\
& \quad {} + 16 y_\tau^4 \left ( m_{H_d}^2 + m_{L_3}^2 + m_{e_3}^2 +
2 A_\tau^2 \right ) \nonumber \\
& \quad {} - 32 g_3^2 y_b^2 \left ( m_{H_d}^2 + m_{Q_3}^2 + m_{d_3}^2 + A_b^2
- 2 M_3 A_b + 2 M_3^2 \right ) \nonumber \\
& \quad {} - 18 g_2^2 y_b^2 \left ( m_{H_d}^2 + m_{Q_3}^2 + m_{d_3}^2 + A_b^2
- 2 M_2 A_b + 2 M_2^2 \right ) \nonumber \\
& \quad {} - 6 g_2^2 y_\tau^2 \left ( m_{H_d}^2 + m_{L_3}^2 + m_{e_3}^2 +
A_\tau^2 - 2 M_2 A_\tau + 2 M_2^2 \right ) \nonumber \\
& \quad {} - \frac{14}{5} g_1^2 y_b^2 \left ( m_{H_d}^2 + m_{Q_3}^2 +
m_{d_3}^2 + A_b^2 - 2 M_1 A_b + 2 M_1^2 \right ) \nonumber \\
& \quad {} - \frac{18}{5} g_1^2 y_\tau^2 \left ( m_{H_d}^2 + m_{L_3}^2 +
m_{e_3}^2 + A_\tau^2 - 2 M_1 A_\tau + 2 M_1^2 \right ) \nonumber \\
& \quad {} - 18 g_2^4 M_2^2 - \frac{198}{25} g_1^4 \left ( \mathcal{S} +
3 M_1^2 \right ) , \label{eq:MSSMmHd2b2} \\
b_{m_{H_u}^2}^{(2)} &= 72 y_t^4 \left ( m_{H_u}^2 + m_{Q_3}^2 + m_{u_3}^2 +
2 A_t^2 \right ) \nonumber \\
& \quad {} + 6 y_t^2 y_b^2 \left ( m_{H_u}^2 + m_{H_d}^2 + 2 m_{Q_3}^2 +
m_{u_3}^2 + m_{d_3}^2 + ( A_t + A_b )^2 \right ) \nonumber \\
& \quad {} - 32 g_3^2 y_t^2 \left ( m_{H_u}^2 + m_{Q_3}^2 + m_{u_3}^2 + A_t^2
- 2 A_t M_3 + 2 M_3^2 \right ) \nonumber \\
& \quad{} - 18 g_2^2 y_t^2 \left ( m_{H_u}^2 + m_{Q_3}^2 + m_{u_3}^2 + A_t^2
- 2 A_t M_2 + 2 M_2^2 \right ) \nonumber \\
& \quad {} - \frac{26}{5} g_1^2 y_t^2 \left ( m_{H_u}^2 + m_{Q_3}^2 +
m_{u_3}^2 + A_t^2 - 2 A_t M_1 + 2 M_1^2 \right ) \nonumber \\
& \quad {} + \frac{198}{25} g_1^4 \left ( \mathcal{S} - 3 M_1^2 \right )
- 18 g_2^4 M_2^2 . \label{eq:MSSMmHu2b2}
\end{align}
\end{subequations}
In these expressions the quantity $\mathcal{S}$ is defined by
\begin{equation} \label{eq:MSSMgaugeBetaContribution}
\mathcal{S} = m_{H_u}^2 - m_{H_d}^2 + \sum_{i=1}^3 \left ( m_{Q_i}^2 -
m_{L_i}^2 - 2 m_{u_i}^2 + m_{d_i}^2 + m_{e_i}^2 \right ) .
\end{equation}
If, in addition, the one-loop contributions to the effective potential
from top and stop loops are included, it is also necessary to
construct the expansions for $m_{Q_3}^2$, $m_{u_3}^2$ and $A_t$. The
coefficients read
\begin{subequations} \label{eq:MSSMsEffPotCoeffs}
\begin{align}
b_{m_{Q_3}^2}^{(2)} &= 24 y_t^4 \left ( m_{H_u}^2 + m_{Q_3}^2 + m_{u_3}^2 +
2 A_t^2 \right ) + 24 y_b^4 \left ( m_{H_d}^2 + m_{Q_3}^2 + m_{d_3}^2 +
2 A_b^2 \right ) \nonumber \\
& \quad {} + 4 y_t^2 y_b^2 \left ( m_{H_u}^2 + m_{H_d}^2 + 2 m_{Q_3}^2 +
m_{u_3}^2 + m_{d_3}^2 + ( A_t + A_b )^2 \right ) \nonumber \\
& \quad {} + 2 y_b^2 y_\tau^2 \left ( 2 m_{H_d}^2 + m_{Q_3}^2 + m_{L_3}^2 +
m_{d_3}^2 + m_{e_3}^2 + ( A_b + A_\tau )^2 \right ) \nonumber \\
& \quad {} - \frac{32}{3} g_3^2 y_t^2 \left ( m_{H_u}^2 + m_{Q_3}^2 +
m_{u_3}^2 + A_t^2 - 2 M_3 A_t + 2 M_3^2 \right ) \nonumber \\
& \quad {} - \frac{32}{3} g_3^2 y_b^2 \left ( m_{H_d}^2 + m_{Q_3}^2 +
m_{d_3}^2 + A_b^2 - 2 M_3 A_b + 2 M_3^2 \right ) \nonumber \\
& \quad {} - 6 g_2^2 y_t^2 \left ( m_{H_u}^2 + m_{Q_3}^2 + m_{u_3}^2 + A_t^2
- 2 M_2 A_t + 2 M_2^2 \right ) \nonumber \\
& \quad {} - 6 g_2^2 y_b^2 \left ( m_{H_d}^2 + m_{Q_3}^2 + m_{d_3}^2 + A_b^2
- 2 M_2 A_b + 2 M_2^2 \right ) \nonumber \\
& \quad {} - \frac{26}{15} g_1^2 y_t^2 \left ( m_{H_u}^2 + m_{Q_3}^2 +
m_{u_3}^2 + A_t^2 - 2 M_1 A_t + 2 M_1^2 \right ) \nonumber \\
& \quad {} - \frac{14}{15} g_1^2 y_b^2 \left ( m_{H_d}^2 + m_{Q_3}^2 +
m_{d_3}^2 + A_b^2 - 2 M_1 A_b + 2 M_1^2 \right ) \nonumber \\
& \quad {} + 96 g_3^4 M_3^2 - 18 g_2^4 M_2^2 + \frac{66}{25} g_1^4
( \mathcal{S} - M_1^2 ) , \label{eq:MSSMmQ32b2} \\
b_{m_{u_3}^2}^{(2)} &= 48 y_t^4 \left ( m_{H_u}^2 + m_{Q_3}^2 + m_{u_3}^2 +
2 A_t^2 \right ) \nonumber \\
& \quad {} + 4 y_t^2 y_b^2 \left ( m_{H_u}^2 + m_{H_d}^2 + 2 m_{Q_3}^2 +
m_{u_3}^2 + m_{d_3}^2 + ( A_t + A_b )^2 \right ) \nonumber \\
& \quad {} - \frac{64}{3} g_3^2 y_t^2 \left ( m_{H_u}^2 + m_{Q_3}^2 +
m_{u_3}^2 + A_t^2 - 2 M_3 A_t + 2 M_3^2 \right ) \nonumber \\
& \quad {} - 12 g_2^2 y_t^2 \left ( m_{H_u}^2 + m_{Q_3}^2 + m_{u_3}^2 +
A_t^2 - 2 M_2 A_t + 2 M_2^2 \right ) \nonumber \\
& \quad {} - \frac{52}{15} g_1^2 y_t^2 \left ( m_{H_u}^2 + m_{Q_3}^2 +
m_{u_3}^2 + A_t^2 - 2 M_1 A_t + 2 M_1^2 \right ) \nonumber \\
& \quad {} + 96 g_3^4 M_3^2 - \frac{264}{25} g_1^4 \left ( \mathcal{S} +
4 M_1^2 \right ) , \label{eq:MSSMmu32b2} \\
b_{A_t}^{(2)} &= 144 y_t^4 A_t + 24 y_b^4 A_b + 14 y_t^2 y_b^2 ( A_t + A_b )
+ 2 y_b^2 y_\tau^2 ( A_b + A_\tau ) \nonumber \\
& \quad {} - 64 g_3^2 y_t^2 ( A_t - M_3 ) - 36 g_2^2 y_t^2 ( A_t - M_2) -
\frac{52}{5} g_1^2 y_t^2 ( A_t - M_1 ) \nonumber \\
& \quad {} - \frac{32}{3} g_3^2 y_b^2 ( A_b - M_3 ) - 6 g_2^2 y_b^2
( A_b - M_2 ) - \frac{14}{15} g_1^2 y_b^2 ( A_b - M_1 ) \nonumber \\
& \quad {} - 64 g_3^4 M_3 + 12 g_2^4 M_2 + \frac{572}{25} g_1^4 M_1
\label{eq:MSSMAtb2}
\end{align}
\end{subequations}
We can similarly obtain the two-loop $\beta$ functions and
coefficients $b_p^{(2)}$ for a general set of $U(1)'$
charges.  Two-loop RGEs for the gauge and Yukawa couplings, gaugino masses and
soft trilinears, along with the one-loop RGEs for the soft scalar
masses, were originally obtained in Ref.~\cite{Athron:2009bs} for the
particular case of the E$_6$SSM.  FlexibleSUSY uses full one- and
two-loop RGEs from SARAH, which for the models considered here are
based on Ref.~\cite{Martin:1993zk} and the recent extension\footnote{In
the version of SARAH which we used the extra terms from this
extension were included for all terms except the trilinear and
bilinear soft masses.  We have been in contact with the SARAH author
about this and understand they will be included in future
versions.} in Ref.~\cite{Fonseca:2011vn} to include models with multiple
$U(1)$ gauge groups, in the most general case where the trace of the
matrix formed from the charges $Q^Y_i$ of the $U(1)_Y$ gauge symmetry
and $Q_i$ of the extra $U(1)^\prime$ symmetry does not vanish,
i.e.~$\sum_i Q_i^YQ_i \neq 0$.

When this trace is nonzero it will also induce gauge kinetic mixing
to be generated during RGE evolution and this is the case in the
models we consider here.\footnote{In $E_6$ inspired models with only
complete $27$-plet matter multiplets this trace would vanish.
However since we assume some incomplete multiplets so that our
models are consistent with gauge coupling unification this trace
doesn't vanish.}  However, when these models are evolved down from
the GUT scale, the radiatively generated gauge kinetic mixing gives an
off-diagonal gauge coupling, $g_{11}$, of just $\approx 0.02$ at the
electroweak scale \cite{King:2005jy} and so it does not play a large
role.  At the same time if gauge kinetic mixing is included the RGE
expressions become very large and unmanageable, so we neglected the
gauge kinetic mixing by setting the SARAH flag \texttt{NoU1Mixing} to true.

At tree level in the EWSB conditions the parameters that must be considered
are $\lambda$, $A_\lambda$, $m_{H_u}^2$, $m_{H_d}^2$, $m_S^2$ and $g_1$, $g_2$
and $g_1'$.  Neglecting kinetic mixing, the two-loop $\beta$ functions for
the relevant gauge couplings read
\begin{subequations} \label{eq:USSMGaugeBetas}
\begin{align}
\beta_{g_1}^{(1)} &= \frac{48}{5} g_1^3 , \label{eq:USSMg1BetaOneLoop} \\
\beta_{g_1}^{(2)} &= \frac{2}{25} g_1^3 \left ( 30 g_1'^2 \Pi_{Q}^Y +
117 g_1^2 + 135 g_2^2 + 300 g_3^2 - 10 \Sigma_{\kappa } - 15 \Sigma_{\lambda }
- 65 y_t^2 - 35 y_b^2 - 45 y_{\tau }^2 \right ) , \label{eq:USSMg1BetaTwoLoop}\\
\beta_{g_2}^{(1)} &= 4 g_2^3 , \label{eq:USSMg2BetaOneLoop} \\
\beta_{g_2}^{(2)} &= \frac{2}{5} g_2^3 \left ( 5 g_1'^2 \Pi _{Q}^L + 9 g_1^2
+ 115 g_2^2 + 60 g_3^2 - 5 \Sigma_{\lambda } - 15 y_t^2 - 15 y_b^2 -
5 y_{\tau }^2 \right) , \label{eq:USSMg2BetaTwoLoop} \\
\beta_{g_1'}^{(1)} &= g_1'^3 \Sigma_{Q} , \label{eq:USSMg1pBetaOneLoop} \\
\beta_{g_1'}^{(2)} &= \frac{2}{5} g_1'^3 \left ( -15 \Sigma_{\kappa } \left (
Q_{\bar{D}}^2 + Q_D^2 + Q_S^2 \right ) - 30 y_b^2 \left ( Q_d^2 + Q_1^2 +
Q_Q^2 \right ) + 120 g_3^2 \Pi_{Q}^C \right . \nonumber \\
& \quad {} - 10 y_{\tau }^2 \left ( Q_e^2 + Q_L^2 + Q_1^2 \right ) +
15 g_2^2 \Pi_{Q}^L + 10 g_1'^2 \Pi_{Q} + 6 g_1^2 \Pi_{Q}^Y \nonumber \\
& \quad \left . {} - 10 \Sigma_{\lambda } \left ( Q_S^2 + Q_1^2 + Q_2^2
\right ) - 30 y_t^2 \left ( Q_u^2 + Q_2^2 + Q_Q^2 \right ) \right ) .
\label{eq:USSMg1pBetaTwoLoop}
\end{align}
\end{subequations}
In order to keep these expressions compact, we have used the notation
\begin{align*}
\Sigma_{Q} = \sum_i Q_i^2 = \frac{321}{40} \cos^2\theta +
\frac{217}{24} \sin^2\theta + \frac{27}{8 \sqrt{15}} \sin 2\theta
\end{align*}
to denote the trace over the $U(1)'$ charges, along
with\footnote{The first of these is the trace which is assumed to vanish in
Ref.~\cite{Martin:1993zk}.  Although we use the \texttt{NoU1Mixing} flag to
neglect gauge kinetic mixing, SARAH does this by removing the RGE for
the off diagonal gauge couplings and effectively setting them to
zero at all scales by removing all terms involving them from the
RGEs.  Therefore, some terms with this trace remain and the RGEs shown here
do not reduce to those which one would obtain from
Ref.~\cite{Martin:1993zk} or Ref.~\cite{Athron:2009bs} unless
$\Sigma_{Q}^Y = 0$.  Note, however, these contributions do not appear in the
corresponding trilinears due to the version of SARAH used.}
\begin{align*}
\Sigma_{Q}^Y &= \sum_i \sqrt{\frac{5}{3}} Q_i^Y Q_i =
-\frac{3}{\sqrt{10}} \cos\theta - \frac{1}{\sqrt{6}} \sin\theta , \\
\Pi_{Q} &= \sum_i Q_i^4 \\
&= \frac{2049}{1600} \cos^4\theta +
\frac{483}{80\sqrt{15}} \cos^3\theta \sin\theta +
\frac{681}{160} \cos^2\theta \sin^2\theta +
\frac{9}{16\sqrt{15}} \cos\theta \sin^3\theta +
\frac{1297}{576}\sin^4\theta , \\
\Pi_{Q}^Y &= \sum_i \left ( \sqrt{\frac{5}{3}} Q_i^Y \right )^2 Q_i^2 =
\frac{59}{40} \cos^2\theta + \frac{31}{24} \sin^2\theta +
\frac{3}{8\sqrt{15}} \sin 2\theta , \\
\Pi_{Q}^L &= 3 Q_1^2 + 3Q_2^2 + Q_{H'}^2 + Q_{\overline{H'}}^2 + 3 Q_L^2 +
9 Q_Q^2 = \frac{39}{20} \cos^2\theta + \frac{19}{12} \sin^2\theta +
\frac{3}{4\sqrt{15}} \sin 2\theta , \\
\Pi_{Q}^C &= Q_d^2 + Q_D^2 + Q_{\overline{D}}^2 + 2 Q_Q^2 + Q_u^2 =
\frac{1}{2} .
\end{align*}
Note that in these expressions the $U(1)_Y$ and $U(1)'$ charges are assumed to
be GUT-normalized.  The expressions in terms of the $E_6$ mixing angle
$\theta$ follow from the charge assignments given in Table \ref{tab:E6charges}
and hold provided that $U(1)$ mixing is neglected.  Similarly, we write
\begin{align*}
&\Sigma_\lambda = \lambda_1^2 + \lambda_2^2 + \lambda_3^2 ,
\qquad \Sigma_\kappa = \kappa_1^2 + \kappa_2^2 + \kappa_3^2 , \\
&\Pi_\lambda = \lambda_1^4 + \lambda_2^4 + \lambda_3^4 ,
\qquad \Pi_\kappa = \kappa_1^4 + \kappa_2^4 + \kappa_3^4 .
\end{align*}
The corresponding $O(t^2)$ coefficients for the gauge couplings are simply
\begin{subequations} \label{eq:USSMGaugeOt2Coeffs}
\begin{align}
b_{g_1}^{(2)} &= \frac{3456}{25} g_1^5 , \label{eq:USSMg1Ot2Coeff} \\
b_{g_2}^{(2)} &= 24 g_2^5 , \label{eq:USSMg2Ot2Coeff} \\
b_{g_1'}^{(2)} &= \frac{3}{2} g_1'^5 \Sigma_{Q}^2 . \label{eq:USSMg1pOt2Coeff}
\end{align}
\end{subequations}
The one- and two-loop contributions to the $\beta$ function for $\lambda$
and the $O(t^2)$ coefficient in the series expansion are
\begin{subequations} \label{eq:USSMLambdaBetas}
\begin{align}
\beta_\lambda^{(1)} &= \lambda \left [ 2 \lambda^2 + 2 \Sigma_\lambda +
3 \Sigma_\kappa + 3 y_t^2 + 3 y_b^2 + y_{\tau }^2 - 3 g_2^2 - \frac{3}{5}
g_1^2 - 2 \left ( Q_1^2 + Q_2^2 + Q_S^2 \right ) g_1'^2 \right ] ,
\label{eq:USSMLambdaBetaOneLoop} \\
\beta_\lambda^{(2)} &= \lambda \bigg \{ -2 \lambda^2 \left ( \lambda^2 +
2 \Sigma_\lambda + 3 \Sigma_\kappa \right ) - 4 \Pi_\lambda - 6 \Pi_\kappa
- 3 \lambda^2 \left ( 3 y_t^2 + 3 y_b^2 + y_\tau^2 \right ) \nonumber \\
& {} - 3 \left ( 3 y_t^4 + 3 y_b^4 + 2 y_t^2 y_b^2 + y_\tau^4 \right ) +
6 g_2^2 \Sigma_\lambda + \frac{2}{5} g_1^2 \left ( 2 y_t^2 - y_b^2 +
3 y_\tau^2 + 2 \Sigma_\kappa + 3 \Sigma_\lambda \right ) \nonumber \\
& {} \left . + g_1'^2 \left [ 4 Q_S^2 \lambda^2 - 6 \left ( Q_2^2 - Q_Q^2
- Q_u^2 \right ) y_t^2 - 6 \left ( Q_1^2 - Q_Q^2 - Q_d^2 \right ) y_b^2
\right . \right . \nonumber \\
& {} \left . - 2 \left ( Q_1^2 - Q_L^2 - Q_e^2 \right ) y_\tau^2 -
6 \left ( Q_S^2 - Q_D^2 - Q_{\overline{D}}^2 \right ) \Sigma_\kappa -
4 \left ( Q_S^2 - Q_1^2 - Q_2^2 \right ) \Sigma_\lambda \right ] \nonumber \\
& {} + 16 g_3^2 \left ( y_t^2 + y_b^2 + \Sigma_\kappa \right ) +
\frac{33}{2} g_2^4 + \frac{297}{50} g_1^4 + 2 g_1'^4 \left [ 2 Q_1^4 +
2 Q_2^4 + 2 Q_S^4 + \left ( Q_1^2 + Q_2^2 + Q_S^2 \right ) \Sigma_{Q}
\right ] \nonumber \\
& {} + \frac{9}{5} g_1^2 g_2^2 + 6 g_1'^2 g_2^2 \left ( Q_1^2 + Q_2^2 \right )
+ \frac{6}{5} g_1^2 g_1'^2 \left [ Q_1^2 + Q_2^2 + \left ( Q_2 - Q_1 \right )
\Sigma_Q^Y \right ] \bigg \} , \label{eq:USSMLambdaBetaTwoLoop} \\
b_\lambda^{(2)} &= \lambda \bigg \{ 2 \lambda^2 \left ( 3 \lambda^2 +
4 \Sigma_\lambda + 6 \Sigma_\kappa \right ) + 4 \Pi_\lambda + 6\Pi_\kappa +
6 \left ( \Sigma_\lambda + \frac{3}{2} \Sigma_\kappa \right )^2 +
7 \lambda^2 \left ( 3 y_t^2 + 3 y_b^2 + y_\tau^2 \right ) \nonumber \\
& {} + \left ( 3 y_t^2 + 3 y_b^2 + y_\tau^2 \right ) \left ( 2 \Sigma_\lambda
+ 3 \Sigma_\kappa \right ) + 3 \left ( \frac{15}{2} y_t^4 + \frac{15}{2} y_b^4
+ \frac{3}{2} y_\tau^4 + 5 y_b^2 y_t^2 + 3 y_b^2 y_\tau^2 + y_t^2 y_\tau^2
\right ) \nonumber \\
& {} - \frac{1}{5} g_1^2 \left ( 12 \lambda^2 + 16 y_b^2 + 22 y_t^2 +
12 y_\tau^2 + 12 \Sigma_\lambda + 13 \Sigma_\kappa \right ) \nonumber \\
& {} - 3 g_2^2 \left ( 4 \lambda^2 + 6 y_t^2 + 6 y_b^2 + 2 y_\tau^2 +
4 \Sigma_\lambda + 3 \Sigma_\kappa \right ) - 16 g_3^2 \left ( y_t^2 + y_b^2
+ \Sigma_\kappa \right ) \nonumber \\
& {} - 2 g_1'^2 \left [ 4 \left ( Q_1^2 + Q_2^2 + Q_S^2 \right )
\left ( \lambda^2 + \Sigma_\lambda \right ) + 3 \Sigma_\kappa
\left ( Q_1^2 + Q_2^2 + Q_S^2 + Q_D^2 + Q_{\overline{D}}^2 \right )
\right . \nonumber \\
& {} + 3y_b^2 \left ( 2 Q_1^2 + Q_2^2 + Q_d^2 + Q_Q^2 + Q_S^2 \right ) +
3 y_t^2 \left ( Q_1^2 + 2 Q_2^2 + Q_Q^2 + Q_S^2 + Q_u^2 \right ) \nonumber \\
& {} \left . + y_\tau^2 \left ( 2 Q_1^2 + Q_2^2 + Q_e^2 + Q_L^2 + Q_S^2
\right ) \right ] - \frac{15}{2} g_2^4 - \frac{279}{50} g_1^4 \nonumber \\
& {} + 2 g_1'^4 \left ( Q_1^2 + Q_2^2 + Q_S^2 \right )
\left ( Q_1^2 + Q_2^2 + Q_S^2 - \Sigma_{Q} \right ) + \frac{9}{5} g_1^2 g_2^2
+ 6 g_1'^2 g_2^2 \left ( Q_1^2 + Q_2^2 + Q_S^2 \right ) \nonumber \\
& {} + \frac{6}{5} g_1^2 g_1'^2 \left ( Q_1^2 + Q_2^2 + Q_S^2 \right )
\bigg \} . \label{eq:USSMLambdaOt2Coeff}
\end{align}
\end{subequations}
It is sufficient for our purposes to consider the trilinear coupling
$a_\lambda \equiv \lambda A_\lambda$, rather than $A_\lambda$, for which the
relevant expressions read
\begin{subequations}
\begin{align}
\beta_{a_\lambda}^{(1)} &= a_\lambda \left [ 2 \lambda ^2 + 2 \Sigma_\lambda +
3 \Sigma_\kappa + 3 y_t^2 + 3 y_b^2 + y_{\tau }^2 - 3 g_2^2 -
\frac{3}{5} g_1^2 - 2 \left ( Q_1^2 + Q_2^2 + Q_S^2 \right ) g_1'^2 \right ]
\nonumber \\
& {} + \lambda \bigg [ 4 \lambda a_\lambda + 4 \Sigma_{a_\lambda} +
6 \Sigma_{a_\kappa} + 6 y_t a_t + 6 y_b a_b + 2 y_\tau a_\tau + 6 g_2^2 M_2 +
\frac{6}{5} g_1^2 M_1 \nonumber \\
& {} + 4 g_1'^2 M_1' \left ( Q_1^2 + Q_2^2 + Q_S^2 \right ) \bigg ] ,
\label{eq:USSMTLambdaBetaOneLoop} \\
\beta_{a_\lambda}^{(2)} &= a_\lambda \bigg \{ -2 \lambda^2 \left ( \lambda^2
+ 2 \Sigma_\lambda + 3 \Sigma_\kappa \right ) - 4 \Pi_\lambda - 6 \Pi_\kappa
- 3 \lambda^2 \left ( 3 y_t^2 + 3 y_b^2 + y_\tau^2 \right ) \nonumber \\
& {} - 3 \left ( 3 y_t^4 + 3 y_b^4 + 2 y_t^2 y_b^2 + y_\tau^4 \right ) +
6 g_2^2 \Sigma_\lambda + \frac{2}{5} g_1^2 \left ( 2 y_t^2 - y_b^2 +
3 y_\tau^2 + 2 \Sigma_\kappa + 3 \Sigma_\lambda \right ) \nonumber \\
& {} \left . + g_1'^2 \left [ 4 Q_S^2 \lambda^2 - 6 \left ( Q_2^2 - Q_Q^2 -
Q_u^2 \right ) y_t^2 - 6 \left ( Q_1^2 - Q_Q^2 - Q_d^2 \right ) y_b^2 \right .
\right . \nonumber \\
& {} \left . - 2 \left ( Q_1^2 - Q_L^2 - Q_e^2 \right ) y_\tau^2 -
6 \left ( Q_S^2 - Q_D^2 - Q_{\overline{D}}^2 \right ) \Sigma_\kappa -
4 \left ( Q_S^2 - Q_1^2 - Q_2^2 \right ) \Sigma_\lambda \right ] \nonumber \\
& {} + 16 g_3^2 \left ( y_t^2 + y_b^2 + \Sigma_\kappa \right ) +
\frac{33}{2} g_2^4 + \frac{297}{50} g_1^4 + 2 g_1'^4 \left [ 2 Q_1^4 +
2 Q_2^4 + 2 Q_S^4 + \left ( Q_1^2 + Q_2^2 + Q_S^2 \right ) \Sigma_{Q} \right ]
\nonumber \\
& {} + \frac{9}{5} g_1^2 g_2^2 +6 g_1'^2 g_2^2 \left ( Q_1^2 + Q_2^2 \right )
+ \frac{6}{5} g_1^2 g_1'^2 \left ( Q_1^2 + Q_2^2 \right ) \bigg \} +
\lambda \bigg \{ -4 \lambda a_\lambda \left ( \lambda^2 + 2 \Sigma_\lambda +
3 \Sigma_\kappa \right ) \nonumber \\
& {} - 4 \lambda^2 \left ( \lambda a_\lambda + 2 \Sigma_{a_\lambda} +
3 \Sigma_{a_\kappa} \right ) - 16 \Pi_{a_\lambda} - 24 \Pi_{a_\kappa} -
6 \lambda a_\lambda \left ( 3 y_t^2 + 3 y_b^2 + y_\tau^2 \right ) \nonumber \\
& {} - 6 \lambda^2 \left ( 3 y_t a_t + 3 y_b a_b + y_\tau a_\tau \right )
- 12 \left [ 3 y_t^3 a_t + 3 y_b^3 a_b + y_t y_b \left ( y_t a_b + y_b a_t
\right ) + y_\tau^3 a_\tau \right ] \nonumber \\
& {} + 32 g_3^2 \left [ y_t a_t + y_b a_b + \Sigma_{a_\kappa} - \left ( y_t^2
+ y_b^2 + \Sigma_\kappa \right ) M_3 \right ] + 12 g_2^2 \left (
\Sigma_{a_\lambda} - \Sigma_\lambda M_2 \right ) \nonumber \\
& {} + \frac{2}{5} g_1^2 \left [ 4 y_t a_t - 2 y_b a_b + 6 y_\tau a_\tau +
4 \Sigma_{a_\kappa} + 6 \Sigma_{a_\lambda} - 2 \left ( 2 y_t^2 - y_b^2 +
3 y_\tau^2 + 2 \Sigma_\kappa + 3 \Sigma_\lambda \right ) M_1 \right ]
\nonumber \\
& {} + 4 g_1'^2 \big [ 3 y_t \left ( Q_Q^2 - Q_2^2 + Q_u^2 \right )
\left ( a_t - y_t M_1' \right ) + 3 y_b \left ( Q_Q^2 - Q_1^2 + Q_d^2 \right )
\left ( a_b - y_b M_1' \right ) \nonumber \\
& {} + y_\tau \left ( Q_L^2 - Q_1^2 + Q_e^2 \right ) \left ( a_\tau -
y_\tau M_1' \right ) + 2Q_S^2 \lambda \left ( a_\lambda - \lambda M_1'
\right ) \nonumber \\
& {} + 2 \left ( Q_1^2 + Q_2^2 - Q_S^2 \right ) \left ( \Sigma_{a_\lambda} -
\Sigma_\lambda M_1' \right ) + 3 \left ( Q_D^2 + Q_{\overline{D}}^2 - Q_S^2
\right ) \left ( \Sigma_{a_\kappa} - \Sigma_\kappa M_1' \right ) \big ]
\nonumber \\
& {} - 66 g_2^4 M_2 - \frac{594}{25} g_1^4 M_1 - 8 g_1'^4 M_1' \left [ 2
\left ( Q_1^4 + Q_2^4 + Q_S^4 \right ) + \left ( Q_1^2 + Q_2^2 + Q_S^2
\right ) \Sigma_Q \right ] \nonumber \\
& {} - \frac{18}{5} g_2^2 g_1^2 \left ( M_2 + M_1 \right ) - 12 g_2^2 g_1'^2
\left ( Q_1^2 + Q_2^2 \right ) \left ( M_2 + M_1' \right ) \nonumber \\
& {} - \frac{12}{5} g_1^2 g_1'^2 \left ( Q_1^2 + Q_2^2 \right ) \left ( M_1 +
M_1' \right ) \bigg \} , \label{eq:USSMTLambdaBetaTwoLoop} \\
b_{a_\lambda}^{(2)} &= \lambda \bigg [ 4 \lambda a_\lambda +
4 \Sigma_{a_\lambda} + 6 \Sigma_{a_\kappa} + 6 y_t a_t + 6 y_b a_b +
2 y_\tau a_\tau + 6 g_2^2 M_2 + \frac{6}{5} g_1^2 M_1 \nonumber \\
& {} + 4 g_1'^2 M_1' \left ( Q_1^2 + Q_2^2 + Q_S^2 \right ) \bigg ] \times
\bigg [ 2 \lambda^2 + 2 \Sigma_\lambda + 3 \Sigma_\kappa + 3 y_t^2 +
3 y_b^2 + y_{\tau }^2 - 3 g_2^2 \nonumber \\
& {} - \frac{3}{5} g_1^2 - 2 \left ( Q_1^2 + Q_2^2 + Q_S^2 \right ) g_1'^2
\bigg ] + a_\lambda \bigg \{ 2 \lambda^2 \left ( 3 \lambda^2 +
4 \Sigma_\lambda + 6 \Sigma_\kappa \right ) + 4 \Pi_\lambda + 6 \Pi_\kappa
\nonumber \\
& {} + 6 \left ( \Sigma_\lambda + \frac{3}{2} \Sigma_\kappa \right )^2 +
7 \lambda^2 \left ( 3 y_t^2 + 3 y_b^2 + y_\tau^2 \right ) + \left ( 3 y_t^2 +
3 y_b^2 + y_\tau^2 \right ) \left ( 2 \Sigma_\lambda + 3 \Sigma_\kappa
\right ) \nonumber \\
& {} + 3 \left ( \frac{15}{2} y_t^4 + \frac{15}{2} y_b^4 +
\frac{3}{2} y_\tau^4 + 5 y_b^2 y_t^2 + 3 y_b^2 y_\tau^2 + y_t^2 y_\tau^2
\right ) \nonumber \\
& {} - \frac{1}{5} g_1^2 \left ( 12 \lambda^2 + 16 y_b^2 + 22 y_t^2 +
12 y_\tau^2 + 12 \Sigma_\lambda + 13 \Sigma_\kappa \right ) \nonumber \\
& {} - 3 g_2^2 \left ( 4 \lambda^2 + 6 y_t^2 + 6 y_b^2 + 2 y_\tau^2 +
4 \Sigma_\lambda + 3 \Sigma_\kappa \right ) - 16 g_3^2 \left ( y_t^2 + y_b^2
+ \Sigma_\kappa \right ) \nonumber \\
& {} - 2 g_1'^2 \left [ 4 \left ( Q_1^2 + Q_2^2 + Q_S^2 \right )
\left ( \lambda^2 + \Sigma_\lambda \right ) + 3 \Sigma_\kappa \left ( Q_1^2 +
Q_2^2 + Q_S^2 + Q_D^2 + Q_{\overline{D}}^2 \right ) \right . \nonumber \\
& {} + 3 y_b^2 \left ( 2 Q_1^2 + Q_2^2 + Q_d^2 + Q_Q^2 + Q_S^2 \right ) +
3 y_t^2 \left ( Q_1^2 + 2 Q_2^2 + Q_Q^2 + Q_S^2 + Q_u^2 \right ) \nonumber \\
& {} \left . + y_\tau^2 \left ( 2 Q_1^2 + Q_2^2 + Q_e^2 + Q_L^2 + Q_S^2
\right ) \right ] - \frac{15}{2} g_2^4 - \frac{279}{50} g_1^4 \nonumber \\
& {} + 2 g_1'^4 \left ( Q_1^2 + Q_2^2 + Q_S^2 \right ) \left (Q_1^2 + Q_2^2
+ Q_S^2 - \Sigma_{Q} \right ) + \frac{9}{5} g_1^2 g_2^2 + 6 g_1'^2 g_2^2
\left ( Q_1^2 + Q_2^2 + Q_S^2 \right ) \nonumber \\
& {} + \frac{6}{5} g_1^2 g_1'^2 \left ( Q_1^2 + Q_2^2 + Q_S^2 \right )
\bigg \} + \lambda \bigg [ 16 \lambda^3 a_{\lambda} + 16\Pi_{a_\lambda} +
24 \Pi_{a_\kappa} + 8 \lambda \sum_{i=1}^3 \lambda_i \left ( \lambda_i
a_{\lambda} + a_{\lambda_i} \lambda \right ) \nonumber \\
& {} + 12 \lambda \sum_{i=1}^3 \kappa_i \left ( \kappa_i a_\lambda +
a_{\kappa_i} \lambda \right ) + 8 \sum_{i=1}^3 \sum_{j=1}^3 \lambda_i
\lambda_j \left ( a_{\lambda_i} \lambda_j + \lambda_i a_{\lambda_j} \right )
\nonumber \\
& {} + 18 \sum_{i=1}^3 \sum_{j=1}^3 \kappa_i \kappa_j \left ( \kappa_i
a_{\kappa_j} + a_{\kappa_i} \kappa_j \right ) + 24 \sum_{i=1}^3 \sum_{j=1}^3
\lambda_i \kappa_j \left ( \lambda_i a_{\kappa_j} + a_{\lambda_i} \kappa_j
\right ) + 72 y_t^3 a_t + 72 y_b^3 a_b \nonumber \\
& {} + 16 y_\tau^3 a_\tau + 12 y_t y_b \left ( a_t y_b + y_t a_b \right ) +
12 y_b y_\tau \left ( a_b y_\tau + y_b a_\tau \right ) + 30 \lambda y_t
\left ( a_{\lambda} y_t + \lambda a_t \right ) \nonumber \\
& {} + 30 \lambda y_b \left ( a_{\lambda} y_b + \lambda a_b \right ) +
10 \lambda y_\tau \left ( a_{\lambda} y_\tau + \lambda a_{\tau} \right ) -
32 g_3^2 y_t \left ( a_t - y_t M_3 \right ) - 32 g_3^2 y_b \left ( a_b -
y_b M_3 \right ) \nonumber \\
& {} - 32 g_3^2 \left ( \Sigma_{a_\kappa} - M_3 \Sigma_\kappa \right ) -
12 g_2^2 \lambda \left ( a_\lambda - \lambda M_2 \right ) - 18 g_2^2 y_t
\left ( a_t - y_t M_2 \right ) - 18 g_2^2 y_b \left ( a_b - y_b M_2 \right )
\nonumber \\
& {} - 6 g_2^2 y_\tau \left ( a_\tau - y_\tau M_2 \right ) - 12 g_2^2
\left ( \Sigma_{a_\lambda} - M_2 \Sigma_\lambda \right ) -
\frac{12}{5} g_1^2 \lambda \left ( a_{\lambda} - \lambda M_1 \right )
- \frac{26}{5} g_1^2 y_t \left ( a_t - y_t M_1 \right ) \nonumber \\
& {} - \frac{14}{5} g_1^2 y_b \left ( a_b - y_b M_1 \right ) -
\frac{18}{5} g_1^2 y_\tau \left ( a_\tau - y_\tau M_1 \right ) -
\frac{12}{5} g_1^2 \left ( \Sigma_{a_\lambda} - M_1 \Sigma_\lambda \right )
- \frac{8}{5}g_1^2 \left ( \Sigma_{a_\kappa} - M_1 \Sigma_\kappa \right )
\nonumber \\
& {} - 8 g_1'^2 \lambda \left ( Q_1^2 + Q_2^2 + Q_S^2 \right )
\left ( a_\lambda - \lambda M_1' \right ) - 12 g_1'^2 y_t \left ( Q_2^2 +
Q_Q^2 + Q_u^2 \right ) \left ( a_t - y_t M_1' \right ) \nonumber \\
& {} - 12 g_1'^2 y_b \left ( Q_1^2 + Q_Q^2 + Q_d^2 \right )
\left ( a_b - y_b M_1' \right ) - 4 g_1'^2 y_\tau \left ( Q_1^2 + Q_L^2 +
Q_e^2 \right ) \left ( a_\tau - y_\tau M_1' \right ) \nonumber \\
& {} - 8 g_1'^2 \left ( Q_1^2 + Q_2^2 + Q_S^2 \right )
\left ( \Sigma_{a_\lambda} - M_1' \Sigma_\lambda \right ) - 12 g_1'^2
\left ( Q_S^2 + Q_D^2 + Q_{\overline{D}}^2 \right ) \left (\Sigma_{a_\kappa}
- M_1' \Sigma_{\kappa} \right ) \nonumber \\
& {} + 48 g_2^4 M_2 + \frac{576}{25} g_1^4 M_1 + 8 g_1'^4 M_1' \Sigma_Q
\left ( Q_1^2 + Q_2^2 + Q_S^2 \right ) \bigg ] ,
\label{eq:USSMTLambdaOt2Coeff}
\end{align}
\end{subequations}
where
\begin{align*}
&\Sigma_{a_\lambda} = \lambda_1 a_{\lambda_1} + \lambda_2 a_{\lambda_2} +
\lambda_3 a_{\lambda_3} , \qquad \Sigma_{a_\kappa} = \kappa_1 a_{\kappa_1}
+ \kappa_2 a_{\kappa_2} + \kappa_3 a_{\kappa_3} , \\
&\Pi_{a_\lambda} = \lambda_1^3 a_{\lambda_1} + \lambda_2^3 a_{\lambda_2} +
\lambda_3^3 a_{\lambda_3} , \qquad \Pi_{a_\kappa} = \kappa_1^3 a_{\kappa_1}
+ \kappa_2^3 a_{\kappa_2} + \kappa_3^3 a_{\kappa_3} ,
\end{align*}
Note that $a_{\lambda_i} \equiv \lambda_i A_{\lambda_i}$, $a_{\kappa_i}
\equiv \kappa_i A_{\kappa_i}$, $a_t \equiv y_t A_t$, $a_b \equiv y_b A_b$ and
$a_\tau \equiv y_\tau A_\tau$.  Defining
\begin{align}
\Sigma_1 &= \sum_{i=1}^3 \left ( m_{Q_i}^2 - 2 m_{u_i}^2 + m_{d_i}^2 +
m_{e_i}^2 - m_{L_i}^2 + m_{H_i^u}^2 - m_{H_i^d}^2 + m_{\overline{D}_i}^2 -
m_{D_i}^2 \right ) - m_{H'}^2 + m_{\overline{H'}}^2 ,
\label{eq:USSMGaugeSigmaTerm1} \\
\Sigma_1' &= \sum_{i=1}^3 \left ( 6 Q_Q m_{Q_i}^2 + 3 Q_u m_{u_i}^2 +
3 Q_d m_{d_i}^2 + Q_e m_{e_i}^2 + 2 Q_L m_{L_i}^2 + 2 Q_2 m_{H_i^u}^2 +
2 Q_1 m_{H_i^d}^2 \right . \nonumber \\
& \quad {} \left . + Q_S m_{S_i}^2 + 3 Q_{\overline{D}} m_{\overline{D}_i}^2 +
3 Q_D m_{D_i}^2 \right ) + 2 Q_{H'} m_{H'}^2 +
2 Q_{\overline{H'}} m_{\overline{H'}}^2 , \label{eq:USSMGaugeSigmaTerm1Pr}
\end{align}
the one- and two-loop $\beta$ functions and the $O(t^2)$ coefficients for
$m_{H_d}^2$ are
\begin{subequations}
\begin{align}
\beta_{m_{H_d}^2}^{(1)} &= 2 \lambda^2 \left ( m_{H_d}^2 + m_{H_u}^2 +
m_S^2 \right ) + 2 a_\lambda^2 + 6 y_b^2 \left ( m_{H_d}^2 + m_{Q_3}^2 +
m_{d_3}^2 \right ) + 6 a_b^2 \nonumber \\
& {} + 2 y_\tau^2 \left ( m_{H_d}^2 + m_{L_3}^2 + m_{e_3}^2 \right ) +
2 a_\tau^2 - 6 g_2^2 M_2^2 - \frac{6}{5} g_1^2 M_1^2 - 8 Q_1^2 g_1'^2 M_1'^2
\nonumber \\
& {} - \frac{3}{5} g_1^2 \Sigma_1 + 2 Q_1 g_1'^2 \Sigma_1' ,
\label{eq:USSMmHd2BetaOneLoop} \\
\beta_{m_{H_d}^2}^{(2)} &= -36 y_b^4 \left ( m_{H_d}^2 + m_{Q_3}^2 +
m_{d_3}^2 \right ) - 12 y_\tau^4 \left ( m_{H_d}^2 + m_{L_3}^2 + m_{e_3}^2
\right ) - 72 y_b^2 a_b^2 - 24 y_\tau^2 a_\tau^2 \nonumber \\
& {} - 6 y_t^2 y_b^2 \left ( m_{H_d}^2 + m_{H_u}^2 + 2 m_{Q_3}^2 + m_{u_3}^2
+ m_{d_3}^2 \right ) - 6 \left ( y_t a_b + y_b a_t \right )^2 \nonumber \\
& {} - 4 \lambda^4 \left ( m_{H_d}^2 + m_{H_u}^2 + m_S^2 \right ) -
8 \lambda^2 a_\lambda^2 - \lambda^2 \sum_{i=1}^3 \left [ 4 \lambda_i^2
\left ( m_{H_d}^2 + m_{H_u}^2 + 2 m_S^2 + m_{H_i^d}^2 \right . \right .
\nonumber \\
& {} \left . \left . + m_{H_i^u}^2 \right ) + 6 \kappa_i^2 \left ( m_{H_d}^2
+ m_{H_u}^2 + 2 m_S^2 + m_{D_i}^2 + m_{\overline{D}_i}^2 \right ) \right ]
- \sum_{i=1}^3 \left [ 4 \left ( \lambda_i a_{\lambda} + \lambda a_{\lambda_i}
\right )^2 \right . \nonumber \\
& {} \left . + 6 \left ( \kappa_i a_\lambda + \lambda a_{\kappa_i} \right )^2
\right ] - 6 \lambda^2 y_t^2 \left ( m_{H_d}^2 + 2 m_{H_u}^2 + m_S^2 +
m_{Q_3}^2 + m_{u_3}^2 \right ) - 6 \left ( \lambda a_t + y_t a_\lambda
\right )^2 \nonumber \\
& {} + 32 g_3^2 y_b^2 \left ( m_{H_d}^2 + m_{Q_3}^2 + m_{d_3}^2 + 2 M_3^2
\right ) + 32 g_3^2 \left ( a_b^2 - 2 y_b a_b M_3 \right ) \nonumber \\
& {} + \frac{6}{5} g_1^2 y_t^2 \left ( 3 m_{H_u}^2 + m_{Q_3}^2 -
4 m_{u_3}^2 \right ) - \frac{2}{5} g_1^2 y_b^2 \left ( 11 m_{H_d}^2 -
m_{Q_3}^2 - 4 m_{d_3}^2 + 4 M_1^2 \right ) \nonumber \\
& {} - \frac{4}{5} g_1^2 \left ( a_b^2 - 2 y_b a_b M_1 \right ) +
\frac{6}{5} g_1^2 y_\tau^2 \left ( m_{H_d}^2 + m_{L_3}^2 + 4 m_{e_3}^2 +
4 M_1^2 \right ) + \frac{12}{5} g_1^2 \left ( a_\tau^2 - 2 y_\tau a_\tau M_1
\right ) \nonumber \\
& {} + \frac{6}{5} g_1^2 \sum_{i=1}^3 \left [ \lambda_i^2 \left ( m_{H_i^d}^2
- m_{H_i^u}^2 \right ) + \kappa_i^2 \left ( m_{\overline{D}_i}^2 - m_{D_i}^2
\right ) \right ] + 12 g_1'^2 y_b^2 \left ( Q_Q^2 + Q_d^2 - Q_1^2 \right )
\nonumber \\
& {} \times \left ( m_{H_d}^2 + m_{Q_3}^2 + m_{d_3}^2 + 2 M_1'^2 \right ) +
12 g_1'^2 \left ( Q_Q^2 + Q_d^2 - Q_1^2 \right ) \left ( a_b^2 -
2 y_b a_b M_1' \right ) \nonumber \\
& {} - 24 Q_1 g_1'^2 y_b^2 \left ( Q_1 m_{H_d}^2 + Q_Q m_{Q_3}^2 +
Q_d m_{d_3}^2 \right ) + 4 g_1'^2 y_\tau^2 \left ( Q_L^2 + Q_e^2 - Q_1^2
\right ) \nonumber \\
& {} \times \left ( m_{H_d}^2 + m_{L_3}^2 + m_{e_3}^2 + 2 M_1'^2 \right ) +
4 g_1'^2 \left ( Q_L^2 + Q_e^2 - Q_1^2 \right ) \left ( a_\tau^2 -
2 y_\tau a_\tau M_1' \right ) \nonumber \\
& {} - 8 Q_1 g_1'^2 y_\tau^2 \left ( Q_1 m_{H_d}^2 + Q_L m_{L_3}^2 +
Q_e m_{e_3}^2 \right ) -24 Q_1 g_1'^2 y_t^2 \left ( Q_2 m_{H_u}^2 +
Q_Q m_{Q_3}^2 + Q_u m_{u_3}^2 \right ) \nonumber \\
& {} + 4 g_1'^2 \lambda^2 \left ( Q_2^2 + Q_S^2 - Q_1^2 \right )
\left ( m_{H_d}^2 + m_{H_u}^2 + m_S^2 + 2 M_1'^2 \right ) + 4 g_1'^2
\left ( Q_2^2 + Q_S^2 - Q_1^2 \right ) \nonumber \\
& {} \times \left ( a_\lambda^2 - 2 \lambda a_\lambda M_1' \right ) -
4 Q_1 g_1'^2 \sum_{i=1}^3 \left [ 2 \lambda_i^2 \left ( Q_1 m_{H_i^d}^2 +
Q_2 m_{H_i^u}^2 + Q_S m_S^2 \right ) \right . \nonumber \\
& {} \left . + 3 \kappa_i^2 \left ( m_S^2 + m_{D_i}^2 + m_{\overline{D}_i}^2
\right ) \right ] - \frac{16}{5} g_3^2 g_1^2 \sum_{i=1}^3 \left ( m_{Q_i}^2
- 2 m_{u_i}^2 + m_{d_i}^2 + m_{\overline{D}_i}^2 - m_{D_i}^2 \right )
\nonumber \\
& {} + 32 Q_1 g_3^2 g_1'^2 \sum_{i=1}^3 \left ( 2 Q_Q m_{Q_i}^2 +
Q_u m_{u_i}^2 + Q_d m_{d_i}^2 + Q_{\overline{D}} m_{\overline{D}_i}^2 +
Q_D m_{D_i}^2 \right ) \nonumber \\
& {} + 3 g_2^4 \left [ 29 M_2^2 + m_{H'}^2 + m_{\overline{H'}}^2 +
\sum_{i=1}^3 \left ( 3 m_{Q_i}^2 + m_{L_i}^2 + m_{H_i^d}^2 + m_{H_i^u}^2
\right ) \right ] \nonumber \\
& {} + \frac{9}{5} g_2^2 g_1^2 \left [ 2 \left ( M_1^2 + M_1 M_2 + M_2^2
\right ) + m_{H'}^2 - m_{\overline{H'}}^2 - \sum_{i=1}^3 \left ( m_{Q_i}^2 -
m_{L_i}^2 + m_{H_i^u}^2 - m_{H_i^d}^2 \right ) \right ] \nonumber \\
& {} + 12 Q_1 g_2^2 g_1'^2 \bigg [ 2 Q_1 \left ( M_1'^2 + M_1' M_2 + M_2^2
\right ) + Q_{H'} m_{H'}^2 + Q_{\overline{H'}} m_{\overline{H'}}^2 \nonumber \\
& {} + \sum_{i=1}^3 \left ( 3 Q_Q m_{Q_i}^2 + Q_L m_{L_i}^2 + Q_1 m_{H_i^d}^2
+ Q_2 m_{H_i^u}^2 \right ) \bigg ] + \frac{1}{25} g_1^4 \bigg [ 891 M_1^2 +
18 m_{H'}^2 \nonumber \\
& {} + 2 \sum_{i=1}^3 \left ( m_{d_i}^2 + 5 m_{D_i}^2 + m_{\overline{D}_i}^2
- 9 m_{e_i}^2 + 9 m_{H_i^d}^2 + 9 m_{L_i}^2 + m_{Q_i}^2 + 28 m_{u_i}^2
\right ) \bigg ] \nonumber \\
& {} - \frac{4}{5} g_1^2 g_1'^2 \bigg [ 6 Q_1 \big ( 3 Q_d +
3 Q_{\overline{D}} - 3 Q_D + 3 Q_e - 4 Q_1 + 3 Q_2 + Q_{\overline{H'}} -
Q_{H'} - 3 Q_L \nonumber \\
& {} + 3 Q_Q - 6 Q_u \big ) \left ( M_1^2 + M_1 M_1' + M_1'^2 \right ) +
3 Q_{\overline{H'}}^2 m_{\overline{H'}}^2 - 3 Q_{H'}^2 m_{H'}^2 \nonumber \\
& {} + 3 \sum_{i=1}^3 \big ( Q_d^2 m_{d_i}^2 +
Q_{\overline{D}}^2 m_{\overline{D}_i}^2 - Q_D^2 m_{D_i}^2 + Q_e^2 m_{e_i}^2
- Q_1^2 m_{H_i^d}^2 + Q_2^2 m_{H_i^u}^2 - Q_L^2 m_{L_i}^2 \nonumber \\
& {} + Q_Q^2 m_{Q_i}^2 - 2 Q_u^2 m_{u_i}^2 \big ) +
3 Q_1 Q_{\overline{H'}} m_{\overline{H'}}^2 - 9 Q_1 Q_{H'} m_{H'}^2 +
Q_1 \sum_{i=1}^3 \bigg ( 4 Q_d m_{d_i}^2 +
4 Q_{\overline{D}} m_{\overline{D}_i}^2 \nonumber \\
& {} - 8 Q_D m_{D_i}^2 - 9 Q_1 m_{H_i^d}^2 + 3 Q_2 m_{H_i^u}^2 -
9 Q_L m_{L_i}^2 + 5 Q_Q m_{Q_i}^2 - 20 Q_u m_{u_i}^2 \big ) \bigg ] \nonumber \\
& {} + 8 Q_1 g_1'^4 \bigg [ 3 Q_1 M_1'^2 \big ( 9 Q_d^2 + 9 Q_{\overline{D}}^2
+ 9 Q_D^2 + 3 Q_e^2 + 8 Q_1^2 + 6 Q_2^2 + 2 Q_{\overline{H'}}^2 + 2 Q_{H'}^2
+ 6 Q_L^2 \nonumber \\
& {} + 18 Q_Q^2 + 3 Q_S^2 + 9 Q_u^2 \big ) +
2 Q_{\overline{H'}}^3 m_{\overline{H'}}^2 + 2 Q_{H'}^3 m_{H'}^2 +
\sum_{i=1}^3 \big ( 3 Q_d^3 m_{d_i}^2 +
3 Q_{\overline{D}}^3 m_{\overline{D}_i}^2 + 3 Q_D^3 m_{D_i}^2 \nonumber \\
& {} + Q_e^3 m_{e_i}^2 + 2 Q_1^3 m_{H_i^d}^2 + 2 Q_2^3 m_{H_i^u}^2 +
2 Q_L^3 m_{L_i}^2 + 6 Q_Q^3 m_{Q_i}^2 + Q_S^3 m_{S_i}^2 + 3 Q_u^3 m_{u_i}^2
\big ) \nonumber \\
& {} + 2 Q_1 Q_{\overline{H'}}^2 m_{\overline{H'}}^2 + 2 Q_1 Q_{H'}^2 m_{H'}^2
+ Q_1 \sum_{i=1}^3 \big ( 3 Q_d^2 m_{d_i}^2 +
3 Q_{\overline{D}}^2 m_{\overline{D}_i}^2 + 3 Q_D^2 m_{D_i}^2 +
Q_e^2 m_{e_i}^2 \nonumber \\
& {} + 2 Q_1^2 m_{H_i^d}^2 + 2 Q_2^2 m_{H_i^u}^2 + 2 Q_L^2 m_{L_i}^2 +
6 Q_Q^2 m_{Q_i}^2 + Q_S^2 m_{S_i}^2 + 3 Q_u^2 m_{u_i}^2 \big ) \bigg ] ,
\label{eq:USSMmHd2BetaTwoLoop} \\
b_{m_{H_d}^2}^{(2)} &= 72 y_b^4 \left ( m_{H_d}^2 + m_{Q_3}^2 + m_{d_3}^2
\right ) + 144 y_b^2 a_b^2 + 16 y_\tau^4 \left ( m_{H_d}^2 + m_{L_3}^2 +
m_{e_3}^2 \right ) + 32 y_\tau^2 a_\tau^2 \nonumber \\
& {} + 8 \lambda^4 \left ( m_{H_d}^2 + m_{H_u}^2 + m_S^2 \right ) +
16 \lambda^2 a_\lambda^2 + 6 y_t^2 y_b^2 \left ( m_{H_d}^2 + m_{H_u}^2 +
2 m_{Q_3}^2 + m_{u_3}^2 + m_{d_3}^2 \right ) \nonumber \\
& {} + 6 \left ( y_t a_b + y_b a_t \right )^2 + 12 y_b^2 y_\tau^2
\left ( 2 m_{H_d}^2 + m_{Q_3}^2 + m_{d_3}^2 + m_{L_3}^2 + m_{e_3}^2 \right )
+ 12 \left ( y_b a_\tau + y_\tau a_b \right )^2 \nonumber \\
& {} + 6 \lambda^2 y_t^2 \left ( m_{H_d}^2 + 2 m_{H_u}^2 + m_S^2 + m_{Q_3}^2
+ m_{u_3}^2 \right ) + 6 \left ( \lambda a_t + y_t a_\lambda \right )^2
\nonumber \\
& {} + 12 \lambda^2 y_b^2 \left ( 2 m_{H_d}^2 + m_{H_u}^2 + m_S^2 +
m_{Q_3}^2 + m_{d_3}^2 \right ) + 12 \left ( \lambda a_b + y_b a_\lambda
\right )^2 \nonumber \\
& {} + 4 \lambda^2 y_\tau^2 \left ( 2 m_{H_d}^2 + m_{H_u}^2 + m_S^2 +
m_{L_3}^2 + m_{e_3}^2 \right ) + 4 \left ( \lambda a_\tau + y_\tau a_\lambda
\right )^2 \nonumber \\
& {} + 4 \sum_{i=1}^3 \left [ \lambda^2 \lambda_i^2 \left ( m_{H_i^d}^2 +
m_{H_i^u}^2 + m_{H_d}^2 + m_{H_u}^2 + 2 m_S^2 \right ) +
\left ( \lambda a_{\lambda_i} + \lambda_i a_\lambda \right )^2 \right ]
\nonumber \\
& {} + 6 \sum_{i=1}^3 \left [ \lambda^2 \kappa_i^2 \left ( m_{H_d}^2 +
m_{H_u}^2 + 2 m_S^2 + m_{D_i}^2 + m_{\overline{D}_i}^2 \right ) +
\left ( \lambda a_{\kappa_i} + \kappa_i a_\lambda \right )^2 \right ]
\nonumber \\
& {} - 32 g_3^2 y_b^2 \left ( m_{H_d}^2 + m_{Q_3}^2 + m_{d_3}^2 + 2 M_3^2
\right ) - 32 g_3^2 \left ( a_b^2 - 2 y_b a_b M_3 \right ) \nonumber \\
& {} - 18 g_2^2 y_b^2 \left ( m_{H_d}^2 + m_{Q_3}^2 + m_{d_3}^2 + 2 M_2^2
\right ) - 18 g_2^2 \left ( a_b^2 - 2 y_b a_b M_2 \right ) \nonumber \\
& {} - 6 g_2^2 y_\tau^2 \left ( m_{H_d}^2 + m_{L_3}^2 + m_{e_3}^2 + 2 M_2^2
\right ) - 6 g_2^2 \left ( a_\tau^2 - 2 y_\tau a_\tau M_2 \right ) \nonumber \\
& {} - 6 g_2^2 \lambda^2 \left ( m_{H_d}^2 + m_{H_u}^2 + m_S^2 + 2 M_2^2
\right ) - 6 g_2^2 \left ( a_\lambda^2 - 2 \lambda a_\lambda M_2 \right )
\nonumber \\
& {} - \frac{14}{5} g_1^2 y_b^2 \left ( m_{H_d}^2 + m_{Q_3}^2 + m_{d_3}^2 +
2 M_1^2 \right ) - \frac{14}{5} g_1^2 \left ( a_b^2 - 2 y_b a_b M_1 \right )
\nonumber \\
& {} - \frac{18}{5} g_1^2 y_\tau^2 \left ( m_{H_d}^2 + m_{L_3}^2 + m_{e_3}^2 +
 2 M_1^2 \right ) - \frac{18}{5} g_1^2 \left ( a_\tau^2 - 2 y_\tau a_\tau M_1
\right ) \nonumber \\
& {} - \frac{6}{5} g_1^2 \lambda^2 \left ( m_{H_d}^2 + m_{H_u}^2 + m_S^2 +
2 M_1^2 \right ) - \frac{6}{5} g_1^2 \left ( a_\lambda^2 -
2 \lambda a_\lambda M_1 \right ) \nonumber \\
& {} - 12 g_1'^2 y_b^2 \left ( Q_1^2 + Q_Q^2 + Q_d^2 \right )
\left ( m_{H_d}^2 + m_{Q_3}^2 + m_{d_3}^2 + 2 M_1'^2 \right ) -12 g_1'^2
\left ( Q_1^2 + Q_Q^2 + Q_d^2 \right ) \nonumber \\
& {} \times \left ( a_b^2 - 2 y_b a_b M_1'\right ) + 6 g_1'^2 y_b^2
\left ( Q_1 + Q_Q + Q_d \right ) \left ( 2 Q_1 m_{H_d}^2 + 2 Q_1 m_{Q_3}^2 +
2 Q_1 m_{d_3}^2 + \Sigma_1' \right ) \nonumber \\
& {} + 12 Q_1 g_1'^2 a_b^2 \left ( Q_1 + Q_Q + Q_d \right ) -
4 g_1'^2 y_\tau^2 \left ( Q_1^2 + Q_L^2 + Q_e^2 \right ) \left ( m_{H_d}^2 +
m_{L_3}^2 + m_{e_3}^2 + 2 M_1'^2 \right ) \nonumber \\
& {} - 4 g_1'^2 \left ( Q_1^2 + Q_L^2 + Q_e^2 \right ) \left ( a_\tau^2 -
2 y_\tau a_\tau M_1' \right ) + 2 g_1'^2 y_\tau^2 \left ( Q_1 + Q_L +Q_e
\right ) \nonumber \\
& {} \times \left ( 2 Q_1 m_{H_d}^2 + 2 Q_1 m_{L_3}^2 + 2 Q_1 m_{e_3}^2 +
\Sigma_1' \right ) + 4 Q_1 g_1'^2 a_\tau^2 \left ( Q_1 + Q_L + Q_e \right )
\nonumber \\
& {} + 12 Q_1 g_1'^2 y_t^2 \left ( Q_2 + Q_Q + Q_u \right ) \left ( m_{H_u}^2
+ m_{Q_3}^2 + m_{u_3}^2 \right ) + 12 Q_1 g_1'^2 a_t^2 \left ( Q_2 + Q_Q +
Q_u \right ) \nonumber \\
& {} - 4 g_1'^2 \lambda^2 \left ( Q_1^2 + Q_2^2 + Q_S^2 \right )
\left ( m_{H_d}^2 + m_{H_u}^2 + m_S^2 + 2 M_1'^2 \right ) - 4 g_1'^2
\left ( Q_1^2 + Q_2^2 + Q_S^2 \right ) \nonumber \\
& {} \times \left ( a_\lambda^2 - 2 \lambda a_\lambda M_1' \right ) +
2 g_1'^2 \lambda^2 \left ( Q_1 + Q_2 + Q_S \right ) \Sigma_1' +
4 Q_1 g_1'^2 \left ( Q_1 + Q_2 + Q_S \right ) \nonumber \\
& {} \times \sum_{i=1}^3 \left [ \lambda_i^2 \left ( m_{H_i^d}^2 + m_{H_i^u}^2
+ m_S^2 \right ) + a_{\lambda_i}^2 \right ] + 6 Q_1 g_1'^2 \left ( Q_S + Q_D
+ Q_{\overline{D}} \right ) \nonumber \\
& {} \times \sum_{i=1}^3 \left [ \kappa_i^2 \left ( m_S^2 + m_{D_i}^2 +
m_{\overline{D}_i}^2 \right ) + a_{\kappa_i}^2 \right ] -
96 Q_1 g_3^2 g_1'^2 M_3^2 \left ( 2 Q_Q + Q_u + Q_d + Q_D + Q_{\overline{D}}
\right ) \nonumber \\
& {} - 72 g_2^4 M_2^2 - 12 Q_1 g_2^2 g_1'^2 M_2^2 \left ( 9 Q_Q + 3 Q_L +
3 Q_1 + 3 Q_2 + Q_{\overline{H'}} + Q_{H'} \right ) \nonumber \\
& {} - \frac{288}{25} g_1^4 \left ( \Sigma_1 + 3 M_1^2 \right ) -
\frac{3}{5} g_1^2 g_1'^2 \Big [ 4 Q_1 M_1^2 \big ( 2 Q_d + 2 Q_{\overline{D}}
+ 2 Q_D + 6 Q_e + 3 Q_1 + 3 Q_2 \nonumber \\
& {} + Q_{\overline{H'}} + Q_{H'} + 3 Q_L + Q_Q + 8 Q_u \big ) - 4 M_1'^2
\big ( 3 Q_d^2 + 3 Q_{\overline{D}}^2 - 3 Q_D^2 + 3 Q_e^2 - 3 Q_1^2 +
3 Q_2^2 \nonumber \\
& {} + Q_{\overline{H'}}^2 - Q_{H'}^2 - 3 Q_L^2 + 3 Q_Q^2 - 6 Q_u^2 \big ) +
\left ( \Sigma_1' - 2 Q_1 \Sigma_1 \right ) \Sigma_Q^Y \Big ] -
4 Q_1 g_1'^4 \Big [ 2 M_1'^2 \big ( 9 Q_d^3 \nonumber \\
& {} + 9 Q_{\overline{D}}^3 + 9 Q_D^3 + 3 Q_e^3 + 6 Q_1^3 + 6 Q_2^3 +
2 Q_{\overline{H'}}^3 + 2 Q_{H'}^3 + 6 Q_L^3 + 18 Q_Q^3 + 3 Q_S^3 +
9 Q_u^3 \big ) \nonumber \\
& {} + \left ( 6 Q_1 M_1'^2 - \Sigma_1' \right ) \Sigma_Q \Big ] .
\label{eq:USSMmHd2Ot2Coeff}
\end{align}
\end{subequations}
Similarly, those for $m_{H_u}^2$ read
\begin{subequations}
\begin{align}
\beta_{m_{H_u}^2}^{(1)} &= 2 \lambda^2 \left ( m_{H_d}^2 + m_{H_u}^2 +
m_S^2 \right ) + 2 a_\lambda^2 + 6 y_t^2 \left ( m_{H_u}^2 + m_{Q_3}^2 +
m_{u_3}^2 \right ) + 6 a_t^2 \nonumber \\
& {} - 6 g_2^2 M_2^2 - \frac{6}{5} g_1^2 M_1^2 - 8 Q_2^2 g_1'^2 M_1'^2 +
\frac{3}{5} g_1^2 \Sigma_1 + 2 Q_2 g_1'^2 \Sigma_1' ,
\label{eq:USSMmHu2BetaOneLoop} \\
\beta_{m_{H_u}^2}^{(2)} &= -36 y_t^4 \left ( m_{H_u}^2 + m_{Q_3}^2 +
m_{u_3}^2 \right ) - 6 y_t^2 y_b^2 \left ( m_{H_u}^2 + m_{H_d}^2 + 2 m_{Q_3}^2
+ m_{u_3}^2 + m_{d_3}^2 \right ) \nonumber \\
& {} - 72 y_t^2 a_t^2 - 6 \left ( y_t a_b + y_b a_t \right )^2 -
4 \lambda^4 \left ( m_{H_d}^2 + m_{H_u}^2 + m_S^2 \right ) -
8 \lambda^2 a_{\lambda}^2 \nonumber \\
& {} - \lambda^2 \sum_{i=1}^3 \bigg [ 4 \lambda_i^2 \left ( m_{H_d}^2 +
m_{H_u}^2 + 2 m_S^2 + m_{H_i^d}^2 + m_{H_i^u}^2 \right ) \nonumber \\
& {} + 6 \kappa_i^2 \left ( m_{H_d}^2 + m_{H_u}^2 + 2 m_S^2 + m_{D_i}^2 +
m_{\overline{D}_i}^2 \right ) \bigg ] - \sum_{i=1}^3 \big [
4 \left ( \lambda_i a_{\lambda} + \lambda a_{\lambda_i} \right )^2
\nonumber \\
& {} + 6 \left ( \kappa_i a_{\lambda} + \lambda a_{\kappa_i} \right )^2
\big ] - 6 \lambda^2 y_b^2 \left ( 2 m_{H_d}^2 + m_{H_u}^2 + m_S^2 +
m_{Q_3}^2 + m_{d_3}^2 \right ) \nonumber \\
& {} - 2 \lambda^2 y_\tau^2 \left ( 2 m_{H_d}^2 + m_{H_u}^2 + m_S^2 +
m_{L_3}^2 + m_{e_3}^2 \right ) - 6 \left ( \lambda a_b + y_b a_{\lambda}
\right )^2 - 2 \left ( \lambda a_\tau + y_\tau a_{\lambda} \right )^2
\nonumber \\
& {} + 32 g_3^2 y_t^2 \left ( m_{H_u}^2 + m_{Q_3}^2 + m_{u_3}^2 + 2 M_3^2
\right ) + 32 g_3^2 \left ( a_t^2 - 2 y_t a_t M_3 \right ) \nonumber \\
& {} + \frac{2}{5} g_1^2 y_t^2 \left ( -5 m_{H_u}^2 + m_{Q_3}^2 +
16 m_{u_3}^2 + 8 M_1^2 \right ) + \frac{8}{5} g_1^2 \left ( a_t^2 -
2 y_t a_t M_1 \right ) \nonumber \\
& {} + \frac{6}{5} g_1^2 y_b^2 \left ( 3 m_{H_d}^2 - m_{Q_3}^2 - 2 m_{d_3}^2
\right ) + \frac{6}{5} g_1^2 y_\tau^2 \left ( m_{H_d}^2 + m_{L_3}^2 -
2 m_{e_3}^2 \right ) \nonumber \\
& {} + \frac{6}{5} g_1^2 \sum_{i=1}^3 \left [ \lambda_i^2 \left ( m_{H_i^d}^2
- m_{H_i^u}^2 \right ) + \kappa_i^2 \left ( m_{D_i}^2 - m_{\overline{D}_i}^2
\right ) \right ] \nonumber \\
& {} + 12 g_1'^2 y_t^2 \left ( Q_Q^2 + Q_u^2 - Q_2^2 \right )
\left ( m_{H_u}^2 + m_{Q_3}^2 + m_{u_3}^2 + 2 M_1'^2 \right ) \nonumber \\
& {} + 12 g_1'^2 \left ( Q_Q^2 + Q_u^2 - Q_2^2 \right ) \left ( a_t^2 -
2 y_t a_t M_1' \right ) - 24 Q_2 g_1'^2 y_t^2 \left ( Q_2 m_{H_u}^2 +
Q_Q m_{Q_3}^2 + Q_u m_{u_3}^2 \right ) \nonumber \\
& {} - 24 Q_2 g_1'^2 y_b^2 \left ( Q_1 m_{H_d}^2 + Q_Q m_{Q_3}^2 +
Q_d m_{d_3}^2 \right ) - 8 Q_2 g_1'^2 y_\tau^2 \left ( Q_1 m_{H_d}^2 +
Q_L m_{L_3}^2 + Q_e m_{e_3}^2 \right ) \nonumber \\
& {} + 4 g_1'^2 \lambda^2 \left ( Q_1^2 - Q_2^2 + Q_S^2 \right )
\left ( m_{H_d}^2 + m_{H_u}^2 + m_S^2 + 2 M_1'^2 \right ) \nonumber \\
& {} + 4 g_1'^2 \left ( Q_1^2 - Q_2^2 + Q_S^2 \right ) \left ( a_\lambda^2 -
2 \lambda a_\lambda M_1' \right ) \nonumber \\
& {} - 4 Q_2 g_1'^2 \sum_{i=1}^3 \left [ 2 \lambda_i^2 \left ( Q_1 m_{H_i^d}^2
+ Q_2 m_{H_i^u}^2 + Q_S m_S^2 \right ) + 3 \kappa_i^2 \left ( Q_S m_S^2 +
Q_D m_{D_i}^2 + Q_{\overline{D}} m_{\overline{D}_i}^2 \right ) \right ]
\nonumber \\
& {} + \frac{16}{5} g_3^2 g_1^2 \sum_{i=1}^3 \left ( m_{Q_i}^2 - 2 m_{u_i}^2
+ m_{d_i}^2 + m_{\overline{D}_i}^2 - m_{D_i}^2 \right ) \nonumber \\
& {} + 32 Q_2 g_3^2 g_1'^2 \sum_{i=1}^3 \left ( 2 Q_Q m_{Q_i}^2 +
Q_u m_{u_i}^2 + Q_d m_{d_i}^2 + Q_{\overline{D}} m_{\overline{D}_i}^2
+ Q_D m_{D_i}^2 \right ) \nonumber \\
& {} + 3 g_2^4 \left [ 29 M_2^2 + m_{H'}^2 + m_{\overline{H'}}^2 +
\sum_{i=1}^3 \left ( 3 m_{Q_i}^2 + m_{L_i}^2 + m_{H_i^d}^2 + m_{H_i^u}^2
\right ) \right ] \nonumber \\
& {} + \frac{9}{5} g_2^2 g_1^2 \left [ 2 \left ( M_1^2 + M_1 M_2 + M_2^2
\right ) + m_{\overline{H'}}^2 - m_{H'}^2 + \sum_{i=1}^3 \left ( m_{Q_i}^2 -
m_{L_i}^2 + m_{H_i^u}^2 - m_{H_i^d}^2 \right ) \right ] \nonumber \\
& {} + 12 Q_2 g_2^2 g_1'^2 \bigg [ 2 Q_2 \left ( M_1'^2 + M_1' M_2 + M_2^2
\right ) + Q_{H'} m_{H'}^2 + Q_{\overline{H'}} m_{\overline{H'}}^2
\nonumber \\
& {} + \sum_{i=1}^3 \left ( 3 Q_Q m_{Q_i}^2 + Q_L m_{L_i}^2 + Q_1 m_{H_i^d}^2
+ Q_2 m_{H_i^u}^2 \right ) \bigg ] + \frac{1}{25} g_1^4 \bigg [ 891 M_1^2 +
18 m_{\overline{H'}}^2 \nonumber \\
& {} + \sum_{i=1}^3 \left ( 10 m_{d_i}^2 + 2 m_{D_i}^2 +
10 m_{\overline{D}_i}^2 + 54 m_{e_i}^2 + 18 m_{H_i^u}^2 + 4 m_{Q_i}^2 -
8 m_{u_i}^2 \right ) \bigg ] \nonumber \\
& {} + \frac{4}{5} g_1^2 g_1'^2 \bigg [ 6 Q_2 \big ( 3 Q_d + Q_{\overline{D}}
- 3 Q_D + 3 Q_e - 3 Q_1 + 4 Q_2 + Q_{\overline{H'}} - Q_{H'} - 3 Q_L
\nonumber \\
& {} + 3 Q_Q - 6 Q_u \big ) \left ( M_1^2 + M_1 M_1' + M_1'^2 \right ) +
3 Q_{\overline{H'}}^2 m_{\overline{H'}}^2 - 3 Q_{H'}^2 m_{H'}^2 \nonumber \\
& {} + 3 \sum_{i=1}^3 \left ( Q_d^2 m_{d_i}^2 +
Q_{\overline{D}}^2 m_{\overline{D}_i}^2 - Q_D^2 m_{D_i}^2 + Q_e^2 m_{e_i}^2
- Q_1^2 m_{H_i^d}^2 + Q_2^2 m_{H_i^u}^2 - Q_L^2 m_{L_i}^2 \right .
\nonumber \\
& {} \left . + Q_Q^2 m_{Q_i}^2 - 2 Q_u^2 m_{u_i}^2 \right ) +
9 Q_2 Q_{\overline{H'}} m_{\overline{H'}}^2 - 3 Q_2 Q_{H'} m_{H'}^2 +
Q_2 \sum_{i=1}^3 \left ( 8 Q_d m_{d_i}^2 +
8 Q_{\overline{D}} m_{\overline{D}_i}^2 \right . \nonumber \\
& {} \left . - 4 Q_D m_{D_i}^2 + 12 Q_e m_{e_i}^2 - 3 Q_1 m_{H_i^d}^2 +
9 Q_2 m_{H_i^u}^2 - 3 Q_L m_{L_i}^2 + 7 Q_Q m_{Q_i}^2 - 4 Q_u m_{u_i}^2
\right ) \bigg ] \nonumber \\
& {} + 8 Q_2 g_1'^4 \bigg [ 3 Q_2 M_1'^2 \big ( 9 Q_d^2 +
9 Q_{\overline{D}}^2 + 9 Q_D^2 + 3 Q_e^2 + 6 Q_1^2 + 8 Q_2^2 +
2 Q_{\overline{H'}}^2 + 2 Q_{H'}^2 + 6 Q_L^2 \nonumber \\
& {} + 18 Q_Q^2 + 3 Q_S^2 + 9 Q_u^2 \big ) +
2 Q_{\overline{H'}}^3 m_{\overline{H'}}^2 + 2 Q_{H'}^3 m_{H'}^2 +
\sum_{i=1}^3 \big ( 3 Q_d^3 m_{d_i}^2 +
3 Q_{\overline{D}}^3 m_{\overline{D}_i}^2 \nonumber \\
& {} + 3 Q_D^3 m_{D_i}^2 + Q_e^3 m_{e_i}^2 + 2 Q_1^3 m_{H_i^d}^2 +
2 Q_2^3 m_{H_i^u}^2 + 2 Q_L^3 m_{L_i}^2 + 6 Q_Q^3 m_{Q_i}^2 + Q_S^3 m_{S_i}^2
+ 3 Q_u^3 m_{u_i}^2 \big ) \nonumber \\
& {} + 2 Q_2 Q_{\overline{H'}}^2 m_{\overline{H'}}^2 +
2 Q_2 Q_{H'}^2 m_{H'}^2 + Q_2 \sum_{i=1}^3 \big ( 3 Q_d^2 m_{d_i}^2 +
3 Q_{\overline{D}}^2 m_{\overline{D}_i}^2 + 3 Q_D^2 m_{D_i}^2 +
Q_e^2 m_{e_i}^2 \nonumber \\
& {} + 2 Q_1^2 m_{H_i^d}^2 + 2 Q_2^2 m_{H_i^u}^2 + 2 Q_L^2 m_{L_i}^2 +
6 Q_Q^2 m_{Q_i}^2 + Q_S^2 m_{S_i}^2 + 3 Q_u^2 m_{u_i}^2 \big ) \bigg ] ,
\label{eq:USSMmHu2BetaTwoLoop} \\
b_{m_{H_u}^2}^{(2)} &= 72 y_t^4 \left ( m_{H_u}^2 + m_{Q_3}^2 + m_{u_3}^2
\right ) + 144 y_t^2 a_t^2 + 8 \lambda^4 \left ( m_{H_d}^2 + m_{H_u}^2 +
m_S^2 \right ) + 16 \lambda^2 a_\lambda^2 \nonumber \\
& {} + 6 y_t^2 y_b^2 \left ( m_{H_d}^2 + m_{H_u}^2 + 2 m_{Q_3}^2 +
m_{u_3}^2 + m_{d_3}^2 \right ) + 6 \left ( y_t a_b + y_b a_t \right )^2
\nonumber \\
& {} + 12 \lambda^2 y_t^2 \left ( m_{H_d}^2 + 2 m_{H_u}^2 + m_S^2 + m_{Q_3}^2
+ m_{u_3}^2 \right ) + 12 \left ( \lambda a_t + y_t a_\lambda \right )^2
\nonumber \\
& {} + 6 \lambda^2 y_b^2 \left ( 2 m_{H_d}^2 + m_{H_u}^2 + m_S^2 + m_{Q_3}^2
+ m_{d_3}^2 \right ) + 6 \left ( \lambda a_b + y_b a_\lambda \right )^2
\nonumber \\
& {} + 2 \lambda^2 y_\tau^2 \left ( 2 m_{H_d}^2 + m_{H_u}^2 + m_S^2 +
m_{L_3}^2 + m_{e_3}^2 \right ) + 2 \left ( \lambda a_\tau + y_\tau a_\lambda
\right )^2 \nonumber \\
& {} + 4 \sum_{i=1}^3 \left [ \lambda^2 \lambda_i^2 \left ( m_{H_i^d}^2 +
m_{H_i^u}^2 + m_{H_d}^2 + m_{H_u}^2 + 2 m_S^2 \right ) + \left (
\lambda a_{\lambda_i} + \lambda_i a_{\lambda} \right )^2 \right ] \nonumber \\
& {} + 6 \sum_{i=1}^3 \left [ \lambda^2 \kappa_i^2 \left ( m_{H_d}^2 +
m_{H_u}^2 + 2 m_S^2 + m_{D_i}^2 + m_{\overline{D}_i}^2 \right )
+ \left ( \lambda a_{\kappa_i} + \kappa_i a_\lambda \right )^2 \right ]
\nonumber \\
& {} - 32 g_3^2 y_t^2 \left ( m_{H_u}^2 + m_{Q_3}^2 + m_{u_3}^2 + 2 M_3^2
\right ) - 32 g_3^2 \left ( a_t^2 - 2 y_t a_t M_3 \right ) \nonumber \\
& {} - 18 g_2^2 y_t^2 \left ( m_{H_u}^2 + m_{Q_3}^2 + m_{u_3}^2 + 2 M_2^2
\right ) - 18 g_2^2 \left ( a_t^2 - 2 y_t a_t M_2 \right ) \nonumber \\
& {} - 6 g_2^2 \lambda^2 \left ( m_{H_d}^2 + m_{H_u}^2 + m_S^2 + 2 M_2^2
\right ) - 6 g_2^2 \left ( a_\lambda^2 - 2 \lambda a_\lambda M_2\right )
\nonumber \\
& {} - \frac{26}{5} g_1^2 y_t^2 \left ( m_{H_u}^2 + m_{Q_3}^2 + m_{u_3}^2 +
2 M_1^2 \right ) - \frac{26}{5} g_1^2 \left ( a_t^2 - 2 y_t a_t M_1 \right )
\nonumber \\
& {} - \frac{6}{5} g_1^2 \lambda^2 \left ( m_{H_d}^2 + m_{H_u}^2 + m_S^2 +
2 M_1^2 \right ) - \frac{6}{5} g_1^2 \left ( a_\lambda^2 -
2 \lambda a_\lambda M_1 \right ) \nonumber \\
& {} - 12 g_1'^2 y_t^2 \left ( Q_2^2 + Q_Q^2 + Q_u^2 \right )
\left ( m_{H_u}^2 + m_{Q_3}^2 + m_{u_3}^2 + 2 M_1'^2 \right ) -
12 g_1'^2 \left ( Q_2^2 + Q_Q^2 + Q_u^2 \right ) \nonumber \\
& {} \times \left ( a_t^2 - 2 y_t a_t M_1' \right ) +
6 g_1'^2 y_t^2 \left ( Q_2 + Q_Q + Q_u \right )
\left ( 2 Q_2 m_{H_u}^2 + 2 Q_2 m_{Q_3}^2 + 2 Q_2 m_{u_3}^2 + \Sigma_1'
\right ) \nonumber \\
& {} + 12 Q_2 g_1'^2 a_t^2 \left ( Q_2 + Q_Q + Q_u \right ) +
12 Q_2 g_1'^2 y_b^2 \left ( Q_1 + Q_Q + Q_d \right ) \left ( m_{H_d}^2 +
m_{Q_3}^2 + m_{d_3}^2 \right ) \nonumber \\
& {} + 12 Q_2 g_1'^2 a_b^2 \left ( Q_1 + Q_Q + Q_d \right ) +
4 Q_2 g_1'^2 y_\tau^2 \left ( Q_1 + Q_L + Q_e \right ) \left ( m_{H_d}^2 +
m_{L_3}^2 + m_{e_3}^2 \right ) \nonumber \\
& {} + 4 Q_2 g_1'^2 a_\tau^2 \left ( Q_1 + Q_L + Q_e \right ) -
4 g_1'^2 \lambda^2 \left ( Q_1^2 + Q_2^2 + Q_S^2 \right ) \left ( m_{H_d}^2 +
m_{H_u}^2 + m_S^2 + 2 M_1'^2 \right ) \nonumber \\
& {} - 4 g_1'^2 \left ( Q_1^2 + Q_2^2 + Q_S^2 \right ) \left ( a_\lambda^2 -
2 \lambda a_\lambda M_1' \right ) + 2 g_1'^2 \lambda^2 \left ( Q_1 + Q_2 +
Q_S \right ) \Sigma_1' \nonumber \\
& {} + 4 Q_2 g_1'^2 \left ( Q_1 + Q_2 + Q_S \right ) \sum_{i=1}^3 \left [
\lambda_i^2 \left ( m_{H_i^d}^2 + m_{H_i^u}^2 + m_S^2 \right ) +
a_{\lambda_i}^2 \right ] \nonumber \\
& {} + 6 Q_2 g_1'^2 \left ( Q_S + Q_D + Q_{\overline{D}} \right ) \sum_{i=1}^3
\left [ \kappa_i^2 \left ( m_S^2 + m_{D_i}^2 + m_{\overline{D}_i}^2 \right )
+ a_{\kappa_i}^2 \right ] \nonumber \\
& {} - 96 Q_2 g_3^2 g_1'^2 M_3^2 \left ( 2 Q_Q + Q_u + Q_d + Q_D +
Q_{\overline{D}} \right ) - 72 g_2^4 M_2^2 \nonumber \\
& {} - 12 Q_2 g_2^2 g_1'^2 M_2^2 \left ( 9 Q_Q + 3 Q_L + 3 Q_1 + 3 Q_2 +
Q_{\overline{H'}} + Q_{H'} \right ) + \frac{288}{25} g_1^4 \left ( \Sigma_1 -
3 M_1^2 \right ) \nonumber \\
& {} - \frac{3}{5} g_1^2 g_1'^2 \Big [ 4 Q_2 M_1^2 \big ( 2 Q_d +
2 Q_{\overline{D}} + 2 Q_D + 6 Q_e + 3 Q_1 + 3 Q_2 + Q_{\overline{H'}} +
Q_{H'} + 3 Q_L \nonumber \\
& {} + Q_Q + 8 Q_u \big ) + 4 M_1'^2 \big ( 3 Q_d^2 + 3 Q_{\overline{D}}^2 -
3 Q_D^2 + 3 Q_e^2 - 3 Q_1^2 + 3 Q_2^2 + Q_{\overline{H'}}^2 - Q_{H'}^2
\nonumber \\
& {} - 3 Q_L^2 + 3 Q_Q^2 - 6 Q_u^2 \big ) - \left ( 2 Q_2 \Sigma_1 + \Sigma_1'
\right ) \Sigma_Q^Y \Big ] - 4 Q_2 g_1'^4 \Big [ 2 M_1'^2 \big ( 9 Q_d^3 +
9 Q_{\overline{D}}^3 + 9 Q_D^3 \nonumber \\
& {} + 3 Q_e^3 + 6 Q_1^3 + 6 Q_2^3 + 2 Q_{\overline{H'}}^3 + 2 Q_{H'}^3 +
6 Q_L^3 + 18 Q_Q^3 + 3 Q_S^3 + 9 Q_u^3 \big ) \nonumber \\
& {} + \left ( 6 Q_2 M_1'^2 - \Sigma_1' \right ) \Sigma_Q \Big ] ,
\label{eq:USSMmHu2Ot2Coeff}
\end{align}
\end{subequations}
while those for $m_S^2$ are
\begin{subequations}
\begin{align}
\beta_{m_S^2}^{(1)} &= \sum_{i=1}^3 \left [ 4 \lambda_i^2 \left ( m_{H_i^d}^2
+ m_{H_i^u}^2 + m_S^2 \right ) + 4 a_{\lambda_i}^2 + 6\kappa_i^2 \left ( m_S^2
+ m_{D_i}^2 + m_{\overline{D}_i}^2 \right ) + 6 a_{\kappa_i}^2 \right ]
\nonumber \\
& {} - 8 Q_S^2 g_1'^2 M_1'^2 + 2 Q_S g_1'^2 \Sigma_1' ,
\label{eq:USSMms2BetaOneLoop} \\
\beta_{m_S^2}^{(2)} &= \sum_{i=1}^3 \bigg [ -16 \lambda_i^4
\left ( m_{H_i^u}^2 + m_{H_i^d}^2 + m_S^2 \right ) - 24 \kappa_i^4
\left ( m_{D_i}^2 + m_{\overline{D}_i}^2 + m_S^2 \right ) \nonumber \\
& {} - 32 \lambda_i^2 a_{\lambda_i}^2 - 48 \kappa_i^2 a_{\kappa_i}^2 \bigg ]
- 12 \lambda^2 y_t^2 \left ( 2 m_{H_u}^2 + m_{H_d}^2 + m_S^2 + m_{Q_3}^2 +
m_{u_3}^2 \right ) \nonumber \\
& {} - 12 \lambda^2 y_b^2 \left ( m_{H_u}^2 + 2 m_{H_d}^2 + m_S^2 + m_{Q_3}^2
+ m_{d_3}^2 \right ) \nonumber \\
& {} - 4 \lambda^2 y_\tau^2 \left ( m_{H_u}^2 + 2 m_{H_d}^2 + m_S^2 +
m_{L_3}^2 + m_{e_3}^2 \right ) - 12 \left ( \lambda a_t + y_t a_\lambda
\right )^2 - 12 \left ( \lambda a_b + y_b a_\lambda \right )^2 \nonumber \\
& {} - 4 \left ( \lambda a_\tau + y_\tau a_\lambda \right )^2 +
32 g_3^2 \sum_{i=1}^3 \left [ \kappa_i^2 \left ( m_S^2 + m_{D_i}^2 +
m_{\overline{D}_i}^2 + 2 M_3^2 \right ) + a_{\kappa_i}^2 -
2 \kappa_i a_{\kappa_i} M_3 \right ] \nonumber \\
& {} + 12 g_2^2 \sum_{i=1}^3 \left [ \lambda_i^2 \left ( m_{H_i^d}^2 +
m_{H_i^u}^2 + m_S^2 + 2 M_2^2 \right ) + a_{\lambda_i}^2 -
2 \lambda_i a_{\lambda_i} M_2 \right ] \nonumber \\
& {} + \frac{4}{5} g_1^2 \sum_{i=1}^3 \bigg [ 3 \lambda_i^2
\left ( m_{H_i^d}^2 + m_{H_i^u}^2 + m_S^2 + 2 M_1^2 \right ) + 2 \kappa_i^2
\left ( m_S^2 + m_{D_i}^2 + m_{\overline{D}_i}^2 + 2 M_1^2 \right )
\nonumber \\
& {} + 3 \left ( a_{\lambda_i}^2 - 2 \lambda_i a_{\lambda_i} M_1 \right ) +
2 \left ( a_{\kappa_i}^2 - 2 \kappa_i a_{\kappa_i} M_1 \right ) \bigg ]
\nonumber \\
& {} + 4 g_1'^2 \sum_{i=1}^3 \bigg [ 2 \lambda_i^2 \left ( Q_1^2 + Q_2^2 -
Q_S^2 \right ) \left ( m_{H_i^d}^2 + m_{H_i^u}^2 + m_S^2 + 2 M_1'^2 \right )
\nonumber \\
& {} - 2 Q_S \lambda_i^2 \left ( Q_1 m_{H_i^d}^2 + Q_2 m_{H_i^u}^2 + Q_S m_S^2
\right ) + 3 \kappa_i^2 \left ( Q_D^2 + Q_{\overline{D}}^2 - Q_S^2 \right )
\nonumber \\
& {} \times \left ( m_S^2 + m_{D_i}^2 + m_{\overline{D}_i}^2 + 2 M_1'^2
\right ) - 3 Q_S \kappa_i^2 \left ( Q_S m_S^2 + Q_D m_{D_i}^2 +
Q_{\overline{D}} m_{\overline{D}_i}^2 \right ) \nonumber \\
& {} + 2 \left ( a_{\lambda_i}^2 - 2 \lambda_i a_{\lambda_i} M_1' \right )
\left ( Q_1^2 + Q_2^2 - Q_S^2 \right ) + 3 \left ( a_{\kappa_i}^2 -
2 \kappa_i a_{\kappa_i} M_1' \right ) \left ( Q_D^2 + Q_{\overline{D}}^2 -
Q_S^2 \right ) \bigg ] \nonumber \\
& {} - 24 Q_S g_1'^2 y_t^2 \left ( Q_2 m_{H_u}^2 + Q_Q m_{Q_3}^2 +
Q_u m_{u_3}^2 \right ) - 24 Q_S g_1'^2 y_b^2 \left ( Q_1 m_{H_d}^2 +
Q_Q m_{Q_3}^2 + Q_d m_{d_3}^2 \right ) \nonumber \\
& {} - 8 Q_S g_1'^2 y_\tau^2 \left ( Q_1 m_{H_d}^2 + Q_L m_{L_3}^2 +
Q_e m_{e_3}^2 \right ) + 32 Q_S g_3^2 g_1'^2 \sum_{i=1}^3
\big ( 2 Q_Q m_{Q_i}^2 + Q_u m_{u_i}^2 \nonumber \\
& {} + Q_d m_{d_i}^2 + Q_D m_{D_i}^2 + Q_{\overline{D}} m_{\overline{D}_i}^2
\big ) + 12 Q_S g_2^2 g_1'^2 \bigg [ Q_{\overline{H'}} m_{\overline{H'}}^2 +
Q_{H'} m_{H'}^2 \nonumber \\
& {} + \sum_{i=1}^3 \left ( 3 Q_Q m_{Q_i}^2 + Q_L m_{L_i}^2 + Q_1 m_{H_i^d}^2
+ Q_2 m_{H_i^u}^2 \right ) \bigg ] + \frac{4}{5} Q_S g_1^2 g_1'^2
\bigg [ 3 Q_{\overline{H'}} m_{\overline{H'}}^2 + 3 Q_{H'} m_{H'}^2
\nonumber \\
& {} + \sum_{i=1}^3 \big ( 2 Q_d m_{d_i}^2 +
2 Q_{\overline{D}} m_{\overline{D}_i}^2 + 2 Q_D m_{D_i}^2 + 6 Q_e m_{e_i}^2 +
3 Q_1 m_{H_i^d}^2 + 3 Q_2 m_{H_i^u}^2 + 3 Q_L m_{L_i}^2 \nonumber \\
& {} + Q_Q m_{Q_i}^2 + 8 Q_u m_{u_i}^2 \big ) \bigg ] + 8 Q_S g_1'^4 \bigg [
3 Q_S M_1'^2 \big ( 9 Q_d^2 + 9 Q_{\overline{D}}^2 + 9 Q_D^2 + 3 Q_e^2 +
6 Q_1^2 + 6 Q_2^2 \nonumber \\
& {} + 2 Q_{\overline{H'}}^2 + 2 Q_{H'}^2 + 6 Q_L^2 + 18 Q_Q^2 + 5 Q_S^2 +
9 Q_u^2 \big ) + 2 Q_{\overline{H'}}^3 m_{\overline{H'}}^2 +
2 Q_{H'}^3 m_{H'}^2 \nonumber \\
& {} + \sum_{i=1}^3 \big ( 3 Q_d^3 m_{d_i}^2 +
3 Q_{\overline{D}}^3 m_{\overline{D}_i}^2 + 3 Q_D^3 m_{D_i}^2 +
Q_e^3 m_{e_i}^2 + 2 Q_1^3 m_{H_i^d}^2 + 2 Q_2^3 m_{H_i^u}^2 +
2 Q_L^3 m_{L_i}^2 \nonumber \\
& {} + 6 Q_Q^3 m_{Q_i}^2 + Q_S^3 m_{S_i}^2 + 3 Q_u^3 m_{u_i}^2 \big ) +
2 Q_S Q_{\overline{H'}}^2 m_{\overline{H'}}^2 + 2 Q_S Q_{H'}^2 m_{H'}^2
\nonumber \\
& {} + Q_S \sum_{i=1}^3 \big ( 3 Q_d^2 m_{d_i}^2 +
3 Q_{\overline{D}}^2 m_{\overline{D}_i}^2 + 3 Q_D^2 m_{D_i}^2 +
Q_e^2 m_{e_i}^2 + 2 Q_1^2 m_{H_i^d}^2 + 2 Q_2^2 m_{H_i^u}^2 +
2 Q_L^2 m_{L_i}^2 \nonumber \\
& {} + 6 Q_Q^2 m_{Q_i}^2 + Q_S^2 m_{S_i}^2 + 3 Q_u^2 m_{u_i}^2 \big ) \bigg ]
, \label{eq:USSMms2BetaTwoLoop} \\
b_{m_S^2}^{(2)} &= 8 \sum_{i=1}^3 \left [ 2 \lambda_i^4 \left ( m_{H_i^d}^2 +
m_{H_i^u}^2 + m_S^2 \right ) + 4 \lambda_i^2 a_{\lambda_i}^2 + 3 \kappa_i^4
\left ( m_S^2 + m_{D_i}^2 + m_{\overline{D}_i}^2 \right ) +
6 \kappa_i^2 a_{\kappa_i}^2 \right ] \nonumber \\
& {} + 8 \sum_{i=1}^3 \sum_{j=1}^3 \left [ \lambda_i^2 \lambda_j^2
\left ( m_{H_i^d}^2 + m_{H_i^u}^2 + m_{H_j^d}^2 + m_{H_j^u}^2 + 2 m_S^2
\right ) + \left ( \lambda_i a_{\lambda_j} + \lambda_j a_{\lambda_i}
\right )^2 \right ] \nonumber \\
& {} + 24 \sum_{i=1}^3 \sum_{j=1}^3 \left [ \lambda_i^2 \kappa_j^2
\left ( m_{H_i^d}^2 + m_{H_i^u}^2 + 2 m_S^2 + m_{D_i}^2 + m_{\overline{D}_i}^2
\right ) + \left ( \lambda_i a_{\kappa_j} + \kappa_j a_{\lambda_i} \right )^2
\right ] \nonumber \\
& {} + 18 \sum_{i=1}^3 \sum_{j=1}^3 \left [ \kappa_i^2 \kappa_j^2
\left ( 2 m_S^2 + m_{D_i}^2 + m_{\overline{D}_i}^2 + m_{D_j}^2 +
m_{\overline{D}_j}^2 \right ) + \left ( \kappa_i a_{\kappa_j} +
\kappa_j a_{\kappa_i} \right )^2 \right ] \nonumber \\
& {} + 12 \lambda^2 y_t^2 \left ( 2 m_{H_u}^2 + m_{H_d}^2 + m_S^2 + m_{Q_3}^2
+ m_{u_3}^2 \right ) + 12 \lambda^2 y_b^2 \left ( m_{H_u}^2 + 2 m_{H_d}^2 +
m_S^2 \right . \nonumber \\
& {} \left . + m_{Q_3}^2 + m_{d_3}^2 \right ) + 4 \lambda^2 y_\tau^2
\left ( 2 m_{H_d}^2 + m_{H_u}^2 + m_S^2 + m_{L_3}^2 + m_{e_3}^2 \right ) +
12 \left ( \lambda a_t + y_t a_{\lambda} \right )^2 \nonumber \\
& {} + 12 \left ( \lambda a_b + y_b a_{\lambda} \right )^2 +
4 \left ( \lambda a_\tau + y_\tau a_{\lambda} \right )^2 -
32 g_3^2 \sum_{i=1}^3 \left [ \kappa_i^2 \left ( m_S^2 + m_{D_i}^2 +
m_{\overline{D}_i}^2 + 2 M_3^2 \right ) \right . \nonumber \\
& {} \left . + a_{\kappa_i}^2 - 2 \kappa_i a_{\kappa_i} M_3 \right ] -
12 g_2^2 \sum_{i=1}^3 \left [ \lambda_i^2 \left ( m_{H_i^d}^2 + m_{H_i^u}^2 +
m_S^2 + 2 M_2^2 \right ) + a_{\lambda_i}^2 - 2 \lambda_i a_{\lambda_i} M_2
\right ] \nonumber \\
& {} - \frac{4}{5} g_1^2 \sum_{i=1}^3 \Big [ 3 \lambda_i^2 \left ( m_{H_i^d}^2
+ m_{H_i^u}^2 + m_S^2 + 2 M_1^2 \right ) + 3 a_{\lambda_i}^2 -
6 \lambda_i a_{\lambda_i} M_1 \nonumber \\
& {} + 2 \kappa_i^2 \left ( m_S^2 + m_{D_i}^2 + m_{\overline{D}_i}^2 +
2 M_1^2 \right ) + 2 a_{\kappa_i}^2 - 4 \kappa_i a_{\kappa_i} M_1 \Big ]
\nonumber \\
& {} + 2 g_1'^2 \sum_{i=1}^3 \Big [ -4 \lambda_i^2 \left ( Q_1^2 + Q_2^2 +
Q_S^2 \right ) \left ( m_{H_i^d}^2 + m_{H_i^u}^2 + m_S^2 + 2 M_1'^2 \right )
\nonumber \\
& {} + 2 \lambda_i^2 \left ( Q_1 + Q_2 + Q_S \right ) \left ( Q_S m_{H_i^d}^2
+ Q_S m_{H_i^u}^2 + Q_S m_S^2 + \Sigma_1' \right ) - 4 \left ( Q_1^2 + Q_2^2
+ Q_S^2 \right ) \nonumber \\
& {} \times \left ( a_{\lambda_i}^2 - 2 \lambda_i a_{\lambda_i} M_1' \right )
+ 2 Q_S a_{\lambda_i}^2 \left ( Q_1 + Q_2 + Q_S \right ) -
6 \kappa_i^2 \left ( Q_S^2 + Q_D^2 + Q_{\overline{D}}^2 \right ) \nonumber \\
& {} \times \left ( m_S^2 + m_{D_i}^2 + m_{\overline{D}_i}^2 + 2 M_1'^2
\right ) + 3 \kappa_i^2 \left ( Q_S + Q_D + Q_{\overline{D}} \right )
\left ( Q_S m_S^2 + Q_S m_{D_i}^2 + Q_S m_{\overline{D}_i}^2 + \Sigma_1'
\right ) \nonumber \\
& {} - 6 \left ( Q_S^2 + Q_D^2 + Q_{\overline{D}}^2 \right )
\left ( a_{\kappa_i}^2 - 2 \kappa_i a_{\kappa_i} M_1' \right ) +
3 Q_S a_{\kappa_i}^2 \left ( Q_S + Q_D + Q_{\overline{D}} \right ) \Big ]
\nonumber \\
& {} + 12 Q_S g_1'^2 y_t^2 \left ( Q_2 + Q_Q + Q_u \right )
\left ( m_{H_u}^2 + m_{Q_3}^2 + m_{u_3}^2 \right ) + 12 Q_S g_1'^2 a_t^2
\left ( Q_2 + Q_Q + Q_u \right ) \nonumber \\
& {} + 12 Q_S g_1'^2 y_b^2 \left ( Q_1 + Q_Q + Q_d \right )
\left ( m_{H_d}^2 + m_{Q_3}^2 + m_{d_3}^2 \right ) + 12 Q_S g_1'^2 a_b^2
\left ( Q_1 + Q_Q + Q_d \right ) \nonumber \\
& {} + 4 Q_S g_1'^2 y_\tau^2 \left ( Q_1 + Q_L + Q_e \right )
\left ( m_{H_d}^2 + m_{L_3}^2 + m_{e_3}^2 \right ) + 4 Q_S g_1'^2 a_\tau^2
\left ( Q_1 + Q_L + Q_e \right ) \nonumber \\
& {} - 96 Q_S g_3^2 g_1'^2 M_3^2 \left ( 2 Q_Q + Q_u + Q_d + Q_D +
Q_{\overline{D}} \right ) - 12 Q_S g_2^2 g_1'^2 M_2^2 \big ( 9 Q_Q + 3 Q_L +
3 Q_1 \nonumber \\
& {} + 3 Q_2 + Q_{\overline{H'}} + Q_{H'} \big ) -
\frac{6}{5} Q_S g_1^2 g_1'^2 \Big [ 2 M_1^2 \big ( 2 Q_d + 2 Q_{\overline{D}}
+ 2 Q_D + 6 Q_e + 3 Q_1 + 3 Q_2 + Q_{\overline{H'}} \nonumber \\
& {} + Q_{H'} + 3 Q_L + Q_Q + 8 Q_u \big ) - \Sigma_1 \Sigma_Q^Y \Big ] -
4 Q_S g_1'^4 \Big [ 2 M_1'^2 \big ( 9 Q_d^3 + 9 Q_{\overline{D}}^3 + 9 Q_D^3 +
3 Q_e^3 + 6 Q_1^3 \nonumber \\
& {} + 6 Q_2^3 + 2 Q_{\overline{H'}}^3 + 2 Q_{H'}^3 + 6 Q_L^3 + 18 Q_Q^3 +
3 Q_S^3 + 9 Q_u^3 \big ) + \left ( 6 Q_S M_1'^2 - \Sigma_1' \right ) \Sigma_Q
\Big ] . \label{eq:USSMms2Ot2Coeff}
\end{align}
\end{subequations}
If the one-loop contributions to the effective potential from top and stop
loops are also included, it is necessary to consider the expansions for $y_t$,
$a_t$, $m_{Q_3}^2$ and $m_{u_3}^2$. The required expressions for $y_t$ read
\begin{subequations}
\begin{align}
\beta_{y_t}^{(1)} &= y_t \left [ \lambda^2 + 6 y_t^2 + y_b^2 -
\frac{16}{3} g_3^2 - 3 g_2^2 - \frac{13}{15} g_1^2 - 2 g_1'^2 \left ( Q_2^2 +
Q_Q^2 + Q_u^2 \right ) \right ] , \label{eq:USSMYu22BetaOneLoop} \\
\beta_{y_t}^{(2)} &= y_t \bigg \{ -22 y_t^4 - 5 y_b^4 - 5 y_t^2 y_b^2 -
y_b^2 y_\tau^2 - \lambda^2 \left ( \lambda^2 + 3 y_t^2 + 4 y_b^2 + y_\tau^2 +
2 \Sigma_\lambda + 3 \Sigma_\kappa \right ) \nonumber \\
& {} + 2 g_1'^2 \left [ \lambda^2 \left ( Q_1^2 - Q_2^2 + Q_S^2 \right ) +
2 y_t^2 \left ( 2 Q_Q^2 + Q_u^2 \right ) + y_b^2 \left ( Q_1^2 - Q_Q^2 +
Q_d^2 \right ) \right ] \nonumber \\
& {} + 16 g_3^2 y_t^2 + 6 g_2^2 y_t^2 + g_1^2 \left ( \frac{6}{5} y_t^2 +
\frac{2}{5} y_b^2 \right ) + \frac{128}{9} g_3^4 + \frac{33}{2} g_2^4 +
\frac{3913}{450} g_1^4 \nonumber \\
& {} + 2 g_1'^4 \left [ 2 \left ( Q_2^4 + Q_Q^4 + Q_u^4 \right ) +
\left ( Q_2^2 + Q_Q^2 + Q_u^2 \right ) \Sigma_{Q} \right ] + 8 g_3^2 g_2^2 +
\frac{136}{45} g_3^2 g_1^2 \nonumber \\
& {} + \frac{32}{3} g_3^2 g_1'^2 \left ( Q_Q^2 + Q_u^2 \right ) + g_2^2 g_1^2
+ 6 g_2^2 g_1'^2 \left ( Q_2^2 + Q_Q^2 \right ) \nonumber \\
& {} + \frac{2}{5} g_1^2 g_1'^2 \left [ 3 Q_2^2 + \frac{1}{3} Q_Q^2 +
\frac{16}{3} Q_u^2 + \left ( 3 Q_2 + Q_Q - 4 Q_u \right ) \Sigma_Q^Y \right ]
\bigg \} , \label{eq:USSMYu22BetaTwoLoop} \\
b_{y_t}^{(2)} &= y_t \bigg \{ 54 y_t^4 + \frac{13}{2} y_b^4 + 13 y_t^2 y_b^2
+ y_b^2 y_\tau^2 + \lambda^2 \left ( \frac{5}{2} \lambda^2 + 15 y_t^2 +
5 y_b^2 + y_\tau^2 + 2 \Sigma_\lambda + 3 \Sigma_\kappa \right ) \nonumber \\
& {} - \frac{16}{3} g_3^2 \left ( \lambda^2 + 2 y_b^2 + 12 y_t^2 \right ) -
6 g_2^2 \left ( \lambda^2 + y_b^2 + 6 y_t^2 \right ) -
g_1^2 \left ( \frac{22}{15} \lambda^2 + \frac{4}{3} y_b^2 + \frac{52}{5} y_t^2
\right ) \nonumber \\
& {} - 2 g_1'^2 \left [ \lambda^2 \left ( Q_1^2 + 2 Q_2^2 + Q_S^2 + Q_Q^2 +
Q_u^2 \right ) + y_b^2 \left ( Q_1^2 + Q_2^2 + 2 Q_Q^2 + Q_u^2 + Q_d^2
\right ) \right . \nonumber \\
& {} \left . + 12 y_t^2 \left ( Q_2^2 + Q_Q^2 + Q_u^2 \right ) \right ] +
\frac{128}{9} g_3^4 - \frac{15}{2} g_2^4 - \frac{143}{18} g_1^4 \nonumber \\
& {} + 2 g_1'^4 \left ( Q_2^2 + Q_Q^2 + Q_u^2 \right ) \left ( Q_2^2 + Q_Q^2
+ Q_u^2 - \Sigma_{Q} \right ) + 16 g_3^2 g_2^2 + \frac{208}{45} g_3^2 g_1^2
\nonumber \\
& {} + \frac{32}{3} g_3^2 g_1'^2 \left ( Q_2^2 + Q_Q^2 + Q_u^2 \right ) +
\frac{13}{5} g_2^2 g_1^2 + 6 g_2^2 g_1'^2 \left ( Q_2^2 + Q_Q^2 + Q_u^2
\right ) \nonumber \\
& {} + \frac{26}{15} g_1^2 g_1'^2 \left ( Q_2^2 + Q_Q^2 + Q_u^2 \right )
\bigg \} , \label{eq:USSMYu22Ot2Coeff}
\end{align}
\end{subequations}
and those for $a_t$ read
\begin{subequations}
\begin{align}
\beta_{a_t}^{(1)} &= a_t \left [ \lambda^2 + 6 y_t^2 + y_b^2 -
\frac{16}{3} g_3^2 - 3 g_2^2 - \frac{13}{15} g_1^2 - 2 g_1'^2 \left ( Q_2^2 +
Q_Q^2 + Q_u^2 \right ) \right ] \nonumber \\
& {} + y_t \bigg [ 2 \lambda a_\lambda + 12 y_t a_t + 2 y_b a_b +
\frac{32}{3} g_3^2 M_3 + 6 g_2^2 M_2 + \frac{26}{15} g_1^2 M_1 \nonumber \\
& {} + 4 g_1'^2 M_1' \left ( Q_2^2 + Q_Q^2 + Q_u^2 \right ) \bigg ] ,
\label{eq:USSMTYu22BetaOneLoop} \\
\beta_{a_t}^{(2)} &= a_t \bigg \{ -22 y_t^4 - 5 y_b^4 - 5 y_t^2 y_b^2 -
y_b^2 y_\tau^2 - \lambda^2 \left ( \lambda^2 + 3 y_t^2 + 4 y_b^2 + y_\tau^2 +
2 \Sigma_\lambda + 3 \Sigma_\kappa \right ) \nonumber \\
& {} + 2 g_1'^2 \left [ \lambda^2 \left ( Q_1^2 - Q_2^2 + Q_S^2 \right ) +
2 y_t^2 \left ( 2 Q_Q^2 + Q_u^2 \right ) + y_b^2 \left ( Q_1^2 - Q_Q^2 +
Q_d^2 \right ) \right ] \nonumber \\
& {} + 16 g_3^2 y_t^2 + 6 g_2^2 y_t^2 + g_1^2 \left ( \frac{6}{5} y_t^2 +
\frac{2}{5} y_b^2 \right ) + \frac{128}{9} g_3^4 + \frac{33}{2} g_2^4 +
\frac{3913}{450} g_1^4 \nonumber \\
& {} + 2 g_1'^4 \left [ 2 \left ( Q_2^4 + Q_Q^4 + Q_u^4 \right ) +
\left ( Q_2^2 + Q_Q^2 + Q_u^2 \right ) \Sigma_{Q} \right ] + 8 g_3^2 g_2^2 +
\frac{136}{45} g_3^2 g_1^2 \nonumber \\
& {} + \frac{32}{3} g_3^2 g_1'^2 \left ( Q_Q^2 + Q_u^2 \right ) + g_2^2 g_1^2
+ 6 g_2^2 g_1'^2 \left ( Q_2^2 + Q_Q^2 \right ) \nonumber \\
& {} + \frac{2}{5} g_1^2 g_1'^2 \left [ 3 Q_2^2 + \frac{1}{3} Q_Q^2 +
\frac{16}{3} Q_u^2 \right ] \bigg \} + y_t \bigg \{ -88 y_t^3 a_t -
20 y_b^3 a_b - 10 y_t y_b \left ( y_b a_t + y_t a_b \right ) \nonumber \\
& {} - 2 y_b y_\tau \left ( y_b a_\tau + y_\tau a_b \right ) -
2 \lambda a_\lambda \left ( 2 \lambda^2 + 3 y_t^2 + 4 y_b^2 + y_\tau^2 +
2 \Sigma_\lambda + 3 \Sigma_\kappa \right ) \nonumber \\
& {} - 2 \lambda^2 \left ( 3 y_t a_t + 4 y_b a_b + y_\tau a_\tau +
2 \Sigma_{a_\lambda} + 3 \Sigma_{a_\kappa} \right ) +
32 g_3^2 y_t \left ( a_t - y_t M_3 \right ) \nonumber \\
& {} + 12 g_2^2 y_t \left ( a_t - y_t M_2 \right ) + \frac{2}{5} g_1^2 \left [
6 y_t a_t + 2 y_b a_b - \left ( 6 y_t^2 + 2 y_b^2 \right ) M_1 \right ]
\nonumber \\
& {} + 4 g_1'^2 \big [ \lambda \left ( Q_1^2 - Q_2^2 + Q_S^2 \right )
\left ( a_\lambda - \lambda M_1' \right ) + 2 y_t \left ( 2 Q_Q^2 + Q_u^2
\right ) \left ( a_t - y_t M_1' \right ) \nonumber \\
& {} + y_b \left ( Q_1^2 - Q_Q^2 + Q_d^2 \right ) \left ( a_b - y_b M_1'
\right ) \big ] - \frac{512}{9} g_3^4 M_3 - 66 g_2^4 M_2 -
\frac{7826}{225} g_1^4 M_1 \nonumber \\
& {} - 8 g_1'^4 M_1' \left [ 2 \left ( Q_2^4 + Q_Q^4 + Q_u^4 \right ) +
\left ( Q_2^2 + Q_Q^2 + Q_u^2 \right ) \Sigma_{Q} \right ] -
16 g_3^2 g_2^2 \left ( M_3 + M_2 \right ) \nonumber \\
& {} - \frac{272}{45} g_3^2 g_1^2 \left ( M_3 + M_1 \right ) -
\frac{64}{3} g_3^2 g_1'^2 \left ( Q_Q^2 + Q_u^2 \right )
\left ( M_3 + M_1' \right ) - 2 g_2^2 g_1^2 \left ( M_2 + M_1 \right )
\nonumber \\
& {} - 12 g_2^2 g_1'^2 \left ( Q_2^2 + Q_Q^2 \right ) \left ( M_2 + M_1'
\right ) - \frac{4}{15} g_1^2 g_1'^2 \left ( 9 Q_2^2 + Q_Q^2 + 16 Q_u^2
\right ) \left ( M_1 + M_1' \right ) \bigg \} ,
\label{eq:USSMTYu22BetaTwoLoop} \\
b_{a_t}^{(2)} &= y_t \bigg [ 2 \lambda a_\lambda + 12 y_t a_t + 2 y_b a_b +
\frac{32}{3} g_3^2 M_3 + 6 g_2^2 M_2 + \frac{26}{15} g_1^2 M_1 \nonumber \\
& {} + 4 g_1'^2 M_1' \left ( Q_2^2 + Q_Q^2 + Q_u^2 \right ) \bigg ] \times
\bigg [ \lambda^2 + 6 y_t^2 + y_b^2 - \frac{16}{3} g_3^2 - 3 g_2^2 -
\frac{13}{15} g_1^2 \nonumber \\
& {} - 2 g_1'^2 \left ( Q_2^2 + Q_Q^2 + Q_u^2 \right ) \bigg ] +
a_t \bigg \{ 54 y_t^4 + \frac{13}{2} y_b^4 + 13 y_t^2 y_b^2 + y_b^2 y_\tau^2
\nonumber \\
& {} + \lambda^2 \left ( \frac{5}{2} \lambda^2 + 15 y_t^2 + 5 y_b^2 + y_\tau^2
+ 2 \Sigma_\lambda + 3 \Sigma_\kappa \right ) -
\frac{16}{3} g_3^2 \left ( \lambda^2 + 2 y_b^2 + 12 y_t^2 \right ) \nonumber \\
& {} - 6 g_2^2 \left ( \lambda^2 + y_b^2 + 6 y_t^2 \right ) - g_1^2 \left (
\frac{22}{15} \lambda^2 + \frac{4}{3} y_b^2 + \frac{52}{5} y_t^2 \right )
\nonumber \\
& {} - 2 g_1'^2 \left [ \lambda^2 \left ( Q_1^2 + 2 Q_2^2 + Q_S^2 + Q_Q^2 +
Q_u^2 \right ) + y_b^2 \left ( Q_1^2 + Q_2^2 + 2 Q_Q^2 + Q_u^2 + Q_d^2
\right ) \right . \nonumber \\
& {} \left . + 12 y_t^2 \left ( Q_2^2 + Q_Q^2 + Q_u^2 \right ) \right ] +
\frac{128}{9} g_3^4 - \frac{15}{2} g_2^4 - \frac{143}{18} g_1^4 \nonumber \\
& {} + 2 g_1'^4 \left ( Q_2^2 + Q_Q^2 + Q_u^2 \right ) \left ( Q_2^2 + Q_Q^2
+ Q_u^2 - \Sigma_{Q} \right ) + 16 g_3^2 g_2^2 + \frac{208}{45} g_3^2 g_1^2
\nonumber \\
& {} + \frac{32}{3} g_3^2 g_1'^2 \left ( Q_2^2 + Q_Q^2 + Q_u^2 \right ) +
\frac{13}{5} g_2^2 g_1^2 + 6 g_2^2 g_1'^2 \left ( Q_2^2 + Q_Q^2 + Q_u^2
\right ) \nonumber \\
& {} + \frac{26}{15} g_1^2 g_1'^2 \left ( Q_2^2 + Q_Q^2 + Q_u^2 \right )
\bigg \} + y_t \bigg [ 144 y_t^3 a_t + 24 y_b^3 a_b + 14 y_t y_b \left (
y_t a_b + a_t y_b \right ) \nonumber \\
& {} + 2 y_b y_\tau \left ( y_b a_\tau + a_b y_\tau \right ) +
18 \lambda y_t \left ( \lambda a_t + a_{\lambda} y_t \right ) +
8 \lambda y_b \left ( \lambda a_b + a_{\lambda} y_b \right ) +
2 \lambda y_\tau \left ( \lambda a_\tau + a_{\lambda} y_\tau \right )
\nonumber \\
& {} + 8 \lambda^3 a_{\lambda} + 4 \lambda \sum_{i=1}^3 \lambda_i \left (
\lambda a_{\lambda_i} + a_{\lambda} \lambda_i \right ) +
6 \lambda \sum_{i=1}^3 \kappa_i \left ( \lambda a_{\kappa_i} +
a_{\lambda} \kappa_i \right ) - 64 g_3^2 y_t \left ( a_t - y_t M_3 \right )
\nonumber \\
& {} - \frac{32}{3} g_3^2 y_b \left ( a_b - y_b M_3 \right ) -
36 g_2^2 y_t \left ( a_t - y_t M_2 \right ) - 6 g_2^2 y_b \left ( a_b -
y_b M_2 \right ) - 6 g_2^2 \lambda \left ( a_\lambda - \lambda M_2 \right )
\nonumber \\
& {} - \frac{52}{5} g_1^2 y_t \left ( a_t - y_t M_1 \right ) -
\frac{14}{15} g_1^2 y_b \left ( a_b - y_b M_1 \right ) -
\frac{18}{15} g_1^2 \lambda \left ( a_\lambda - \lambda M_1 \right )
\nonumber \\
& {} - 24 g_1'^2 y_t \left ( Q_2^2 + Q_Q^2 + Q_u^2 \right ) \left ( a_t -
y_t M_1' \right ) - 4 g_1'^2 y_b \left ( Q_1^2 + Q_Q^2 + Q_d^2 \right )
\left ( a_b - y_b M_1' \right ) \nonumber \\
& {} - 4 g_1'^2 \lambda \left ( Q_1^2 + Q_2^2 + Q_S^2 \right )
\left ( a_\lambda - \lambda M_1' \right ) + 48 g_2^4 M_2 +
\frac{832}{25} g_1^4 M_1 \nonumber \\
& {} + 8 g_1'^4 M_1' \Sigma_Q \left ( Q_2^2 + Q_Q^2 + Q_u^2 \right ) \bigg ] .
\label{eq:USSMTYu22Ot2Coeff}
\end{align}
\end{subequations}
The one- and two-loop $\beta$ functions and the resulting $O(t^2)$ coefficient
for $m_{Q_3}^2$ are
\begin{subequations}
\begin{align}
\beta_{m_{Q_3}^2}^{(1)} &= 2 y_t^2 \left ( m_{H_u}^2 + m_{Q_3}^2 + m_{u_3}^2
\right ) + 2 a_t^2 + 2 y_b^2 \left ( m_{H_d}^2 + m_{Q_3}^2 + m_{d_3}^2
\right ) + 2 a_b^2 \nonumber \\
& {} - \frac{32}{3} g_3^2 M_3^2 - 6 g_2^2 M_2^2 - \frac{2}{15} g_1^2 M_1^2 -
8 Q_Q^2 g_1'^2 M_1'^2 + \frac{1}{5} g_1^2 \Sigma_1 + 2 Q_Q g_1'^2 \Sigma_1' ,
\label{eq:USSMmq222BetaOneLoop} \\
\beta_{m_{Q_3}^2}^{(2)} &= -20 y_t^4 \left ( m_{H_u}^2 + m_{Q_3}^2 + m_{u_3}^2
 \right ) - 20 y_b^4 \left ( m_{H_d}^2 + m_{Q_3}^2 + m_{d_3}^2 \right )
\nonumber \\
& {} - 2 y_\tau^2 y_b^2 \left ( 2 m_{H_d}^2 + m_{Q_3}^2 + m_{L_3}^2 +
m_{d_3}^2 + m_{e_3}^2 \right ) - 40 y_t^2 a_t^2 - 40 y_b^2 a_b^2 \nonumber \\
& {} - 2 \left ( y_b a_\tau + y_\tau a_b \right )^2 - 2 \lambda^2 y_t^2
\left ( 2 m_{H_u}^2 + m_{H_d}^2 + m_S^2 + m_{Q_3}^2 + m_{u_3}^2 \right )
\nonumber \\
& {} - 2 \lambda^2 y_b^2 \left ( m_{H_u}^2 + 2 m_{H_d}^2 + m_S^2 + m_{Q_3}^2
+ m_{d_3}^2 \right ) - 2 \left ( \lambda a_t + y_t a_\lambda \right )^2 -
2 \left ( \lambda a_b + y_b a_\lambda \right )^2 \nonumber \\
& {} + \frac{2}{5} g_1^2 y_t^2 \left ( m_{H_u}^2 + 3m_{Q_3}^2 + 8 m_{u_3}^2
+ 8 M_1^2 \right ) + \frac{8}{5} g_1^2 \left ( a_t^2 - 2 y_t a_t M_1 \right )
\nonumber \\
& {} + \frac{2}{5} g_1^2 y_b^2 \left ( 5 m_{H_d}^2 + m_{Q_3}^2 + 4 M_1^2
\right ) + \frac{4}{5} g_1^2 \left ( a_b^2 - 2 y_b a_b M_1 \right ) +
\frac{2}{5} g_1^2 y_\tau^2 \left ( m_{H_d}^2 + m_{L_3}^2 - 2 m_{e_3}^2
\right ) \nonumber \\
& {} + \frac{2}{5} g_1^2 \sum_{i=1}^3 \left [ \kappa_i^2 \left ( m_{D_i}^2 -
m_{\overline{D}_i}^2 \right ) + \lambda_i^2 \left ( m_{H_i^d}^2 - m_{H_i^u}^2
\right ) \right ] \nonumber \\
& {} + 4 g_1'^2 y_t^2 \left ( Q_2^2 - Q_Q^2 + Q_u^2 \right )
\left ( m_{H_u}^2 + m_{Q_3}^2 + m_{u_3}^2 + 2 M_1'^2 \right ) \nonumber \\
& {} + 4 g_1'^2 \left ( Q_2^2 - Q_Q^2 + Q_u^2 \right ) \left ( a_t^2 -
2 y_t a_t M_1' \right ) - 24 Q_Q g_1'^2 y_t^2 \left ( Q_2 m_{H_u}^2 +
Q_Q m_{Q_3}^2 + Q_um_{u_3}^2 \right ) \nonumber \\
& {} + 4 g_1'^2 y_b^2 \left ( Q_1^2 - Q_Q^2 + Q_d^2 \right ) \left ( m_{H_d}^2
+ m_{Q_3}^2 + m_{d_3}^2 + 2 M_1'^2 \right ) \nonumber \\
& {} + 4 g_1'^2 \left ( Q_1^2 - Q_Q^2 + Q_d^2 \right ) \left ( a_b^2 -
2 y_b a_b M_1' \right ) - 24 Q_Q g_1'^2 y_b^2 \left ( Q_1 m_{H_d}^2 +
Q_Q m_{Q_3}^2 + Q_d m_{d_3}^2 \right ) \nonumber \\
& {} - 8 Q_Q g_1'^2 y_\tau^2 \left ( Q_1 m_{H_d}^2 + Q_L m_{L_3}^2 +
Q_e m_{e_3}^2 \right ) \nonumber \\
& {} - 4 Q_Q g_1'^2 \sum_{i=1}^3 \left [ 3 \kappa_i^2 \left ( Q_S m_S^2 +
Q_D m_{D_i}^2 + Q_{\overline{D}} m_{\overline{D}_i}^2 \right ) +
2 \lambda_i^2 \left ( Q_1 m_{H_i^d}^2 + Q_2 m_{H_i^u}^2 + Q_S m_S^2 \right )
\right ] \nonumber \\
& {} + \frac{16}{3} g_3^4 \left [ 10 M_3^2 + \sum_{i=1}^3 \left ( 2 m_{Q_i}^2
+ m_{u_i}^2 + m_{d_i}^2 + m_{D_i}^2 + m_{\overline{D}_i}^2 \right ) \right ]
+ 32 g_3^2 g_2^2 \left ( M_2^2 + M_2 M_3 + M_3^2 \right ) \nonumber \\
& {} + \frac{16}{45} g_3^2 g_1^2 \left [ 2 \left ( M_1^2 + M_1 M_3 + M_3^2
\right ) + 3 \sum_{i=1}^3 \left ( m_{Q_i}^2 + m_{d_i}^2 - 2 m_{u_i}^2 -
m_{D_i}^2 + m_{\overline{D}_i}^2 \right ) \right ] \nonumber \\
& {} + \frac{32}{3} Q_Q g_3^2 g_1'^2 \bigg [ 4 Q_Q \left ( M_1'^2 + M_1' M_3
+ M_3^2 \right ) + 3 \sum_{i=1}^3 \big ( 2 Q_Q m_{Q_i}^2 + Q_u m_{u_i}^2 +
Q_d m_{d_i}^2 \nonumber \\
& {} + Q_D m_{D_i}^2 + Q_{\overline{D}} m_{\overline{D}_i}^2 \big ) \bigg ]
+ 3 g_2^4 \left [ 29 M_2^2 + m_{\overline{H'}}^2 + m_{H'}^2 +
\sum_{i=1}^3 \left ( 3 m_{Q_i}^2 + m_{L_i}^2 + m_{H_i^d}^2 + m_{H_i^u}^2
\right ) \right ] \nonumber \\
& {} + \frac{1}{5} g_2^2 g_1^2 \left [ 2 \left ( M_1^2 + M_1 M_2 + M_2^2
\right ) + 3 m_{\overline{H'}}^2 - 3 m_{H'}^2 +
3 \sum_{i=1}^3 \left ( m_{Q_i}^2 - m_{L_i}^2 + m_{H_i^u}^2 - m_{H_i^d}^2
\right ) \right ] \nonumber \\
& {} + 12 Q_Q g_2^2 g_1'^2 \bigg [ 2 Q_Q \left ( M_1'^2 + M_1' M_2 + M_2^2
\right ) + Q_{\overline{H'}} m_{\overline{H'}}^2 + Q_{H'} m_{H'}^2 \nonumber \\
& {} + \sum_{i=1}^3 \left ( 3 Q_Q m_{Q_i}^2 + Q_L m_{L_i}^2 + Q_1 m_{H_i^d}^2
+ Q_2 m_{H_i^u}^2 \right ) \bigg ] + \frac{1}{75} g_1^4 \bigg [ 289 M_1^2 +
12 m_{\overline{H'}}^2 - 6 m_{H'}^2 \nonumber \\
& {} + \sum_{i=1}^3 \left ( 6 m_{d_i}^2 - 2 m_{D_i}^2 + 6 m_{\overline{D}_i}^2
+ 42 m_{e_i}^2 - 6 m_{H_i^d}^2 + 12 m_{H_i^u}^2 - 6 m_{L_i}^2 + 2 m_{Q_i}^2
- 24 m_{u_i}^2 \right ) \bigg ] \nonumber \\
& {} + \frac{4}{15} g_1^2 g_1'^2 \bigg [ 2 Q_Q \big ( 9 Q_d +
9 Q_{\overline{D}} - 9 Q_D + 9 Q_e - 9 Q_1 + 9 Q_2 + 3 Q_{\overline{H'}} -
3 Q_{H'} - 9 Q_L \nonumber \\
& {} + 10 Q_Q - 18 Q_u \big ) \left ( M_1^2 + M_1 M_1' + M_1'^2 \right ) +
3 Q_{\overline{H'}}^2 m_{\overline{H'}}^2 - 3 Q_{H'}^2 m_{H'}^2 \nonumber \\
& {} + 3 \sum_{i=1}^3 \big ( Q_d^2 m_{d_i}^2 +
Q_{\overline{D}}^2 m_{\overline{D}_i}^2 - Q_D^2 m_{D_i}^2 + Q_e^2 m_{e_i}^2 -
Q_1^2 m_{H_i^d}^2 + Q_2^2 m_{H_i^u}^2 \nonumber \\
& {} - Q_L^2 m_{L_i}^2 - Q_Q^2 m_{Q_i}^2 - 2 Q_u^2 m_{u_i}^2 \big ) +
15 Q_Q Q_{\overline{H'}} m_{\overline{H'}}^2 + 3 Q_Q Q_{H'} m_{H'}^2
\nonumber \\
& {} + Q_Q \sum_{i=1}^3 \big ( 12 Q_d m_{d_i}^2 +
12 Q_{\overline{D}} m_{\overline{D}_i}^2 + 24 Q_e m_{e_i}^2 +
3 Q_1 m_{H_i^d}^2 + 15 Q_2 m_{H_i^u}^2 + 3 Q_L m_{L_i}^2 \nonumber \\
& {} + 15 Q_Q m_{Q_i}^2 + 12 m_{u_i}^2 \big ) \bigg ] + 8 Q_Q g_1'^4 \bigg [
3 Q_Q M_1'^2 \big ( 9 Q_d^2 + 9 Q_{\overline{D}}^2 + 9 Q_D^2 + 3 Q_e^2 +
6 Q_1^2 \nonumber \\
& {} + 6 Q_2^2 + 2 Q_{\overline{H'}}^2 + 2 Q_{H'}^2 + 6 Q_L^2 + 20 Q_Q^2 +
3 Q_S^2 + 9 Q_u^2 \big ) + 2 Q_{\overline{H'}}^3 m_{\overline{H'}}^2 +
2 Q_{H'}^3 m_{H'}^2 \nonumber \\
& {} + \sum_{i=1}^3 \big ( 3 Q_d^3 m_{d_i}^2 +
3 Q_{\overline{D}}^3 m_{\overline{D}_i}^2 + 3 Q_D^3 m_{D_i}^2 +
Q_e^3 m_{e_i}^2 + 2 Q_1^3 m_{H_i^d}^2 + 2 Q_2^3 m_{H_i^u}^2 +
2 Q_L^3 m_{L_i}^2 \nonumber \\
& {} + 6 Q_Q^3 m_{Q_i}^2 + Q_S^3 m_{S_i}^2 + 3 Q_u^3 m_{u_i}^2 \big ) +
2 Q_Q Q_{\overline{H'}}^2 m_{\overline{H'}}^2 + 2 Q_Q Q_{H'}^2 m_{H'}^2
\nonumber \\
& {} + Q_Q \sum_{i=1}^3 \big ( 3 Q_d^2 m_{d_i}^2 +
3 Q_{\overline{D}}^2 m_{\overline{D}_i}^2 + 3 Q_D^2 m_{D_i}^2 +
Q_e^2 m_{e_i}^2 + 2 Q_1^2 m_{H_i^d}^2 + 2 Q_2^2 m_{H_i^u}^2 \nonumber \\
& {} + 2 Q_L^2 m_{L_i}^2 + 6 Q_Q^2 m_{Q_i}^2 + Q_S^2 m_{S_i}^2 +
3 Q_u^2 m_{u_i}^2 \big ) \bigg ] , \label{eq:USSMmq222BetaTwoLoop} \\
b_{m_{Q_3}^2}^{(2)} &= 24 y_t^4 \left ( m_{H_u^2} + m_{Q_3}^2 + m_{u_3}^2
\right ) + 24 y_b^4 \left ( m_{H_d}^2 + m_{Q_3}^2 + m_{d_3}^2 \right ) +
48 y_t^2 a_t^2 + 48 y_b^2 a_b^2 \nonumber \\
& {} + 4 y_t^2 y_b^2 \left ( m_{H_u}^2 + m_{H_d}^2 + 2 m_{Q_3}^2 + m_{u_3}^2
+ m_{d_3}^2 \right ) + 4 \left ( y_t a_b + y_b a_t \right )^2 \nonumber \\
& {} + 2 y_b^2 y_\tau^2 \left ( 2 m_{H_d}^2 + m_{Q_3}^2 + m_{d_3}^2 +
m_{L_3}^2 + m_{e_3}^2 \right ) + 2 \left ( y_\tau a_b + y_b a_\tau \right )^2
\nonumber \\
& {} + 2 \lambda^2 y_t^2 \left ( m_{H_d}^2 + 2 m_{H_u}^2 + m_S^2 + m_{Q_3}^2
+ m_{u_3}^2 \right ) + 2 \left ( \lambda a_t + y_t a_\lambda \right )^2
\nonumber \\
& {} + 2 \lambda^2 y_b^2 \left ( 2 m_{H_d}^2 + m_{H_u}^2 + m_S^2 + m_{Q_3}^2
+ m_{d_3}^2 \right ) + 2 \left ( \lambda a_b + y_b a_\lambda \right )^2
\nonumber \\
& {} - \frac{32}{3} g_3^2 y_t^2 \left ( m_{H_u}^2 + m_{Q_3}^2 + m_{u_3}^2 +
2 M_3^2 \right ) - \frac{32}{3} g_3^2 \left ( a_t^2 - 2 y_t a_t M_3 \right )
\nonumber \\
& {} - \frac{32}{3} g_3^2 y_b^2 \left ( m_{H_d}^2 + m_{Q_3}^2 + m_{d_3}^2 +
2 M_3^2 \right ) - \frac{32}{3} g_3^2 \left ( a_b^2 - 2 y_b a_b M_3 \right )
\nonumber \\
& {} - 6 g_2^2 y_t^2 \left ( m_{H_u}^2 + m_{Q_3}^2 + m_{u_3}^2 + 2 M_2^2
\right ) - 6 g_2^2 \left ( a_t^2 - 2 y_t a_t M_2 \right ) \nonumber \\
& {} - 6 g_2^2 y_b^2 \left ( m_{H_d}^2 + m_{Q_3}^2 + m_{d_3}^2 + 2 M_2^2
\right ) - 6 g_2^2 \left ( a_b^2 - 2 y_b a_b M_2 \right ) \nonumber \\
& {} - \frac{26}{15} g_1^2 y_t^2 \left ( m_{H_u}^2 + m_{Q_3}^2 + m_{u_3}^2 +
2 M_1^2 \right ) - \frac{26}{15} g_1^2 \left ( a_t^2 - 2 y_t a_t M_1 \right )
\nonumber \\
& {} - \frac{14}{15} g_1^2 y_b^2 \left ( m_{H_d}^2 + m_{Q_3}^2 + m_{d_3}^2 +
2 M_1^2 \right ) - \frac{14}{15} g_1^2 \left ( a_b^2 - 2 y_b a_b M_1 \right )
\nonumber \\
& {} - 4 g_1'^2 y_t^2 \left ( Q_2^2 + Q_Q^2 + Q_u^2 \right ) \left ( m_{H_u}^2
+ m_{Q_3}^2 + m_{u_3}^2 + 2 M_1'^2 \right ) - 4 g_1'^2 \left ( Q_2^2 + Q_Q^2
+ Q_u^2 \right ) \nonumber \\
& {} \times \left ( a_t^2 - 2 y_t a_t M_1' \right ) + 2 g_1'^2 y_t^2 \left (
Q_2 + Q_Q + Q_u \right ) \left ( 6 Q_Q m_{H_u}^2 + 6 Q_Q m_{Q_3}^2 +
6 Q_Q m_{u_3}^2 + \Sigma_1' \right ) \nonumber \\
& {} + 12 Q_Q g_1'^2 a_t^2 \left ( Q_2 + Q_Q + Q_u \right ) -
4 g_1'^2 y_b^2 \left ( Q_1^2 + Q_Q^2 + Q_d^2 \right ) \left ( m_{H_d}^2 +
m_{Q_3}^2 + m_{d_3}^2 + 2 M_1'^2 \right ) \nonumber \\
& {} - 4 g_1'^2 \left ( Q_1^2 + Q_Q^2 + Q_d^2 \right ) \left ( a_b^2 -
2 y_b a_b M_1' \right ) + 2 g_1'^2 y_b^2 \left ( Q_1 + Q_Q + Q_d \right )
\nonumber \\
& {} \times \left ( 6 Q_Q m_{H_d}^2 + 6 Q_Q m_{Q_3}^2 + 6 Q_Q m_{d_3}^2 +
\Sigma_1' \right ) + 12 Q_Q g_1'^2 a_b^2 \left ( Q_1 + Q_Q + Q_d \right )
\nonumber \\
& {} + 4 Q_Q g_1'^2 y_\tau^2 \left ( Q_1 + Q_L + Q_e \right ) \left (
m_{H_d}^2 + m_{L_3}^2 + m_{e_3}^2 \right ) + 4 Q_Q g_1'^2 a_\tau^2 \left (
Q_1 + Q_L + Q_e \right ) \nonumber \\
& {} + 2 Q_Q g_1'^2 \sum_{i=1}^3 \Big [ 2 \lambda_i^2 \left ( Q_1 + Q_2 + Q_S
\right ) \left ( m_{H_i^d}^2 + m_{H_i^u}^2 + m_S^2 \right ) +
2 a_{\lambda_i}^2 \left ( Q_1 + Q_2 + Q_S \right ) \nonumber \\
& {} + 3 \kappa_i^2 \left ( Q_S + Q_D + Q_{\overline{D}} \right ) \left (
m_S^2 + m_{D_i}^2 + m_{\overline{D}_i}^2 \right ) + 3 a_{\kappa_i}^2 \left (
Q_S + Q_D + Q_{\overline{D}} \right ) \Big ] \nonumber \\
& {} - 96 Q_Q g_3^2 g_1'^2 M_3^2 \left ( 2 Q_Q + Q_u + Q_d + Q_D +
Q_{\overline{D}} \right ) - 72 g_2^4 M_2^2 \nonumber \\
& {} - 12 Q_Q g_2^2 g_1'^2 M_2^2 \left ( 9 Q_Q + 3 Q_L + 3 Q_1 + 3 Q_2 +
Q_{\overline{H'}} + Q_{H'} \right ) + \frac{96}{25} g_1^4 \left ( \Sigma_1 -
M_1^2 \right ) \nonumber \\
& {} - \frac{1}{5} g_1^2 g_1'^2 \big [ 12 Q_Q M_1^2 \big ( 2 Q_d +
2 Q_{\overline{D}} + 2 Q_D + 6 Q_e + 3 Q_1 + 3 Q_2 + Q_{\overline{H'}} +
Q_{H'} + 3 Q_L \nonumber \\
& {} + Q_Q + 8 Q_u \big ) + 4 M_1'^2 \big ( 3 Q_d^2 + 3 Q_{\overline{D}}^2 -
3 Q_D^2 + 3 Q_e^2 - 3 Q_1^2 + 3 Q_2^2 + Q_{\overline{H'}}^2 - Q_{H'}^2 -
3 Q_L^2 \nonumber \\
& {} + 3 Q_Q^2 - 6 Q_u^2 \big ) - \left ( 6 Q_Q \Sigma_1 + \Sigma_1' \right )
\Sigma_Q^Y \big ] - 4 Q_Q g_1'^4 \big [ 2 M_1'^2 \big ( 9 Q_d^3 +
9 Q_{\overline{D}}^3 + 9 Q_D^3 + 3 Q_e^3 + 6 Q_1^3 \nonumber \\
& {} + 6 Q_2^3 + 2 Q_{\overline{H'}}^3 + 2 Q_{H'}^3 + 6 Q_L^3 + 18 Q_Q^3 +
3 Q_S^3 + 9 Q_u^3 \big ) + \left ( 6 Q_Q M_1'^2 - \Sigma_1' \right )
\Sigma_Q \big ] . \label{eq:USSMmq222Ot2Coeff}
\end{align}
\end{subequations}
Finally, the relevant expressions for the soft mass $m_{u_3}^2$ read
\begin{subequations}
\begin{align}
\beta_{m_{u_3}^2}^{(1)} &= 4 y_t^2 \left ( m_{H_u}^2 + m_{Q_3}^2 + m_{u_3}^2
\right ) + 4 a_t^2 - \frac{32}{3} g_3^2 M_3^2 - \frac{32}{15} g_1^2 M_1^2
- 8 Q_u^2 g_1'^2 M_1'^2 \nonumber \\
& {} - \frac{4}{5} g_1^2 \Sigma_1 + 2 Q_u g_1'^2 \Sigma_1' ,
\label{eq:USSMmu222BetaOneLoop} \\
\beta_{m_{u_3}^2}^{(2)} &= -32 y_t^4 \left ( m_{H_u}^2 + m_{Q_3}^2 + m_{u_3}^2
\right ) - 4 y_t^2 y_b^2 \left ( m_{H_u}^2 + m_{H_d}^2 + 2 m_{Q_3}^2 +
m_{u_3}^2 + m_{d_3}^2 \right ) \nonumber \\
& {} - 64 y_t^2 a_t^2 -4 \left ( y_t a_b + y_b a_t \right )^2 -
4 \lambda^2 y_t^2 \left ( 2 m_{H_u}^2 + m_{H_d}^2 + m_S^2 + m_{Q_3}^2 +
m_{u_3}^2 \right ) \nonumber \\
& {} - 4 \left ( \lambda a_t + y_t a_\lambda \right )^2 + 12 g_2^2 y_t^2
\left ( m_{H_u}^2 + m_{Q_3}^2 + m_{u_3}^2 + 2 M_2^2 \right ) + 12 g_2^2
\left ( a_t^2 - 2 y_t a_t M_2 \right ) \nonumber \\
& {} - \frac{4}{5} g_1^2 y_t^2 \left ( -5 m_{H_u}^2 - m_{Q_3}^2 + 9 m_{u_3}^2
+ 2 M_1^2 \right ) - \frac{4}{5} g_1^2 \left ( a_t^2 - 2 y_t a_t M_1 \right )
\nonumber \\
& {} - \frac{8}{5} g_1^2 y_b^2 \left ( 3 m_{H_d}^2 - m_{Q_3}^2 - 2 m_{d_3}^2
\right ) - \frac{8}{5} g_1^2 y_\tau^2 \left ( m_{H_d}^2 + m_{L_3}^2 -
2 m_{e_3}^2 \right ) \nonumber \\
& {} - \frac{8}{5} g_1^2 \sum_{i=1}^3 \left [ \kappa_i^2 \left ( m_{D_i}^2 -
m_{\overline{D}_i}^2 \right ) + \lambda_i^2 \left ( m_{H_i^d}^2 - m_{H_i^u}^2
\right ) \right ] \nonumber \\
& {} + 8 g_1'^2 y_t^2 \left ( Q_2^2 + Q_Q^2 - Q_u^2 \right ) \left ( m_{H_u}^2
+ m_{Q_3}^2 + m_{u_3}^2 + 2 M_1'^2 \right ) \nonumber \\
& {} + 8 g_1'^2 \left ( Q_2^2 + Q_Q^2 - Q_u^2 \right ) \left ( a_t^2 -
2 y_t a_t M_1' \right ) - 24 Q_u g_1'^2 y_t^2 \left ( Q_2 m_{H_u}^2 +
Q_Q m_{Q_3}^2 + Q_u m_{u_3}^2 \right ) \nonumber \\
& {} - 24 Q_u g_1'^2 y_b^2 \left ( Q_1 m_{H_d}^2 + Q_Q m_{Q_3}^2 +
Q_d m_{d_3}^2 \right ) - 8 Q_u g_1'^2 y_\tau^2 \left ( Q_1 m_{H_d}^2 +
Q_L m_{L_3}^2 + Q_e m_{e_3}^2 \right ) \nonumber \\
& {} - 12 Q_u g_1'^2 \sum_{i=1}^3 \kappa_i^2 \left ( Q_S m_S^2 + Q_D m_{D_i}^2
+ Q_{\overline{D}} m_{\overline{D}_i}^2 \right ) \nonumber \\
& {} - 8 Q_u g_1'^2 \sum_{i=1}^3 \lambda_i^2 \left ( Q_1 m_{H_d}^2 +
Q_2 m_{H_u}^2 + Q_S m_S^2 \right ) \nonumber \\
& {} + \frac{16}{3} g_3^4 \left [ 10 M_3^2 + \sum_{i=1}^3 \left ( m_{d_i}^2
+ m_{D_i}^2 + m_{\overline{D}_i}^2 + 2 m_{Q_i}^2 + m_{u_i}^2 \right )
\right ] \nonumber \\
& {} + \frac{64}{45} g_3^2 g_1^2 \left [ 8 \left ( M_1^2 + M_1 M_3 + M_3^2
\right ) - 3 \sum_{i=1}^3 \left ( m_{d_i}^2 - m_{D_i}^2 +
m_{\overline{D}_i}^2 + m_{Q_i}^2 - 2 m_{u_i}^2 \right ) \right ] \nonumber \\
& {} + \frac{8}{3} Q_u g_3^2 g_1'^2 \bigg [ 16 Q_u \left ( M_1'^2 + M_1' M_3 +
M_3^2 \right ) \nonumber \\
& {} + 12 \sum_{i=1}^3 \left ( Q_d m_{d_i}^2 + Q_D m_{D_i}^2 +
Q_{\overline{D}} m_{\overline{D}_i}^2 + 2 Q_Q m_{Q_i}^2 + Q_u m_{u_i}^2
\right ) \bigg ] \nonumber \\
& {} + \frac{12}{5} g_2^2 g_1^2 \left [ m_{H'}^2 - m_{\overline{H'}}^2 +
\sum_{i=1}^3 \left ( m_{H_i^d}^2 - m_{H_i^u}^2 + m_{L_i}^2 - m_{Q_i}^2
\right ) \right ] \nonumber \\
& {} + 12 Q_u g_2^2 g_1'^2 \left [ Q_{H'} m_{H'}^2 +
Q_{\overline{H'}} m_{\overline{H'}}^2 + \sum_{i=1}^3 \left ( Q_1 m_{H_i^d}^2
+ Q_2 m_{H_i^u}^2 + Q_L m_{L_i}^2 + 3 Q_Q m_{Q_i}^2 \right ) \right ]
\nonumber \\
& {} + \frac{4}{75} g_1^4 \bigg [ 1261 M_1^2 + 21 m_{H'}^2 +
3 m_{\overline{H'}}^2 \nonumber \\
& {} + \sum_{i=1}^3 \left ( 4 m_{d_i}^2 + 12 m_{D_i}^2 +
4 m_{\overline{D}_i}^2 - 12 m_{e_i}^2 + 21 m_{H_i^d}^2 + 3 m_{H_i^u}^2 +
21 m_{L_i}^2 + 3 m_{Q_i}^2 + 64 m_{u_i}^2 \right ) \bigg ] \nonumber \\
& {} + \frac{4}{15} g_1^2g_1'^2 \bigg [ -8 Q_u \big ( 9 Q_d +
9 Q_{\overline{D}} - 9 Q_D + 9 Q_e - 9 Q_1 + 9 Q_2 + 3Q_{\overline{H'}} -
3 Q_{H'} - 9 Q_L \nonumber \\
& {} + 9 Q_Q - 22 Q_u \big ) \left ( M_1^2 + M_1 M_1' + M_1'^2 \right ) +
12 Q_{H'}^2 m_{H'}^2 - 12 Q_{\overline{H'}}^2 m_{\overline{H'}}^2 \nonumber \\
& {} + 12 \sum_{i=1}^3 \big ( -Q_d^2 m_{d_i}^2 -
Q_{\overline{D}}^2 m_{\overline{D}_i}^2 + Q_D^2 m_{D_i}^2 - Q_e^2 m_{e_i}^2 +
Q_1^2 m_{H_i^d}^2 - Q_2^2 m_{H_i^u}^2 + Q_L^2 m_{L_i}^2 \nonumber \\
& {} - Q_Q^2 m_{Q_i}^2 + Q_u^2 m_{u_i}^2 \big ) -
15 Q_{\overline{H'}} Q_u m_{\overline{H'}}^2 + 33 Q_{H'} Q_u m_{H'}^2 +
Q_u \sum_{i=1}^3 \big ( -18 Q_d m_{d_i}^2 \nonumber \\
& {} - 18 Q_{\overline{D}} m_{\overline{D}_i}^2 + 30 Q_D m_{D_i}^2 -
6 Q_e m_{e_i}^2 + 33 Q_1 m_{H_i^d}^2 - 15 Q_2 m_{H_i^u}^2 + 33 Q_L m_{L_i}^2
\nonumber \\
& {} - 21 Q_Q m_{Q_i}^2 + 84 Q_u m_{u_i}^2 \big ) \bigg ] +
8 Q_u g_1'^4 \bigg [ 3 Q_u M_1'^2 \big ( 9 Q_d^2 + 9 Q_{\overline{D}}^2 +
9 Q_D^2 + 3 Q_e^2 + 6 Q_1^2 \nonumber \\
& {} + 6 Q_2^2 + 2 Q_{\overline{H'}}^2 + 2 Q_{H'}^2 + 6 Q_L^2 + 18 Q_Q^2 +
3 Q_S^2 + 11 Q_u^2 \big ) + 2 Q_{H'}^3 m_{H'}^2 +
2 Q_{\overline{H'}}^3 m_{\overline{H'}}^2 \nonumber \\
& {} + \sum_{i=1}^3 \big ( 3 Q_d^3 m_{d_i}^2 +
3 Q_{\overline{D}}^3 m_{\overline{D}_i}^2 + 3 Q_D^3 m_{D_i}^2 +
Q_e^3 m_{e_i}^2 + 2 Q_1^3 m_{H_i^d}^2 + 2 Q_2^3 m_{H_i^u}^2 +
2 Q_L^3 m_{L_i}^2 \nonumber \\
& {} + 6 Q_Q^3 m_{Q_i}^2 + Q_S^3 m_{S_i}^2 + 3 Q_u^3 m_{u_i}^2 \big ) +
2 Q_u Q_{H'}^2 m_{H'}^2 + 2 Q_u Q_{\overline{H'}}^2 m_{\overline{H'}}^2
\nonumber \\
& {} + Q_u \sum_{i=1}^3 \big ( 3 Q_d^2 m_{d_i}^2 +
3 Q_{\overline{D}}^2 m_{\overline{D}_i}^2 + 3 Q_D^2 m_{D_i}^2 +
Q_e^2 m_{e_i}^2 + 2 Q_1^2 m_{H_i^d}^2 + 2 Q_2^2 m_{H_i^u}^2 +
2 Q_L^2 m_{L_i}^2 \nonumber \\
& {} + 6 Q_Q^2 m_{Q_i}^2 + Q_S^2 m_{S_i}^2 + 3 Q_u^2 m_{u_i}^2 \big )
\bigg ] , \label{eq:USSMmu222BetaTwoLoop} \\
b_{m_{u_3}^2}^{(2)} &= 48 y_t^4 \left ( m_{H_u}^2 + m_{Q_3}^2 + m_{u_3}^2
\right ) + 4 y_t^2 y_b^2 \left ( m_{H_u}^2 + m_{H_d}^2 + 2 m_{Q_3}^2 +
m_{u_3}^2 + m_{d_3}^2 \right ) + 96 y_t^2 a_t^2 \nonumber \\
& {} + 4 \left ( y_t a_b + y_b a_t \right )^2 + 4 \lambda^2 y_t^2 \left (
2 m_{H_u}^2 + m_{H_d}^2 + m_S^2 + m_{Q_3}^2 + m_{u_3}^2 \right ) +
4 \left ( \lambda a_t + y_t a_\lambda \right )^2 \nonumber \\
& {} - \frac{64}{3} g_3^2 y_t^2 \left ( m_{H_u}^2 + m_{Q_3}^2 + m_{u_3}^2 +
2 M_3^2 \right ) - \frac{64}{3} g_3^2 \left ( a_t^2 - 2 y_t a_t M_3 \right )
\nonumber \\
& {} - 12 g_2^2 y_t^2 \left ( m_{H_u}^2 + m_{Q_3}^2 + m_{u_3}^2 + 2 M_2^2
\right ) - 12 g_2^2 \left ( a_t^2 - 2 y_t a_t M_2 \right ) \nonumber \\
& {} - \frac{52}{15} g_1^2 y_t^2 \left ( m_{H_u}^2 + m_{Q_3}^2 + m_{u_3}^2 +
2 M_1^2 \right ) - \frac{52}{15} g_1^2 \left ( a_t^2 - 2 y_t a_t M_1 \right )
\nonumber \\
& {} - 8 g_1'^2 y_t^2 \left ( Q_2^2 + Q_Q^2 + Q_u^2 \right ) \left ( m_{H_u}^2
+ m_{Q_3}^2 + m_{u_3}^2 + 2 M_1'^2 \right ) \nonumber \\
& {} - 8 g_1'^2 \left ( Q_2^2 + Q_Q^2 + Q_u^2 \right ) \left ( a_t^2 -
2 y_t a_t M_1' \right ) \nonumber \\
& {} + 4 g_1'^2 y_t^2 \left ( Q_2 + Q_Q + Q_u \right ) \left ( 3 Q_u m_{H_u}^2
+ 3 Q_u m_{Q_3}^2 + 3 Q_u m_{u_3}^2 + \Sigma_1' \right ) \nonumber \\
& {} + 12 Q_u g_1'^2 a_t^2 \left ( Q_2 + Q_Q + Q_u \right ) +
12 Q_u g_1'^2 y_b^2 \left ( Q_1 + Q_Q + Q_d \right ) \left ( m_{H_d}^2 +
m_{Q_3}^2 + m_{d_3}^2 \right ) \nonumber \\
& {} + 12 Q_u g_1'^2 a_b^2 \left ( Q_1 + Q_Q + Q_d \right ) \nonumber \\
& {} + 4 Q_u g_1'^2 y_\tau^2 \left ( Q_1 + Q_L + Q_e \right ) \left (
m_{H_d}^2 + m_{L_3}^2 + m_{e_3}^2 \right ) + 4 Q_u g_1'^2 a_\tau^2 \left (
Q_1 + Q_L + Q_e \right ) \nonumber \\
& {} + 2 Q_u g_1'^2 \sum_{i=1}^3 \bigg [ 3 \kappa_i^2 \left ( Q_S + Q_D +
Q_{\overline{D}} \right ) \left ( m_S^2 + m_{D_i}^2 + m_{\overline{D}_i}^2
\right ) + 3 a_{\kappa_i}^2 \left ( Q_S + Q_D + Q_{\overline{D}} \right )
\nonumber \\
& {} + 2 \lambda_i^2 \left ( Q_1 + Q_2 + Q_S \right ) \left ( m_{H_i^d}^2 +
m_{H_i^u}^2 + m_S^2 \right ) + 2 a_{\lambda_i}^2 \left ( Q_1 + Q_2 + Q_S
\right ) \bigg ] \nonumber \\
& {} - 96 Q_u g_3^2 g_1'^2 M_3^2 \left ( 2 Q_Q + Q_u + Q_d + Q_D +
Q_{\overline{D}} \right ) \nonumber \\
& {} - 12 Q_u g_2^2 g_1'^2 M_2^2 \left ( 9 Q_Q + 3 Q_L + 3 Q_1 + 3 Q_2 +
Q_{H'} + Q_{\overline{H'}} \right ) - \frac{384}{25} g_1^4 \left ( \Sigma_1 +
4 M_1^2 \right ) \nonumber \\
& {} + \frac{2}{5} g_1^2 g_1'^2 \big [ 8 M_1'^2 ( 3 Q_d^2 +
3 Q_{\overline{D}}^2 - 3 Q_D^2 + 3 Q_e^2 - 3 Q_1^2 + 3 Q_2^2 +
Q_{\overline{H'}}^2 - Q_{H'}^2 - 3 Q_L^2 \nonumber \\
& {} + 3 Q_Q^2 - 6 Q_u^2 ) + 6 Q_u M_1^2 ( 2 Q_d + 2 Q_{\overline{D}} + 2 Q_D
+ 6 Q_e + 3 Q_1 + 3 Q_2 + Q_{\overline{H'}} + Q_{H'} \nonumber \\
& {} + 3 Q_L + Q_Q + 8 Q_u ) + \left ( 3 Q_u \Sigma_1 - 2 \Sigma_1^\prime
\right ) \Sigma_Q^Y \big ] - 4 Q_u g_1'^4 \big [ 2 M_1'^2 ( 9 Q_d^3 + 9 Q_D^3
+ 9 Q_{\overline{D}}^3 \nonumber \\
& {} + 3 Q_e^3 + 6 Q_1^3 + 6 Q_2^3 + 2 Q_{\overline{H'}}^3 + 2 Q_{H'}^3 +
6 Q_L^3 + 18 Q_Q^3 + 3 Q_S^3 + 9 Q_u^3 ) \nonumber \\
& {} + \left ( 6 Q_u M_1'^2 - \Sigma_1^\prime \right ) \Sigma_Q \big ] .
\label{eq:USSMmu222Ot2Coeff}
\end{align}
\end{subequations}

\bibliographystyle{apsrev4-1}
\bibliography{bibliography}{}

\begin{thebibliography}{160}%
\makeatletter
\providecommand \@ifxundefined [1]{%
 \@ifx{#1\undefined}
}%
\providecommand \@ifnum [1]{%
 \ifnum #1\expandafter \@firstoftwo
 \else \expandafter \@secondoftwo
 \fi
}%
\providecommand \@ifx [1]{%
 \ifx #1\expandafter \@firstoftwo
 \else \expandafter \@secondoftwo
 \fi
}%
\providecommand \natexlab [1]{#1}%
\providecommand \enquote  [1]{``#1''}%
\providecommand \bibnamefont  [1]{#1}%
\providecommand \bibfnamefont [1]{#1}%
\providecommand \citenamefont [1]{#1}%
\providecommand \href@noop [0]{\@secondoftwo}%
\providecommand \href [0]{\begingroup \@sanitize@url \@href}%
\providecommand \@href[1]{\@@startlink{#1}\@@href}%
\providecommand \@@href[1]{\endgroup#1\@@endlink}%
\providecommand \@sanitize@url [0]{\catcode `\\12\catcode `\$12\catcode
  `\&12\catcode `\#12\catcode `\^12\catcode `\_12\catcode `\%12\relax}%
\providecommand \@@startlink[1]{}%
\providecommand \@@endlink[0]{}%
\providecommand \url  [0]{\begingroup\@sanitize@url \@url }%
\providecommand \@url [1]{\endgroup\@href {#1}{\urlprefix }}%
\providecommand \urlprefix  [0]{URL }%
\providecommand \Eprint [0]{\href }%
\providecommand \doibase [0]{http://dx.doi.org/}%
\providecommand \selectlanguage [0]{\@gobble}%
\providecommand \bibinfo  [0]{\@secondoftwo}%
\providecommand \bibfield  [0]{\@secondoftwo}%
\providecommand \translation [1]{[#1]}%
\providecommand \BibitemOpen [0]{}%
\providecommand \bibitemStop [0]{}%
\providecommand \bibitemNoStop [0]{.\EOS\space}%
\providecommand \EOS [0]{\spacefactor3000\relax}%
\providecommand \BibitemShut  [1]{\csname bibitem#1\endcsname}%
\let\auto@bib@innerbib\@empty
%</preamble>
\bibitem [{\citenamefont {Aad}\ \emph {et~al.}(2012)\citenamefont {Aad} \emph
  {et~al.}}]{Aad:2012tfa}%
  \BibitemOpen
  \bibfield  {author} {\bibinfo {author} {\bibfnamefont {G.}~\bibnamefont
  {Aad}} \emph {et~al.} (\bibinfo {collaboration} {ATLAS Collaboration}),\
  }\href {\doibase 10.1016/j.physletb.2012.08.020} {\bibfield  {journal}
  {\bibinfo  {journal} {Phys. Lett. B}\ }\textbf {\bibinfo {volume} {716}},\
  \bibinfo {pages} {1} (\bibinfo {year} {2012})},\ \Eprint
  {http://arxiv.org/abs/1207.7214} {arXiv:1207.7214 [hep-ex]} \BibitemShut
  {NoStop}%
%%CITATION = ARXIV:1207.7214;%%
\bibitem [{\citenamefont {Chatrchyan}\ \emph {et~al.}(2012)\citenamefont
  {Chatrchyan} \emph {et~al.}}]{Chatrchyan:2012ufa}%
  \BibitemOpen
  \bibfield  {author} {\bibinfo {author} {\bibfnamefont {S.}~\bibnamefont
  {Chatrchyan}} \emph {et~al.} (\bibinfo {collaboration} {CMS Collaboration}),\
  }\href {\doibase 10.1016/j.physletb.2012.08.021} {\bibfield  {journal}
  {\bibinfo  {journal} {Phys. Lett. B}\ }\textbf {\bibinfo {volume} {716}},\
  \bibinfo {pages} {30} (\bibinfo {year} {2012})},\ \Eprint
  {http://arxiv.org/abs/1207.7235} {arXiv:1207.7235 [hep-ex]} \BibitemShut
  {NoStop}%
%%CITATION = ARXIV:1207.7235;%%
\bibitem [{\citenamefont {Cassel}\ and\ \citenamefont
  {Ghilencea}(2012)}]{Cassel:2011zd}%
  \BibitemOpen
  \bibfield  {author} {\bibinfo {author} {\bibfnamefont {S.}~\bibnamefont
  {Cassel}}\ and\ \bibinfo {author} {\bibfnamefont {D.~M.}\ \bibnamefont
  {Ghilencea}},\ }\href {\doibase 10.1142/S0217732312300030} {\bibfield
  {journal} {\bibinfo  {journal} {Mod. Phys. Lett. A}\ }\textbf {\bibinfo
  {volume} {27}},\ \bibinfo {pages} {1230003} (\bibinfo {year} {2012})},\
  \Eprint {http://arxiv.org/abs/1103.4793} {arXiv:1103.4793 [hep-ph]}
  \BibitemShut {NoStop}%
%%CITATION = ARXIV:1103.4793;%%
\bibitem [{\citenamefont {Ghilencea}\ \emph {et~al.}(2012)\citenamefont
  {Ghilencea}, \citenamefont {Lee},\ and\ \citenamefont
  {Park}}]{Ghilencea:2012gz}%
  \BibitemOpen
  \bibfield  {author} {\bibinfo {author} {\bibfnamefont {D.~M.}\ \bibnamefont
  {Ghilencea}}, \bibinfo {author} {\bibfnamefont {H.~M.}\ \bibnamefont {Lee}},
  \ and\ \bibinfo {author} {\bibfnamefont {M.}~\bibnamefont {Park}},\ }\href
  {\doibase 10.1007/JHEP07(2012)046} {\bibfield  {journal} {\bibinfo  {journal}
  {J. High Energy Phys.}\ }\textbf {\bibinfo {volume} {1207}},\ \bibinfo
  {pages} {046} (\bibinfo {year} {2012})},\ \Eprint
  {http://arxiv.org/abs/1203.0569} {arXiv:1203.0569 [hep-ph]} \BibitemShut
  {NoStop}%
%%CITATION = ARXIV:1203.0569;%%
\bibitem [{\citenamefont {Fayet}(1977)}]{Fayet:1977yc}%
  \BibitemOpen
  \bibfield  {author} {\bibinfo {author} {\bibfnamefont {P.}~\bibnamefont
  {Fayet}},\ }\href {\doibase 10.1016/0370-2693(77)90852-8} {\bibfield
  {journal} {\bibinfo  {journal} {Phys. Lett. B}\ }\textbf {\bibinfo {volume}
  {69}},\ \bibinfo {pages} {489} (\bibinfo {year} {1977})}\BibitemShut
  {NoStop}%
%%CITATION = PHLTA,B69,489;%%
\bibitem [{\citenamefont {Kim}\ and\ \citenamefont
  {Nilles}(1984)}]{Kim:1983dt}%
  \BibitemOpen
  \bibfield  {author} {\bibinfo {author} {\bibfnamefont {J.~E.}\ \bibnamefont
  {Kim}}\ and\ \bibinfo {author} {\bibfnamefont {H.~P.}\ \bibnamefont
  {Nilles}},\ }\href {\doibase 10.1016/0370-2693(84)91890-2} {\bibfield
  {journal} {\bibinfo  {journal} {Phys. Lett. B}\ }\textbf {\bibinfo {volume}
  {138}},\ \bibinfo {pages} {150} (\bibinfo {year} {1984})}\BibitemShut
  {NoStop}%
%%CITATION = PHLTA,B138,150;%%
\bibitem [{\citenamefont {Suematsu}\ and\ \citenamefont
  {Yamagishi}(1995)}]{Suematsu:1994qm}%
  \BibitemOpen
  \bibfield  {author} {\bibinfo {author} {\bibfnamefont {D.}~\bibnamefont
  {Suematsu}}\ and\ \bibinfo {author} {\bibfnamefont {Y.}~\bibnamefont
  {Yamagishi}},\ }\href {\doibase 10.1142/S0217751X95002096} {\bibfield
  {journal} {\bibinfo  {journal} {Int. J. Mod. Phys. A}\ }\textbf {\bibinfo
  {volume} {10}},\ \bibinfo {pages} {4521} (\bibinfo {year} {1995})},\ \Eprint
  {http://arxiv.org/abs/hep-ph/9411239} {arXiv:hep-ph/9411239 [hep-ph]}
  \BibitemShut {NoStop}%
%%CITATION = HEP-PH/9411239;%%
\bibitem [{\citenamefont {Cveti\v{c}}\ and\ \citenamefont
  {Langacker}(1996{\natexlab{a}})}]{Cvetic:1995rj}%
  \BibitemOpen
  \bibfield  {author} {\bibinfo {author} {\bibfnamefont {M.}~\bibnamefont
  {Cveti\v{c}}}\ and\ \bibinfo {author} {\bibfnamefont {P.}~\bibnamefont
  {Langacker}},\ }\href {\doibase 10.1103/PhysRevD.54.3570} {\bibfield
  {journal} {\bibinfo  {journal} {Phys. Rev. D}\ }\textbf {\bibinfo {volume}
  {54}},\ \bibinfo {pages} {3570} (\bibinfo {year} {1996}{\natexlab{a}})},\
  \Eprint {http://arxiv.org/abs/hep-ph/9511378} {arXiv:hep-ph/9511378 [hep-ph]}
  \BibitemShut {NoStop}%
%%CITATION = HEP-PH/9511378;%%
\bibitem [{\citenamefont {Cveti\v{c}}\ and\ \citenamefont
  {Langacker}(1996{\natexlab{b}})}]{Cvetic:1996mf}%
  \BibitemOpen
  \bibfield  {author} {\bibinfo {author} {\bibfnamefont {M.}~\bibnamefont
  {Cveti\v{c}}}\ and\ \bibinfo {author} {\bibfnamefont {P.}~\bibnamefont
  {Langacker}},\ }\href {\doibase 10.1142/S0217732396001260} {\bibfield
  {journal} {\bibinfo  {journal} {Mod. Phys. Lett. A}\ }\textbf {\bibinfo
  {volume} {11}},\ \bibinfo {pages} {1247} (\bibinfo {year}
  {1996}{\natexlab{b}})},\ \Eprint {http://arxiv.org/abs/hep-ph/9602424}
  {arXiv:hep-ph/9602424 [hep-ph]} \BibitemShut {NoStop}%
%%CITATION = HEP-PH/9602424;%%
\bibitem [{\citenamefont {Jain}\ and\ \citenamefont {Shrock}()}]{Jain:1995cb}%
  \BibitemOpen
  \bibfield  {author} {\bibinfo {author} {\bibfnamefont {V.}~\bibnamefont
  {Jain}}\ and\ \bibinfo {author} {\bibfnamefont {R.}~\bibnamefont {Shrock}},\
  }\href@noop {} {\ }\Eprint {http://arxiv.org/abs/hep-ph/9507238}
  {arXiv:hep-ph/9507238 [hep-ph]} \BibitemShut {NoStop}%
%%CITATION = HEP-PH/9507238;%%
\bibitem [{\citenamefont {Nir}(1995)}]{Nir:1995bu}%
  \BibitemOpen
  \bibfield  {author} {\bibinfo {author} {\bibfnamefont {Y.}~\bibnamefont
  {Nir}},\ }\href {\doibase 10.1016/0370-2693(95)00619-V} {\bibfield  {journal}
  {\bibinfo  {journal} {Phys. Lett. B}\ }\textbf {\bibinfo {volume} {354}},\
  \bibinfo {pages} {107} (\bibinfo {year} {1995})},\ \Eprint
  {http://arxiv.org/abs/hep-ph/9504312} {arXiv:hep-ph/9504312 [hep-ph]}
  \BibitemShut {NoStop}%
%%CITATION = HEP-PH/9504312;%%
\bibitem [{\citenamefont {Cveti\v{c}}\ \emph {et~al.}(1997)\citenamefont
  {Cveti\v{c}}, \citenamefont {Demir}, \citenamefont {Espinosa}, \citenamefont
  {Everett},\ and\ \citenamefont {Langacker}}]{Cvetic:1997ky}%
  \BibitemOpen
  \bibfield  {author} {\bibinfo {author} {\bibfnamefont {M.}~\bibnamefont
  {Cveti\v{c}}}, \bibinfo {author} {\bibfnamefont {D.~A.}\ \bibnamefont
  {Demir}}, \bibinfo {author} {\bibfnamefont {J.~R.}\ \bibnamefont {Espinosa}},
  \bibinfo {author} {\bibfnamefont {L.~L.}\ \bibnamefont {Everett}}, \ and\
  \bibinfo {author} {\bibfnamefont {P.}~\bibnamefont {Langacker}},\ }\href
  {\doibase 10.1103/PhysRevD.56.2861} {\bibfield  {journal} {\bibinfo
  {journal} {Phys. Rev. D}\ }\textbf {\bibinfo {volume} {56}},\ \bibinfo
  {pages} {2861} (\bibinfo {year} {1997})},\ \Eprint
  {http://arxiv.org/abs/hep-ph/9703317} {arXiv:hep-ph/9703317 [hep-ph]}
  \BibitemShut {NoStop}%
%%CITATION = HEP-PH/9703317;%%
\bibitem [{\citenamefont {King}\ \emph
  {et~al.}(2006{\natexlab{a}})\citenamefont {King}, \citenamefont {Moretti},\
  and\ \citenamefont {Nevzorov}}]{King:2005jy}%
  \BibitemOpen
  \bibfield  {author} {\bibinfo {author} {\bibfnamefont {S.~F.}\ \bibnamefont
  {King}}, \bibinfo {author} {\bibfnamefont {S.}~\bibnamefont {Moretti}}, \
  and\ \bibinfo {author} {\bibfnamefont {R.}~\bibnamefont {Nevzorov}},\ }\href
  {\doibase 10.1103/PhysRevD.73.035009} {\bibfield  {journal} {\bibinfo
  {journal} {Phys. Rev. D}\ }\textbf {\bibinfo {volume} {73}},\ \bibinfo
  {pages} {035009} (\bibinfo {year} {2006}{\natexlab{a}})},\ \Eprint
  {http://arxiv.org/abs/hep-ph/0510419} {arXiv:hep-ph/0510419 [hep-ph]}
  \BibitemShut {NoStop}%
%%CITATION = HEP-PH/0510419;%%
\bibitem [{\citenamefont {King}\ \emph
  {et~al.}(2006{\natexlab{b}})\citenamefont {King}, \citenamefont {Moretti},\
  and\ \citenamefont {Nevzorov}}]{King:2005my}%
  \BibitemOpen
  \bibfield  {author} {\bibinfo {author} {\bibfnamefont {S.~F.}\ \bibnamefont
  {King}}, \bibinfo {author} {\bibfnamefont {S.}~\bibnamefont {Moretti}}, \
  and\ \bibinfo {author} {\bibfnamefont {R.}~\bibnamefont {Nevzorov}},\ }\href
  {\doibase 10.1016/j.physletb.2005.12.070} {\bibfield  {journal} {\bibinfo
  {journal} {Phys. Lett. B}\ }\textbf {\bibinfo {volume} {634}},\ \bibinfo
  {pages} {278} (\bibinfo {year} {2006}{\natexlab{b}})},\ \Eprint
  {http://arxiv.org/abs/hep-ph/0511256} {arXiv:hep-ph/0511256 [hep-ph]}
  \BibitemShut {NoStop}%
%%CITATION = HEP-PH/0511256;%%
\bibitem [{\citenamefont {Athron}\ \emph {et~al.}(2010)\citenamefont {Athron},
  \citenamefont {Hall}, \citenamefont {Howl}, \citenamefont {King},
  \citenamefont {Miller}, \citenamefont {Moretti},\ and\ \citenamefont
  {Nevzorov}}]{Athron:2010zz}%
  \BibitemOpen
  \bibfield  {author} {\bibinfo {author} {\bibfnamefont {P.}~\bibnamefont
  {Athron}}, \bibinfo {author} {\bibfnamefont {J.~P.}\ \bibnamefont {Hall}},
  \bibinfo {author} {\bibfnamefont {R.}~\bibnamefont {Howl}}, \bibinfo {author}
  {\bibfnamefont {S.~F.}\ \bibnamefont {King}}, \bibinfo {author}
  {\bibfnamefont {D.~J.}\ \bibnamefont {Miller}}, \bibinfo {author}
  {\bibfnamefont {S.}~\bibnamefont {Moretti}}, \ and\ \bibinfo {author}
  {\bibfnamefont {R.}~\bibnamefont {Nevzorov}},\ }\href {\doibase
  10.1016/j.nuclphysbps.2010.02.074} {\bibfield  {journal} {\bibinfo  {journal}
  {Nucl. Phys. Proc. Suppl.}\ }\textbf {\bibinfo {volume} {200-202}},\ \bibinfo
  {pages} {120} (\bibinfo {year} {2010})}\BibitemShut {NoStop}%
%%CITATION = NUPHZ,200-202,120;%%
\bibitem [{\citenamefont {Athron}\ \emph {et~al.}(2013)\citenamefont {Athron},
  \citenamefont {Binjonaid},\ and\ \citenamefont {King}}]{Athron:2013ipa}%
  \BibitemOpen
  \bibfield  {author} {\bibinfo {author} {\bibfnamefont {P.}~\bibnamefont
  {Athron}}, \bibinfo {author} {\bibfnamefont {M.}~\bibnamefont {Binjonaid}}, \
  and\ \bibinfo {author} {\bibfnamefont {S.~F.}\ \bibnamefont {King}},\ }\href
  {\doibase 10.1103/PhysRevD.87.115023} {\bibfield  {journal} {\bibinfo
  {journal} {Phys. Rev. D}\ }\textbf {\bibinfo {volume} {87}},\ \bibinfo
  {pages} {115023} (\bibinfo {year} {2013})},\ \Eprint
  {http://arxiv.org/abs/1302.5291} {arXiv:1302.5291 [hep-ph]} \BibitemShut
  {NoStop}%
%%CITATION = ARXIV:1302.5291;%%
\bibitem [{\citenamefont {Athron}\ \emph
  {et~al.}(2009{\natexlab{a}})\citenamefont {Athron}, \citenamefont {King},
  \citenamefont {Miller}, \citenamefont {Moretti},\ and\ \citenamefont
  {Nevzorov}}]{Athron:2009ue}%
  \BibitemOpen
  \bibfield  {author} {\bibinfo {author} {\bibfnamefont {P.}~\bibnamefont
  {Athron}}, \bibinfo {author} {\bibfnamefont {S.~F.}\ \bibnamefont {King}},
  \bibinfo {author} {\bibfnamefont {D.~J.}\ \bibnamefont {Miller}}, \bibinfo
  {author} {\bibfnamefont {S.}~\bibnamefont {Moretti}}, \ and\ \bibinfo
  {author} {\bibfnamefont {R.}~\bibnamefont {Nevzorov}},\ }\href {\doibase
  10.1016/j.physletb.2009.10.051} {\bibfield  {journal} {\bibinfo  {journal}
  {Phys. Lett. B}\ }\textbf {\bibinfo {volume} {681}},\ \bibinfo {pages} {448}
  (\bibinfo {year} {2009}{\natexlab{a}})},\ \Eprint
  {http://arxiv.org/abs/0901.1192} {arXiv:0901.1192 [hep-ph]} \BibitemShut
  {NoStop}%
%%CITATION = ARXIV:0901.1192;%%
\bibitem [{\citenamefont {Athron}\ \emph
  {et~al.}(2009{\natexlab{b}})\citenamefont {Athron}, \citenamefont {King},
  \citenamefont {Miller}, \citenamefont {Moretti},\ and\ \citenamefont
  {Nevzorov}}]{Athron:2009bs}%
  \BibitemOpen
  \bibfield  {author} {\bibinfo {author} {\bibfnamefont {P.}~\bibnamefont
  {Athron}}, \bibinfo {author} {\bibfnamefont {S.~F.}\ \bibnamefont {King}},
  \bibinfo {author} {\bibfnamefont {D.~J.}\ \bibnamefont {Miller}}, \bibinfo
  {author} {\bibfnamefont {S.}~\bibnamefont {Moretti}}, \ and\ \bibinfo
  {author} {\bibfnamefont {R.}~\bibnamefont {Nevzorov}},\ }\href {\doibase
  10.1103/PhysRevD.80.035009} {\bibfield  {journal} {\bibinfo  {journal} {Phys.
  Rev. D}\ }\textbf {\bibinfo {volume} {80}},\ \bibinfo {pages} {035009}
  (\bibinfo {year} {2009}{\natexlab{b}})},\ \Eprint
  {http://arxiv.org/abs/0904.2169} {arXiv:0904.2169 [hep-ph]} \BibitemShut
  {NoStop}%
%%CITATION = ARXIV:0904.2169;%%
\bibitem [{\citenamefont {Drees}\ \emph {et~al.}(1986)\citenamefont {Drees},
  \citenamefont {Falck},\ and\ \citenamefont {Gluck}}]{Drees:1985js}%
  \BibitemOpen
  \bibfield  {author} {\bibinfo {author} {\bibfnamefont {M.}~\bibnamefont
  {Drees}}, \bibinfo {author} {\bibfnamefont {N.~K.}\ \bibnamefont {Falck}}, \
  and\ \bibinfo {author} {\bibfnamefont {M.}~\bibnamefont {Gluck}},\ }\href
  {\doibase 10.1016/0370-2693(86)90596-4} {\bibfield  {journal} {\bibinfo
  {journal} {Phys. Lett. B}\ }\textbf {\bibinfo {volume} {167}},\ \bibinfo
  {pages} {187} (\bibinfo {year} {1986})}\BibitemShut {NoStop}%
%%CITATION = PHLTA,B167,187;%%
\bibitem [{\citenamefont {Athron}\ \emph
  {et~al.}(2015{\natexlab{a}})\citenamefont {Athron}, \citenamefont
  {M{\"u}hlleitner}, \citenamefont {Nevzorov},\ and\ \citenamefont
  {Williams}}]{Athron:2014pua}%
  \BibitemOpen
  \bibfield  {author} {\bibinfo {author} {\bibfnamefont {P.}~\bibnamefont
  {Athron}}, \bibinfo {author} {\bibfnamefont {M.}~\bibnamefont
  {M{\"u}hlleitner}}, \bibinfo {author} {\bibfnamefont {R.}~\bibnamefont
  {Nevzorov}}, \ and\ \bibinfo {author} {\bibfnamefont {A.~G.}\ \bibnamefont
  {Williams}},\ }\href {\doibase 10.1007/JHEP01(2015)153} {\bibfield  {journal}
  {\bibinfo  {journal} {J. High Energy Phys.}\ }\textbf {\bibinfo {volume}
  {1501}},\ \bibinfo {pages} {153} (\bibinfo {year} {2015}{\natexlab{a}})},\
  \Eprint {http://arxiv.org/abs/1410.6288} {arXiv:1410.6288 [hep-ph]}
  \BibitemShut {NoStop}%
%%CITATION = ARXIV:1410.6288;%%
\bibitem [{\citenamefont {Ellis}\ \emph {et~al.}(1986)\citenamefont {Ellis},
  \citenamefont {Enqvist}, \citenamefont {Nanopoulos},\ and\ \citenamefont
  {Zwirner}}]{Ellis:1986yg}%
  \BibitemOpen
  \bibfield  {author} {\bibinfo {author} {\bibfnamefont {J.~R.}\ \bibnamefont
  {Ellis}}, \bibinfo {author} {\bibfnamefont {K.}~\bibnamefont {Enqvist}},
  \bibinfo {author} {\bibfnamefont {D.~V.}\ \bibnamefont {Nanopoulos}}, \ and\
  \bibinfo {author} {\bibfnamefont {F.}~\bibnamefont {Zwirner}},\ }\href
  {\doibase 10.1142/S0217732386000105} {\bibfield  {journal} {\bibinfo
  {journal} {Mod. Phys. Lett. A}\ }\textbf {\bibinfo {volume} {01}},\ \bibinfo
  {pages} {57} (\bibinfo {year} {1986})}\BibitemShut {NoStop}%
%%CITATION = MPLAE,A1,57;%%
\bibitem [{\citenamefont {Barbieri}\ and\ \citenamefont
  {Giudice}(1988)}]{Barbieri:1987fn}%
  \BibitemOpen
  \bibfield  {author} {\bibinfo {author} {\bibfnamefont {R.}~\bibnamefont
  {Barbieri}}\ and\ \bibinfo {author} {\bibfnamefont {G.~F.}\ \bibnamefont
  {Giudice}},\ }\href {\doibase 10.1142/S0217732386000105} {\bibfield
  {journal} {\bibinfo  {journal} {Nucl. Phys.}\ }\textbf {\bibinfo {volume}
  {B306}},\ \bibinfo {pages} {63} (\bibinfo {year} {1988})}\BibitemShut
  {NoStop}%
%%CITATION = NUPHA,B306,63;%%
\bibitem [{\citenamefont {de~Carlos}\ and\ \citenamefont
  {Casas}(1993)}]{deCarlos:1993yy}%
  \BibitemOpen
  \bibfield  {author} {\bibinfo {author} {\bibfnamefont {B.}~\bibnamefont
  {de~Carlos}}\ and\ \bibinfo {author} {\bibfnamefont {J.~A.}\ \bibnamefont
  {Casas}},\ }\href {\doibase 10.1016/0370-2693(93)90940-J} {\bibfield
  {journal} {\bibinfo  {journal} {Phys. Lett. B}\ }\textbf {\bibinfo {volume}
  {309}},\ \bibinfo {pages} {320} (\bibinfo {year} {1993})},\ \Eprint
  {http://arxiv.org/abs/hep-ph/9303291} {arXiv:hep-ph/9303291 [hep-ph]}
  \BibitemShut {NoStop}%
%%CITATION = HEP-PH/9303291;%%
\bibitem [{\citenamefont {de~Carlos}\ and\ \citenamefont
  {Casas}()}]{deCarlos:1993ca}%
  \BibitemOpen
  \bibfield  {author} {\bibinfo {author} {\bibfnamefont {B.}~\bibnamefont
  {de~Carlos}}\ and\ \bibinfo {author} {\bibfnamefont {J.~A.}\ \bibnamefont
  {Casas}},\ }\href@noop {} {\ }\Eprint {http://arxiv.org/abs/hep-ph/9310232}
  {arXiv:hep-ph/9310232 [hep-ph]} \BibitemShut {NoStop}%
%%CITATION = HEP-PH/9310232;%%
\bibitem [{\citenamefont {Chankowski}\ \emph {et~al.}(1998)\citenamefont
  {Chankowski}, \citenamefont {Ellis},\ and\ \citenamefont
  {Pokorski}}]{Chankowski:1997zh}%
  \BibitemOpen
  \bibfield  {author} {\bibinfo {author} {\bibfnamefont {P.~H.}\ \bibnamefont
  {Chankowski}}, \bibinfo {author} {\bibfnamefont {J.~R.}\ \bibnamefont
  {Ellis}}, \ and\ \bibinfo {author} {\bibfnamefont {S.}~\bibnamefont
  {Pokorski}},\ }\href {\doibase 10.1016/S0370-2693(98)00060-4} {\bibfield
  {journal} {\bibinfo  {journal} {Phys. Lett. B}\ }\textbf {\bibinfo {volume}
  {423}},\ \bibinfo {pages} {327} (\bibinfo {year} {1998})},\ \Eprint
  {http://arxiv.org/abs/hep-ph/9712234} {arXiv:hep-ph/9712234 [hep-ph]}
  \BibitemShut {NoStop}%
%%CITATION = HEP-PH/9712234;%%
\bibitem [{\citenamefont {Agashe}\ and\ \citenamefont
  {Graesser}(1997)}]{Agashe:1997kn}%
  \BibitemOpen
  \bibfield  {author} {\bibinfo {author} {\bibfnamefont {K.}~\bibnamefont
  {Agashe}}\ and\ \bibinfo {author} {\bibfnamefont {M.}~\bibnamefont
  {Graesser}},\ }\href {\doibase 10.1016/S0550-3213(97)00569-5} {\bibfield
  {journal} {\bibinfo  {journal} {Nucl. Phys.}\ }\textbf {\bibinfo {volume}
  {B507}},\ \bibinfo {pages} {3} (\bibinfo {year} {1997})},\ \Eprint
  {http://arxiv.org/abs/hep-ph/9704206} {arXiv:hep-ph/9704206 [hep-ph]}
  \BibitemShut {NoStop}%
%%CITATION = HEP-PH/9704206;%%
\bibitem [{\citenamefont {Wright}()}]{Wright:1998mk}%
  \BibitemOpen
  \bibfield  {author} {\bibinfo {author} {\bibfnamefont {D.}~\bibnamefont
  {Wright}},\ }\href@noop {} {\ }\Eprint {http://arxiv.org/abs/hep-ph/9801449}
  {arXiv:hep-ph/9801449 [hep-ph]} \BibitemShut {NoStop}%
%%CITATION = HEP-PH/9801449;%%
\bibitem [{\citenamefont {Kane}\ and\ \citenamefont
  {King}(1999)}]{Kane:1998im}%
  \BibitemOpen
  \bibfield  {author} {\bibinfo {author} {\bibfnamefont {G.~L.}\ \bibnamefont
  {Kane}}\ and\ \bibinfo {author} {\bibfnamefont {S.~F.}\ \bibnamefont
  {King}},\ }\href {\doibase 10.1016/S0370-2693(99)00190-2} {\bibfield
  {journal} {\bibinfo  {journal} {Phys. Lett. B}\ }\textbf {\bibinfo {volume}
  {451}},\ \bibinfo {pages} {113} (\bibinfo {year} {1999})},\ \Eprint
  {http://arxiv.org/abs/hep-ph/9810374} {arXiv:hep-ph/9810374 [hep-ph]}
  \BibitemShut {NoStop}%
%%CITATION = HEP-PH/9810374;%%
\bibitem [{\citenamefont {Bastero-Gil}\ \emph {et~al.}(2000)\citenamefont
  {Bastero-Gil}, \citenamefont {Kane},\ and\ \citenamefont
  {King}}]{BasteroGil:1999gu}%
  \BibitemOpen
  \bibfield  {author} {\bibinfo {author} {\bibfnamefont {M.}~\bibnamefont
  {Bastero-Gil}}, \bibinfo {author} {\bibfnamefont {G.~L.}\ \bibnamefont
  {Kane}}, \ and\ \bibinfo {author} {\bibfnamefont {S.~F.}\ \bibnamefont
  {King}},\ }\href {\doibase 10.1016/S0370-2693(00)00002-2} {\bibfield
  {journal} {\bibinfo  {journal} {Phys. Lett. B}\ }\textbf {\bibinfo {volume}
  {474}},\ \bibinfo {pages} {103} (\bibinfo {year} {2000})},\ \Eprint
  {http://arxiv.org/abs/hep-ph/9910506} {arXiv:hep-ph/9910506 [hep-ph]}
  \BibitemShut {NoStop}%
%%CITATION = HEP-PH/9910506;%%
\bibitem [{\citenamefont {Feng}\ \emph {et~al.}(2000)\citenamefont {Feng},
  \citenamefont {Matchev},\ and\ \citenamefont {Moroi}}]{Feng:1999zg}%
  \BibitemOpen
  \bibfield  {author} {\bibinfo {author} {\bibfnamefont {J.~L.}\ \bibnamefont
  {Feng}}, \bibinfo {author} {\bibfnamefont {K.~T.}\ \bibnamefont {Matchev}}, \
  and\ \bibinfo {author} {\bibfnamefont {T.}~\bibnamefont {Moroi}},\ }\href
  {\doibase 10.1103/PhysRevD.61.075005} {\bibfield  {journal} {\bibinfo
  {journal} {Phys. Rev. D}\ }\textbf {\bibinfo {volume} {61}},\ \bibinfo
  {pages} {075005} (\bibinfo {year} {2000})},\ \Eprint
  {http://arxiv.org/abs/hep-ph/9909334} {arXiv:hep-ph/9909334 [hep-ph]}
  \BibitemShut {NoStop}%
%%CITATION = HEP-PH/9909334;%%
\bibitem [{\citenamefont {Allanach}\ \emph {et~al.}(2000)\citenamefont
  {Allanach}, \citenamefont {Hetherington}, \citenamefont {Parker},\ and\
  \citenamefont {Webber}}]{Allanach:2000ii}%
  \BibitemOpen
  \bibfield  {author} {\bibinfo {author} {\bibfnamefont {B.~C.}\ \bibnamefont
  {Allanach}}, \bibinfo {author} {\bibfnamefont {J.~P.~J.}\ \bibnamefont
  {Hetherington}}, \bibinfo {author} {\bibfnamefont {M.~A.}\ \bibnamefont
  {Parker}}, \ and\ \bibinfo {author} {\bibfnamefont {B.~R.}\ \bibnamefont
  {Webber}},\ }\href@noop {} {\bibfield  {journal} {\bibinfo  {journal} {J.
  High Energy Phys.}\ }\textbf {\bibinfo {volume} {0008}},\ \bibinfo {pages}
  {017} (\bibinfo {year} {2000})},\ \Eprint
  {http://arxiv.org/abs/hep-ph/0005186} {arXiv:hep-ph/0005186 [hep-ph]}
  \BibitemShut {NoStop}%
%%CITATION = HEP-PH/0005186;%%
\bibitem [{\citenamefont {Derm\'\i\v{s}ek}\ and\ \citenamefont
  {Gunion}(2005)}]{Dermisek:2005ar}%
  \BibitemOpen
  \bibfield  {author} {\bibinfo {author} {\bibfnamefont {R.}~\bibnamefont
  {Derm\'\i\v{s}ek}}\ and\ \bibinfo {author} {\bibfnamefont {J.~F.}\
  \bibnamefont {Gunion}},\ }\href {\doibase 10.1103/PhysRevLett.95.041801}
  {\bibfield  {journal} {\bibinfo  {journal} {Phys. Rev. Lett.}\ }\textbf
  {\bibinfo {volume} {95}},\ \bibinfo {pages} {041801} (\bibinfo {year}
  {2005})},\ \Eprint {http://arxiv.org/abs/hep-ph/0502105}
  {arXiv:hep-ph/0502105 [hep-ph]} \BibitemShut {NoStop}%
%%CITATION = HEP-PH/0502105;%%
\bibitem [{\citenamefont {Barbieri}\ and\ \citenamefont
  {Hall}()}]{Barbieri:2005kf}%
  \BibitemOpen
  \bibfield  {author} {\bibinfo {author} {\bibfnamefont {R.}~\bibnamefont
  {Barbieri}}\ and\ \bibinfo {author} {\bibfnamefont {L.~J.}\ \bibnamefont
  {Hall}},\ }\href@noop {} {\ }\Eprint {http://arxiv.org/abs/hep-ph/0510243}
  {arXiv:hep-ph/0510243 [hep-ph]} \BibitemShut {NoStop}%
%%CITATION = HEP-PH/0510243;%%
\bibitem [{\citenamefont {Allanach}(2006)}]{Allanach:2006jc}%
  \BibitemOpen
  \bibfield  {author} {\bibinfo {author} {\bibfnamefont {B.~C.}\ \bibnamefont
  {Allanach}},\ }\href {\doibase 10.1016/j.physletb.2006.02.052} {\bibfield
  {journal} {\bibinfo  {journal} {Phys. Lett. B}\ }\textbf {\bibinfo {volume}
  {635}},\ \bibinfo {pages} {123} (\bibinfo {year} {2006})},\ \Eprint
  {http://arxiv.org/abs/hep-ph/0601089} {arXiv:hep-ph/0601089 [hep-ph]}
  \BibitemShut {NoStop}%
%%CITATION = HEP-PH/0601089;%%
\bibitem [{\citenamefont {Gripaios}\ and\ \citenamefont
  {West}(2006)}]{Gripaios:2006nn}%
  \BibitemOpen
  \bibfield  {author} {\bibinfo {author} {\bibfnamefont {B.}~\bibnamefont
  {Gripaios}}\ and\ \bibinfo {author} {\bibfnamefont {S.~M.}\ \bibnamefont
  {West}},\ }\href {\doibase 10.1103/PhysRevD.74.075002} {\bibfield  {journal}
  {\bibinfo  {journal} {Phys. Rev. D}\ }\textbf {\bibinfo {volume} {74}},\
  \bibinfo {pages} {075002} (\bibinfo {year} {2006})},\ \Eprint
  {http://arxiv.org/abs/hep-ph/0603229} {arXiv:hep-ph/0603229 [hep-ph]}
  \BibitemShut {NoStop}%
%%CITATION = HEP-PH/0603229;%%
\bibitem [{\citenamefont {Derm\'\i\v{s}ek}\ \emph {et~al.}(2007)\citenamefont
  {Derm\'\i\v{s}ek}, \citenamefont {Gunion},\ and\ \citenamefont
  {McElrath}}]{Dermisek:2006py}%
  \BibitemOpen
  \bibfield  {author} {\bibinfo {author} {\bibfnamefont {R.}~\bibnamefont
  {Derm\'\i\v{s}ek}}, \bibinfo {author} {\bibfnamefont {J.~F.}\ \bibnamefont
  {Gunion}}, \ and\ \bibinfo {author} {\bibfnamefont {B.}~\bibnamefont
  {McElrath}},\ }\href {\doibase 10.1103/PhysRevD.76.051105} {\bibfield
  {journal} {\bibinfo  {journal} {Phys. Rev. D}\ }\textbf {\bibinfo {volume}
  {76}},\ \bibinfo {pages} {051105} (\bibinfo {year} {2007})},\ \Eprint
  {http://arxiv.org/abs/hep-ph/0612031} {arXiv:hep-ph/0612031 [hep-ph]}
  \BibitemShut {NoStop}%
%%CITATION = HEP-PH/0612031;%%
\bibitem [{\citenamefont {Barbieri}\ \emph {et~al.}(2006)\citenamefont
  {Barbieri}, \citenamefont {Hall},\ and\ \citenamefont
  {Rychkov}}]{Barbieri:2006dq}%
  \BibitemOpen
  \bibfield  {author} {\bibinfo {author} {\bibfnamefont {R.}~\bibnamefont
  {Barbieri}}, \bibinfo {author} {\bibfnamefont {L.~J.}\ \bibnamefont {Hall}},
  \ and\ \bibinfo {author} {\bibfnamefont {V.~S.}\ \bibnamefont {Rychkov}},\
  }\href {\doibase 10.1103/PhysRevD.74.015007} {\bibfield  {journal} {\bibinfo
  {journal} {Phys. Rev. D}\ }\textbf {\bibinfo {volume} {74}},\ \bibinfo
  {pages} {015007} (\bibinfo {year} {2006})},\ \Eprint
  {http://arxiv.org/abs/hep-ph/0603188} {arXiv:hep-ph/0603188 [hep-ph]}
  \BibitemShut {NoStop}%
%%CITATION = HEP-PH/0603188;%%
\bibitem [{\citenamefont {Kobayashi}\ \emph {et~al.}(2006)\citenamefont
  {Kobayashi}, \citenamefont {Terao},\ and\ \citenamefont
  {Tsuchiya}}]{Kobayashi:2006fh}%
  \BibitemOpen
  \bibfield  {author} {\bibinfo {author} {\bibfnamefont {T.}~\bibnamefont
  {Kobayashi}}, \bibinfo {author} {\bibfnamefont {H.}~\bibnamefont {Terao}}, \
  and\ \bibinfo {author} {\bibfnamefont {A.}~\bibnamefont {Tsuchiya}},\ }\href
  {\doibase 10.1103/PhysRevD.74.015002} {\bibfield  {journal} {\bibinfo
  {journal} {Phys. Rev. D}\ }\textbf {\bibinfo {volume} {74}},\ \bibinfo
  {pages} {015002} (\bibinfo {year} {2006})},\ \Eprint
  {http://arxiv.org/abs/hep-ph/0604091} {arXiv:hep-ph/0604091 [hep-ph]}
  \BibitemShut {NoStop}%
%%CITATION = HEP-PH/0604091;%%
\bibitem [{\citenamefont {Perelstein}\ and\ \citenamefont
  {Shakya}(2013)}]{Perelstein:2012qg}%
  \BibitemOpen
  \bibfield  {author} {\bibinfo {author} {\bibfnamefont {M.}~\bibnamefont
  {Perelstein}}\ and\ \bibinfo {author} {\bibfnamefont {B.}~\bibnamefont
  {Shakya}},\ }\href {\doibase 10.1103/PhysRevD.88.075003} {\bibfield
  {journal} {\bibinfo  {journal} {Phys. Rev. D}\ }\textbf {\bibinfo {volume}
  {88}},\ \bibinfo {pages} {075003} (\bibinfo {year} {2013})},\ \Eprint
  {http://arxiv.org/abs/1208.0833} {arXiv:1208.0833 [hep-ph]} \BibitemShut
  {NoStop}%
%%CITATION = ARXIV:1208.0833;%%
\bibitem [{\citenamefont {Antusch}\ \emph {et~al.}(2013)\citenamefont
  {Antusch}, \citenamefont {Calibbi}, \citenamefont {Maurer}, \citenamefont
  {Monaco},\ and\ \citenamefont {Spinrath}}]{Antusch:2012gv}%
  \BibitemOpen
  \bibfield  {author} {\bibinfo {author} {\bibfnamefont {S.}~\bibnamefont
  {Antusch}}, \bibinfo {author} {\bibfnamefont {L.}~\bibnamefont {Calibbi}},
  \bibinfo {author} {\bibfnamefont {V.}~\bibnamefont {Maurer}}, \bibinfo
  {author} {\bibfnamefont {M.}~\bibnamefont {Monaco}}, \ and\ \bibinfo {author}
  {\bibfnamefont {M.}~\bibnamefont {Spinrath}},\ }\href {\doibase
  10.1007/JHEP01(2013)187} {\bibfield  {journal} {\bibinfo  {journal} {J. High
  Energy Phys.}\ }\textbf {\bibinfo {volume} {1301}},\ \bibinfo {pages} {187}
  (\bibinfo {year} {2013})},\ \Eprint {http://arxiv.org/abs/1207.7236}
  {arXiv:1207.7236} \BibitemShut {NoStop}%
%%CITATION = ARXIV:1207.7236;%%
\bibitem [{\citenamefont {Cheng}\ \emph {et~al.}()\citenamefont {Cheng},
  \citenamefont {Li}, \citenamefont {Li}, \citenamefont {Wan}, \citenamefont
  {Wang} \emph {et~al.}}]{Cheng:2012pe}%
  \BibitemOpen
  \bibfield  {author} {\bibinfo {author} {\bibfnamefont {T.}~\bibnamefont
  {Cheng}}, \bibinfo {author} {\bibfnamefont {J.}~\bibnamefont {Li}}, \bibinfo
  {author} {\bibfnamefont {T.}~\bibnamefont {Li}}, \bibinfo {author}
  {\bibfnamefont {X.}~\bibnamefont {Wan}}, \bibinfo {author} {\bibfnamefont
  {Y.~k.}\ \bibnamefont {Wang}},  \emph {et~al.},\ }\href@noop {} {\ }\Eprint
  {http://arxiv.org/abs/1207.6392} {arXiv:1207.6392 [hep-ph]} \BibitemShut
  {NoStop}%
%%CITATION = ARXIV:1207.6392;%%
\bibitem [{\citenamefont {Cahill-Rowley}\ \emph {et~al.}(2012)\citenamefont
  {Cahill-Rowley}, \citenamefont {Hewett}, \citenamefont {Ismail},\ and\
  \citenamefont {Rizzo}}]{CahillRowley:2012rv}%
  \BibitemOpen
  \bibfield  {author} {\bibinfo {author} {\bibfnamefont {M.~W.}\ \bibnamefont
  {Cahill-Rowley}}, \bibinfo {author} {\bibfnamefont {J.~L.}\ \bibnamefont
  {Hewett}}, \bibinfo {author} {\bibfnamefont {A.}~\bibnamefont {Ismail}}, \
  and\ \bibinfo {author} {\bibfnamefont {T.~G.}\ \bibnamefont {Rizzo}},\ }\href
  {\doibase 10.1103/PhysRevD.86.075015} {\bibfield  {journal} {\bibinfo
  {journal} {Phys. Rev. D}\ }\textbf {\bibinfo {volume} {86}},\ \bibinfo
  {pages} {075015} (\bibinfo {year} {2012})},\ \Eprint
  {http://arxiv.org/abs/1206.5800} {arXiv:1206.5800 [hep-ph]} \BibitemShut
  {NoStop}%
%%CITATION = ARXIV:1206.5800;%%
\bibitem [{\citenamefont {Ross}\ \emph {et~al.}(2012)\citenamefont {Ross},
  \citenamefont {Schmidt-Hoberg},\ and\ \citenamefont {Staub}}]{Ross:2012nr}%
  \BibitemOpen
  \bibfield  {author} {\bibinfo {author} {\bibfnamefont {G.~G.}\ \bibnamefont
  {Ross}}, \bibinfo {author} {\bibfnamefont {K.}~\bibnamefont
  {Schmidt-Hoberg}}, \ and\ \bibinfo {author} {\bibfnamefont {F.}~\bibnamefont
  {Staub}},\ }\href {\doibase 10.1007/JHEP08(2012)074} {\bibfield  {journal}
  {\bibinfo  {journal} {J. High Energy Phys.}\ }\textbf {\bibinfo {volume}
  {1208}},\ \bibinfo {pages} {074} (\bibinfo {year} {2012})},\ \Eprint
  {http://arxiv.org/abs/1205.1509} {arXiv:1205.1509 [hep-ph]} \BibitemShut
  {NoStop}%
%%CITATION = ARXIV:1205.1509;%%
\bibitem [{\citenamefont {Basak}\ and\ \citenamefont
  {Mohanty}(2012)}]{Basak:2012bd}%
  \BibitemOpen
  \bibfield  {author} {\bibinfo {author} {\bibfnamefont {T.}~\bibnamefont
  {Basak}}\ and\ \bibinfo {author} {\bibfnamefont {S.}~\bibnamefont
  {Mohanty}},\ }\href {\doibase 10.1103/PhysRevD.86.075031} {\bibfield
  {journal} {\bibinfo  {journal} {Phys. Rev. D}\ }\textbf {\bibinfo {volume}
  {86}},\ \bibinfo {pages} {075031} (\bibinfo {year} {2012})},\ \Eprint
  {http://arxiv.org/abs/1204.6592} {arXiv:1204.6592 [hep-ph]} \BibitemShut
  {NoStop}%
%%CITATION = ARXIV:1204.6592;%%
\bibitem [{\citenamefont {Kang}\ \emph {et~al.}(2012)\citenamefont {Kang},
  \citenamefont {Li},\ and\ \citenamefont {Li}}]{Kang:2012sy}%
  \BibitemOpen
  \bibfield  {author} {\bibinfo {author} {\bibfnamefont {Z.}~\bibnamefont
  {Kang}}, \bibinfo {author} {\bibfnamefont {J.}~\bibnamefont {Li}}, \ and\
  \bibinfo {author} {\bibfnamefont {T.}~\bibnamefont {Li}},\ }\href {\doibase
  10.1007/JHEP11(2012)024} {\bibfield  {journal} {\bibinfo  {journal} {J. High
  Energy Phys.}\ }\textbf {\bibinfo {volume} {1211}},\ \bibinfo {pages} {024}
  (\bibinfo {year} {2012})},\ \Eprint {http://arxiv.org/abs/1201.5305}
  {arXiv:1201.5305 [hep-ph]} \BibitemShut {NoStop}%
%%CITATION = ARXIV:1201.5305;%%
\bibitem [{\citenamefont {Boehm}\ \emph {et~al.}(2013)\citenamefont {Boehm},
  \citenamefont {Dev}, \citenamefont {Mazumdar},\ and\ \citenamefont
  {Pukartas}}]{Boehm:2013qva}%
  \BibitemOpen
  \bibfield  {author} {\bibinfo {author} {\bibfnamefont {C.}~\bibnamefont
  {Boehm}}, \bibinfo {author} {\bibfnamefont {P.~S.~B.}\ \bibnamefont {Dev}},
  \bibinfo {author} {\bibfnamefont {A.}~\bibnamefont {Mazumdar}}, \ and\
  \bibinfo {author} {\bibfnamefont {E.}~\bibnamefont {Pukartas}},\ }\href
  {\doibase 10.1007/JHEP06(2013)113} {\bibfield  {journal} {\bibinfo  {journal}
  {J. High Energy Phys.}\ }\textbf {\bibinfo {volume} {1306}},\ \bibinfo
  {pages} {113} (\bibinfo {year} {2013})},\ \Eprint
  {http://arxiv.org/abs/1303.5386} {arXiv:1303.5386 [hep-ph]} \BibitemShut
  {NoStop}%
%%CITATION = ARXIV:1303.5386;%%
\bibitem [{\citenamefont {Miller}\ and\ \citenamefont
  {Morais}(2013)}]{Miller:2013jra}%
  \BibitemOpen
  \bibfield  {author} {\bibinfo {author} {\bibfnamefont {D.~J.}\ \bibnamefont
  {Miller}}\ and\ \bibinfo {author} {\bibfnamefont {A.~P.}\ \bibnamefont
  {Morais}},\ }\href {\doibase 10.1007/JHEP10(2013)226} {\bibfield  {journal}
  {\bibinfo  {journal} {J. High Energy Phys.}\ }\textbf {\bibinfo {volume}
  {1310}},\ \bibinfo {pages} {226} (\bibinfo {year} {2013})},\ \Eprint
  {http://arxiv.org/abs/1307.1373} {arXiv:1307.1373 [hep-ph]} \BibitemShut
  {NoStop}%
%%CITATION = ARXIV:1307.1373;%%
\bibitem [{\citenamefont {Binjonaid}\ and\ \citenamefont
  {King}(2014)}]{Binjonaid:2014oga}%
  \BibitemOpen
  \bibfield  {author} {\bibinfo {author} {\bibfnamefont {M.~Y.}\ \bibnamefont
  {Binjonaid}}\ and\ \bibinfo {author} {\bibfnamefont {S.~F.}\ \bibnamefont
  {King}},\ }\href {\doibase 10.1103/PhysRevD.90.055020} {\bibfield  {journal}
  {\bibinfo  {journal} {Phys. Rev. D}\ }\textbf {\bibinfo {volume} {90}},\
  \bibinfo {pages} {055020} (\bibinfo {year} {2014})},\ \Eprint
  {http://arxiv.org/abs/1403.2088} {arXiv:1403.2088 [hep-ph]} \BibitemShut
  {NoStop}%
%%CITATION = ARXIV:1403.2088;%%
\bibitem [{\citenamefont {Miller}\ and\ \citenamefont
  {Morais}(2014)}]{Miller:2014jza}%
  \BibitemOpen
  \bibfield  {author} {\bibinfo {author} {\bibfnamefont {D.~J.}\ \bibnamefont
  {Miller}}\ and\ \bibinfo {author} {\bibfnamefont {A.~P.}\ \bibnamefont
  {Morais}},\ }\href {\doibase 10.1007/JHEP12(2014)132} {\bibfield  {journal}
  {\bibinfo  {journal} {J. High Energy Phys.}\ }\textbf {\bibinfo {volume}
  {1412}},\ \bibinfo {pages} {132} (\bibinfo {year} {2014})},\ \Eprint
  {http://arxiv.org/abs/1408.3013} {arXiv:1408.3013 [hep-ph]} \BibitemShut
  {NoStop}%
%%CITATION = ARXIV:1408.3013;%%
\bibitem [{\citenamefont {Anderson}\ and\ \citenamefont
  {Castano}(1995{\natexlab{a}})}]{Anderson:1994dz}%
  \BibitemOpen
  \bibfield  {author} {\bibinfo {author} {\bibfnamefont {G.~W.}\ \bibnamefont
  {Anderson}}\ and\ \bibinfo {author} {\bibfnamefont {D.~J.}\ \bibnamefont
  {Castano}},\ }\href {\doibase 10.1016/0370-2693(95)00051-L} {\bibfield
  {journal} {\bibinfo  {journal} {Phys. Lett. B}\ }\textbf {\bibinfo {volume}
  {347}},\ \bibinfo {pages} {300} (\bibinfo {year} {1995}{\natexlab{a}})},\
  \Eprint {http://arxiv.org/abs/hep-ph/9409419} {arXiv:hep-ph/9409419 [hep-ph]}
  \BibitemShut {NoStop}%
%%CITATION = HEP-PH/9409419;%%
\bibitem [{\citenamefont {Anderson}\ and\ \citenamefont
  {Castano}(1995{\natexlab{b}})}]{Anderson:1994tr}%
  \BibitemOpen
  \bibfield  {author} {\bibinfo {author} {\bibfnamefont {G.~W.}\ \bibnamefont
  {Anderson}}\ and\ \bibinfo {author} {\bibfnamefont {D.~J.}\ \bibnamefont
  {Castano}},\ }\href {\doibase 10.1103/PhysRevD.52.1693} {\bibfield  {journal}
  {\bibinfo  {journal} {Phys. Rev. D}\ }\textbf {\bibinfo {volume} {52}},\
  \bibinfo {pages} {1693} (\bibinfo {year} {1995}{\natexlab{b}})},\ \Eprint
  {http://arxiv.org/abs/hep-ph/9412322} {arXiv:hep-ph/9412322 [hep-ph]}
  \BibitemShut {NoStop}%
%%CITATION = HEP-PH/9412322;%%
\bibitem [{\citenamefont {Anderson}\ and\ \citenamefont
  {Castano}(1996)}]{Anderson:1995cp}%
  \BibitemOpen
  \bibfield  {author} {\bibinfo {author} {\bibfnamefont {G.~W.}\ \bibnamefont
  {Anderson}}\ and\ \bibinfo {author} {\bibfnamefont {D.~J.}\ \bibnamefont
  {Castano}},\ }\href {\doibase 10.1103/PhysRevD.53.2403} {\bibfield  {journal}
  {\bibinfo  {journal} {Phys. Rev. D}\ }\textbf {\bibinfo {volume} {53}},\
  \bibinfo {pages} {2403} (\bibinfo {year} {1996})},\ \Eprint
  {http://arxiv.org/abs/hep-ph/9509212} {arXiv:hep-ph/9509212 [hep-ph]}
  \BibitemShut {NoStop}%
%%CITATION = HEP-PH/9509212;%%
\bibitem [{\citenamefont {Anderson}\ \emph {et~al.}(1997)\citenamefont
  {Anderson}, \citenamefont {Castano},\ and\ \citenamefont
  {Riotto}}]{Anderson:1996ew}%
  \BibitemOpen
  \bibfield  {author} {\bibinfo {author} {\bibfnamefont {G.~W.}\ \bibnamefont
  {Anderson}}, \bibinfo {author} {\bibfnamefont {D.~J.}\ \bibnamefont
  {Castano}}, \ and\ \bibinfo {author} {\bibfnamefont {A.}~\bibnamefont
  {Riotto}},\ }\href {\doibase 10.1103/PhysRevD.55.2950} {\bibfield  {journal}
  {\bibinfo  {journal} {Phys. Rev. D}\ }\textbf {\bibinfo {volume} {55}},\
  \bibinfo {pages} {2950} (\bibinfo {year} {1997})},\ \Eprint
  {http://arxiv.org/abs/hep-ph/9609463} {arXiv:hep-ph/9609463 [hep-ph]}
  \BibitemShut {NoStop}%
%%CITATION = HEP-PH/9609463;%%
\bibitem [{\citenamefont {Ciafaloni}\ and\ \citenamefont
  {Strumia}(1997)}]{Ciafaloni:1996zh}%
  \BibitemOpen
  \bibfield  {author} {\bibinfo {author} {\bibfnamefont {P.}~\bibnamefont
  {Ciafaloni}}\ and\ \bibinfo {author} {\bibfnamefont {A.}~\bibnamefont
  {Strumia}},\ }\href {\doibase 10.1016/S0550-3213(97)00138-7} {\bibfield
  {journal} {\bibinfo  {journal} {Nucl. Phys.}\ }\textbf {\bibinfo {volume}
  {B494}},\ \bibinfo {pages} {41} (\bibinfo {year} {1997})},\ \Eprint
  {http://arxiv.org/abs/hep-ph/9611204} {arXiv:hep-ph/9611204 [hep-ph]}
  \BibitemShut {NoStop}%
%%CITATION = HEP-PH/9611204;%%
\bibitem [{\citenamefont {Chan}\ \emph {et~al.}(1998)\citenamefont {Chan},
  \citenamefont {Chattopadhyay},\ and\ \citenamefont {Nath}}]{Chan:1997bi}%
  \BibitemOpen
  \bibfield  {author} {\bibinfo {author} {\bibfnamefont {K.~L.}\ \bibnamefont
  {Chan}}, \bibinfo {author} {\bibfnamefont {U.}~\bibnamefont {Chattopadhyay}},
  \ and\ \bibinfo {author} {\bibfnamefont {P.}~\bibnamefont {Nath}},\ }\href
  {\doibase 10.1103/PhysRevD.58.096004} {\bibfield  {journal} {\bibinfo
  {journal} {Phys. Rev. D}\ }\textbf {\bibinfo {volume} {58}},\ \bibinfo
  {pages} {096004} (\bibinfo {year} {1998})},\ \Eprint
  {http://arxiv.org/abs/hep-ph/9710473} {arXiv:hep-ph/9710473 [hep-ph]}
  \BibitemShut {NoStop}%
%%CITATION = HEP-PH/9710473;%%
\bibitem [{\citenamefont {Barbieri}\ and\ \citenamefont
  {Strumia}(1998)}]{Barbieri:1998uv}%
  \BibitemOpen
  \bibfield  {author} {\bibinfo {author} {\bibfnamefont {R.}~\bibnamefont
  {Barbieri}}\ and\ \bibinfo {author} {\bibfnamefont {A.}~\bibnamefont
  {Strumia}},\ }\href {\doibase 10.1016/S0370-2693(98)00577-2} {\bibfield
  {journal} {\bibinfo  {journal} {Phys. Lett. B}\ }\textbf {\bibinfo {volume}
  {433}},\ \bibinfo {pages} {63} (\bibinfo {year} {1998})},\ \Eprint
  {http://arxiv.org/abs/hep-ph/9801353} {arXiv:hep-ph/9801353 [hep-ph]}
  \BibitemShut {NoStop}%
%%CITATION = HEP-PH/9801353;%%
\bibitem [{\citenamefont {Giusti}\ \emph {et~al.}(1999)\citenamefont {Giusti},
  \citenamefont {Romanino},\ and\ \citenamefont {Strumia}}]{Giusti:1998gz}%
  \BibitemOpen
  \bibfield  {author} {\bibinfo {author} {\bibfnamefont {L.}~\bibnamefont
  {Giusti}}, \bibinfo {author} {\bibfnamefont {A.}~\bibnamefont {Romanino}}, \
  and\ \bibinfo {author} {\bibfnamefont {A.}~\bibnamefont {Strumia}},\ }\href
  {\doibase 10.1016/S0550-3213(99)00153-4} {\bibfield  {journal} {\bibinfo
  {journal} {Nucl. Phys.}\ }\textbf {\bibinfo {volume} {B550}},\ \bibinfo
  {pages} {3} (\bibinfo {year} {1999})},\ \Eprint
  {http://arxiv.org/abs/hep-ph/9811386} {arXiv:hep-ph/9811386 [hep-ph]}
  \BibitemShut {NoStop}%
%%CITATION = HEP-PH/9811386;%%
\bibitem [{\citenamefont {Casas}\ \emph
  {et~al.}(2004{\natexlab{a}})\citenamefont {Casas}, \citenamefont {Espinosa},\
  and\ \citenamefont {Hidalgo}}]{Casas:2003jx}%
  \BibitemOpen
  \bibfield  {author} {\bibinfo {author} {\bibfnamefont {J.~A.}\ \bibnamefont
  {Casas}}, \bibinfo {author} {\bibfnamefont {J.~R.}\ \bibnamefont {Espinosa}},
  \ and\ \bibinfo {author} {\bibfnamefont {I.}~\bibnamefont {Hidalgo}},\ }\href
  {\doibase 10.1088/1126-6708/2004/01/008} {\bibfield  {journal} {\bibinfo
  {journal} {J. High Energy Phys.}\ }\textbf {\bibinfo {volume} {0401}},\
  \bibinfo {pages} {008} (\bibinfo {year} {2004}{\natexlab{a}})},\ \Eprint
  {http://arxiv.org/abs/hep-ph/0310137} {arXiv:hep-ph/0310137 [hep-ph]}
  \BibitemShut {NoStop}%
%%CITATION = HEP-PH/0310137;%%
\bibitem [{\citenamefont {Casas}\ \emph
  {et~al.}(2004{\natexlab{b}})\citenamefont {Casas}, \citenamefont {Espinosa},\
  and\ \citenamefont {Hidalgo}}]{Casas:2004uu}%
  \BibitemOpen
  \bibfield  {author} {\bibinfo {author} {\bibfnamefont {J.~A.}\ \bibnamefont
  {Casas}}, \bibinfo {author} {\bibfnamefont {J.~R.}\ \bibnamefont {Espinosa}},
  \ and\ \bibinfo {author} {\bibfnamefont {I.}~\bibnamefont {Hidalgo}},\ }in\
  \href {\doibase 10.1142/9789812702463_0011} {\emph {\bibinfo {booktitle}
  {{String Phenomenology 2003}}}},\ \bibinfo {editor} {edited by\ \bibinfo
  {editor} {\bibfnamefont {V.}~\bibnamefont {Sanz}}, \bibinfo {editor}
  {\bibfnamefont {S.}~\bibnamefont {Abel}}, \bibinfo {editor} {\bibfnamefont
  {J.}~\bibnamefont {Santiago}}, \ and\ \bibinfo {editor} {\bibfnamefont
  {A.}~\bibnamefont {Fraraggi}}}\ (\bibinfo  {publisher} {World Scientific,
  Singapore},\ \bibinfo {year} {2004})\ p.~\bibinfo {pages} {76},\ \Eprint
  {http://arxiv.org/abs/hep-ph/0402017} {arXiv:hep-ph/0402017 [hep-ph]}
  \BibitemShut {NoStop}%
%%CITATION = HEP-PH/0402017;%%
\bibitem [{\citenamefont {Casas}\ \emph
  {et~al.}(2004{\natexlab{c}})\citenamefont {Casas}, \citenamefont {Espinosa},\
  and\ \citenamefont {Hidalgo}}]{Casas:2004gh}%
  \BibitemOpen
  \bibfield  {author} {\bibinfo {author} {\bibfnamefont {J.~A.}\ \bibnamefont
  {Casas}}, \bibinfo {author} {\bibfnamefont {J.~R.}\ \bibnamefont {Espinosa}},
  \ and\ \bibinfo {author} {\bibfnamefont {I.}~\bibnamefont {Hidalgo}},\ }\href
  {\doibase 10.1088/1126-6708/2004/11/057} {\bibfield  {journal} {\bibinfo
  {journal} {J. High Energy Phys.}\ }\textbf {\bibinfo {volume} {0411}},\
  \bibinfo {pages} {057} (\bibinfo {year} {2004}{\natexlab{c}})},\ \Eprint
  {http://arxiv.org/abs/hep-ph/0410298} {arXiv:hep-ph/0410298 [hep-ph]}
  \BibitemShut {NoStop}%
%%CITATION = HEP-PH/0410298;%%
\bibitem [{\citenamefont {Casas}\ \emph {et~al.}(2007)\citenamefont {Casas},
  \citenamefont {Espinosa},\ and\ \citenamefont {Hidalgo}}]{Casas:2006bd}%
  \BibitemOpen
  \bibfield  {author} {\bibinfo {author} {\bibfnamefont {J.~A.}\ \bibnamefont
  {Casas}}, \bibinfo {author} {\bibfnamefont {J.~R.}\ \bibnamefont {Espinosa}},
  \ and\ \bibinfo {author} {\bibfnamefont {I.}~\bibnamefont {Hidalgo}},\ }\href
  {\doibase 10.1016/j.nuclphysb.2007.04.024} {\bibfield  {journal} {\bibinfo
  {journal} {Nucl. Phys.}\ }\textbf {\bibinfo {volume} {B777}},\ \bibinfo
  {pages} {226} (\bibinfo {year} {2007})},\ \Eprint
  {http://arxiv.org/abs/hep-ph/0607279} {arXiv:hep-ph/0607279 [hep-ph]}
  \BibitemShut {NoStop}%
%%CITATION = HEP-PH/0607279;%%
\bibitem [{\citenamefont {Kitano}\ and\ \citenamefont
  {Nomura}(2005)}]{Kitano:2005wc}%
  \BibitemOpen
  \bibfield  {author} {\bibinfo {author} {\bibfnamefont {R.}~\bibnamefont
  {Kitano}}\ and\ \bibinfo {author} {\bibfnamefont {Y.}~\bibnamefont
  {Nomura}},\ }\href {\doibase 10.1016/j.physletb.2005.10.003} {\bibfield
  {journal} {\bibinfo  {journal} {Phys. Lett. B}\ }\textbf {\bibinfo {volume}
  {631}},\ \bibinfo {pages} {58} (\bibinfo {year} {2005})},\ \Eprint
  {http://arxiv.org/abs/hep-ph/0509039} {arXiv:hep-ph/0509039 [hep-ph]}
  \BibitemShut {NoStop}%
%%CITATION = HEP-PH/0509039;%%
\bibitem [{\citenamefont {Athron}\ and\ \citenamefont
  {Miller}(2007{\natexlab{a}})}]{Athron:2007ry}%
  \BibitemOpen
  \bibfield  {author} {\bibinfo {author} {\bibfnamefont {P.}~\bibnamefont
  {Athron}}\ and\ \bibinfo {author} {\bibfnamefont {D.~J.}\ \bibnamefont
  {Miller}},\ }\href {\doibase 10.1103/PhysRevD.76.075010} {\bibfield
  {journal} {\bibinfo  {journal} {Phys. Rev. D}\ }\textbf {\bibinfo {volume}
  {76}},\ \bibinfo {pages} {075010} (\bibinfo {year} {2007}{\natexlab{a}})},\
  \Eprint {http://arxiv.org/abs/0705.2241} {arXiv:0705.2241 [hep-ph]}
  \BibitemShut {NoStop}%
%%CITATION = ARXIV:0705.2241;%%
\bibitem [{\citenamefont {Athron}\ and\ \citenamefont
  {Miller}(2007{\natexlab{b}})}]{Athron:2007qr}%
  \BibitemOpen
  \bibfield  {author} {\bibinfo {author} {\bibfnamefont {P.}~\bibnamefont
  {Athron}}\ and\ \bibinfo {author} {\bibfnamefont {D.~J.}\ \bibnamefont
  {Miller}},\ }in\ \href
  {http://www.susy07.uni-karlsruhe.de/PROC/ProcSUSY07_FINAL.pdf} {\emph
  {\bibinfo {booktitle} {{Proceedings of the 15th International Conference on
  Supersymmetry and the Unification of Fundamental Interactions}}}},\ \bibinfo
  {editor} {edited by\ \bibinfo {editor} {\bibfnamefont {W.}~\bibnamefont
  {de~Boer}}\ and\ \bibinfo {editor} {\bibfnamefont {I.}~\bibnamefont
  {Gebauer}}}\ (\bibinfo  {publisher} {University of Karlsruhe, Karlsruhe},\
  \bibinfo {year} {2007})\ p.\ \bibinfo {pages} {554},\ \Eprint
  {http://arxiv.org/abs/0710.2486} {arXiv:0710.2486 [hep-ph]} \BibitemShut
  {NoStop}%
%%CITATION = ARXIV:0710.2486;%%
\bibitem [{\citenamefont {Baer}\ \emph {et~al.}(2012)\citenamefont {Baer},
  \citenamefont {Barger}, \citenamefont {Huang}, \citenamefont {Mustafayev},\
  and\ \citenamefont {Tata}}]{Baer:2012up}%
  \BibitemOpen
  \bibfield  {author} {\bibinfo {author} {\bibfnamefont {H.}~\bibnamefont
  {Baer}}, \bibinfo {author} {\bibfnamefont {V.}~\bibnamefont {Barger}},
  \bibinfo {author} {\bibfnamefont {P.}~\bibnamefont {Huang}}, \bibinfo
  {author} {\bibfnamefont {A.}~\bibnamefont {Mustafayev}}, \ and\ \bibinfo
  {author} {\bibfnamefont {X.}~\bibnamefont {Tata}},\ }\href {\doibase
  10.1103/PhysRevLett.109.161802} {\bibfield  {journal} {\bibinfo  {journal}
  {Phys. Rev. Lett.}\ }\textbf {\bibinfo {volume} {109}},\ \bibinfo {pages}
  {161802} (\bibinfo {year} {2012})},\ \Eprint {http://arxiv.org/abs/1207.3343}
  {arXiv:1207.3343 [hep-ph]} \BibitemShut {NoStop}%
%%CITATION = ARXIV:1207.3343;%%
\bibitem [{\citenamefont {Allanach}\ \emph {et~al.}(2007)\citenamefont
  {Allanach}, \citenamefont {Cranmer}, \citenamefont {Lester},\ and\
  \citenamefont {Weber}}]{Allanach:2007qk}%
  \BibitemOpen
  \bibfield  {author} {\bibinfo {author} {\bibfnamefont {B.~C.}\ \bibnamefont
  {Allanach}}, \bibinfo {author} {\bibfnamefont {K.}~\bibnamefont {Cranmer}},
  \bibinfo {author} {\bibfnamefont {C.~G.}\ \bibnamefont {Lester}}, \ and\
  \bibinfo {author} {\bibfnamefont {A.~M.}\ \bibnamefont {Weber}},\ }\href
  {\doibase 10.1088/1126-6708/2007/08/023} {\bibfield  {journal} {\bibinfo
  {journal} {J. High Energy Phys.}\ }\textbf {\bibinfo {volume} {0708}},\
  \bibinfo {pages} {023} (\bibinfo {year} {2007})},\ \Eprint
  {http://arxiv.org/abs/0705.0487} {arXiv:0705.0487 [hep-ph]} \BibitemShut
  {NoStop}%
%%CITATION = ARXIV:0705.0487;%%
\bibitem [{\citenamefont {Cabrera}\ \emph {et~al.}(2009)\citenamefont
  {Cabrera}, \citenamefont {Casas},\ and\ \citenamefont {Ruiz~de
  Austri}}]{Cabrera:2008tj}%
  \BibitemOpen
  \bibfield  {author} {\bibinfo {author} {\bibfnamefont {M.~E.}\ \bibnamefont
  {Cabrera}}, \bibinfo {author} {\bibfnamefont {J.~A.}\ \bibnamefont {Casas}},
  \ and\ \bibinfo {author} {\bibfnamefont {R.}~\bibnamefont {Ruiz~de Austri}},\
  }\href {\doibase 10.1088/1126-6708/2009/03/075} {\bibfield  {journal}
  {\bibinfo  {journal} {J. High Energy Phys.}\ }\textbf {\bibinfo {volume}
  {0903}},\ \bibinfo {pages} {075} (\bibinfo {year} {2009})},\ \Eprint
  {http://arxiv.org/abs/0812.0536} {arXiv:0812.0536 [hep-ph]} \BibitemShut
  {NoStop}%
%%CITATION = ARXIV:0812.0536;%%
\bibitem [{\citenamefont {Ghilencea}\ and\ \citenamefont
  {Ross}(2013)}]{Ghilencea:2012qk}%
  \BibitemOpen
  \bibfield  {author} {\bibinfo {author} {\bibfnamefont {D.~M.}\ \bibnamefont
  {Ghilencea}}\ and\ \bibinfo {author} {\bibfnamefont {G.~G.}\ \bibnamefont
  {Ross}},\ }\href {\doibase 10.1016/j.nuclphysb.2012.11.007} {\bibfield
  {journal} {\bibinfo  {journal} {Nucl. Phys.}\ }\textbf {\bibinfo {volume}
  {B868}},\ \bibinfo {pages} {65} (\bibinfo {year} {2013})},\ \Eprint
  {http://arxiv.org/abs/1208.0837} {arXiv:1208.0837 [hep-ph]} \BibitemShut
  {NoStop}%
%%CITATION = ARXIV:1208.0837;%%
\bibitem [{\citenamefont {Fichet}(2012)}]{Fichet:2012sn}%
  \BibitemOpen
  \bibfield  {author} {\bibinfo {author} {\bibfnamefont {S.}~\bibnamefont
  {Fichet}},\ }\href {\doibase 10.1103/PhysRevD.86.125029} {\bibfield
  {journal} {\bibinfo  {journal} {Phys. Rev. D}\ }\textbf {\bibinfo {volume}
  {86}},\ \bibinfo {pages} {125029} (\bibinfo {year} {2012})},\ \Eprint
  {http://arxiv.org/abs/1204.4940} {arXiv:1204.4940 [hep-ph]} \BibitemShut
  {NoStop}%
%%CITATION = ARXIV:1204.4940;%%
\bibitem [{\citenamefont {Kim}\ \emph {et~al.}(2014)\citenamefont {Kim},
  \citenamefont {Athron}, \citenamefont {Bal{\'a}zs}, \citenamefont {Farmer},\
  and\ \citenamefont {Hutchison}}]{Kim:2013uxa}%
  \BibitemOpen
  \bibfield  {author} {\bibinfo {author} {\bibfnamefont {D.}~\bibnamefont
  {Kim}}, \bibinfo {author} {\bibfnamefont {P.}~\bibnamefont {Athron}},
  \bibinfo {author} {\bibfnamefont {C.}~\bibnamefont {Bal{\'a}zs}}, \bibinfo
  {author} {\bibfnamefont {B.}~\bibnamefont {Farmer}}, \ and\ \bibinfo {author}
  {\bibfnamefont {E.}~\bibnamefont {Hutchison}},\ }\href {\doibase
  10.1103/PhysRevD.90.055008} {\bibfield  {journal} {\bibinfo  {journal} {Phys.
  Rev. D}\ }\textbf {\bibinfo {volume} {90}},\ \bibinfo {pages} {055008}
  (\bibinfo {year} {2014})},\ \Eprint {http://arxiv.org/abs/1312.4150}
  {arXiv:1312.4150 [hep-ph]} \BibitemShut {NoStop}%
%%CITATION = ARXIV:1312.4150;%%
\bibitem [{\citenamefont {Langacker}(2009)}]{Langacker:2008yv}%
  \BibitemOpen
  \bibfield  {author} {\bibinfo {author} {\bibfnamefont {P.}~\bibnamefont
  {Langacker}},\ }\href {\doibase 10.1103/RevModPhys.81.1199} {\bibfield
  {journal} {\bibinfo  {journal} {Rev. Mod. Phys.}\ }\textbf {\bibinfo {volume}
  {81}},\ \bibinfo {pages} {1199} (\bibinfo {year} {2009})},\ \Eprint
  {http://arxiv.org/abs/0801.1345} {arXiv:0801.1345 [hep-ph]} \BibitemShut
  {NoStop}%
%%CITATION = ARXIV:0801.1345;%%
\bibitem [{\citenamefont {Gunion}\ \emph {et~al.}(2000)\citenamefont {Gunion},
  \citenamefont {Haber}, \citenamefont {Kane},\ and\ \citenamefont
  {Dawson}}]{Gunion:1989we}%
  \BibitemOpen
  \bibfield  {author} {\bibinfo {author} {\bibfnamefont {J.~F.}\ \bibnamefont
  {Gunion}}, \bibinfo {author} {\bibfnamefont {H.~E.}\ \bibnamefont {Haber}},
  \bibinfo {author} {\bibfnamefont {G.~L.}\ \bibnamefont {Kane}}, \ and\
  \bibinfo {author} {\bibfnamefont {S.}~\bibnamefont {Dawson}},\ }\href@noop {}
  {\emph {\bibinfo {title} {{The Higgs Hunter's Guide}}}}\ (\bibinfo
  {publisher} {Westview Press, Boulder},\ \bibinfo {year} {2000})\BibitemShut
  {NoStop}%
%%CITATION = FRPHA,80,1;%%
\bibitem [{\citenamefont {Gunion}\ \emph {et~al.}()\citenamefont {Gunion},
  \citenamefont {Haber}, \citenamefont {Kane},\ and\ \citenamefont
  {Dawson}}]{Gunion:1992hs}%
  \BibitemOpen
  \bibfield  {author} {\bibinfo {author} {\bibfnamefont {J.~F.}\ \bibnamefont
  {Gunion}}, \bibinfo {author} {\bibfnamefont {H.~E.}\ \bibnamefont {Haber}},
  \bibinfo {author} {\bibfnamefont {G.~L.}\ \bibnamefont {Kane}}, \ and\
  \bibinfo {author} {\bibfnamefont {S.}~\bibnamefont {Dawson}},\ }\href@noop {}
  {\ }\Eprint {http://arxiv.org/abs/hep-ph/9302272} {arXiv:hep-ph/9302272
  [hep-ph]} \BibitemShut {NoStop}%
%%CITATION = HEP-PH/9302272;%%
\bibitem [{\citenamefont {Binetruy}\ \emph {et~al.}(1986)\citenamefont
  {Binetruy}, \citenamefont {Dawson}, \citenamefont {Hinchliffe},\ and\
  \citenamefont {Sher}}]{Binetruy:1985xm}%
  \BibitemOpen
  \bibfield  {author} {\bibinfo {author} {\bibfnamefont {P.}~\bibnamefont
  {Binetruy}}, \bibinfo {author} {\bibfnamefont {S.}~\bibnamefont {Dawson}},
  \bibinfo {author} {\bibfnamefont {I.}~\bibnamefont {Hinchliffe}}, \ and\
  \bibinfo {author} {\bibfnamefont {M.}~\bibnamefont {Sher}},\ }\href {\doibase
  10.1016/0550-3213(86)90374-3} {\bibfield  {journal} {\bibinfo  {journal}
  {Nucl. Phys.}\ }\textbf {\bibinfo {volume} {B273}},\ \bibinfo {pages} {501}
  (\bibinfo {year} {1986})}\BibitemShut {NoStop}%
%%CITATION = NUPHA,B273,501;%%
\bibitem [{\citenamefont {Ibanez}\ and\ \citenamefont
  {Mas}(1987)}]{Ibanez:1986si}%
  \BibitemOpen
  \bibfield  {author} {\bibinfo {author} {\bibfnamefont {L.~E.}\ \bibnamefont
  {Ibanez}}\ and\ \bibinfo {author} {\bibfnamefont {J.}~\bibnamefont {Mas}},\
  }\href {\doibase 10.1016/0550-3213(87)90434-2} {\bibfield  {journal}
  {\bibinfo  {journal} {Nucl. Phys.}\ }\textbf {\bibinfo {volume} {B286}},\
  \bibinfo {pages} {107} (\bibinfo {year} {1987})}\BibitemShut {NoStop}%
%%CITATION = NUPHA,B286,107;%%
\bibitem [{\citenamefont {Gunion}\ \emph {et~al.}(1987)\citenamefont {Gunion},
  \citenamefont {Roszkowski},\ and\ \citenamefont {Haber}}]{Gunion:1986ky}%
  \BibitemOpen
  \bibfield  {author} {\bibinfo {author} {\bibfnamefont {J.~F.}\ \bibnamefont
  {Gunion}}, \bibinfo {author} {\bibfnamefont {L.}~\bibnamefont {Roszkowski}},
  \ and\ \bibinfo {author} {\bibfnamefont {H.~E.}\ \bibnamefont {Haber}},\
  }\href {\doibase 10.1016/0370-2693(87)90651-4} {\bibfield  {journal}
  {\bibinfo  {journal} {Phys. Lett. B}\ }\textbf {\bibinfo {volume} {189}},\
  \bibinfo {pages} {409} (\bibinfo {year} {1987})}\BibitemShut {NoStop}%
%%CITATION = PHLTA,B189,409;%%
\bibitem [{\citenamefont {Haber}\ and\ \citenamefont
  {Sher}(1987)}]{Haber:1986gz}%
  \BibitemOpen
  \bibfield  {author} {\bibinfo {author} {\bibfnamefont {H.~E.}\ \bibnamefont
  {Haber}}\ and\ \bibinfo {author} {\bibfnamefont {M.}~\bibnamefont {Sher}},\
  }\href {\doibase 10.1103/PhysRevD.35.2206} {\bibfield  {journal} {\bibinfo
  {journal} {Phys. Rev. D}\ }\textbf {\bibinfo {volume} {35}},\ \bibinfo
  {pages} {2206} (\bibinfo {year} {1987})}\BibitemShut {NoStop}%
%%CITATION = PHRVA,D35,2206;%%
\bibitem [{\citenamefont {Baer}\ \emph {et~al.}(1987)\citenamefont {Baer},
  \citenamefont {Dicus}, \citenamefont {Drees},\ and\ \citenamefont
  {Tata}}]{Baer:1987eb}%
  \BibitemOpen
  \bibfield  {author} {\bibinfo {author} {\bibfnamefont {H.}~\bibnamefont
  {Baer}}, \bibinfo {author} {\bibfnamefont {D.}~\bibnamefont {Dicus}},
  \bibinfo {author} {\bibfnamefont {M.}~\bibnamefont {Drees}}, \ and\ \bibinfo
  {author} {\bibfnamefont {X.}~\bibnamefont {Tata}},\ }\href {\doibase
  10.1103/PhysRevD.36.1363} {\bibfield  {journal} {\bibinfo  {journal} {Phys.
  Rev. D}\ }\textbf {\bibinfo {volume} {36}},\ \bibinfo {pages} {1363}
  (\bibinfo {year} {1987})}\BibitemShut {NoStop}%
%%CITATION = PHRVA,D36,1363;%%
\bibitem [{\citenamefont {Gunion}\ \emph {et~al.}(1988)\citenamefont {Gunion},
  \citenamefont {Roszkowski},\ and\ \citenamefont {Haber}}]{Gunion:1987jd}%
  \BibitemOpen
  \bibfield  {author} {\bibinfo {author} {\bibfnamefont {J.~F.}\ \bibnamefont
  {Gunion}}, \bibinfo {author} {\bibfnamefont {L.}~\bibnamefont {Roszkowski}},
  \ and\ \bibinfo {author} {\bibfnamefont {H.~E.}\ \bibnamefont {Haber}},\
  }\href {\doibase 10.1103/PhysRevD.38.105} {\bibfield  {journal} {\bibinfo
  {journal} {Phys. Rev. D}\ }\textbf {\bibinfo {volume} {38}},\ \bibinfo
  {pages} {105} (\bibinfo {year} {1988})}\BibitemShut {NoStop}%
%%CITATION = PHRVA,D38,105;%%
\bibitem [{\citenamefont {Grifols}\ \emph {et~al.}(1986)\citenamefont
  {Grifols}, \citenamefont {M{\'e}ndez},\ and\ \citenamefont
  {Sol{\`a}}}]{Grifols:1986vr}%
  \BibitemOpen
  \bibfield  {author} {\bibinfo {author} {\bibfnamefont {J.~A.}\ \bibnamefont
  {Grifols}}, \bibinfo {author} {\bibfnamefont {A.}~\bibnamefont {M{\'e}ndez}},
  \ and\ \bibinfo {author} {\bibfnamefont {J.}~\bibnamefont {Sol{\`a}}},\
  }\href {\doibase 10.1103/PhysRevLett.57.2348} {\bibfield  {journal} {\bibinfo
   {journal} {Phys. Rev. Lett.}\ }\textbf {\bibinfo {volume} {57}},\ \bibinfo
  {pages} {2348} (\bibinfo {year} {1986})}\BibitemShut {NoStop}%
%%CITATION = PRLTA,57,2348;%%
\bibitem [{\citenamefont {Ellis}\ \emph {et~al.}(1987)\citenamefont {Ellis},
  \citenamefont {Nanopoulos}, \citenamefont {Petcov},\ and\ \citenamefont
  {Zwirner}}]{Ellis:1986ip}%
  \BibitemOpen
  \bibfield  {author} {\bibinfo {author} {\bibfnamefont {J.~R.}\ \bibnamefont
  {Ellis}}, \bibinfo {author} {\bibfnamefont {D.~V.}\ \bibnamefont
  {Nanopoulos}}, \bibinfo {author} {\bibfnamefont {S.~T.}\ \bibnamefont
  {Petcov}}, \ and\ \bibinfo {author} {\bibfnamefont {F.}~\bibnamefont
  {Zwirner}},\ }\href {\doibase 10.1016/0550-3213(87)90263-X} {\bibfield
  {journal} {\bibinfo  {journal} {Nucl. Phys.}\ }\textbf {\bibinfo {volume}
  {B283}},\ \bibinfo {pages} {93} (\bibinfo {year} {1987})}\BibitemShut
  {NoStop}%
%%CITATION = NUPHA,B283,93;%%
\bibitem [{\citenamefont {Morris}(1988)}]{Morris:1987fm}%
  \BibitemOpen
  \bibfield  {author} {\bibinfo {author} {\bibfnamefont {D.~A.}\ \bibnamefont
  {Morris}},\ }\href {\doibase 10.1103/PhysRevD.37.2012} {\bibfield  {journal}
  {\bibinfo  {journal} {Phys. Rev. D}\ }\textbf {\bibinfo {volume} {37}},\
  \bibinfo {pages} {2012} (\bibinfo {year} {1988})}\BibitemShut {NoStop}%
%%CITATION = PHRVA,D37,2012;%%
\bibitem [{\citenamefont {Drees}(1987)}]{Drees:1987tp}%
  \BibitemOpen
  \bibfield  {author} {\bibinfo {author} {\bibfnamefont {M.}~\bibnamefont
  {Drees}},\ }\href {\doibase 10.1103/PhysRevD.35.2910} {\bibfield  {journal}
  {\bibinfo  {journal} {Phys. Rev. D}\ }\textbf {\bibinfo {volume} {35}},\
  \bibinfo {pages} {2910} (\bibinfo {year} {1987})}\BibitemShut {NoStop}%
%%CITATION = PHRVA,D35,2910;%%
\bibitem [{\citenamefont {Ma}(1996)}]{Ma:1995xk}%
  \BibitemOpen
  \bibfield  {author} {\bibinfo {author} {\bibfnamefont {E.}~\bibnamefont
  {Ma}},\ }\href {\doibase 10.1016/0370-2693(96)00524-2} {\bibfield  {journal}
  {\bibinfo  {journal} {Phys. Lett. B}\ }\textbf {\bibinfo {volume} {380}},\
  \bibinfo {pages} {286} (\bibinfo {year} {1996})},\ \Eprint
  {http://arxiv.org/abs/hep-ph/9507348} {arXiv:hep-ph/9507348 [hep-ph]}
  \BibitemShut {NoStop}%
%%CITATION = HEP-PH/9507348;%%
\bibitem [{\citenamefont {Suematsu}(1997)}]{Suematsu:1997tv}%
  \BibitemOpen
  \bibfield  {author} {\bibinfo {author} {\bibfnamefont {D.}~\bibnamefont
  {Suematsu}},\ }\href {\doibase 10.1142/S0217732397001746} {\bibfield
  {journal} {\bibinfo  {journal} {Mod. Phys. Lett. A}\ }\textbf {\bibinfo
  {volume} {12}},\ \bibinfo {pages} {1709} (\bibinfo {year} {1997})},\ \Eprint
  {http://arxiv.org/abs/hep-ph/9705412} {arXiv:hep-ph/9705412 [hep-ph]}
  \BibitemShut {NoStop}%
%%CITATION = HEP-PH/9705412;%%
\bibitem [{\citenamefont {Suematsu}(1998{\natexlab{a}})}]{Suematsu:1997qt}%
  \BibitemOpen
  \bibfield  {author} {\bibinfo {author} {\bibfnamefont {D.}~\bibnamefont
  {Suematsu}},\ }\href {\doibase 10.1016/S0370-2693(97)01298-7} {\bibfield
  {journal} {\bibinfo  {journal} {Phys. Lett. B}\ }\textbf {\bibinfo {volume}
  {416}},\ \bibinfo {pages} {108} (\bibinfo {year} {1998}{\natexlab{a}})},\
  \Eprint {http://arxiv.org/abs/hep-ph/9705405} {arXiv:hep-ph/9705405 [hep-ph]}
  \BibitemShut {NoStop}%
%%CITATION = HEP-PH/9705405;%%
\bibitem [{\citenamefont {Suematsu}(1998{\natexlab{b}})}]{Suematsu:1997au}%
  \BibitemOpen
  \bibfield  {author} {\bibinfo {author} {\bibfnamefont {D.}~\bibnamefont
  {Suematsu}},\ }\href {\doibase 10.1103/PhysRevD.57.1738} {\bibfield
  {journal} {\bibinfo  {journal} {Phys. Rev. D}\ }\textbf {\bibinfo {volume}
  {57}},\ \bibinfo {pages} {1738} (\bibinfo {year} {1998}{\natexlab{b}})},\
  \Eprint {http://arxiv.org/abs/hep-ph/9708413} {arXiv:hep-ph/9708413 [hep-ph]}
  \BibitemShut {NoStop}%
%%CITATION = HEP-PH/9708413;%%
\bibitem [{\citenamefont {Keith}\ and\ \citenamefont
  {Ma}(1996)}]{Keith:1996fv}%
  \BibitemOpen
  \bibfield  {author} {\bibinfo {author} {\bibfnamefont {E.}~\bibnamefont
  {Keith}}\ and\ \bibinfo {author} {\bibfnamefont {E.}~\bibnamefont {Ma}},\
  }\href {\doibase 10.1103/PhysRevD.54.3587} {\bibfield  {journal} {\bibinfo
  {journal} {Phys. Rev. D}\ }\textbf {\bibinfo {volume} {54}},\ \bibinfo
  {pages} {3587} (\bibinfo {year} {1996})},\ \Eprint
  {http://arxiv.org/abs/hep-ph/9603353} {arXiv:hep-ph/9603353 [hep-ph]}
  \BibitemShut {NoStop}%
%%CITATION = HEP-PH/9603353;%%
\bibitem [{\citenamefont {Keith}\ and\ \citenamefont
  {Ma}(1997)}]{Keith:1997zb}%
  \BibitemOpen
  \bibfield  {author} {\bibinfo {author} {\bibfnamefont {E.}~\bibnamefont
  {Keith}}\ and\ \bibinfo {author} {\bibfnamefont {E.}~\bibnamefont {Ma}},\
  }\href {\doibase 10.1103/PhysRevD.56.7155} {\bibfield  {journal} {\bibinfo
  {journal} {Phys. Rev. D}\ }\textbf {\bibinfo {volume} {56}},\ \bibinfo
  {pages} {7155} (\bibinfo {year} {1997})},\ \Eprint
  {http://arxiv.org/abs/hep-ph/9704441} {arXiv:hep-ph/9704441 [hep-ph]}
  \BibitemShut {NoStop}%
%%CITATION = HEP-PH/9704441;%%
\bibitem [{\citenamefont {Gherghetta}\ \emph {et~al.}(1998)\citenamefont
  {Gherghetta}, \citenamefont {Kaeding},\ and\ \citenamefont
  {Kane}}]{Gherghetta:1996yr}%
  \BibitemOpen
  \bibfield  {author} {\bibinfo {author} {\bibfnamefont {T.}~\bibnamefont
  {Gherghetta}}, \bibinfo {author} {\bibfnamefont {T.~A.}\ \bibnamefont
  {Kaeding}}, \ and\ \bibinfo {author} {\bibfnamefont {G.~L.}\ \bibnamefont
  {Kane}},\ }\href {\doibase 10.1103/PhysRevD.57.3178} {\bibfield  {journal}
  {\bibinfo  {journal} {Phys. Rev. D}\ }\textbf {\bibinfo {volume} {57}},\
  \bibinfo {pages} {3178} (\bibinfo {year} {1998})},\ \Eprint
  {http://arxiv.org/abs/hep-ph/9701343} {arXiv:hep-ph/9701343 [hep-ph]}
  \BibitemShut {NoStop}%
%%CITATION = HEP-PH/9701343;%%
\bibitem [{\citenamefont {Demir}\ and\ \citenamefont
  {Pak}(1998)}]{Demir:1998dk}%
  \BibitemOpen
  \bibfield  {author} {\bibinfo {author} {\bibfnamefont {D.~A.}\ \bibnamefont
  {Demir}}\ and\ \bibinfo {author} {\bibfnamefont {N.~K.}\ \bibnamefont
  {Pak}},\ }\href {\doibase 10.1103/PhysRevD.57.6609} {\bibfield  {journal}
  {\bibinfo  {journal} {Phys. Rev. D}\ }\textbf {\bibinfo {volume} {57}},\
  \bibinfo {pages} {6609} (\bibinfo {year} {1998})},\ \Eprint
  {http://arxiv.org/abs/hep-ph/9809357} {arXiv:hep-ph/9809357 [hep-ph]}
  \BibitemShut {NoStop}%
%%CITATION = HEP-PH/9809357;%%
\bibitem [{\citenamefont {Langacker}\ and\ \citenamefont
  {Wang}(1998)}]{Langacker:1998tc}%
  \BibitemOpen
  \bibfield  {author} {\bibinfo {author} {\bibfnamefont {P.}~\bibnamefont
  {Langacker}}\ and\ \bibinfo {author} {\bibfnamefont {J.}~\bibnamefont
  {Wang}},\ }\href {\doibase 10.1103/PhysRevD.58.115010} {\bibfield  {journal}
  {\bibinfo  {journal} {Phys. Rev. D}\ }\textbf {\bibinfo {volume} {58}},\
  \bibinfo {pages} {115010} (\bibinfo {year} {1998})},\ \Eprint
  {http://arxiv.org/abs/hep-ph/9804428} {arXiv:hep-ph/9804428 [hep-ph]}
  \BibitemShut {NoStop}%
%%CITATION = HEP-PH/9804428;%%
\bibitem [{\citenamefont {Hambye}\ \emph {et~al.}(2001)\citenamefont {Hambye},
  \citenamefont {Ma}, \citenamefont {Raidal},\ and\ \citenamefont
  {Sarkar}}]{Hambye:2000bn}%
  \BibitemOpen
  \bibfield  {author} {\bibinfo {author} {\bibfnamefont {T.}~\bibnamefont
  {Hambye}}, \bibinfo {author} {\bibfnamefont {E.}~\bibnamefont {Ma}}, \bibinfo
  {author} {\bibfnamefont {M.}~\bibnamefont {Raidal}}, \ and\ \bibinfo {author}
  {\bibfnamefont {U.}~\bibnamefont {Sarkar}},\ }\href {\doibase
  10.1016/S0370-2693(01)00577-9} {\bibfield  {journal} {\bibinfo  {journal}
  {Phys. Lett. B}\ }\textbf {\bibinfo {volume} {512}},\ \bibinfo {pages} {373}
  (\bibinfo {year} {2001})},\ \Eprint {http://arxiv.org/abs/hep-ph/0011197}
  {arXiv:hep-ph/0011197 [hep-ph]} \BibitemShut {NoStop}%
%%CITATION = HEP-PH/0011197;%%
\bibitem [{\citenamefont {Ma}\ and\ \citenamefont {Raidal}(2002)}]{Ma:2000jf}%
  \BibitemOpen
  \bibfield  {author} {\bibinfo {author} {\bibfnamefont {E.}~\bibnamefont
  {Ma}}\ and\ \bibinfo {author} {\bibfnamefont {M.}~\bibnamefont {Raidal}},\
  }\href {\doibase 10.1088/0954-3899/28/1/307} {\bibfield  {journal} {\bibinfo
  {journal} {J. Phys. G}\ }\textbf {\bibinfo {volume} {28}},\ \bibinfo {pages}
  {95} (\bibinfo {year} {2002})},\ \Eprint
  {http://arxiv.org/abs/hep-ph/0012366} {arXiv:hep-ph/0012366 [hep-ph]}
  \BibitemShut {NoStop}%
%%CITATION = HEP-PH/0012366;%%
\bibitem [{\citenamefont {Hewett}\ and\ \citenamefont
  {Rizzo}(1989)}]{Hewett:1988xc}%
  \BibitemOpen
  \bibfield  {author} {\bibinfo {author} {\bibfnamefont {J.~L.}\ \bibnamefont
  {Hewett}}\ and\ \bibinfo {author} {\bibfnamefont {T.~G.}\ \bibnamefont
  {Rizzo}},\ }\href {\doibase 10.1016/0370-1573(89)90071-9} {\bibfield
  {journal} {\bibinfo  {journal} {Phys. Rep.}\ }\textbf {\bibinfo {volume}
  {183}},\ \bibinfo {pages} {193} (\bibinfo {year} {1989})}\BibitemShut
  {NoStop}%
%%CITATION = PRPLC,183,193;%%
\bibitem [{\citenamefont {Hesselbach}\ \emph {et~al.}(2002)\citenamefont
  {Hesselbach}, \citenamefont {Franke},\ and\ \citenamefont
  {Fraas}}]{Hesselbach:2001ri}%
  \BibitemOpen
  \bibfield  {author} {\bibinfo {author} {\bibfnamefont {S.}~\bibnamefont
  {Hesselbach}}, \bibinfo {author} {\bibfnamefont {F.}~\bibnamefont {Franke}},
  \ and\ \bibinfo {author} {\bibfnamefont {H.}~\bibnamefont {Fraas}},\ }\href
  {\doibase 10.1007/s100520100865} {\bibfield  {journal} {\bibinfo  {journal}
  {Eur. Phys. J. C}\ }\textbf {\bibinfo {volume} {23}},\ \bibinfo {pages} {149}
  (\bibinfo {year} {2002})},\ \Eprint {http://arxiv.org/abs/hep-ph/0107080}
  {arXiv:hep-ph/0107080 [hep-ph]} \BibitemShut {NoStop}%
%%CITATION = HEP-PH/0107080;%%
\bibitem [{\citenamefont {Barger}\ \emph {et~al.}(2005)\citenamefont {Barger},
  \citenamefont {Langacker},\ and\ \citenamefont {Lee}}]{Barger:2005hb}%
  \BibitemOpen
  \bibfield  {author} {\bibinfo {author} {\bibfnamefont {V.}~\bibnamefont
  {Barger}}, \bibinfo {author} {\bibfnamefont {P.}~\bibnamefont {Langacker}}, \
  and\ \bibinfo {author} {\bibfnamefont {H.-S.}\ \bibnamefont {Lee}},\ }\href
  {\doibase 10.1016/j.physletb.2005.09.023} {\bibfield  {journal} {\bibinfo
  {journal} {Phys. Lett. B}\ }\textbf {\bibinfo {volume} {630}},\ \bibinfo
  {pages} {85} (\bibinfo {year} {2005})},\ \Eprint
  {http://arxiv.org/abs/hep-ph/0508027} {arXiv:hep-ph/0508027 [hep-ph]}
  \BibitemShut {NoStop}%
%%CITATION = HEP-PH/0508027;%%
\bibitem [{\citenamefont {Choi}\ \emph {et~al.}(2007)\citenamefont {Choi},
  \citenamefont {Haber}, \citenamefont {Kalinowski},\ and\ \citenamefont
  {Zerwas}}]{Choi:2006fz}%
  \BibitemOpen
  \bibfield  {author} {\bibinfo {author} {\bibfnamefont {S.~Y.}\ \bibnamefont
  {Choi}}, \bibinfo {author} {\bibfnamefont {H.~E.}\ \bibnamefont {Haber}},
  \bibinfo {author} {\bibfnamefont {J.}~\bibnamefont {Kalinowski}}, \ and\
  \bibinfo {author} {\bibfnamefont {P.~M.}\ \bibnamefont {Zerwas}},\ }\href
  {\doibase 10.1016/j.nuclphysb.2007.04.019} {\bibfield  {journal} {\bibinfo
  {journal} {Nucl. Phys.}\ }\textbf {\bibinfo {volume} {B778}},\ \bibinfo
  {pages} {85} (\bibinfo {year} {2007})},\ \Eprint
  {http://arxiv.org/abs/hep-ph/0612218} {arXiv:hep-ph/0612218 [hep-ph]}
  \BibitemShut {NoStop}%
%%CITATION = HEP-PH/0612218;%%
\bibitem [{\citenamefont {Barger}\ \emph {et~al.}(2007)\citenamefont {Barger},
  \citenamefont {Langacker}, \citenamefont {Lewis}, \citenamefont {McCaskey},
  \citenamefont {Shaughnessy},\ and\ \citenamefont {Yencho}}]{Barger:2007nv}%
  \BibitemOpen
  \bibfield  {author} {\bibinfo {author} {\bibfnamefont {V.}~\bibnamefont
  {Barger}}, \bibinfo {author} {\bibfnamefont {P.}~\bibnamefont {Langacker}},
  \bibinfo {author} {\bibfnamefont {I.}~\bibnamefont {Lewis}}, \bibinfo
  {author} {\bibfnamefont {M.}~\bibnamefont {McCaskey}}, \bibinfo {author}
  {\bibfnamefont {G.}~\bibnamefont {Shaughnessy}}, \ and\ \bibinfo {author}
  {\bibfnamefont {B.}~\bibnamefont {Yencho}},\ }\href {\doibase
  10.1103/PhysRevD.75.115002} {\bibfield  {journal} {\bibinfo  {journal} {Phys.
  Rev. D}\ }\textbf {\bibinfo {volume} {75}},\ \bibinfo {pages} {115002}
  (\bibinfo {year} {2007})},\ \Eprint {http://arxiv.org/abs/hep-ph/0702036}
  {arXiv:hep-ph/0702036 [HEP-PH]} \BibitemShut {NoStop}%
%%CITATION = HEP-PH/0702036;%%
\bibitem [{\citenamefont {Kalinowski}\ \emph {et~al.}(2009)\citenamefont
  {Kalinowski}, \citenamefont {King},\ and\ \citenamefont
  {Roberts}}]{Kalinowski:2008iq}%
  \BibitemOpen
  \bibfield  {author} {\bibinfo {author} {\bibfnamefont {J.}~\bibnamefont
  {Kalinowski}}, \bibinfo {author} {\bibfnamefont {S.~F.}\ \bibnamefont
  {King}}, \ and\ \bibinfo {author} {\bibfnamefont {J.~P.}\ \bibnamefont
  {Roberts}},\ }\href {\doibase 10.1088/1126-6708/2009/01/066} {\bibfield
  {journal} {\bibinfo  {journal} {J. High Energy Phys.}\ }\textbf {\bibinfo
  {volume} {0901}},\ \bibinfo {pages} {066} (\bibinfo {year} {2009})},\ \Eprint
  {http://arxiv.org/abs/0811.2204} {arXiv:0811.2204 [hep-ph]} \BibitemShut
  {NoStop}%
%%CITATION = ARXIV:0811.2204;%%
\bibitem [{\citenamefont {Stech}\ and\ \citenamefont
  {Tavartkiladze}(2008)}]{Stech:2008wd}%
  \BibitemOpen
  \bibfield  {author} {\bibinfo {author} {\bibfnamefont {B.}~\bibnamefont
  {Stech}}\ and\ \bibinfo {author} {\bibfnamefont {Z.}~\bibnamefont
  {Tavartkiladze}},\ }\href {\doibase 10.1103/PhysRevD.77.076009} {\bibfield
  {journal} {\bibinfo  {journal} {Phys. Rev. D}\ }\textbf {\bibinfo {volume}
  {77}},\ \bibinfo {pages} {076009} (\bibinfo {year} {2008})},\ \Eprint
  {http://arxiv.org/abs/0802.0894} {arXiv:0802.0894 [hep-ph]} \BibitemShut
  {NoStop}%
%%CITATION = ARXIV:0802.0894;%%
\bibitem [{\citenamefont {Kang}\ \emph
  {et~al.}(2005{\natexlab{a}})\citenamefont {Kang}, \citenamefont {Langacker},\
  and\ \citenamefont {Li}}]{Kang:2004ix}%
  \BibitemOpen
  \bibfield  {author} {\bibinfo {author} {\bibfnamefont {J.-h.}\ \bibnamefont
  {Kang}}, \bibinfo {author} {\bibfnamefont {P.}~\bibnamefont {Langacker}}, \
  and\ \bibinfo {author} {\bibfnamefont {T.-j.}\ \bibnamefont {Li}},\ }\href
  {\doibase 10.1103/PhysRevD.71.015012} {\bibfield  {journal} {\bibinfo
  {journal} {Phys. Rev. D}\ }\textbf {\bibinfo {volume} {71}},\ \bibinfo
  {pages} {015012} (\bibinfo {year} {2005}{\natexlab{a}})},\ \Eprint
  {http://arxiv.org/abs/hep-ph/0411404} {arXiv:hep-ph/0411404 [hep-ph]}
  \BibitemShut {NoStop}%
%%CITATION = HEP-PH/0411404;%%
\bibitem [{\citenamefont {Kang}\ \emph
  {et~al.}(2005{\natexlab{b}})\citenamefont {Kang}, \citenamefont {Langacker},
  \citenamefont {Li},\ and\ \citenamefont {Liu}}]{Kang:2004pp}%
  \BibitemOpen
  \bibfield  {author} {\bibinfo {author} {\bibfnamefont {J.}~\bibnamefont
  {Kang}}, \bibinfo {author} {\bibfnamefont {P.}~\bibnamefont {Langacker}},
  \bibinfo {author} {\bibfnamefont {T.-j.}\ \bibnamefont {Li}}, \ and\ \bibinfo
  {author} {\bibfnamefont {T.}~\bibnamefont {Liu}},\ }\href {\doibase
  10.1103/PhysRevLett.94.061801} {\bibfield  {journal} {\bibinfo  {journal}
  {Phys. Rev. Lett.}\ }\textbf {\bibinfo {volume} {94}},\ \bibinfo {pages}
  {061801} (\bibinfo {year} {2005}{\natexlab{b}})},\ \Eprint
  {http://arxiv.org/abs/hep-ph/0402086} {arXiv:hep-ph/0402086 [hep-ph]}
  \BibitemShut {NoStop}%
%%CITATION = HEP-PH/0402086;%%
\bibitem [{\citenamefont {Kang}\ and\ \citenamefont
  {Langacker}(2005)}]{Kang:2004bz}%
  \BibitemOpen
  \bibfield  {author} {\bibinfo {author} {\bibfnamefont {J.}~\bibnamefont
  {Kang}}\ and\ \bibinfo {author} {\bibfnamefont {P.}~\bibnamefont
  {Langacker}},\ }\href {\doibase 10.1103/PhysRevD.71.035014} {\bibfield
  {journal} {\bibinfo  {journal} {Phys. Rev. D}\ }\textbf {\bibinfo {volume}
  {71}},\ \bibinfo {pages} {035014} (\bibinfo {year} {2005})},\ \Eprint
  {http://arxiv.org/abs/hep-ph/0412190} {arXiv:hep-ph/0412190 [hep-ph]}
  \BibitemShut {NoStop}%
%%CITATION = HEP-PH/0412190;%%
\bibitem [{\citenamefont {Baumgart}\ \emph {et~al.}(2007)\citenamefont
  {Baumgart}, \citenamefont {Hartman}, \citenamefont {Kilic},\ and\
  \citenamefont {Wang}}]{Baumgart:2006pa}%
  \BibitemOpen
  \bibfield  {author} {\bibinfo {author} {\bibfnamefont {M.}~\bibnamefont
  {Baumgart}}, \bibinfo {author} {\bibfnamefont {T.}~\bibnamefont {Hartman}},
  \bibinfo {author} {\bibfnamefont {C.}~\bibnamefont {Kilic}}, \ and\ \bibinfo
  {author} {\bibfnamefont {L.-T.}\ \bibnamefont {Wang}},\ }\href {\doibase
  10.1088/1126-6708/2007/11/084} {\bibfield  {journal} {\bibinfo  {journal} {J.
  High Energy Phys.}\ }\textbf {\bibinfo {volume} {0711}},\ \bibinfo {pages}
  {084} (\bibinfo {year} {2007})},\ \Eprint
  {http://arxiv.org/abs/hep-ph/0608172} {arXiv:hep-ph/0608172 [hep-ph]}
  \BibitemShut {NoStop}%
%%CITATION = HEP-PH/0608172;%%
\bibitem [{\citenamefont {Athron}\ \emph {et~al.}(2011)\citenamefont {Athron},
  \citenamefont {King}, \citenamefont {Miller}, \citenamefont {Moretti},\ and\
  \citenamefont {Nevzorov}}]{Athron:2011wu}%
  \BibitemOpen
  \bibfield  {author} {\bibinfo {author} {\bibfnamefont {P.}~\bibnamefont
  {Athron}}, \bibinfo {author} {\bibfnamefont {S.~F.}\ \bibnamefont {King}},
  \bibinfo {author} {\bibfnamefont {D.~J.}\ \bibnamefont {Miller}}, \bibinfo
  {author} {\bibfnamefont {S.}~\bibnamefont {Moretti}}, \ and\ \bibinfo
  {author} {\bibfnamefont {R.}~\bibnamefont {Nevzorov}},\ }\href {\doibase
  10.1103/PhysRevD.84.055006} {\bibfield  {journal} {\bibinfo  {journal} {Phys.
  Rev. D}\ }\textbf {\bibinfo {volume} {84}},\ \bibinfo {pages} {055006}
  (\bibinfo {year} {2011})},\ \Eprint {http://arxiv.org/abs/1102.4363}
  {arXiv:1102.4363 [hep-ph]} \BibitemShut {NoStop}%
%%CITATION = ARXIV:1102.4363;%%
\bibitem [{\citenamefont {Chang}\ \emph {et~al.}(2011)\citenamefont {Chang},
  \citenamefont {Cheung},\ and\ \citenamefont {Yuan}}]{Chang:2011be}%
  \BibitemOpen
  \bibfield  {author} {\bibinfo {author} {\bibfnamefont {C.-F.}\ \bibnamefont
  {Chang}}, \bibinfo {author} {\bibfnamefont {K.}~\bibnamefont {Cheung}}, \
  and\ \bibinfo {author} {\bibfnamefont {T.-C.}\ \bibnamefont {Yuan}},\ }\href
  {\doibase 10.1007/JHEP09(2011)058} {\bibfield  {journal} {\bibinfo  {journal}
  {J. High Energy Phys.}\ }\textbf {\bibinfo {volume} {1109}},\ \bibinfo
  {pages} {058} (\bibinfo {year} {2011})},\ \Eprint
  {http://arxiv.org/abs/1107.1133} {arXiv:1107.1133 [hep-ph]} \BibitemShut
  {NoStop}%
%%CITATION = ARXIV:1107.1133;%%
\bibitem [{\citenamefont {Gutierrez-Rodriguez}\ \emph
  {et~al.}(2007)\citenamefont {Gutierrez-Rodriguez}, \citenamefont
  {Hernandez-Ruiz},\ and\ \citenamefont {Perez}}]{GutierrezRodriguez:2006hb}%
  \BibitemOpen
  \bibfield  {author} {\bibinfo {author} {\bibfnamefont {A.}~\bibnamefont
  {Gutierrez-Rodriguez}}, \bibinfo {author} {\bibfnamefont {M.~A.}\
  \bibnamefont {Hernandez-Ruiz}}, \ and\ \bibinfo {author} {\bibfnamefont
  {M.~A.}\ \bibnamefont {Perez}},\ }\href {\doibase 10.1142/S0217751X07036865}
  {\bibfield  {journal} {\bibinfo  {journal} {Int. J. Mod. Phys. A}\ }\textbf
  {\bibinfo {volume} {22}},\ \bibinfo {pages} {3493} (\bibinfo {year}
  {2007})},\ \Eprint {http://arxiv.org/abs/hep-ph/0611235}
  {arXiv:hep-ph/0611235 [hep-ph]} \BibitemShut {NoStop}%
%%CITATION = HEP-PH/0611235;%%
\bibitem [{\citenamefont {Asano}\ \emph {et~al.}()\citenamefont {Asano},
  \citenamefont {Kikuchi},\ and\ \citenamefont {Kim}}]{Asano:2008ju}%
  \BibitemOpen
  \bibfield  {author} {\bibinfo {author} {\bibfnamefont {M.}~\bibnamefont
  {Asano}}, \bibinfo {author} {\bibfnamefont {T.}~\bibnamefont {Kikuchi}}, \
  and\ \bibinfo {author} {\bibfnamefont {S.-G.}\ \bibnamefont {Kim}},\
  }\href@noop {} {\ }\Eprint {http://arxiv.org/abs/0807.5084} {arXiv:0807.5084
  [hep-ph]} \BibitemShut {NoStop}%
%%CITATION = ARXIV:0807.5084;%%
\bibitem [{\citenamefont {Daikoku}\ and\ \citenamefont
  {Suematsu}(2000)}]{Daikoku:2000ep}%
  \BibitemOpen
  \bibfield  {author} {\bibinfo {author} {\bibfnamefont {Y.}~\bibnamefont
  {Daikoku}}\ and\ \bibinfo {author} {\bibfnamefont {D.}~\bibnamefont
  {Suematsu}},\ }\href {\doibase 10.1103/PhysRevD.62.095006} {\bibfield
  {journal} {\bibinfo  {journal} {Phys. Rev. D}\ }\textbf {\bibinfo {volume}
  {62}},\ \bibinfo {pages} {095006} (\bibinfo {year} {2000})},\ \Eprint
  {http://arxiv.org/abs/hep-ph/0003205} {arXiv:hep-ph/0003205 [hep-ph]}
  \BibitemShut {NoStop}%
%%CITATION = HEP-PH/0003205;%%
\bibitem [{\citenamefont {Ham}\ \emph {et~al.}(2008)\citenamefont {Ham},
  \citenamefont {Hur}, \citenamefont {Ko},\ and\ \citenamefont
  {Oh}}]{Ham:2008xf}%
  \BibitemOpen
  \bibfield  {author} {\bibinfo {author} {\bibfnamefont {S.~W.}\ \bibnamefont
  {Ham}}, \bibinfo {author} {\bibfnamefont {T.}~\bibnamefont {Hur}}, \bibinfo
  {author} {\bibfnamefont {P.}~\bibnamefont {Ko}}, \ and\ \bibinfo {author}
  {\bibfnamefont {S.~K.}\ \bibnamefont {Oh}},\ }\href {\doibase
  10.1088/0954-3899/35/9/095007} {\bibfield  {journal} {\bibinfo  {journal} {J.
  Phys. G}\ }\textbf {\bibinfo {volume} {35}},\ \bibinfo {pages} {095007}
  (\bibinfo {year} {2008})},\ \Eprint {http://arxiv.org/abs/0801.2361}
  {arXiv:0801.2361 [hep-ph]} \BibitemShut {NoStop}%
%%CITATION = ARXIV:0801.2361;%%
\bibitem [{\citenamefont {Howl}\ and\ \citenamefont
  {King}(2008)}]{Howl:2007zi}%
  \BibitemOpen
  \bibfield  {author} {\bibinfo {author} {\bibfnamefont {R.}~\bibnamefont
  {Howl}}\ and\ \bibinfo {author} {\bibfnamefont {S.~F.}\ \bibnamefont
  {King}},\ }\href {\doibase 10.1088/1126-6708/2008/01/030} {\bibfield
  {journal} {\bibinfo  {journal} {J. High Energy Phys.}\ }\textbf {\bibinfo
  {volume} {0801}},\ \bibinfo {pages} {030} (\bibinfo {year} {2008})},\ \Eprint
  {http://arxiv.org/abs/0708.1451} {arXiv:0708.1451 [hep-ph]} \BibitemShut
  {NoStop}%
%%CITATION = ARXIV:0708.1451;%%
\bibitem [{\citenamefont {Braam}\ \emph
  {et~al.}(2010{\natexlab{a}})\citenamefont {Braam}, \citenamefont {Reuter},\
  and\ \citenamefont {Wiesler}}]{Braam:2009fi}%
  \BibitemOpen
  \bibfield  {author} {\bibinfo {author} {\bibfnamefont {F.}~\bibnamefont
  {Braam}}, \bibinfo {author} {\bibfnamefont {J.}~\bibnamefont {Reuter}}, \
  and\ \bibinfo {author} {\bibfnamefont {D.}~\bibnamefont {Wiesler}},\ }\href
  {\doibase 10.1063/1.3327620} {\bibfield  {journal} {\bibinfo  {journal} {AIP
  Conf. Proc.}\ }\textbf {\bibinfo {volume} {1200}},\ \bibinfo {pages} {458}
  (\bibinfo {year} {2010}{\natexlab{a}})},\ \Eprint
  {http://arxiv.org/abs/0909.3081} {arXiv:0909.3081 [hep-ph]} \BibitemShut
  {NoStop}%
%%CITATION = ARXIV:0909.3081;%%
\bibitem [{\citenamefont {Braam}\ \emph
  {et~al.}(2010{\natexlab{b}})\citenamefont {Braam}, \citenamefont {Knochel},\
  and\ \citenamefont {Reuter}}]{Braam:2010sy}%
  \BibitemOpen
  \bibfield  {author} {\bibinfo {author} {\bibfnamefont {F.}~\bibnamefont
  {Braam}}, \bibinfo {author} {\bibfnamefont {A.}~\bibnamefont {Knochel}}, \
  and\ \bibinfo {author} {\bibfnamefont {J.}~\bibnamefont {Reuter}},\ }\href
  {\doibase 10.1007/JHEP06(2010)013} {\bibfield  {journal} {\bibinfo  {journal}
  {J. High Energy Phys.}\ }\textbf {\bibinfo {volume} {1006}},\ \bibinfo
  {pages} {013} (\bibinfo {year} {2010}{\natexlab{b}})},\ \Eprint
  {http://arxiv.org/abs/1001.4074} {arXiv:1001.4074 [hep-ph]} \BibitemShut
  {NoStop}%
%%CITATION = ARXIV:1001.4074;%%
\bibitem [{\citenamefont {Hall}\ and\ \citenamefont
  {King}(2011)}]{Hall:2011zq}%
  \BibitemOpen
  \bibfield  {author} {\bibinfo {author} {\bibfnamefont {J.~P.}\ \bibnamefont
  {Hall}}\ and\ \bibinfo {author} {\bibfnamefont {S.~F.}\ \bibnamefont
  {King}},\ }\href {\doibase 10.1007/JHEP06(2011)006} {\bibfield  {journal}
  {\bibinfo  {journal} {J. High Energy Phys.}\ }\textbf {\bibinfo {volume}
  {1106}},\ \bibinfo {pages} {006} (\bibinfo {year} {2011})},\ \Eprint
  {http://arxiv.org/abs/1104.2259} {arXiv:1104.2259 [hep-ph]} \BibitemShut
  {NoStop}%
%%CITATION = ARXIV:1104.2259;%%
\bibitem [{\citenamefont {Nevzorov}(2013)}]{Nevzorov:2012hs}%
  \BibitemOpen
  \bibfield  {author} {\bibinfo {author} {\bibfnamefont {R.}~\bibnamefont
  {Nevzorov}},\ }\href {\doibase 10.1103/PhysRevD.87.015029} {\bibfield
  {journal} {\bibinfo  {journal} {Phys. Rev. D}\ }\textbf {\bibinfo {volume}
  {87}},\ \bibinfo {pages} {015029} (\bibinfo {year} {2013})},\ \Eprint
  {http://arxiv.org/abs/1205.5967} {arXiv:1205.5967 [hep-ph]} \BibitemShut
  {NoStop}%
%%CITATION = ARXIV:1205.5967;%%
\bibitem [{\citenamefont {King}\ \emph {et~al.}(2008)\citenamefont {King},
  \citenamefont {Luo}, \citenamefont {Miller},\ and\ \citenamefont
  {Nevzorov}}]{King:2008qb}%
  \BibitemOpen
  \bibfield  {author} {\bibinfo {author} {\bibfnamefont {S.~F.}\ \bibnamefont
  {King}}, \bibinfo {author} {\bibfnamefont {R.}~\bibnamefont {Luo}}, \bibinfo
  {author} {\bibfnamefont {D.~J.}\ \bibnamefont {Miller}}, \ and\ \bibinfo
  {author} {\bibfnamefont {R.}~\bibnamefont {Nevzorov}},\ }\href {\doibase
  10.1088/1126-6708/2008/12/042} {\bibfield  {journal} {\bibinfo  {journal} {J.
  High Energy Phys.}\ }\textbf {\bibinfo {volume} {0812}},\ \bibinfo {pages}
  {042} (\bibinfo {year} {2008})},\ \Eprint {http://arxiv.org/abs/0806.0330}
  {arXiv:0806.0330 [hep-ph]} \BibitemShut {NoStop}%
%%CITATION = ARXIV:0806.0330;%%
\bibitem [{\citenamefont {King}\ \emph {et~al.}(2009)\citenamefont {King},
  \citenamefont {Luo}, \citenamefont {Miller},\ and\ \citenamefont
  {Nevzorov}}]{King:2008gw}%
  \BibitemOpen
  \bibfield  {author} {\bibinfo {author} {\bibfnamefont {S.~F.}\ \bibnamefont
  {King}}, \bibinfo {author} {\bibfnamefont {R.}~\bibnamefont {Luo}}, \bibinfo
  {author} {\bibfnamefont {D.~J.}\ \bibnamefont {Miller}}, \ and\ \bibinfo
  {author} {\bibfnamefont {R.}~\bibnamefont {Nevzorov}},\ }\href {\doibase
  10.1063/1.3052011} {\bibfield  {journal} {\bibinfo  {journal} {AIP Conf.
  Proc.}\ }\textbf {\bibinfo {volume} {1078}},\ \bibinfo {pages} {509}
  (\bibinfo {year} {2009})},\ \Eprint {http://arxiv.org/abs/0808.3739}
  {arXiv:0808.3739 [hep-ph]} \BibitemShut {NoStop}%
%%CITATION = ARXIV:0808.3739;%%
\bibitem [{\citenamefont {Chao}()}]{Chao:2014hya}%
  \BibitemOpen
  \bibfield  {author} {\bibinfo {author} {\bibfnamefont {W.}~\bibnamefont
  {Chao}},\ }\href@noop {} {\ }\Eprint {http://arxiv.org/abs/1411.5575}
  {arXiv:1411.5575 [hep-ph]} \BibitemShut {NoStop}%
%%CITATION = ARXIV:1411.5575;%%
\bibitem [{\citenamefont {King}\ \emph {et~al.}(2007)\citenamefont {King},
  \citenamefont {Moretti},\ and\ \citenamefont {Nevzorov}}]{King:2007uj}%
  \BibitemOpen
  \bibfield  {author} {\bibinfo {author} {\bibfnamefont {S.~F.}\ \bibnamefont
  {King}}, \bibinfo {author} {\bibfnamefont {S.}~\bibnamefont {Moretti}}, \
  and\ \bibinfo {author} {\bibfnamefont {R.}~\bibnamefont {Nevzorov}},\ }\href
  {\doibase 10.1016/j.physletb.2007.04.061} {\bibfield  {journal} {\bibinfo
  {journal} {Phys. Lett. B}\ }\textbf {\bibinfo {volume} {650}},\ \bibinfo
  {pages} {57} (\bibinfo {year} {2007})},\ \Eprint
  {http://arxiv.org/abs/hep-ph/0701064} {arXiv:hep-ph/0701064 [hep-ph]}
  \BibitemShut {NoStop}%
%%CITATION = HEP-PH/0701064;%%
\bibitem [{\citenamefont {Howl}\ and\ \citenamefont
  {King}(2007)}]{Howl:2007hq}%
  \BibitemOpen
  \bibfield  {author} {\bibinfo {author} {\bibfnamefont {R.}~\bibnamefont
  {Howl}}\ and\ \bibinfo {author} {\bibfnamefont {S.~F.}\ \bibnamefont
  {King}},\ }\href {\doibase 10.1016/j.physletb.2007.07.035} {\bibfield
  {journal} {\bibinfo  {journal} {Phys. Lett. B}\ }\textbf {\bibinfo {volume}
  {652}},\ \bibinfo {pages} {331} (\bibinfo {year} {2007})},\ \Eprint
  {http://arxiv.org/abs/0705.0301} {arXiv:0705.0301 [hep-ph]} \BibitemShut
  {NoStop}%
%%CITATION = ARXIV:0705.0301;%%
\bibitem [{\citenamefont {Hall}\ \emph {et~al.}(2011)\citenamefont {Hall},
  \citenamefont {King}, \citenamefont {Nevzorov}, \citenamefont {Pakvasa},\
  and\ \citenamefont {Sher}}]{Hall:2010ix}%
  \BibitemOpen
  \bibfield  {author} {\bibinfo {author} {\bibfnamefont {J.~P.}\ \bibnamefont
  {Hall}}, \bibinfo {author} {\bibfnamefont {S.~F.}\ \bibnamefont {King}},
  \bibinfo {author} {\bibfnamefont {R.}~\bibnamefont {Nevzorov}}, \bibinfo
  {author} {\bibfnamefont {S.}~\bibnamefont {Pakvasa}}, \ and\ \bibinfo
  {author} {\bibfnamefont {M.}~\bibnamefont {Sher}},\ }\href {\doibase
  10.1103/PhysRevD.83.075013} {\bibfield  {journal} {\bibinfo  {journal} {Phys.
  Rev. D}\ }\textbf {\bibinfo {volume} {83}},\ \bibinfo {pages} {075013}
  (\bibinfo {year} {2011})},\ \Eprint {http://arxiv.org/abs/1012.5114}
  {arXiv:1012.5114 [hep-ph]} \BibitemShut {NoStop}%
%%CITATION = ARXIV:1012.5114;%%
\bibitem [{\citenamefont {Nevzorov}\ and\ \citenamefont
  {Pakvasa}(2014)}]{Nevzorov:2013tta}%
  \BibitemOpen
  \bibfield  {author} {\bibinfo {author} {\bibfnamefont {R.}~\bibnamefont
  {Nevzorov}}\ and\ \bibinfo {author} {\bibfnamefont {S.}~\bibnamefont
  {Pakvasa}},\ }\href {\doibase 10.1016/j.physletb.2013.11.050} {\bibfield
  {journal} {\bibinfo  {journal} {Phys. Lett. B}\ }\textbf {\bibinfo {volume}
  {728}},\ \bibinfo {pages} {210} (\bibinfo {year} {2014})},\ \Eprint
  {http://arxiv.org/abs/1308.1021} {arXiv:1308.1021 [hep-ph]} \BibitemShut
  {NoStop}%
%%CITATION = ARXIV:1308.1021;%%
\bibitem [{\citenamefont {Hall}\ and\ \citenamefont
  {King}(2009)}]{Hall:2009aj}%
  \BibitemOpen
  \bibfield  {author} {\bibinfo {author} {\bibfnamefont {J.~P.}\ \bibnamefont
  {Hall}}\ and\ \bibinfo {author} {\bibfnamefont {S.~F.}\ \bibnamefont
  {King}},\ }\href {\doibase 10.1088/1126-6708/2009/08/088} {\bibfield
  {journal} {\bibinfo  {journal} {J. High Energy Phys.}\ }\textbf {\bibinfo
  {volume} {0908}},\ \bibinfo {pages} {088} (\bibinfo {year} {2009})},\ \Eprint
  {http://arxiv.org/abs/0905.2696} {arXiv:0905.2696 [hep-ph]} \BibitemShut
  {NoStop}%
%%CITATION = ARXIV:0905.2696;%%
\bibitem [{\citenamefont {Aprile}\ \emph {et~al.}(2011)\citenamefont {Aprile}
  \emph {et~al.}}]{2011PhRvL.107m1302A}%
  \BibitemOpen
  \bibfield  {author} {\bibinfo {author} {\bibfnamefont {E.}~\bibnamefont
  {Aprile}} \emph {et~al.} (\bibinfo {collaboration} {XENON100
  Collaboration}),\ }\href {\doibase 10.1103/PhysRevLett.107.131302} {\bibfield
   {journal} {\bibinfo  {journal} {Phys. Rev. Lett.}\ }\textbf {\bibinfo
  {volume} {107}},\ \bibinfo {pages} {131302} (\bibinfo {year} {2011})},\
  \Eprint {http://arxiv.org/abs/1104.2549} {arXiv:1104.2549 [astro-ph.CO]}
  \BibitemShut {NoStop}%
%%CITATION = ARXIV:1104.2549;%%
\bibitem [{\citenamefont {Aprile}\ \emph {et~al.}(2012)\citenamefont {Aprile}
  \emph {et~al.}}]{2012PhRvL.109r1301A}%
  \BibitemOpen
  \bibfield  {author} {\bibinfo {author} {\bibfnamefont {E.}~\bibnamefont
  {Aprile}} \emph {et~al.} (\bibinfo {collaboration} {XENON100
  Collaboration}),\ }\href {\doibase 10.1103/PhysRevLett.109.181301} {\bibfield
   {journal} {\bibinfo  {journal} {Phys. Rev. Lett.}\ }\textbf {\bibinfo
  {volume} {109}},\ \bibinfo {pages} {181301} (\bibinfo {year} {2012})},\
  \Eprint {http://arxiv.org/abs/1207.5988} {arXiv:1207.5988 [astro-ph.CO]}
  \BibitemShut {NoStop}%
%%CITATION = ARXIV:1207.5988;%%
\bibitem [{\citenamefont {Akerib}\ \emph {et~al.}(2014)\citenamefont {Akerib}
  \emph {et~al.}}]{Akerib:2013tjd}%
  \BibitemOpen
  \bibfield  {author} {\bibinfo {author} {\bibfnamefont {D.~S.}\ \bibnamefont
  {Akerib}} \emph {et~al.} (\bibinfo {collaboration} {LUX Collaboration}),\
  }\href {\doibase 10.1103/PhysRevLett.112.091303} {\bibfield  {journal}
  {\bibinfo  {journal} {Phys. Rev. Lett.}\ }\textbf {\bibinfo {volume} {112}},\
  \bibinfo {pages} {091303} (\bibinfo {year} {2014})},\ \Eprint
  {http://arxiv.org/abs/1310.8214} {arXiv:1310.8214 [astro-ph.CO]} \BibitemShut
  {NoStop}%
%%CITATION = ARXIV:1310.8214;%%
\bibitem [{\citenamefont {Rizzo}(2012)}]{Rizzo:2012rf}%
  \BibitemOpen
  \bibfield  {author} {\bibinfo {author} {\bibfnamefont {T.~G.}\ \bibnamefont
  {Rizzo}},\ }\href {\doibase 10.1103/PhysRevD.85.099901,
  10.1103/PhysRevD.85.055010} {\bibfield  {journal} {\bibinfo  {journal} {Phys.
  Rev. D}\ }\textbf {\bibinfo {volume} {85}},\ \bibinfo {pages} {055010}
  (\bibinfo {year} {2012})},\ \Eprint {http://arxiv.org/abs/1201.2898}
  {arXiv:1201.2898 [hep-ph]} \BibitemShut {NoStop}%
%%CITATION = ARXIV:1201.2898;%%
\bibitem [{\citenamefont {Miller}\ \emph {et~al.}(2013)\citenamefont {Miller},
  \citenamefont {Morais},\ and\ \citenamefont {Pandita}}]{Miller:2012vn}%
  \BibitemOpen
  \bibfield  {author} {\bibinfo {author} {\bibfnamefont {D.~J.}\ \bibnamefont
  {Miller}}, \bibinfo {author} {\bibfnamefont {A.~P.}\ \bibnamefont {Morais}},
  \ and\ \bibinfo {author} {\bibfnamefont {P.~N.}\ \bibnamefont {Pandita}},\
  }\href {\doibase 10.1103/PhysRevD.87.015007} {\bibfield  {journal} {\bibinfo
  {journal} {Phys. Rev. D}\ }\textbf {\bibinfo {volume} {87}},\ \bibinfo
  {pages} {015007} (\bibinfo {year} {2013})},\ \Eprint
  {http://arxiv.org/abs/1208.5906} {arXiv:1208.5906 [hep-ph]} \BibitemShut
  {NoStop}%
%%CITATION = ARXIV:1208.5906;%%
\bibitem [{\citenamefont {Sperling}\ \emph {et~al.}(2013)\citenamefont
  {Sperling}, \citenamefont {St{\"o}ckinger},\ and\ \citenamefont
  {Voigt}}]{Sperling:2013eva}%
  \BibitemOpen
  \bibfield  {author} {\bibinfo {author} {\bibfnamefont {M.}~\bibnamefont
  {Sperling}}, \bibinfo {author} {\bibfnamefont {D.}~\bibnamefont
  {St{\"o}ckinger}}, \ and\ \bibinfo {author} {\bibfnamefont {A.}~\bibnamefont
  {Voigt}},\ }\href {\doibase 10.1007/JHEP07(2013)132} {\bibfield  {journal}
  {\bibinfo  {journal} {J. High Energy Phys.}\ }\textbf {\bibinfo {volume}
  {1307}},\ \bibinfo {pages} {132} (\bibinfo {year} {2013})},\ \Eprint
  {http://arxiv.org/abs/1305.1548} {arXiv:1305.1548 [hep-ph]} \BibitemShut
  {NoStop}%
%%CITATION = ARXIV:1305.1548;%%
\bibitem [{\citenamefont {Sperling}\ \emph {et~al.}(2014)\citenamefont
  {Sperling}, \citenamefont {St{\"o}ckinger},\ and\ \citenamefont
  {Voigt}}]{Sperling:2013xqa}%
  \BibitemOpen
  \bibfield  {author} {\bibinfo {author} {\bibfnamefont {M.}~\bibnamefont
  {Sperling}}, \bibinfo {author} {\bibfnamefont {D.}~\bibnamefont
  {St{\"o}ckinger}}, \ and\ \bibinfo {author} {\bibfnamefont {A.}~\bibnamefont
  {Voigt}},\ }\href {\doibase 10.1007/JHEP01(2014)068} {\bibfield  {journal}
  {\bibinfo  {journal} {J. High Energy Phys.}\ }\textbf {\bibinfo {volume}
  {1401}},\ \bibinfo {pages} {068} (\bibinfo {year} {2014})},\ \Eprint
  {http://arxiv.org/abs/1310.7629} {arXiv:1310.7629 [hep-ph]} \BibitemShut
  {NoStop}%
%%CITATION = ARXIV:1310.7629;%%
\bibitem [{\citenamefont {Athron}\ \emph
  {et~al.}(2012{\natexlab{a}})\citenamefont {Athron}, \citenamefont {King},
  \citenamefont {Miller}, \citenamefont {Moretti},\ and\ \citenamefont
  {Nevzorov}}]{Athron:2012sq}%
  \BibitemOpen
  \bibfield  {author} {\bibinfo {author} {\bibfnamefont {P.}~\bibnamefont
  {Athron}}, \bibinfo {author} {\bibfnamefont {S.~F.}\ \bibnamefont {King}},
  \bibinfo {author} {\bibfnamefont {D.~J.}\ \bibnamefont {Miller}}, \bibinfo
  {author} {\bibfnamefont {S.}~\bibnamefont {Moretti}}, \ and\ \bibinfo
  {author} {\bibfnamefont {R.}~\bibnamefont {Nevzorov}},\ }\href {\doibase
  10.1103/PhysRevD.86.095003} {\bibfield  {journal} {\bibinfo  {journal} {Phys.
  Rev. D}\ }\textbf {\bibinfo {volume} {86}},\ \bibinfo {pages} {095003}
  (\bibinfo {year} {2012}{\natexlab{a}})},\ \Eprint
  {http://arxiv.org/abs/1206.5028} {arXiv:1206.5028 [hep-ph]} \BibitemShut
  {NoStop}%
%%CITATION = ARXIV:1206.5028;%%
\bibitem [{\citenamefont {Athron}\ \emph
  {et~al.}(2012{\natexlab{b}})\citenamefont {Athron}, \citenamefont
  {St{\"o}ckinger},\ and\ \citenamefont {Voigt}}]{Athron:2012pw}%
  \BibitemOpen
  \bibfield  {author} {\bibinfo {author} {\bibfnamefont {P.}~\bibnamefont
  {Athron}}, \bibinfo {author} {\bibfnamefont {D.}~\bibnamefont
  {St{\"o}ckinger}}, \ and\ \bibinfo {author} {\bibfnamefont {A.}~\bibnamefont
  {Voigt}},\ }\href {\doibase 10.1103/PhysRevD.86.095012} {\bibfield  {journal}
  {\bibinfo  {journal} {Phys. Rev. D}\ }\textbf {\bibinfo {volume} {86}},\
  \bibinfo {pages} {095012} (\bibinfo {year} {2012}{\natexlab{b}})},\ \Eprint
  {http://arxiv.org/abs/1209.1470} {arXiv:1209.1470 [hep-ph]} \BibitemShut
  {NoStop}%
%%CITATION = ARXIV:1209.1470;%%
\bibitem [{\citenamefont {Aad}\ \emph {et~al.}(2014{\natexlab{a}})\citenamefont
  {Aad} \emph {et~al.}}]{Aad:2014cka}%
  \BibitemOpen
  \bibfield  {author} {\bibinfo {author} {\bibfnamefont {G.}~\bibnamefont
  {Aad}} \emph {et~al.} (\bibinfo {collaboration} {ATLAS Collaboration}),\
  }\href {\doibase 10.1103/PhysRevD.90.052005} {\bibfield  {journal} {\bibinfo
  {journal} {Phys. Rev. D}\ }\textbf {\bibinfo {volume} {90}},\ \bibinfo
  {pages} {052005} (\bibinfo {year} {2014}{\natexlab{a}})},\ \Eprint
  {http://arxiv.org/abs/1405.4123} {arXiv:1405.4123 [hep-ex]} \BibitemShut
  {NoStop}%
%%CITATION = ARXIV:1405.4123;%%
\bibitem [{\citenamefont {Rizzo}(1998)}]{Rizzo:1998ut}%
  \BibitemOpen
  \bibfield  {author} {\bibinfo {author} {\bibfnamefont {T.~G.}\ \bibnamefont
  {Rizzo}},\ }\href {\doibase 10.1103/PhysRevD.59.015020} {\bibfield  {journal}
  {\bibinfo  {journal} {Phys. Rev. D}\ }\textbf {\bibinfo {volume} {59}},\
  \bibinfo {pages} {015020} (\bibinfo {year} {1998})},\ \Eprint
  {http://arxiv.org/abs/hep-ph/9806397} {arXiv:hep-ph/9806397 [hep-ph]}
  \BibitemShut {NoStop}%
%%CITATION = HEP-PH/9806397;%%
\bibitem [{\citenamefont {Salvioni}\ \emph {et~al.}(2009)\citenamefont
  {Salvioni}, \citenamefont {Villadoro},\ and\ \citenamefont
  {Zwirner}}]{Salvioni:2009mt}%
  \BibitemOpen
  \bibfield  {author} {\bibinfo {author} {\bibfnamefont {E.}~\bibnamefont
  {Salvioni}}, \bibinfo {author} {\bibfnamefont {G.}~\bibnamefont {Villadoro}},
  \ and\ \bibinfo {author} {\bibfnamefont {F.}~\bibnamefont {Zwirner}},\ }\href
  {\doibase 10.1088/1126-6708/2009/11/068} {\bibfield  {journal} {\bibinfo
  {journal} {J. High Energy Phys.}\ }\textbf {\bibinfo {volume} {0911}},\
  \bibinfo {pages} {068} (\bibinfo {year} {2009})},\ \Eprint
  {http://arxiv.org/abs/0909.1320} {arXiv:0909.1320 [hep-ph]} \BibitemShut
  {NoStop}%
%%CITATION = ARXIV:0909.1320;%%
\bibitem [{\citenamefont {Krauss}\ \emph {et~al.}(2012)\citenamefont {Krauss},
  \citenamefont {O'Leary}, \citenamefont {Porod},\ and\ \citenamefont
  {Staub}}]{Krauss:2012ku}%
  \BibitemOpen
  \bibfield  {author} {\bibinfo {author} {\bibfnamefont {M.~E.}\ \bibnamefont
  {Krauss}}, \bibinfo {author} {\bibfnamefont {B.}~\bibnamefont {O'Leary}},
  \bibinfo {author} {\bibfnamefont {W.}~\bibnamefont {Porod}}, \ and\ \bibinfo
  {author} {\bibfnamefont {F.}~\bibnamefont {Staub}},\ }\href {\doibase
  10.1103/PhysRevD.86.055017} {\bibfield  {journal} {\bibinfo  {journal} {Phys.
  Rev. D}\ }\textbf {\bibinfo {volume} {86}},\ \bibinfo {pages} {055017}
  (\bibinfo {year} {2012})},\ \Eprint {http://arxiv.org/abs/1206.3513}
  {arXiv:1206.3513 [hep-ph]} \BibitemShut {NoStop}%
%%CITATION = ARXIV:1206.3513;%%
\bibitem [{\citenamefont {Cassel}\ \emph {et~al.}(2010)\citenamefont {Cassel},
  \citenamefont {Ghilencea},\ and\ \citenamefont {Ross}}]{Cassel:2010px}%
  \BibitemOpen
  \bibfield  {author} {\bibinfo {author} {\bibfnamefont {S.}~\bibnamefont
  {Cassel}}, \bibinfo {author} {\bibfnamefont {D.~M.}\ \bibnamefont
  {Ghilencea}}, \ and\ \bibinfo {author} {\bibfnamefont {G.~G.}\ \bibnamefont
  {Ross}},\ }\href {\doibase 10.1016/j.nuclphysb.2010.03.031} {\bibfield
  {journal} {\bibinfo  {journal} {Nucl. Phys.}\ }\textbf {\bibinfo {volume}
  {B835}},\ \bibinfo {pages} {110} (\bibinfo {year} {2010})},\ \Eprint
  {http://arxiv.org/abs/1001.3884} {arXiv:1001.3884 [hep-ph]} \BibitemShut
  {NoStop}%
%%CITATION = ARXIV:1001.3884;%%
\bibitem [{\citenamefont {Ellwanger}\ \emph {et~al.}(2011)\citenamefont
  {Ellwanger}, \citenamefont {Espitalier-Noel},\ and\ \citenamefont
  {Hugonie}}]{Ellwanger:2011mu}%
  \BibitemOpen
  \bibfield  {author} {\bibinfo {author} {\bibfnamefont {U.}~\bibnamefont
  {Ellwanger}}, \bibinfo {author} {\bibfnamefont {G.}~\bibnamefont
  {Espitalier-Noel}}, \ and\ \bibinfo {author} {\bibfnamefont {C.}~\bibnamefont
  {Hugonie}},\ }\href {\doibase 10.1007/JHEP09(2011)105} {\bibfield  {journal}
  {\bibinfo  {journal} {J. High Energy Phys.}\ }\textbf {\bibinfo {volume}
  {1109}},\ \bibinfo {pages} {105} (\bibinfo {year} {2011})},\ \Eprint
  {http://arxiv.org/abs/1107.2472} {arXiv:1107.2472 [hep-ph]} \BibitemShut
  {NoStop}%
%%CITATION = ARXIV:1107.2472;%%
\bibitem [{\citenamefont {Staub}(2010)}]{Staub:2009bi}%
  \BibitemOpen
  \bibfield  {author} {\bibinfo {author} {\bibfnamefont {F.}~\bibnamefont
  {Staub}},\ }\href {\doibase 10.1016/j.cpc.2010.01.011} {\bibfield  {journal}
  {\bibinfo  {journal} {Comput. Phys. Commun.}\ }\textbf {\bibinfo {volume}
  {181}},\ \bibinfo {pages} {1077} (\bibinfo {year} {2010})},\ \Eprint
  {http://arxiv.org/abs/0909.2863} {arXiv:0909.2863 [hep-ph]} \BibitemShut
  {NoStop}%
%%CITATION = ARXIV:0909.2863;%%
\bibitem [{\citenamefont {Staub}(2011)}]{Staub:2010jh}%
  \BibitemOpen
  \bibfield  {author} {\bibinfo {author} {\bibfnamefont {F.}~\bibnamefont
  {Staub}},\ }\href {\doibase 10.1016/j.cpc.2010.11.030} {\bibfield  {journal}
  {\bibinfo  {journal} {Comput. Phys. Commun.}\ }\textbf {\bibinfo {volume}
  {182}},\ \bibinfo {pages} {808} (\bibinfo {year} {2011})},\ \Eprint
  {http://arxiv.org/abs/1002.0840} {arXiv:1002.0840 [hep-ph]} \BibitemShut
  {NoStop}%
%%CITATION = ARXIV:1002.0840;%%
\bibitem [{\citenamefont {Staub}(2013)}]{Staub:2012pb}%
  \BibitemOpen
  \bibfield  {author} {\bibinfo {author} {\bibfnamefont {F.}~\bibnamefont
  {Staub}},\ }\href {\doibase 10.1016/j.cpc.2013.02.019} {\bibfield  {journal}
  {\bibinfo  {journal} {Comput. Phys. Commun.}\ }\textbf {\bibinfo {volume}
  {184}},\ \bibinfo {pages} {pp. 1792} (\bibinfo {year} {2013})},\ \Eprint
  {http://arxiv.org/abs/1207.0906} {arXiv:1207.0906 [hep-ph]} \BibitemShut
  {NoStop}%
%%CITATION = ARXIV:1207.0906;%%
\bibitem [{\citenamefont {Staub}(2014)}]{Staub:2013tta}%
  \BibitemOpen
  \bibfield  {author} {\bibinfo {author} {\bibfnamefont {F.}~\bibnamefont
  {Staub}},\ }\href {\doibase 10.1016/j.cpc.2014.02.018} {\bibfield  {journal}
  {\bibinfo  {journal} {Comput. Phys. Commun.}\ }\textbf {\bibinfo {volume}
  {185}},\ \bibinfo {pages} {1773} (\bibinfo {year} {2014})},\ \Eprint
  {http://arxiv.org/abs/1309.7223} {arXiv:1309.7223 [hep-ph]} \BibitemShut
  {NoStop}%
%%CITATION = ARXIV:1309.7223;%%
\bibitem [{\citenamefont {Athron}\ \emph
  {et~al.}(2015{\natexlab{b}})\citenamefont {Athron}, \citenamefont {Park},
  \citenamefont {St{\"o}ckinger},\ and\ \citenamefont
  {Voigt}}]{Athron:2014yba}%
  \BibitemOpen
  \bibfield  {author} {\bibinfo {author} {\bibfnamefont {P.}~\bibnamefont
  {Athron}}, \bibinfo {author} {\bibfnamefont {J.-h.}\ \bibnamefont {Park}},
  \bibinfo {author} {\bibfnamefont {D.}~\bibnamefont {St{\"o}ckinger}}, \ and\
  \bibinfo {author} {\bibfnamefont {A.}~\bibnamefont {Voigt}},\ }\href
  {\doibase 10.1016/j.cpc.2014.12.020} {\bibfield  {journal} {\bibinfo
  {journal} {Comput. Phys. Commun.}\ }\textbf {\bibinfo {volume} {190}},\
  \bibinfo {pages} {139} (\bibinfo {year} {2015}{\natexlab{b}})},\ \Eprint
  {http://arxiv.org/abs/1406.2319} {arXiv:1406.2319 [hep-ph]} \BibitemShut
  {NoStop}%
%%CITATION = ARXIV:1406.2319;%%
\bibitem [{\citenamefont {Allanach}(2002)}]{Allanach:2001kg}%
  \BibitemOpen
  \bibfield  {author} {\bibinfo {author} {\bibfnamefont {B.~C.}\ \bibnamefont
  {Allanach}},\ }\href {\doibase 10.1016/S0010-4655(01)00460-X} {\bibfield
  {journal} {\bibinfo  {journal} {Comput. Phys. Commun.}\ }\textbf {\bibinfo
  {volume} {143}},\ \bibinfo {pages} {305} (\bibinfo {year} {2002})},\ \Eprint
  {http://arxiv.org/abs/hep-ph/0104145} {arXiv:hep-ph/0104145 [hep-ph]}
  \BibitemShut {NoStop}%
%%CITATION = HEP-PH/0104145;%%
\bibitem [{\citenamefont {Allanach}\ \emph {et~al.}(2014)\citenamefont
  {Allanach}, \citenamefont {Athron}, \citenamefont {Tunstall}, \citenamefont
  {Voigt},\ and\ \citenamefont {Williams}}]{Allanach:2013kza}%
  \BibitemOpen
  \bibfield  {author} {\bibinfo {author} {\bibfnamefont {B.~C.}\ \bibnamefont
  {Allanach}}, \bibinfo {author} {\bibfnamefont {P.}~\bibnamefont {Athron}},
  \bibinfo {author} {\bibfnamefont {L.~C.}\ \bibnamefont {Tunstall}}, \bibinfo
  {author} {\bibfnamefont {A.}~\bibnamefont {Voigt}}, \ and\ \bibinfo {author}
  {\bibfnamefont {A.~G.}\ \bibnamefont {Williams}},\ }\href {\doibase
  10.1016/j.cpc.2014.04.015} {\bibfield  {journal} {\bibinfo  {journal}
  {Comput. Phys. Commun.}\ }\textbf {\bibinfo {volume} {185}},\ \bibinfo
  {pages} {2322} (\bibinfo {year} {2014})},\ \Eprint
  {http://arxiv.org/abs/1311.7659} {arXiv:1311.7659 [hep-ph]} \BibitemShut
  {NoStop}%
%%CITATION = ARXIV:1311.7659;%%
\bibitem [{\citenamefont {Goodsell}\ \emph {et~al.}(2015)\citenamefont
  {Goodsell}, \citenamefont {Nickel},\ and\ \citenamefont
  {Staub}}]{Goodsell:2014pla}%
  \BibitemOpen
  \bibfield  {author} {\bibinfo {author} {\bibfnamefont {M.~D.}\ \bibnamefont
  {Goodsell}}, \bibinfo {author} {\bibfnamefont {K.}~\bibnamefont {Nickel}}, \
  and\ \bibinfo {author} {\bibfnamefont {F.}~\bibnamefont {Staub}},\ }\href
  {\doibase 10.1103/PhysRevD.91.035021} {\bibfield  {journal} {\bibinfo
  {journal} {Phys. Rev. D}\ }\textbf {\bibinfo {volume} {91}},\ \bibinfo
  {pages} {035021} (\bibinfo {year} {2015})},\ \Eprint
  {http://arxiv.org/abs/1411.4665} {arXiv:1411.4665 [hep-ph]} \BibitemShut
  {NoStop}%
%%CITATION = ARXIV:1411.4665;%%
\bibitem [{\citenamefont {Goodsell}\ \emph {et~al.}()\citenamefont {Goodsell},
  \citenamefont {Nickel},\ and\ \citenamefont {Staub}}]{Goodsell:2015ira}%
  \BibitemOpen
  \bibfield  {author} {\bibinfo {author} {\bibfnamefont {M.}~\bibnamefont
  {Goodsell}}, \bibinfo {author} {\bibfnamefont {K.}~\bibnamefont {Nickel}}, \
  and\ \bibinfo {author} {\bibfnamefont {F.}~\bibnamefont {Staub}},\
  }\href@noop {} {\ }\Eprint {http://arxiv.org/abs/1503.03098}
  {arXiv:1503.03098 [hep-ph]} \BibitemShut {NoStop}%
%%CITATION = ARXIV:1503.03098;%%
\bibitem [{\citenamefont {Carena}\ \emph {et~al.}(1996)\citenamefont {Carena},
  \citenamefont {Quiros},\ and\ \citenamefont {Wagner}}]{Carena:1995wu}%
  \BibitemOpen
  \bibfield  {author} {\bibinfo {author} {\bibfnamefont {M.}~\bibnamefont
  {Carena}}, \bibinfo {author} {\bibfnamefont {M.}~\bibnamefont {Quiros}}, \
  and\ \bibinfo {author} {\bibfnamefont {C.}~\bibnamefont {Wagner}},\ }\href
  {\doibase 10.1016/0550-3213(95)00665-6} {\bibfield  {journal} {\bibinfo
  {journal} {Nucl. Phys.}\ }\textbf {\bibinfo {volume} {B461}},\ \bibinfo
  {pages} {407} (\bibinfo {year} {1996})},\ \Eprint
  {http://arxiv.org/abs/hep-ph/9508343} {arXiv:hep-ph/9508343 [hep-ph]}
  \BibitemShut {NoStop}%
%%CITATION = HEP-PH/9508343;%%
\bibitem [{\citenamefont {Ellwanger}\ and\ \citenamefont
  {Hugonie}(2002)}]{Ellwanger:1999ji}%
  \BibitemOpen
  \bibfield  {author} {\bibinfo {author} {\bibfnamefont {U.}~\bibnamefont
  {Ellwanger}}\ and\ \bibinfo {author} {\bibfnamefont {C.}~\bibnamefont
  {Hugonie}},\ }\href {\doibase 10.1007/s10052-002-0980-4} {\bibfield
  {journal} {\bibinfo  {journal} {Eur. Phys. J. C}\ }\textbf {\bibinfo {volume}
  {25}},\ \bibinfo {pages} {297} (\bibinfo {year} {2002})},\ \Eprint
  {http://arxiv.org/abs/hep-ph/9909260} {arXiv:hep-ph/9909260 [hep-ph]}
  \BibitemShut {NoStop}%
%%CITATION = HEP-PH/9909260;%%
\bibitem [{\citenamefont {Godfrey}\ and\ \citenamefont
  {Martin}()}]{Godfrey:2013eta}%
  \BibitemOpen
  \bibfield  {author} {\bibinfo {author} {\bibfnamefont {S.}~\bibnamefont
  {Godfrey}}\ and\ \bibinfo {author} {\bibfnamefont {T.}~\bibnamefont
  {Martin}},\ }\href@noop {} {\ }\Eprint {http://arxiv.org/abs/1309.1688}
  {arXiv:1309.1688 [hep-ph]} \BibitemShut {NoStop}%
%%CITATION = ARXIV:1309.1688;%%
\bibitem [{\citenamefont {Kraan}(2006)}]{Kraan:2005vy}%
  \BibitemOpen
  \bibfield  {author} {\bibinfo {author} {\bibfnamefont {A.~C.}\ \bibnamefont
  {Kraan}},\ }in\ \href@noop {} {\emph {\bibinfo {booktitle} {{Fundamental
  Interactions. Proceedings of the 20th Lake Louise Winter Institute}}}},\
  \bibinfo {editor} {edited by\ \bibinfo {editor} {\bibfnamefont
  {A.}~\bibnamefont {Astbury}} \emph {et~al.}}\ (\bibinfo  {publisher} {World
  Scientific, Singapore},\ \bibinfo {year} {2006})\ p.\ \bibinfo {pages}
  {189},\ \Eprint {http://arxiv.org/abs/hep-ex/0505002} {arXiv:hep-ex/0505002
  [hep-ex]} \BibitemShut {NoStop}%
%%CITATION = HEP-EX/0505002;%%
\bibitem [{\citenamefont {Khachatryan}\ \emph {et~al.}(2014)\citenamefont
  {Khachatryan} \emph {et~al.}}]{Khachatryan:2014qwa}%
  \BibitemOpen
  \bibfield  {author} {\bibinfo {author} {\bibfnamefont {V.}~\bibnamefont
  {Khachatryan}} \emph {et~al.} (\bibinfo {collaboration} {CMS
  Collaboration}),\ }\href {\doibase 10.1140/epjc/s10052-014-3036-7} {\bibfield
   {journal} {\bibinfo  {journal} {Eur. Phys. J. C}\ }\textbf {\bibinfo
  {volume} {74}},\ \bibinfo {pages} {3036} (\bibinfo {year} {2014})},\ \Eprint
  {http://arxiv.org/abs/1405.7570} {arXiv:1405.7570 [hep-ex]} \BibitemShut
  {NoStop}%
%%CITATION = ARXIV:1405.7570;%%
\bibitem [{\citenamefont {Aad}\ \emph {et~al.}(2014{\natexlab{b}})\citenamefont
  {Aad} \emph {et~al.}}]{Aad:2014nua}%
  \BibitemOpen
  \bibfield  {author} {\bibinfo {author} {\bibfnamefont {G.}~\bibnamefont
  {Aad}} \emph {et~al.} (\bibinfo {collaboration} {ATLAS Collaboration}),\
  }\href {\doibase 10.1007/JHEP04(2014)169} {\bibfield  {journal} {\bibinfo
  {journal} {J. High Energy Phys.}\ }\textbf {\bibinfo {volume} {1404}},\
  \bibinfo {pages} {169} (\bibinfo {year} {2014}{\natexlab{b}})},\ \Eprint
  {http://arxiv.org/abs/1402.7029} {arXiv:1402.7029 [hep-ex]} \BibitemShut
  {NoStop}%
%%CITATION = ARXIV:1402.7029;%%
\bibitem [{\citenamefont {Aad}\ \emph {et~al.}(2014{\natexlab{c}})\citenamefont
  {Aad} \emph {et~al.}}]{Aad:2014lra}%
  \BibitemOpen
  \bibfield  {author} {\bibinfo {author} {\bibfnamefont {G.}~\bibnamefont
  {Aad}} \emph {et~al.} (\bibinfo {collaboration} {ATLAS Collaboration}),\
  }\href {\doibase 10.1007/JHEP10(2014)024} {\bibfield  {journal} {\bibinfo
  {journal} {J. High Energy Phys.}\ }\textbf {\bibinfo {volume} {1410}},\
  \bibinfo {pages} {24} (\bibinfo {year} {2014}{\natexlab{c}})},\ \Eprint
  {http://arxiv.org/abs/1407.0600} {arXiv:1407.0600 [hep-ex]} \BibitemShut
  {NoStop}%
%%CITATION = ARXIV:1407.0600;%%
\bibitem [{\citenamefont {Barate}\ \emph {et~al.}(2003)\citenamefont {Barate}
  \emph {et~al.}}]{Barate:2003sz}%
  \BibitemOpen
  \bibfield  {author} {\bibinfo {author} {\bibfnamefont {R.}~\bibnamefont
  {Barate}} \emph {et~al.} (\bibinfo {collaboration} {LEP Working Group for
  Higgs boson searches, ALEPH Collaboration, DELPHI Collaboration, L3
  Collaboration, OPAL Collaboration}),\ }\href {\doibase
  10.1016/S0370-2693(03)00614-2} {\bibfield  {journal} {\bibinfo  {journal}
  {Phys. Lett. B}\ }\textbf {\bibinfo {volume} {565}},\ \bibinfo {pages} {61}
  (\bibinfo {year} {2003})},\ \Eprint {http://arxiv.org/abs/hep-ex/0306033}
  {arXiv:hep-ex/0306033 [hep-ex]} \BibitemShut {NoStop}%
%%CITATION = HEP-EX/0306033;%%
\bibitem [{\citenamefont {Schael}\ \emph {et~al.}(2006)\citenamefont {Schael}
  \emph {et~al.}}]{Schael:2006cr}%
  \BibitemOpen
  \bibfield  {author} {\bibinfo {author} {\bibfnamefont {S.}~\bibnamefont
  {Schael}} \emph {et~al.} (\bibinfo {collaboration} {ALEPH Collaboration,
  DELPHI Collaboration, L3 Collaboration, OPAL Collaboration, LEP Working Group
  for Higgs Boson Searches}),\ }\href {\doibase 10.1140/epjc/s2006-02569-7}
  {\bibfield  {journal} {\bibinfo  {journal} {Eur. Phys. J. C}\ }\textbf
  {\bibinfo {volume} {47}},\ \bibinfo {pages} {547} (\bibinfo {year} {2006})},\
  \Eprint {http://arxiv.org/abs/hep-ex/0602042} {arXiv:hep-ex/0602042 [hep-ex]}
  \BibitemShut {NoStop}%
%%CITATION = HEP-EX/0602042;%%
\bibitem [{\citenamefont {Barger}\ \emph {et~al.}(1994)\citenamefont {Barger},
  \citenamefont {Berger},\ and\ \citenamefont {Ohmann}}]{Barger:1993gh}%
  \BibitemOpen
  \bibfield  {author} {\bibinfo {author} {\bibfnamefont {V.~D.}\ \bibnamefont
  {Barger}}, \bibinfo {author} {\bibfnamefont {M.~S.}\ \bibnamefont {Berger}},
  \ and\ \bibinfo {author} {\bibfnamefont {P.}~\bibnamefont {Ohmann}},\ }\href
  {\doibase 10.1103/PhysRevD.49.4908} {\bibfield  {journal} {\bibinfo
  {journal} {Phys. Rev. D}\ }\textbf {\bibinfo {volume} {49}},\ \bibinfo
  {pages} {4908} (\bibinfo {year} {1994})},\ \Eprint
  {http://arxiv.org/abs/hep-ph/9311269} {arXiv:hep-ph/9311269 [hep-ph]}
  \BibitemShut {NoStop}%
%%CITATION = HEP-PH/9311269;%%
\bibitem [{\citenamefont {Martin}\ and\ \citenamefont
  {Vaughn}(1994)}]{Martin:1993zk}%
  \BibitemOpen
  \bibfield  {author} {\bibinfo {author} {\bibfnamefont {S.~P.}\ \bibnamefont
  {Martin}}\ and\ \bibinfo {author} {\bibfnamefont {M.~T.}\ \bibnamefont
  {Vaughn}},\ }\href {\doibase 10.1103/PhysRevD.50.2282,
  10.1103/PhysRevD.78.039903} {\bibfield  {journal} {\bibinfo  {journal} {Phys.
  Rev. D}\ }\textbf {\bibinfo {volume} {50}},\ \bibinfo {pages} {2282}
  (\bibinfo {year} {1994})},\ \Eprint {http://arxiv.org/abs/hep-ph/9311340}
  {arXiv:hep-ph/9311340 [hep-ph]} \BibitemShut {NoStop}%
%%CITATION = HEP-PH/9311340;%%
\bibitem [{\citenamefont {Fonseca}\ \emph {et~al.}(2012)\citenamefont
  {Fonseca}, \citenamefont {Malinsky}, \citenamefont {Porod},\ and\
  \citenamefont {Staub}}]{Fonseca:2011vn}%
  \BibitemOpen
  \bibfield  {author} {\bibinfo {author} {\bibfnamefont {R.~M.}\ \bibnamefont
  {Fonseca}}, \bibinfo {author} {\bibfnamefont {M.}~\bibnamefont {Malinsky}},
  \bibinfo {author} {\bibfnamefont {W.}~\bibnamefont {Porod}}, \ and\ \bibinfo
  {author} {\bibfnamefont {F.}~\bibnamefont {Staub}},\ }\href {\doibase
  10.1016/j.nuclphysb.2011.08.017} {\bibfield  {journal} {\bibinfo  {journal}
  {Nucl. Phys.}\ }\textbf {\bibinfo {volume} {B854}},\ \bibinfo {pages} {28}
  (\bibinfo {year} {2012})},\ \Eprint {http://arxiv.org/abs/1107.2670}
  {arXiv:1107.2670 [hep-ph]} \BibitemShut {NoStop}%
%%CITATION = ARXIV:1107.2670;%%
\end{thebibliography}%
\end{document}